\documentstyle[11pt,twoside,psfig,assym,lscape]{cardiffthesis}

\DeclareSymbolFont{UPM}{U}{eur}{m}{n}
\SetSymbolFont{UPM}{bold}{U}{eur}{b}{n}
\DeclareMathSymbol{\umu}{0}{UPM}{"16}

\newcommand{\micron}{$\umu$m}

\newcommand{\Kb}{{$K$}}

\newcommand{\attrib}[1]{\newline \rule{0pt}{.5\baselineskip} \hfill --- #1.}

\newcommand{\tpm}{$\pm\ $}

\newcommand{\tbn}[1]{${^{#1}}$}

\newcommand{\ah}[1]{{\hat{a}}_{\mathsf{#1}}}
 
\newcommand{\lums}[1]{{\lambda}_{\rm{#1}}}
 
\newcommand{\Itt}[1]{{F}_{\rm{#1}}}
 
\newcommand{\Iq}[1]{{F}_{\rm{Q,#1}}}

\newcommand{\Ptt}[1]{{P}_{\rm{#1}}}
 
\newcommand{\Phm}[1]{{\phi}_{\rm{#1}}}

\newcommand{\Ptm}[1]{{\Pi}_{\rm{#1}}}

\newcommand{\Stt}[1]{{\sigma}_{\rm{#1}}}
  
\newcommand{\Stm}[1]{{\epsilon}_{\rm{#1}}}
  
\newcommand{\Pq}[1]{{P}_{\rm{Q,#1}}}
 
\newcommand{\Pqi}[1]{{P}_{{\rm{Q,}}#1}}
 
\newcommand{\PHI}[1]{{\Phi}_{{#1}}}
 
\newcommand{\ml}[1]{{m}_{\mathsf{#1}}}

\newcommand{\ccen}[1]{\multicolumn{1}{c}{#1}}

\newcommand{\cduo}[1]{\multicolumn{2}{c}{#1}}

\newcommand{\cduol}[1]{\multicolumn{2}{|c|}{#1}}

\newcommand{\rS}{r_{\it S}}

\newcommand{\ha}{{\frac{1}{2}}}

\newcommand{\util}[3]{{{#1}^{{#2}}_{{\sim}}{#3}}}

\newcommand{\gapeq}[2]{\util{#1\:}{>}{\:#2}}

\newcommand{\lapeq}[2]{\util{#1\:}{<}{\:#2}}

\newcommand{\cdf}[2]{\Phi_{#1}(#2)}

\newcommand{\ecd}[2]{\Xi_{#1}(#2)}

\newcommand{\bi}[1]{{\bf{#1}}}

\newcommand{\dg}{\degr}

\newcommand{\bllt}{${\bullet}\hspace{1em}$}

\newcommand{\mL}[1]{{\mathcal{L}}(#1)}

\newcommand{\mU}[1]{{\mathcal{U}}(#1)}

\newcommand{\mA}{{\mathcal{A}}}

\newcommand{\mD}{{\mathcal{D}}}

\newcommand{\abs}[1]{{|#1|}}

\newcommand{\mh}[1]{{{\underline{m}}_{\mathsf{#1}}}}

\newcommand{\eps}[1]{{{\mathcal E}_{\mathsf{#1}}}}

\newcommand{\vas}[1]{{\varepsilon_{\mathsf{#1}}}}

\newcommand{\sbb}[1]{{\hat{\sigma}_{#1}}}

\newcommand{\cls}[1]{{\bar{#1}_{C_{#1}}}}

\newcommand{\cln}[2]{{\bar{#1}_{#2\%}}}

\begin{document}


\title{The infrared polarizations of high-redshift radio 
galaxies} \author{Gareth James Leyshon}
\degreeyear{1999}
\degree{Doctor of Philosophy}
\chair{Professor Michael Disney (provisional)}
\othermembers{Professor Michael Edmonds (provisional internal)\\
Dr Clive Tadhunter, or Professor Jim Hough (provisional externals)}
\numberofmembers{3}
\prevdegrees{B.A. (Oxon.) 1994}
\field{Astrophysics}
\campus{Cardiff}

\begin{frontmatter}

\maketitle
\approvalpage
\copyrightpage

\begin{dedication}
\null\vfil
{\large
\begin{center}
To Jesus and Mary,\\\vspace{12pt}
{\em Ad Majoram Dei Gloriam},\\\vspace{12pt}
and to my parents and grandparents.
\end{center}}
\vfil\null
\end{dedication}

\begin{abstract}

This thesis reports the $K$-band polarizations of a
representative sample of nine radio galaxies: seven 3C objects at $0.7
< z < 1.3$, and two other distinctive sources.
Careful consideration is given to   
the accurate measurement and `debiasing' of faint
polarizations, with recommendations for the function of
polarimetric software.

\begin{description}

\item[3C 22] has 3\% polarization perpendicular to its radio 
structure, consistent with suggestions that it may be an obscured quasar.

\item[3C 41] also has 3\% polarization perpendicular to its 
radio and may also be an obscured quasar; its scattering medium is 
probably dust rather than electrons.

\item[3C 54] is polarized at 6\%, parallel to its radio structure. 

\item[3C 65] is faint: its noisy measurements give no firm 
evidence for polarization.

\item[3C 114] has a complex structure of four bright knots, one offset 
from the radio structure and three along the axis. There is strong 
evidence for polarization in the source as a whole (12\%) and the brightest 
knot (5\%).

\item[3C 356] is faint: we do not detect any $K$-band continuation of the 
known visible/near-ultraviolet polarization.

\item[3C 441] lies in a rich field; one of its companions appears to be 
18\% polarized. The identification of the knot containing the active 
nucleus has been disputed, and is discussed.

\item[LBDS 53W091] was controversially reported to have a 40\% $H$-band 
polarization. No firm evidence is found for non-zero $K$-band 
polarization in 53W091, though there is some evidence for its companion 
being polarized. 
The object is discussed in the context of other radio-weak galaxies.

\item[MRC 0156$-$252] at $z \sim 2$ is found to be unpolarized in $K$.

\end{description}

 Simple spectral and spatial models for polarization in radio galaxies
are discussed and used to interpret the measurements. The important
cosmological question of the fraction of $K$-band light arising in radio
galaxy nuclei is considered: in particular, the contribution of scattered 
nuclear 
light to the total $K$-band emission is estimated to be of order 7\% in 
3C 22 and 3C 41, 26\% in 3C 114, and tentatively 25\% or 
more in 3C 356.


\end{abstract}

\begin{acknowledgements}

 A three year project in astronomy relies on many factors to come to
fruition: the guidance of one's supervisor; chance remarks from
colleagues; the tedious but very necessary work of those who mount
archives on the World Wide Web; and most importantly, the availability of
observatories, software and funding which makes it all possible! 

 Firstly, I would like to thank Steve Eales for his guidance over the 
last four years, and for his philosophy that `you don't need to do lot 
of work to get a PhD' -- provided that {\em sufficient} work has been
done! Also a big thank-you to Steve Rawlings at Oxford: had I not spent a
month working efficiently on his radio galaxies in 1993 [now published at 
long last! \cite{Lacy+99a}], I might never have come to Cardiff. 

 Many thanks to all who gave constructive comments and advice throughout
the last three years: Bob Thomson, Jim Hough, Chris Packham, Mike Disney
and Mike Edmunds; Bryn Jones, Neal Jackson, Arjun Dey, Clive Tadhunter and
Patrick Leahy. Thanks especially to Buell Jannuzi and Richard Elston for
sharing their polarimetry results, Mark Dickinson for some optical
magnitudes, Megan Urry for allowing me to reproduce a complicated 
diagram, and 
Mark Neeser for having his thesis in the right place at the right time. 

 Particular thanks to Jim Dunlop for help on my second observing trip and
with the 53W091 data; and to Colin Aspin, Antonio Chrysostomou and Tim 
Carroll for help with making observations and data reduction. A special 
mention with many thanks to my A-level Statistics teacher, Eric Lewis!
 
 This research has made use of the {\sc nasa/ipac} extragalactic database
({\sc ned}) which is operated by the Jet Propulsion Laboratory, CalTech,
under contract with the National Aeronautics and Space Administration.
The United Kingdom Infrared Telescope is operated by the Joint Astronomy
Centre on behalf of the U.\,K. Particle Physics and Astronomy Research
Council. Thanks to the Department of Physical Sciences, University
of Hertfordshire for providing IRPOL2 for the UKIRT. 

Data reduction was performed with {\sc starlink} and {\sc iraf} routines. 
Thanks to Rodney Smith and Philip Fayers for their ceaseless efforts to keep Cardiff's computers functional! 
The use of NASA's {\em SkyView} facility
({\tt http://skyview.gsfc.nasa.gov}) located at NASA Goddard
Space Flight Center is acknowledged; as is that of the ADS abstract 
service at Harvard. This work was funded by a
{\sc pparc} postgraduate student research award.

\end{acknowledgements}

\tableofcontents

\listoftables

\listoffigures

\begin{specnotes}

Please note that certain conventions are adopted throughout this thesis:

\begin{itemize}

\item The results in this thesis are often quoted in the form of 
percentages (for polarizations, proportions of light from different 
sources, etc.).  Whenever measurements are presented in the form $a \pm 
b\%$, this should be read as $b$ being the absolute error on $a$ with 
both variables having the `units' in percentages. The format of a 
percentage error on an absolute quantity is {\bf never} used.

\item Occasionally it has been necessary to use the same mathematical 
symbols in different ways in different chapters. Usage is always 
consistent within a chapter and the most mathematical ones conclude with 
a glossary of all symbols used. 

\item Position angles are always denoted $\phi$; 
the symbol $\theta$ is only used in polarization vector phase space.

\item Assumptions about the cosmological parameters of the Universe are 
always explicitly stated where required; $h_0$ denotes
the Hubble constant in units of 100 km\,s$^{-1}$\,Mpc$^{-1}$.
Angular to linear scale conversion factors, when required, are taken from 
Peterson \scite[Figure 9.3]{Peterson-97a}.

\item Throughout this thesis, the term `optical' is used to encompass the 
near infrared, visible light and the near ultraviolet, as opposed to 
`visible', which explicitly means the region of the spectrum covered by 
the $R$, $V$ and $B$ bands.

\item The different classes of active galaxies are defined in Chapter 
\ref{defineAGN}. The term `quasar' is used to cover both radio-quiet and 
radio-loud quasi-stellar objects.

\item Each chapter is self-contained in abbreviations for papers cited. 
Any abbreviations used in a chapter are defined in the introduction to 
that chapter.

\item The work is written in the first person plural, the scientific 
`we', throughout. This does not imply collaboration in authorship except 
where explicitly noted by footnotes.

\end{itemize}

\end{specnotes}

\label{oddtest}

\newpage

\ 

\newpage

\end{frontmatter}

\chapter{Active Galaxies and the Unification Hypothesis}
\label{defineAGN}

\begin{quote}
Meddle first, understand later. You had to meddle a bit before you had
anything to try to understand. And the thing was never, ever, to go
back and hide in the Lavatory of Unreason. You have to try to get your
mind around the Universe before you can give it a twist.
\attrib{Ponder Stibbons, {\em Interesting Times}} 
\end{quote}
 
 The study of the most distant objects in the Universe is a demanding 
task. The maximum amount of information must be gleaned from the minimal 
flux of photons reaching Earth. When a target is so faint that our best 
image consists of a few bright pixels on an infrared array, there seems 
little hope of probing the structure hidden within. Yet even the faintest 
light, limited by diffraction or seeing, carries with it a hidden 
property: polarization. This is the tool which has been investigated and 
used in this work, to reveal new data on nine radio galaxies.

\section{Active Galaxies}
\label{intunimod}

Humanity's understanding of the Universe has developed radically since
Immanuel Kant first speculated about the existence of `island universes'
in the eighteenth century. In 1845, Lord Rosse completed construction of
his great reflecting telescope, and subsequently discovered spiral
structure in many of Messier's nebulae. By 1920, it was seriously argued
that the spiral nebulae were in fact galaxies external to our own -
epitomised by the `Great Debate' of Astronomy between Curtis and Shapley
that year \cite{Hoskin-76a}. Hubble's determination of the distance to the
spiral nebulae resolved the debate, and in the years that followed, the
vast majority of galaxies were found to be of elliptical or spiral
formation. 
 
 The advent of radio astronomy opened up a second waveband through which
the Universe could be studied, and by the late 1960s,
radio astrometry was sufficiently
accurate that radio sources could be identified with their 
optical\footnote{Note the definition in the frontmatter:
the term `optical' is used to encompass the near 
infrared, visible light and the near ultraviolet.}
counterparts. It became apparent that many elliptical galaxies were strong
radio sources -- forming the class of {\em radio galaxies} \cite{McCarthy-93a}. Most of the
sources consisted of double radio lobes spanning a distance 5-10 times the
size of the parent galaxy at the centre.

 At the same time, numerous other classes of unusual galaxies or intense 
radio sources were revealing themselves to new instruments. Earliest to 
become apparent was the class of Seyferts, galaxies
(normally spirals) with unusually bright nuclei whose spectra
included narrow ($\sim 1,000$ km/s) permitted and forbidden emission lines. 
Some Seyferts also exhibited broader ($\sim 10,000$ km/s) permitted 
emission lines, and were branded `Type 1', while those without were 
spectroscopic `Type 2'. Seyferts exhibit radio emission, but this is 
usually weak. The lines were accompanied by a `featureless continuum' 
whose profile was flat rather than the curve characteristic of blackbody 
thermal emission \cite[and references therein]{Robson-96a}.

 Meanwhile, radio surveys had also identified sources whose optical
counterparts were found to be brilliant and pointlike: these were named
quasars, the quasi-stellar radio sources. Like Type 1 Seyferts, quasars
exhibited a flat spectrum optical continuum with strong emission lines, both
narrow and broad. In 1963, quasar emission lines were first identified
with an element: 3C 273's emission lines were found to be characteristic
of hydrogen at high redshift, $z=0.158$ \cite[and references therein]{Peterson-97a}. 

 The radio-loud quasars were found to be excessively luminous in the
$U$-band compared with stars and normal galaxies, which prompted optical surveys to hunt for
more objects with ultraviolet excesses. These surveys discovered many more
quasi-stellar objects with similar spectra, and 90-95\% of all 
quasars\footnote{This thesis will adopt the term
`quasar' for these objects regardless of radio intensity.} are
now thought to be radio-quiet. 

 Collectively, Seyferts, quasars and radio galaxies became known as 
`active galaxies', the {\em Collins Dictionary of Astronomy}\,
definition \cite{Illingworth-94a} being 
`galaxies that are emitting unusually large amounts of energy from a very 
compact central source --- hence the alternative name of {\em active 
galactic nuclei}\, or {\em AGN}\,'. Classification of an object as an AGN
may be made because the active nucleus has been observed 
directly, or be inferred from the presence of radio lobes. Certain extremely
energetic AGN clearly dominated by an optically bright nucleus
became known as `blazars' \cite[\S 3.1]{Antonucci-93a}.

 At first it was unclear whether the wide-ranging class of `AGN' was simply 
phenomenological, or whether the different types of AGN were linked by an 
underlying physical mechanism. All species of AGN demanded a mechanism 
whereby a much greater energy output might be obtained from the heart of 
a galaxy than could be accounted for by stellar nuclear fusion; the 
mechanism would have to be capable of giving rise to a flat spectrum in 
both radio and optical wavelengths, provide for the presence of hot 
clouds of gas emitting radiation at particular wavelengths, and allow for 
the presence or absence of radio jets.

\begin{figure}
{The diagram below shows the postulated structure of an active nucleus according to the Unification Hypothesis. 
The molecular torus is cut away at the front to show the broad line region clouds and core. 
The black hole at the centre is surrounded by an accretion disc. 
This figure is reproduced from Urry \& Padovani \protect\scite{Urry+95a}, \copyright\ PASP, reprinted
with permission of the authors. \vspace{4ex}}
{\centering
\psfig{file=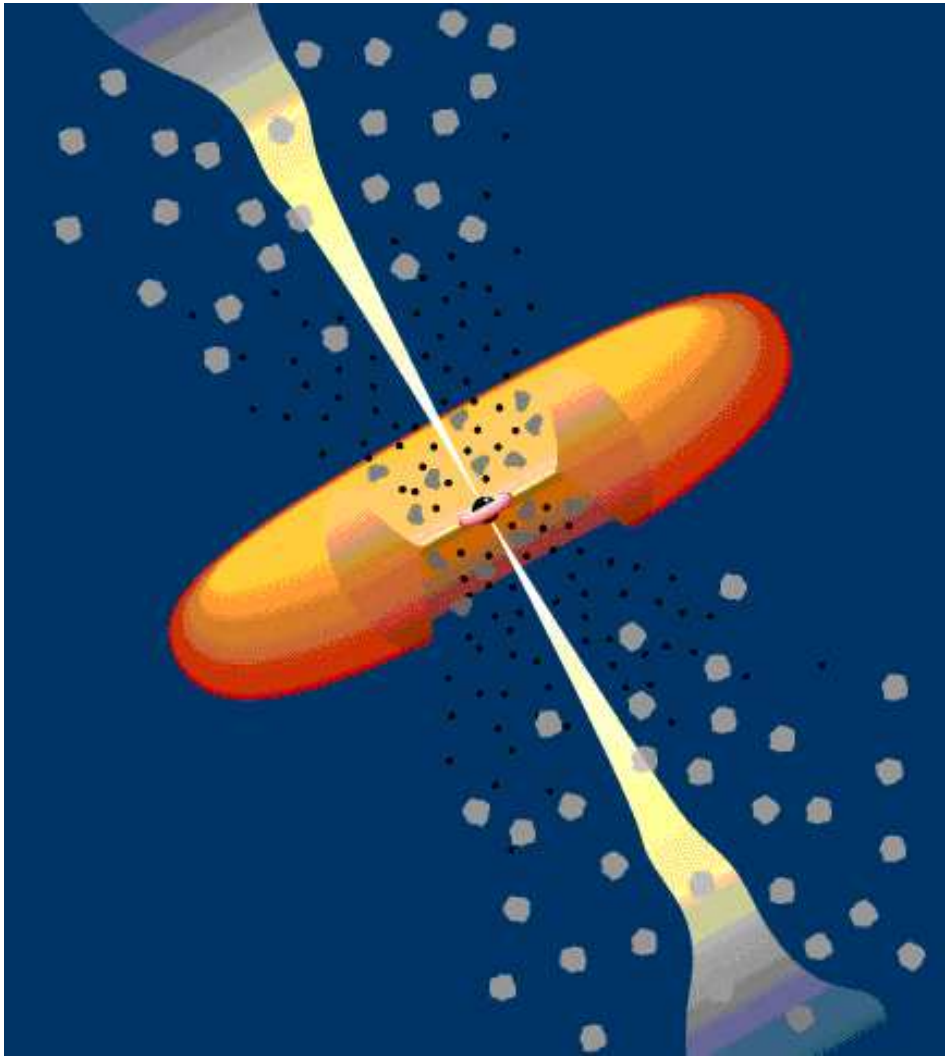,width=125mm} 
\caption{Schematic representation of an active nucleus.}
\label{unifig} }
\end{figure}

 Now it is generally accepted that the underlying mechanism in all these
objects is the accretion of matter on to a black hole
\cite{Antonucci-93a,Urry+95a}. Infalling matter forms an accretion
disk, heated by viscous and/or turbulent processes, which glows in the
ultraviolet and possibly soft X-rays. Hard X-rays are emitted in the
innermost part of the disk. Clouds of gas close to the black hole move
rapidly in its gravitational potential, and produce line emission at
visible and ultraviolet wavelengths --- these form and occupy the Broad
Line Region (BLR). Well beyond the accretion disk, gas and dust forms a
second, warped, disk or torus. This torus
screens the BLR from view in those AGN whose line of sight
to the Earth is not close to the axis of the torus. Energetic particles
escape in well-collimated jets at the poles of the torus.
Gas clouds further
from the active nucleus travel at lower velocities: not obscured by the
torus, such clouds emit light whose emission lines suffer less Doppler
broadening. Hence narrow lines are seen in all forms of AGN, whereas in
those AGN oriented so our line of sight is `down the jet' we see the
otherwise obscured broad line regions and/or continuum light from the
central engine. This model mechanism, illustrated in Figure \ref{unifig}, is generally known as the {\em
Unification Model} of AGN. At present, this stands as the `best buy' 
model for
AGN, but is not universally accepted --- especially in the case of
quasars. 

 Drawing on the spectroscopic classification of Seyferts, AGN generally 
are now classified `Type 1' and `Type 2'. Type 2 objects are those with 
no evidence of a direct view of their central engines: radio galaxies 
exhibiting only narrow emission lines (NLRGs) join Seyfert 2s in this 
category. Type 1 objects are those which do seem to include radiation 
from the central engine, and Seyfert 1s are joined by BLRGs (broad-lined 
radio galaxies -- which also show the narrow lines), and by quasars.

 One important prediction of the Unification Model is that Type 2 objects, 
with their torus axis being aligned roughly in the plane of the sky, 
ought to include broad line regions (BLRs) whose light, though obscured 
from Earth, escapes into the plane of the sky. Dust particles or 
electrons in the host galaxy or in the clouds responsible for the narrow 
lines should scatter some of this light into our line of sight. When 
light becomes scattered, it becomes linearly polarized in the sense 
perpendicular to the pre- and post-scattering flight axes of the photon; 
hence linearly polarised imaging or spectroscopy of AGN should reveal 
light from BLRs polarised perpendicular to the direction of the radio jet 
(presumed to be aligned with the opening of the torus). 

 The motivation of this thesis is to search for evidence of such 
polarization in $K$-band infrared light from radio galaxies -- a waveband 
thought, but not proven, to be dominated by light from the host galaxy's 
stars rather than the active nucleus. Findings of $K$-band polarization 
would set important constraints on the relative strengths of the nuclear 
and stellar components. Accordingly, the next chapter presents a 
review of the {\em status quo}\,\footnote{Since the
observational data recorded in this thesis concerns only
imaging polarimetry, not polarized spectra, I will not review
the finer features of raw and polarised AGN spectra here; the
subject has been extensively treated in the literature.} in our 
knowledge of the 
relationship between radio galaxies' radio structure and their morphological 
and polarization properties. First, however, we must look at the `big 
picture' of the different classes of AGN known to exist, and how they 
might be related to one another if the best-buy Unification Hypothesis is 
correct.

\section{The Unification Hypothesis for AGN}

 Contemporary authors embrace the Unification Model as the
accepted model for AGN with various degrees of enthusiasm:
Antonucci \scite{Antonucci-93a} rushes to set it up as the
`straw person model' against which he reviews the current
observational evidence; for Robson \scite[ch.\ 9]{Robson-96a}
it is a solid foundation, while Peterson's \scite[ch.\
7]{Peterson-97a} approach is more cautious. There is no other
serious contender to explain the wide range of AGN phenomena,
although in certain individual objects \cite[\S \S 2.4,
3.4]{Peterson-97a} starbursts rather than black hole accretion
may form the hidden engine driving the radiation output.

 A fundamental division between the various classes of AGN is
their radio strength. Radio galaxies, by definition, are radio
loud. Seyferts are empirically found to be radio quiet. All
blazars, without known exception, are radio loud; 90--95\% of
quasars are radio quiet. The various classifications of AGN
stem historically from the (often extreme) prototypes of each
class first discovered, and are not always helpful in
classifying less extreme examples: for instance, a radio-loud
galaxy with an obvious active nucleus would now be classified
a radio galaxy rather than a Seyfert, so the absence of
radio-loud Seyferts is a consequence of taxonomy rather than
physics. While some attempt has been made to define the
different AGN classes more rigorously \cite[ch.\
2]{Peterson-97a}, the older literature and human nature
militate against the use of clear-cut terms to distinguish
different classes of object -- objects which are hypothesised
to lie on a continuum of classes in any case!

\subsection{Seyfert Galaxies}
\label{polinSeyferts}

 As summarised earlier, the first key to unification came from
studies of Seyfert galaxy spectra. Seyferts are now defined as
low luminosity AGN with absolute magnitudes $M_B > -21.5 +
\log h_0$ \cite[\S 2.1]{Peterson-97a}; it follows naturally,
therefore, that all known Seyferts are at low redshift and
their morphology is open for study. Nearly all Seyfert AGN are
found to lie within spiral galaxies \cite[\S
8.1.1]{Peterson-97a}, often of type Sa or Sb, and the host
galaxies are more likely than normal spirals to be barred
and/or deformed. Robson \scite[\S 3.2]{Robson-96a} notes that
in the rare cases where Seyferts are radio-loud, they often
have other peculiar characteristics.

 Spectral studies of Seyferts led to the Type 1 / Type 2
classification based on the presence or absence of broad
spectral lines. The discovery of broad lines in Type 2 Seyfert
NGC 1068 in polarised light \cite{Antonucci+85a} prompted the
realization that the Broad Line emitting Region (BLR) must lie
within some geometrical feature which screened it from direct
view. This screening feature -- the postulated molecular torus
-- is typically of diameter $\sim$ 100 pc, and should not be
confused with the accretion disk in the central engine,
measuring perhaps 0.03 pc. A typical schematic diagram is
Figure 7.1 of Peterson \scite{Peterson-97a}; an excellent
cartoon sketching the structure of an AGN at eight different
scales ($10^{-4}$--$10^{+6}$ pc) is borrowed from Blandford by
Robson \scite[Figure 9.9]{Robson-96a}.

 Recent studies of the near-infrared ($H$-band) polarization
of NGC 1068 \cite{Young+96a,Packham+97a} have been found to be
consistent with a scattered light hypothesis and have even
allowed the likely position and orientation of the molecular
torus to be identified: the torus in this case has a diameter
greater than 200 pc.

Miller \& Goodrich \scite{Miller+90a} studied eight further
Seyfert 2s to see if spectropolarimetry would reveal Seyfert 1
features, choosing objects already known to have high
broadband polarizations. Four gave definite positive results,
two gave definite negatives, and two failed to produce
reliable signal-to-noise. The four polarised galaxies revealed
polarized features consistent with Seyfert 1 properties,
albeit with a degree of `bluing' in the spectrum indicative
that dust scattering must be a contributing mechanism. Three
of these, and possibly one of the two low signal-to-noise
sources, exhibited perpendicular alignment between the
polarization orientation and the radio axis. Similar
perpendicular alignments had also been detected in Seyfert 2s
by Antonucci \scite{Antonucci-83a}. Weak {\em parallel}\, polarizations 
have been observed in a few Seyfert 1s \cite[and references 
therein]{Antonucci+90a}.

 The {\em nuclear} polarization levels obtained by Miller \&
Goodrich \scite{Miller+90a} were only of the order of a few
percent --- rather low if the underlying mechanism is the
scattering implied by the perpendicular alignment. Antonucci
\scite{Antonucci-93a} speculates that they may have
underestimated the contribution of host galaxy starlight, and
that the true polarization may be closer to the 16\% level
observed in NGC 1068.

 Further tests for the Unification Model in Seyferts are reviewed
by Antonucci \scite{Antonucci-93a} --- some may indicate
refinements that need to take account of additional parameters
in the model (e.g.\ the opening angle of the molecular torus)
but none fatally wound the principle of Unification. Claims
that some Seyfert 2s contain no BLR emission \cite[\S
2.6.2]{Antonucci-93a} call for more sensitive
spectropolarimetry before they can seriously question
Unification: Robson \scite[\S 3.2]{Robson-96a} and Peterson
\scite[\S 7.4.1]{Peterson-97a} note that this is a hot area of
current research. Until proven otherwise, it can be
safely stated that
Seyfert galaxies fall into two distinct classes: broad line (Type 1) 
objects which sometimes
exhibit weak polarizations parallel to the
radio structure, and narrow line (Type 2)
objects which often display strong perpendicular polarizations.

\subsection{Blazars}

 The precise definition of a blazar seems to depend on which
source is consulted. I will follow Robson's \scite[\S
3.6.4]{Robson-96a} helpful advice that the term refers to a
phenomenon rather than a simple class of object: specifically,
the phenomenon of a relativistic jet beamed roughly in the
direction of terrestrial observers, dominating the radio thru
infrared spectrum with its non-thermal synchrotron emission.
The blazar phenomenon is exhibited by three classes of object:
BL Lacertae objects (BLLs), optically violent variable sources
(OVVs), and highly polarized quasars (HPQs).

 BLLs are distinguished and defined by the lack of emission or
absorption features in their spectra. They often exhibit
variability (changing their output by several magnitudes in
the space of a few weeks) and usually lie in elliptical
galaxies, though spiral hosts are also known. OVVs are AGN
(with spectra including broad emission lines) which exhibit
short timescale luminosity variations ($\ga 0.1$ mag) over
timespans as short as a day \cite[\S 
2.5]{Illingworth-94a,Peterson-97a}. No radio-quiet OVVs or
 BLLs are known \cite{Jannuzi+93a}.

 Antonucci \scite[\S 3.1]{Antonucci-93a} argues strongly that
the distinction between BLLs and OVVs is ill-founded --
especially given that the sources' very variability can switch
them between the two categories -- and that in fact all
radio-loud AGN with radio structures dominated by emission
from the core are part of the same family of objects. The
optical components of core-dominated radio-loud AGN tend to be
quite red, highly variable, and polarized: this is proposed to
be the high-frequency tail of the core synchrotron emission.
Only emission from the core can vary coherently over
timescales of weeks or days.  He proposes that BLLs are simply
the extreme cases where the synchrotron emission utterly
dominates other components of the optical output, and predicts
that more sensitive spectropolarimetry of BL Lacertae objects
would reveal faint unpolarised broad emission lines from the
BLR clouds basking in the synchrotron jets. Conversely, he
also suggests that those core-dominated radio-loud AGN not
classed as blazars would reveal a faint red optical core under
careful scrutiny.

 Robson \scite[\S 3.6.4]{Robson-96a} also includes HPQs in his
phenomenological class of blazars. Scarpa \& Falomo
\scite{Scarpa+97a} recently compared the optical properties of
a sample of HPQs and BLLs, finding that optical properties of
radio-selected BL Lacertae objects were very similar to those
of highly polarised quasars. An earlier survey comparing high-
and low-polarization quasars \cite{Moore+84a} found that all
but two of their HPQs were radio-loud, and the two radio-quiet
quasars had their own peculiarities. The orientation angles of
the HPQ linear polarizations seemed randomly distributed with
respect to the radio axes; this bears out the core emission
hypothesis, as the synchrotron mechanism produces light whose
polarisation orientation has no relationship with the radio
jet geometry. It seems eminently reasonable to accommodate
HPQs between less extreme core-dominated radio-loud AGN and
the OVVs on Antonucci's unified blazar scheme.

\subsection{Quasars}

 Complementing the definition of Seyferts, above, quasars are
now defined as AGN with absolute magnitudes $M_B < -21.5 +
\log h_0$ \cite[\S 2.2]{Peterson-97a}. A small proportion
(5-10\%) are known to be radio-loud; spectroscopically,
quasars exhibit spectra similar to Type 1 Seyferts \cite[\S
7.4.1]{Peterson-97a}. Since Moore \& Stockman
\scite{Moore+84a} found HPQs to be quite distinct from low
polarization quasars (LPQs) we have already dealt with HPQs as
blazars; and Stockman, Moore \& Angel \scite{Stockman+84a}
undertook a specific study of the LPQs. The cut-off cannot be
precisely defined, but 3\% polarization is normally taken as
an effective working threshold in the literature.

Stockman, Moore \& Angel \scite{Stockman+84a} found that the
typical LPQ polarization was around 0.6\% and tended to be
aligned {\em parallel} with the radio axis. There was no
strong evidence for temporal variability in the degree or
orientation of polarization, with upper limits of $\Delta p /
p \leq 0.16$ and $\Delta \phi \leq 8 \degr$. In a sub-sample
of LPQs mostly selected in the radio, the $B$-band
polarization was typically 50\% greater than the $R$-band
value; the equivalent test was not performed on
optically-selected LPQs. The lack of variability and the
tendency for polarization to increase at shorter wavelengths
rules out a blazar origin for the polarized light in LPQs:
models invoking scattering off dust grains or electrons in a
disk or oblate cloud could account for such polarization but
the mechanism is still very unclear. Stockman, Angel \& Miley
\scite{Stockman+79a}, Antonucci
\scite{Antonucci-82a} and Moore \& Stockman \scite{Moore+84a}
provide evidence for a bimodal (parallel/perpendicular)
distribution of scattering angles; perpendicular alignments
can be easily accounted for by the usual mechanism.
Antonucci \& Barvainis \scite{Antonucci+90a} suggest that the parallel
LPQs and the few Seyfert 1s that exhibit weak parallel polarization may
contain disks or tori with very large opening angles, which could produce 
parallel polarization by scattering.

The Stockman, Moore \& Angel \scite{Stockman+84a} survey covered bright
objects from a variety of catalogues, and was not statistically
complete in any meaningful sense. To complement it, Fugmann \&
Meisenheimer \scite{Fugmann+88a} studied a sample of faint 5\,GHz radio
sources, and more recently, Impey, Lawrence \& Tapia \scite{Impey+91a}
studied the optical polarization of a complete sample of radio sources,
also selected at 5\,GHz.

Impey, Lawrence \& Tapia's \scite{Impey+91a} complete 5\,GHz sample
included both radio galaxies and quasars. Since HPQs are known to
exhibit strong variability, it is possible that they may sometimes drop
below the 3\% threshold and are at risk of being labelled LPQs on the
strength of a single measurement. Discovering a trend of polarization
increasing with radio compactness, Impey, Lawrence \& Tapia
\scite{Impey+91a} note that they `cannot exclude the possibility that
{\em all}\, quasars with compact radio emission have $p_{\mathrm
max}>3\%$, at least some of the time'. Again, this would support a
division of radio-loud quasars into those which are part of the blazar
family, aligned such that their radio core would appear compact, and
those whose beaming axis is not so closely aligned with the line of
sight to Earth. Fugmann \& Meisenheimer's \scite{Fugmann+88a} results
also suggest that many compact radio objects not otherwise noted for
optical variability exhibit polarization properties characteristic of
blazars. Robson \scite[\S 3.4.1]{Robson-96a} notes that the
polarization properties of {\em radio-quiet} quasars have not been well
measured, but are tightly constrained in the optical as being very low
--- low enough to be attributed to thermal emission.

One unanswered question for the Unification Model is why Type
2 spectra are not seen in quasars. Peterson \scite[\S
7.4.1]{Peterson-97a} offers two suggestions: that the
molecular torus surrounding such a powerful central engine is
thinned to the point of ineffectiveness; or that `Quasar 2s'
exist but have been misidentified as something else, perhaps
the ultraluminous far infrared galaxies \cite{Sanders+88a}. 

Robson \scite[\S 9.2.3]{Robson-96a} pursues the latter
hypothesis in the shape of IRAS galaxy IRAS FSC 10214+4724.
This remarkable object, at $z=2.286$, appears to be
gravitationally lensed, to be undergoing a starburst phase,
{\em and}\, to contain an active nucleus! Images
\cite{Lawrence+93a} taken through
polarizing filters reveal a polarization of 
about 16\% regardless of aperture, but ambiguous indications of any
Alignment Effect. Polarized spectra \cite{Goodrich+96a} reveal broad 
quasar-like emission lines. 
Dust scattering from an active nucleus is
proposed as the most likely source of the polarization, but
scattered light from a blanketed starburst might also provide
an explanation. IRAS 09104+4109 \cite{Hines+93a} is also notable
as an IRAS galaxy containing a powerful radio source and with a constant
nuclear polarization of $\sim 18 \%$, although the polarization is 
misaligned with its radio structure (possibly due to the geometry of 
thin regions in its blanketing dust).
Both Antonucci \scite{Antonucci-93a} and
Robson \scite[\S 9.2.3]{Robson-96a} speculate that future
analyses of the most luminous IRAS galaxies will reveal some
(perhaps ten percent) of them to be hiding the missing Type 2
quasars.

\section{Radio Galaxies}

 Radio Galaxies form the remaining category of AGN. Most of
the AGN sources considered so far have been radio-quiet,
except for the blazars which are dominated by strongly beamed
radiation over many decades of their spectrum. Radio galaxies
join the 5-10\% of quasars in the distinct class of radio-loud
AGN. Reviewing the status of high redshift radio galaxies,
McCarthy \scite{McCarthy-93a} notes that the distinction
between radio galaxies and radio quasars is becoming blurred
as the host galaxies of quasars have been identified for
quasars with $z \la 0.5$. Classically, the distinction had
been that a powerful radio source was a `quasar' if the host
galaxy could not be seen beneath the active nucleus; an `N
galaxy' if the nucleus was exceptionally bright but did not
wash out all traces of the starlight; or an ordinary radio
galaxy otherwise.

\subsection{The spectra of radio galaxies}
\label{RGspectrum}

 Today, radio galaxies are classified on the basis of two
distinct sets of properties: their optical emission lines and
their radio structure. McCarthy \scite{McCarthy-93a} notes
that Broad Line Radio Galaxies (BLRGs), i.e.\ those with
H\,{\sc i} lines having widths over 2,000 km/s, tend to have
the morphological classification of `N galaxies', and their
broad line spectra are similar to those exhibited by Seyfert
1s. Narrow Line Radio Galaxy (NLRG) spectra have only narrow
lines for both permitted and forbidden transitions; BLRGs have
narrow forbidden line spectra similar to those of NLRGs.

 A distinction is often made between `nuclear' and `extended' emission,
but isolating the nucleus from any extended emission region is not
trivial when a 2\arcsec\ slit encompasses more than 10 kpc of an object
at $z>1$. If spectroscopy can be performed on distinct regions of a
radio galaxy (an operation not possible with unresolved quasars), the
properties of the spectra would enable the composition of the different
parts of the galaxy to be identified. Similarly, imaging polarimetry
has the potential to be an invaluable tool to determine the properties
of different parts of the emission. But both techniques are limited in 
practice by the faintness of the galaxies \cite{Cimatti+96a}.

 The optical radiation emitted by radio galaxies is thought to
be a combination of starlight and nebular emission from the
host galaxy, and a quasar-like (power-law) component
originating in the active nucleus hidden in the heart of the
galaxy. [Note that at this stage we need make no assumptions about the
{\em reason}\, for the shape of the quasar spectrum, but only utilise
its profile. It has been suggested \cite{Binette+88a} that the quasar
spectrum could be synthesised from a suitable combination of blackbody
curves.] 

Manzini \& di Serego Alighieri \scite{Manzini+96a}
tested this three-component hypothesis by modelling radio galaxy
spectra at
rest frame wavelengths from 0.2\,\micron\ to 1.0\,\micron.
 Starlight from the host galaxy was
modelled as the synthetic spectrum of Bruzual \& Charlot
\scite{Bruzual+93a} for a galaxy with an initial burst of
star formation and no subsequent formation. Nebular emission
was modelled with a spectrum selected from Aller
\scite{Aller-87a}. The active nuclear component was modelled
by the composite radio-loud quasar spectrum of Cristiani \&
Vio \scite{Cristiani+90a}, and attenuated according to
different distributions of dust grains which might be present
to scatter nuclear light into the line of sight to Earth.

Manzini \& di Serego Alighieri \scite{Manzini+96a} applied their
modelling to a small sample of radio galaxies at redshifts ranging from
0.11 to 2.63, and have demonstrated that their observed magnitudes (by
multiwaveband photometry) are consistent with artificial spectra
synthesised from three such components. The contribution of the
starlight becomes greater to longer wavelengths, while the nuclear
component decreases. For five out of their six galaxies, the stellar
component of the light has become dominant by a rest-frame wavelength
of 0.5\,\micron; in 3C 277.2\ $(z=0.766)$ the starlight only exceeds the
nuclear component at about 0.85\,\micron. Hence the `galaxy plus quasar'
model predicts that starlight should dominate the infrared output of
radio galaxies, while nuclear emission is predominant in the
ultraviolet. Hammer, LeF\`{e}vre \& Angonin \scite{Hammer+93a} confirm
that the ultraviolet $\lambda < 400$\,nm light from $z \sim 1$ 3C radio
galaxies is dominated by the presence of an active nucleus.

 We have already noted (\S \ref{polinSeyferts}) that the distinction between
BLRGs and NLRGs can be interpreted as a Type 1 / Type 2 orientation
effect, with the BLR obscured by an assumed
molecular torus in those galaxies classed NLRGs. Quasars and radio
galaxies have been shown to have comparable emission line luminosities,
arising in emission line regions less than 1 kpc in diameter
\cite{Spinrad-82a}. If the Unification Model is the correct model to
apply to radio galaxies, then Manzini \& di Serego Alighieri
\scite{Manzini+96a} are correct to model their `quasar' component as
scattered into the line of sight by dust; and their results show that
radio galaxies can be accurately modelled as containing quasar cores
(with molecular tori of dimensions less than 1 kpc), with core light
scattered by plausible (albeit idealized) distributions of dust.

\subsection{The Hubble diagram: radio galaxy evolution}
\label{HubbleK}

 If the infrared emission of radio galaxies is dominated by starlight,
then studies of the variation of their $K$-band magnitudes with
redshift should tell us something about galactic evolution. Lilly \&
Longair \scite{Lilly+84a} produced a $K-z$ plot, or {\em Hubble
diagram}, for 3CR radio galaxies, i.e.~for those radio galaxies with
the most powerful radio emissions. Two features were evident 
in the resulting Hubble diagram: the dispersion of the 
$K$-magnitudes about the average value remained
constant up to $z \ga 1$, but the average magnitude evolved with
redshift such that galaxies at $z \sim 1$ were about 1 mag more
luminous than at $z = 0$. 

 Eales \& Rawlings \scite{Eales+96a} summarise the natural
interpretation of these findings: low dispersion implies that the radio
galaxies were not passing through any transient phase in their
evolution (which would have caused wider variation in their luminosity)
over the span of redshifts covered. The declining luminosity to lower
redshifts is consistent with a period of star formation at $z > 5$,
followed by passive evolution as stars of decreasing mass reach the end
of their lives. This seems perfectly reasonable since nearby radio
galaxies are known to lie within giant ellipticals with a small spread
of absolute magnitude \cite{Laing+83a}, and a similar evolutionary
model has been suggested for radio-quiet elliptical galaxies
\cite{Eggen+62a}. Imaging of the rest-frame visible structure of
radio galaxies out to $z \sim 2$ shows many of them to have dynamically
relaxed structures, suggesting that these are active elliptical
galaxies, too \cite{McCarthy+92a,Rigler+92a,Cimatti+94a}.

 It should be noted, however, that many distant radio galaxies at
$z>0.6$ \cite[see below]{McCarthy-93a,Cimatti+94a} do not have
elliptical morphologies: an evolving elliptical model alone cannot
explain the disturbed morphology of these objects, so at best a
modified evolving elliptical model is needed. [It has also been
suggested that the radio galaxies are in fact young objects which pass
through a radio-loud phase only a few hundred Myr after a rapid
star-forming phase itself lasting of order 100 Myr \cite{Chambers+90a}.]

 Lilly \& Longair's \scite{Lilly+84a} Hubble diagram suffers from the
unavoidable selection effect that 3CR galaxies contain the most
powerful radio sources. Eales et al.\ \scite{Eales+97a} therefore
analysed a 90\% complete set of radio galaxies selected at lower radio
luminosities in the B2 and 6C catalogues, and created a Hubble diagram
allowing the $K$-magnitudes of 3CR galaxies to be compared to those
whose radio output was only one-sixth as strong. Analyzing the diagram
above and below the natural threshold of $z=0.6$, they found that the
low redshift B2/6C galaxies had $K$-magnitudes statistically identical
to the 3CR sample, but at $0.6<z<1.8$, the 3CR sample was brighter by a
median 0.6 magnitudes. Demonstrating that sources of bias have either
been eliminated or would make their result stronger, Eales et al.\ 
\scite{Eales+97a} argue that the $K$-band emission of the brightest
radio galaxies must be contaminated by light from a source whose
luminosity is correlated with the radio strength of the galaxy --
presumably direct or indirect $K$-band emission from the active nucleus
itself (but see below).

Eales \& Rawlings \scite{Eales+96a} note that Hubble diagram for the
6C/B2 galaxies (assumed to be unpolluted by nuclear emission) follows a
curve for {\em no}\, stellar evolution; the 3C Hubble diagram which had
previously been interpreted as indicating stellar evolution rather
represents a series of galaxies showing no evolutionary effects,
hosting nuclear sources which tend to be brighter at higher redshift by
the selection effect of a flux-limited radio sample. It has been argued
\cite{Best+98a} that the apparent `no evolution' result occurs because
of a cosmic conspiracy: the host galaxies of the radio sources evolve
in the same way as radio-quiet Brightest Cluster Galaxies (BCGs). The
$K-z$ Hubble diagram for BCGs also suggests an unphysical no-evolution
scenario, which must be accounted for by postulating evolution in the
galactic structure whose net effect counteracts that of stellar evolution.
The most likely explanation is ongoing formation according to hierarchical
clustering models. Radio galaxies are preferentially found in clusters
at high redshift, so it would not be surprising for their behaviour to
follow that of BCGs; while the fact that low-redshift radio galaxies
are not preferentially found in clusters should make us suspicious of
accepting Lilly \& Longair's \scite{Lilly+84a} continuous Hubble diagram
across the $z=0.6$ morphology break.

Best, Longair \& R\"{o}ttgering \scite{Best+98a} argue that $K$-band
emission from the active nucleus cannot contribute more than 15\%
(typically 4\%) to the brightness of a 3CR radio galaxy, nor cause more
than 0.3 mag of brightening in 3CR objects over 6C objects. Ruling out
the possibility that 3CR objects contain more young stars, they suggest
rather that 3CR objects simply contain greater masses of stars, and
cite evidence \cite{Kormendy+95a} that if the central engine is a black
hole whose accretion rate depends on the material available in the host
galaxy, then the mass of stars in the galaxy will be correlated with
the radio, and hence the optical \cite{Willott+98a,Serjeant+98a}.

\subsection{The radio morphology of radio-loud AGN}

The radio classification of radio-loud AGN distinguishes those
which are `lobe-dominated' and those which are
`core-dominated' \cite[\S 3.7]{Robson-96a}. A more detailed
discussion of the different radio structures observed in radio
galaxies is given by Miley \scite[\S 2.1]{Miley-80a}.

 Measurements of radio flux density, $S$, at different
frequencies allow a power law spectrum to be fitted to the
source, characterised by a spectral index $\alpha$, such that
$S \propto \nu^{-\alpha}$. Core-dominated radio sources tend
to have flat spectra ($\alpha \sim 0$) and often show a single
kpc-scale jet; in fact, most core-dominated radio sources
would fall into the category of blazars rather than radio
galaxies \cite[\S 3.7.2]{Robson-96a} --- and there is recent
evidence that radio galaxies which are core-dominated exhibit 
variable optical polarization due to synchrotron 
emission \cite{Cohen+97a,Tran+98a}.

Lobe-dominated radio structures emit radio waves from a locus
of space which can span many tens of kiloparsecs, up to 3 Mpc
in the case of 3C 236 \cite[\S 3.7.1]{Robson-96a}. The
extended lobes of radiation are zones of synchrotron emission,
and are fed by a stream of relativistic electrons flowing out
of the poles of the central engine. The radio spectra of
lobe-dominated sources tend to be steep, with $\alpha \sim 1$.

 A further division is made according to the criteria of
Fanaroff \& Riley \scite{Fanaroff+74a}, whence sources with
spectral luminosity density $P_{\mathrm 178\,MHz} > 5 \times
10^{25}$ W\,Hz$^{-1}$ are class FR II, and those less luminous
are class FR I. The class FR I sources are usually associated
with radio galaxies alone, and the most prominent parts of the
the radio structure (`hot spots') lie closer to the core than
to the edge of the radio structure.

 Both quasars and radio galaxies can exhibit class FR II
structure, and their hotspots lie closer to the edge of the
emission region than to the core \cite{Illingworth-94a}. Such
sources usually have a compact radio core co-located with the
optical nucleus of the galaxy. All radio-loud quasars and many
class FR II radio galaxies are asymmetric, with a kpc-scale
jet only visible between the core and one of the lobes
\cite[\S 8.1]{Robson-96a}.  The most likely explanation of the
asymmetry again invokes an orientation argument, with those
jets travelling towards us closest to our line of sight being
most Doppler-brightened and the counterjets similarly
Doppler-supressed \cite[\S 8.2.1]{Robson-96a}; hence quasars,
where we are thought to be looking close to `down the jet',
are always asymmetric, while radio galaxies are viewed closer
to `sideways on' and the two jets can appear to be of
comparable brightness. This has been borne out by the
discovery of the Laing-Garrington effect \cite[\S
3.4]{Antonucci-93a}, where the radio emission from the far
side of such a source suffers depolarization by its passage
through the interstellar material in the host galaxy and hence
appears less polarized than the radio emission from the lobe associated
with the jet.

 For nearby radio galaxies, class FR II sources are normally
found in (otherwise normal) giant elliptical galaxies,
although not in galaxies forming part of rich clusters. Nearby FR I
sources, however, are very likely to be hosted by more
luminous ellipticals, often the type D or cD galaxies which
dominate rich clusters \cite[\S 3.7.1]{Robson-96a}. At high
redshift, though, there is no clear evidence for a distinction
between the richness of clusters hosting FR I and FR II
classes; and the morphology of the host galaxy is often
distorted by the presence of knots. We will return to this
in the next chapter, in a discussion of the Alignment Effect.

\subsection{Radio source unification revisited}

 The presence or absence of strong radio emission seems to be a
fundamental characteristic of AGN, and is strongly linked with optical
morphology: Peterson \scite[\S 8.1.3]{Peterson-97a} notes that it is
`true in general' that radio-quiet AGN (Seyferts, most quasars) are
found in spiral galaxies [but see Ridgway \& Stockton
\scite{Ridgway+97a} and references therein for evidence of elliptical
hosts being common in radio-quiet quasars], while strong radio sources
(radio quasars, radio galaxies and blazars) tend to have elliptical
hosts. Further, the host galaxies of radio quasars are, on average, 0.7
mag brighter in absolute $B$-magnitude than radio-quiet quasar hosts.
The absence of FR I class quasars may follow from the strength of the
central engine: if the nucleus is powerful enough to be optically
classified a quasar, then its jets may {\em de facto}\, be strong
enough to produce FR II class radio structure.

 The first indications that orientation effects may be important in
understanding the nature of radio sources came with the discovery of
apparent superluminal motion in four of the brightest radio sources
\cite{Cohen+77a}. Superluminal observations can be understood if the
source is ejecting matter at relativistic velocities along a path close
to the line of sight to the observer \cite{Rees-66a,Blandford+79a}; in
which case superluminal sources must be the beamed subset of some
parent population. This model is now generally accepted as the
explanation of superluminal radio sources and is
consistent with other observed features of the superluminal sources --
including blazar properties and asymmetric jets \cite{Barthel-89a}.

 The Unification Model predicts that radio galaxies should not
exhibit superluminal motion. Of Cohen et al.'s \scite{Cohen+77a} four
original superluminal sources, three were quasars but the fourth was
classed as a radio galaxy. This object, 3C 120, is a nearby $(z=0.033)$
galaxy which has been variously classified as a core-dominated broad-line
radio galaxy, and as a radio-loud Seyfert 1 with disturbed morphology
\cite{Grandi+97a}. If we set aside 3C 120 as an anomalous object, what of
other AGN? A survey of relativistic motion in all sources of known VLBI
core size, appearing in the literature 1986--1992, was assembled by
Ghisellini et al.\ \scite{Ghisellini+93a}. Of the sources for which speeds
were recorded, definite superluminal velocities were observed in all 11
BLLs, 23 out of 29 quasars (with 5 more having superluminal upper limits),
and none of 6 radio galaxies. One radio galaxy (0108+388) displayed an 
apparent speed of $1.0c$, and another (0710+439) had an upper limit of 
$2.5c$ quoted. Neither of these findings are strong enough to prove the 
existence of superluminal motion in a radio galaxy.

 The Unification Model hypothesises that radio-loud quasars are an
oriented subset of some intrinsically radio-loud parent population.
Barthel \scite{Barthel-89a} poses the question, `Is every quasar
beamed?', and reviews the evidence. As discussed above, those quasars
which show superluminal motion are likely to be beamed, and we have
also considered the evidence of the Laing-Garrington effect. A beaming
hypothesis can explain several statistical properties, including the
correlation between brightness and component motion, and more limited
statistics showing that lobe-dominated sources display lower
superluminal velocities than core-dominated (more closely aligned?)
sources. Against that must be set the problem of why certain radio-loud
quasars have very large extended structures, whose deprojected linear
size would be enormous; and the statistical finding that the asymmetric
brightness of jets over counterjets is, on average, larger than can be
accounted for by beaming effects alone if the parent population is of
{\em randomly oriented quasars}.

Barthel \scite{Barthel-89a} goes on to demonstrate that by assuming the
parent population is of powerful radio galaxies, with quasars as the
beamed subset, the statistics of jet/counterjet asymmetries can be
justified; and the sizes of the largest radio galaxies are such that the
largest superluminal quasars are not too large in the context of the
parent population. Radio-loud quasars and FR II radio galaxies at $0.5 < z
<1$ in the 3CRR catalogue were compared; after the exclusion of unsuitable
candidates, 12 quasars and 30 radio galaxies remained, with the linear
dimensions of the galaxy radio structure averaging about twice that of the
quasar mean. The relative numbers suggest that a source will appear as a
quasar if viewed within $44.4\degr$ of its axis, and as a radio galaxy
otherwise. This being the case, the quasars should be foreshortened 1.8
times as much as the radio galaxies, consistent with the observed factor
of 2. [If the radio-quiet unbeamed counterparts of the quasars really are
the far infrared galaxies, Barthel \scite{Barthel-89a} notes that the
number count statistics are consistent with such unification: there are 11
quasars for every 49 far infrared galaxies.]

 Further evidence for radio galaxy / radio quasar Unification comes
from other wavebands \cite{Barthel-89a}: Blazars and quasars are known
to be strong in their X-ray emission, while radio galaxies are weak;
Seyfert 1s are luminous in X-rays while Seyfert 2s are not, suggesting
that the molecular torus is an effective shield of X-ray radiation. On
the other hand, the torus viewed from any angle ought to be bright in
the far infrared because of its own heat, and detection statistics of
both radio galaxies and quasars in this waveband bear this out. Two
pieces of evidence oppose Unification, however: claims that the host
galaxies of radio quasars differ significantly from radio
galaxies \cite{Hutchings-87a}; and that extended radio sources lie in
denser environments than compact sources \cite{Prestage+88a}.

 Robson and Antonucci draw differing conclusions on how blazars fit
into the unification picture. We have already seen how Antonucci
\scite[\S 3.1]{Antonucci-93a} proposes that BLLs and OVVs are
attributed to different luminosities in otherwise similar synchrotron
cores. Recall that under this scheme, BLLs are postulated to be blazars
with the most powerful synchrotron sources, whose emission effectively
drowns out the broad lines from the BLR which lies in the line of
sight.  Robson \scite[\S 8.4.2]{Robson-96a} suggests that BL Lac objects
are not end-on quasars, claiming rather than BLLs are beamed FR I
objects and OVVs are beamed FR II objects --- and noting differences in
the radio polarization properties of BLLs and OVVs which hint that BLLs
are more likely to have shocks in their jets. The absence of emission
lines in BLLs would be related to the weakness of the output of the
line-emitting clouds rather than the overpowering strength of the
optical synchrotron core emission. Ghisellini et al.
\scite{Ghisellini+93a} find that their statistics from a survey of 105
radio-loud AGN support this idea.

Antonucci \scite[\S 3.1]{Antonucci-93a} warns against the
automatic identification of the BLL/OVV classification with
the Fanaroff-Riley class, noting that some famous BL Lacs are
FR class II \cite{Kollgaard+92a}. What is clear, is that many
blazars have sufficient radio output in their diffuse emission
alone to make it into the 3C or 4C catalogues; and so
`misaligned blazars' not beamed towards the Earth must be part
of a unified continuum with some other classes of AGN which
are {\em already known}.

 One modification to the standard Unification Model is that
some putative radio galaxies may be quasars obscured not by
their molecular torus, but by other obscuring material. For
instance, infrared spectroscopy revealed a broad H$\alpha$
line in `NLRG' 3C 22 \cite{Rawlings+95a}, suggesting that this
source may be in the quasar orientation, but with opaque
material obscuring much of the light from the active nucleus.
Similarly polarization measurements of 3C 109
\cite{Goodrich+92a} can be understood if there is a hidden
quasar core whose light suffers polarization by transmission
through dust. If such obscured quasars are 
common \cite[3C 234 could be another example]{Tran+98a}, then many
putative radio galaxies may have their axis closer to their
line of sight to Earth than hitherto thought, and orientation
statistics will be affected accordingly.

 Two major questions remained unanswered by current
Unification models. Why do some galaxies and quasars produce
jets strong enough to drive radio emission, and others not?
Why is radio emission is only found in elliptical galaxies? We
shall not pursue these questions or debate the nature of
blazars here, as such matters lie outside the scope of this
thesis, but pause to note that work is still very much in
progress on the refining of the Unification Model.

\chapter{The Alignment Effect and Polarization in Radio Galaxies}
\label{reviewRG}

\begin{quote}
There comes a time when for every addition of knowledge you forget
something that you knew before. It is of the highest importance,
therefore, not to have useless facts elbowing out the useful ones.
\attrib{Sherlock Holmes, {\em A Study in Scarlet}}
\end{quote}

The current paradigm within which radio galaxies are explored is that
of the {\em Unification Hypothesis}. We have already explored how this
hypothesis can be used to account for the wide range of AGN phenomena
described in the previous chapter; now we look more specifically at
radio galaxies.  Distinctively among AGN, radio galaxies often display
significant alignments between their radio structures and the
orientation of their polarization and/or optical structure. Study of
these alignments can help confirm or refute the appropriateness of the
Unification Hypothesis to describe individual radio galaxies, and
statistically, the class as a whole.

\section{Polarization in AGN}
\label{rguni}
 
 Since active nuclei lie within host galaxies whose stars emit unpolarized
blackbody radiation, the measured polarization of any active galaxy will
be that of the active nucleus diluted by starlight. It is important to
distinguish whether polarization figures quoted in a given case are those
of the raw measurement, or corrected for removing the unpolarized stellar
intensity to yield the polarization of the nucleus. The
contribution of starlight diminishes in the near ultraviolet and at   
shorter wavelengths; rest-frame ultraviolet measurements can be
presumed to give a good indication of the nuclear polarization.
 
 The 1980 review paper of Angel \& Stockman summarised what was then known
of the visible and infrared polarization of extragalactic objects:  the
three classes of active galaxies known to produce polarised light were
blazars, quasars and Seyfert 2s. Although the spectra of BLRGs
suggested that they were related to quasars and Seyfert 2s, their
relative faintness meant that polarimetric studies of low
redshift radio galaxies did not appear in the
literature until the early 1980s \cite{Antonucci-82a,Rudy+83a}, and
high redshift studies a
decade later \cite{Cimatti+93a}.
 
 Angel \& Stockman \scite[\S IV]{Angel+80a} wrote before the
Unification Hypothesis had become popular, and reviewed numerous
mechanisms which might account for the low visible/infrared
polarizations observed in many Seyferts. Originally, Seyfert
polarizations (of order 1\%) were attributed to the visible
high-frequency tail of synchrotron radiation. Multicolour and
spectroscopic studies of Seyfert polarization, however, showed that in
most Seyferts, both the core continuum light and the emission lines
were polarised, and the polarization (corrected for the stellar
contribution) was stronger in the blue than in the red, but with little
rotation of position angle. These facts suggested that the total light
emerging from the central engine was being polarised by some subsequent
interaction, most probably with dust. [We dealt with this in some
detail in \S \ref{polinSeyferts}.]
 
 The discovery of Type 1 features in the polarised spectra of Seyfert 2s
was the key to the first stage of Unification \cite[\S 
7.1]{Peterson-97a}, the realization that
orientation alone might be the distinguishing feature between the two
classes of Seyferts. Antonucci's \scite{Antonucci-93a} review paper
describes how the prototypical Seyfert NGC 1068 was investigated and  
found to be generating visible/ultraviolet light polarised at 16\% in its
nucleus. Since electrons scatter light equally strongly at all wavelengths
whereas dust (via Rayleigh scattering and similar mechanisms)
preferentially scatters blue light, it is implied that the scattering
medium in NGC 1068 may be free electrons. Other Seyferts show evidence of
higher nuclear polarizations at shorter wavelengths, characteristic of  
dust. Most yield a polarization orientation perpendicular to the radio
structure, as would be expected for scattering.
 
 Similarly, by the time of Antonucci's \scite{Antonucci-93a} paper,
evidence was accumulating that BLRGs and NLRGs were the Type 1 and Type
2 classes for radio galaxies analogous to the classification of
Seyferts. The picture seemed to be clearest for radio galaxies at
redshifts $z>0.6$, where light measured in the $V$-filter on Earth
corresponded to rest frame ultraviolet ($\lambda <330$\,nm) emissions
in the AGN, uncontaminated by significant starlight. The first measured
high redshift radio galaxy polarization was reported in {\em Nature}\,
by di Serego Alighieri et al.\ \scite{Alighieri+89a}; and since then,
mounting evidence
\cite{Jannuzi+91a,Tadhunter+92a,Alighieri+93a,Cimatti+97a} has
generally borne out the empirical rule of thumb \cite{Cimatti+93a} that
distant radio galaxies $(z \ga 0.6)$ should display diluted
polarizations of 5\% or more, oriented roughly perpendicular to the
radio lobe structure.
 
In addition to this polarimetric evidence (reviewed in more detail
below, \S \ref{polevi}), the apparent alignment of the knotted
optical
structures of high redshift radio galaxies with the radio axes \cite[\S
5]{Chambers+87a,McCarthy+87a,McCarthy-93a} lent weight to the concept
of the torus and central engine proposed by the Unification Model. As we
shall see in \S \ref{aligneff}, this so-called `Alignment Effect' is
manifested most strongly in the AGN with the most powerful radio
emission, and is intimately linked to the Unification Model and the
presumed scattering mechanism for polarization.
 
 While not proven -- and inevitably suffering from a number
of pathological cases which fit poorly -- the Unification Model is now
generally accepted, to the extent of being the foundation of the first
textbooks on AGN to become available \cite{Robson-96a,Peterson-97a}.
The model must stand or fall, however, according to objective tests, not 
by indications of its popularity among astronomers. As new technology 
becomes available to the astronomical community, it is naturally
the Unification Model which experiments are designed to test, and
reinforce or falsify. With the recent availability of infrared arrays
and polarisers \cite{McLean-97a}, it has become possible to extend
imaging polarimetry into the $K$-band.
The work contained in this thesis represents the first studies of
linear polarization in high redshift radio galaxies in this waveband.
 
\section{The Observational Challenge of Distant Radio Galaxies}

\subsection{Observational techniques}

Modern astronomical detectors \cite{McLean-97a} make a range of
observational techniques available. Light from distant objects can be
imaged on a detector array, and the light intensity measured in a
synthetic aperture covering any part of the image. The light can be
dispersed to form a spectrum or passed through an analyser which
separates orthogonally polarised components.

Both spectroscopy and polarimetry are time-consuming procedures for
faint objects: the former disperses the minimal available light into
its component wavelengths, and the latter requires an accurate
measurement of the intensity difference between the two orthogonal
components of the light. Only recently has technology made it possible
to employ both techniques simultaneously and perform spectropolarimetry
of high-redshift radio galaxies, and our capabilities are limited: even
with the light-gathering capacity of the 10\,m Keck telescope, the
resulting spectra must be rebinned at low resolution to extract a
meaningful signal \cite{Cimatti+96a}. Alternatively, if the orientation
of the polarization is already known, a 4\,m class telescope can obtain
a spectrum of light polarized in the known direction in a reasonable
time \cite{Antonucci+94a}.

 The unique value of spectropolarimetry lies in its ability to identify
the spectrum of scattered light present in the total signal, and so
trace the emission properties of whatever hidden component is
illuminating the scattering medium. (The spectrum will also, of course,
give us an indication of whether polarization is attributable to a
mechanism other than scattering.)

 Images taken through a polarizing filter have their own value;
photopolarimetry (i.e.\ photometry of polarized light) can be performed
in synthetic apertures, yielding a polarization
map or a study of the polarization of individual structures in an
object of complex morphology. But again, the faintest sources are not
amenable to a pixel-by-pixel polarimetric analysis; regions several
pixels square may need to be binned together to obtain an acceptable
signal-to-noise; and many of the radio galaxy figures given in the
literature are simply polarizations integrated over the whole
structure.

 Where photopolarimetry is available in multiple
wavebands, models of the polarized spectra of radio galaxies can be
fitted against these broadband measurements: Manzini \& di Serego
Alighieri \scite{Manzini+96a} used this technique to establish their
result (\S \ref{RGspectrum}) that if the 
radio galaxies sampled consist of stars, nebular emission and an excess
component in the form of a power-law, then starlight is still
their dominant component in the near infrared.

\subsection{The importance of the $K$-band}
 
 While observations of the rest frame ultraviolet have been important in
polarimetric studies of AGN, on the assumption that the host galaxy
contribution to the ultraviolet emission is negligible, the infrared is
important for the opposite reason. Radio galaxies, as a species of AGN
distinguished by their radio properties but [ideally] unremarkable in
their optical emission, could be used as examples of `normal' galaxies at
high redshift (hence having experienced less cosmological evolution than
nearby galaxies). They could be detected at high redshifts by virtue of
their radio emission, and then studied optically in the hope that the
active nucleus has not had too great an influence on their evolution 
(compared to `inactive' galaxies), and is not polluting the light from the
host galaxy. At high redshifts, of course, light originating in the
visible or near infrared arrives at the Earth shifted into longer
infrared wavelengths.
 
 The atmospheric $K$-band window lies at a convenient wavelength for
studying the near infrared properties of objects at $z \sim 1$, and the
previous chapter reviewed how the $K-z$ Hubble diagram for 3CR radio galaxies
\cite{Lilly+84a} suggested that the observed $K$-band light was
essentially stellar emission from passively evolving elliptical
galaxies. So in the late 1980s, $K$-band studies of radio galaxies were
thought to be revealing the properties of young elliptical galaxies. As
we have seen (\S \ref{HubbleK}), this has now been called into question
by studies of fainter radio sources \cite{Eales+97a}; and it seems that
a substantial fraction of the observed $K$-band light in 3CR galaxies
must come from the active nucleus after all.

 While the Hubble diagram is a useful tool to analyse the statistical
properties of a set of galaxies, it tells us nothing about individual
galaxies. Studies of the polarization, morphology and spectra of 3CR
galaxies are needed to determine the properties of their $K$-band
excess; an understanding of the influence of the excess in individual
radio galaxies is essential if we are to salvage their role as probes
of galactic evolution.

 The new observations presented in this thesis are polarized $K$-band
images of nine radio galaxies, including seven from the 3CR catalogue.
The signal-to-noise limit prevents the meaningful analysis of
any structure
finer than lobes of individual bright objects. In some cases,
polarimetry in other optical bands is available and can be used
together with
our findings for comparison with synthetic spectra. In all cases, an
upper limit can be assigned to the maximum contribution of any
scattered component to the $K$-band light, providing an independent
means of determining the influence of the active nucleus on the
apparent luminosity of 3CR sources. 

 In the light of our measurements, and those in the literature, it is
then possible to model the most likely mechanism giving rise to the
observed polarizations. If the strength of the $K$-band emission is
related to the power output of the central engine, the simplest
explanation of this result is that a significant fraction of a radio
galaxy's infrared light emerges from the active nucleus in a restricted
cone, and enters our line of sight after scattering by dust or
electrons. In this case, as for the visible and ultraviolet light, the
scattered infrared light should be polarised perpendicular to the
direction of the radio jet. As we shall see in the rest of this
chapter, the relationship between the polarization and structure of
radio galaxies measured at radio, visible and infrared wavelengths is
already well documented, and we shall discuss the properties of our
representative sample of radio galaxies in this context.
 
 The motivation for performing studies of the $K$-band polarizations of
high redshift radio galaxies, therefore, is to probe the origin of
their $K$-band emission. A finding of no polarization would suggest
that infrared light could be used as a safe indicator of the properties
of the host galaxy (but would make it hard to explain the suspected
infrared Alignment Effect). A finding of infrared light polarised
perpendicular to the radio axis would suggest scattered nuclear light;
and where polarimetry exists in other wavebands, would provide a longer
baseline to test the likely origin of the polarisation --- electron
scattering, dust scattering and direct sight of a synchrotron source
each have a distinct dependence on wavelength. Any other finding would
be an invitation to further scientific study!

\subsection{The need for rigorous statistics}
 
 As in any scientific investigation, a thorough error analysis is
required to give the final data their due weight. Polarimetry, however,
is more demanding than other forms of photometry. Measuring a
polarization is akin to determining the magnitude of a vector, a definite 
positive quantity. While the measurements of the vector's components may
fluctuate about zero for an unpolarised object, the magnitude stubbornly 
remains greater than zero and must be `debiased' accordingly.
 
 The astronomical literature contains not a few papers by statisticians
\cite{Simmons+85a} and careful polarimetrists accusing astronomers of
failing to debias their work adequately, although most 1990s papers on
radio galaxy polarizations do address this issue. Given the need for
debiasing polarization figures and the low signal-to-noise inevitable
when studying objects at high redshift, a great deal of work in this
thesis has been devoted to the accurate debiasing and error estimation
of the data available. Much of the work has been published in the form
of a step-by-step guide to polarimetry \cite{Leyshon-98a}; the format
has been retained, though the work has been refined and updated, in
Chapter \ref{stoch} of this thesis.
 
\section{Observational Evidence for Orientation Correlations in Radio
Galaxies}

\subsection{Evidence for the Alignment Effect}
\label{aligneff}

 CCD technology of the mid-1980s allowed the optical structures of high
redshift radio galaxies to be investigated for the first time. Radio
galaxies at $z \sim 1$ were found to look nothing like the giant
ellipticals associated with lower redshift radio galaxies; rather, the
high redshift galaxies were often elongated and contained two or more
bright `knots' rather than a single identifiable nucleus
\cite{Lilly+84a,Spinrad+84a,Spinrad+84b,McCarthy+87b}.  After further
studies of the most powerful radio sources, it was found that the
major axis of the optical elongated or knotted structures was usually
aligned within a few tens of degrees of the radio axis
\cite{Chambers+87a,McCarthy+87a} -- an association which has become
known as the `Alignment Effect'.

 Subsequent investigations with detectors sensitive to visible
light revealed that the Alignment Effect cuts in at
redshifts $z \geq 0.6$ \cite{McCarthy-93a}, and that the knotted
optical structures are known to be emitting {\em continuum}\,
radiation, ruling out theories that the aligned structures are
attributable to line-emitting gas clouds. More recent observations
\cite{Longair+95a,Best+97a,Ridgway+97a} confirm the alignment of the
rest-frame ultraviolet emitting regions with the radio structure.

 There is no consensus at present about the mechanism which gives rise
to the knotted structure; current observations continue to investigate
the extent to which the Alignment Effect is associated with emission
from the active nucleus. Two key tests of the relationship with the
nuclear emission are whether the effect becomes weaker to longer
wavelengths, and whether it becomes less prominent in less powerful
radio galaxies.

 The $z=0.6$ cut-off suggests that either there is an evolutionary
process at work, or that we are observing a property of the rest-frame
ultraviolet which does not extend to the rest-frame visible. Infrared
observations of twenty 1 Jy galaxies at $z>1.5$ \cite{McCarthy-93a}
show little evidence for extended structure at rest-frame visible
wavelengths. Conversely, $U$-band images of the low-redshift $(z=0.1)$
radio galaxy 3C 195 reveal a distinct aligned ultraviolet structure
\cite{Cimatti+95a}, and a bipolar aligned structure is also seen in
ultraviolet images of Cygnus A \cite{Hurt+99a}.
Recent $U$-band imaging of the nearby radio galaxy NGC 6251
\cite{Crane+97a} has revealed several extended regions of emission: the
most prominent feature of this radiation lies interior to a dust ring,
is nearly {\em perpendicular}\, to the radio jet axis, and has a
polarization below 10\%. 

The first $U$-band survey of low redshift radio 
galaxies (15 3CR objects at $0 < z < 0.6$) \cite{Roche+99a} found 
evidence for $\Delta \phi < 12\degr$ alignment between the radio axis and 
the $U$-band structure in 6 objects (two such alignments would be 
expected by chance alone). Two different mechanisms seemed to be at 
work: three sources showed alignment in the optical structure surrounding 
the radio nucleus, while the other three appeared to have a merging 
companion galaxy close to the radio axis. Of the three sources with an 
elongated nucleus, 3C 348 was also elongated in $V$ but gave no evidence 
for knots in either band, while the other two examples displayed knots in 
$U$ but no elongation or knots in $V$. The most radio-luminous radio 
galaxies, therefore, including those with no obvious aligned structure in 
the $V$-band, are now known to be able to display 
alignment in their near-ultraviolet structure at low redshifts, too.

 The first $K$-band images of 3CR galaxies
\cite{Chambers+88a,Eales+90a,Eisenhardt+90a} revealed that the
near-infrared emissions of the most luminous radio galaxies displayed
structure as knotted and complex as the ultraviolet emissions. As
with visible light detectors, efforts were made to obtain infrared
images of radio galaxies with lower radio luminosity. Many of these
programmes used the $K$-band, in which observations of radio galaxies
at $z \sim 1$ trace emissions at $\lambda \sim 1.1$\,\micron\ in the
rest frame.

 Dunlop \& Peacock \scite{Dunlop+93a} compared $K$-band images of 3CR
and PSR (Parkes Selected Regions) radio galaxies in a narrow bin of
redshifts. The 3CR galaxies were selected at $0.8 < z < 1.3$, and the
PSR galaxies ($S_{\mathrm 2.7\,GHz} > 0.1$\,Jy) were known or estimated
to be in a similar redshift range. A definite infrared-radio alignment
effect was determined in the sample of 19 3CR galaxies, although the
$K$-band structure was, on average, less extended than the optical
structure. In some cases, the infrared structure seemed to be
significantly more closely aligned with the radio axis than structure
observed at visible wavelengths.
 These findings are consistent with
the smaller 3C sample of Rigler et al.\ \scite{Rigler+92a}, which
suggested that an infrared alignment effect was present, but weak.
Best, Longair \& R\"{o}ttgering \scite{Best+97a} imaged 28 3CR
galaxies at $0.6<z<1.2$ and again found distinct alignments at
visible wavelengths, with less complex structure and a weaker
Alignment Effect in the infrared. On average, only about 10\% of the
$K$-band flux of 3CR galaxies at $z \sim 1$ is associated with aligned
structures \cite{Rigler+92a,Best+98a}.

 Lacy et al.\ \scite{Lacy+98b} investigated the Alignment Effect in a
sample of $0.5 < z < 0.82$ 7C radio galaxies, with radio luminosities
of order one-twentieth those of 3C galaxies. The Effect was still
present, albeit very weakly above 400\,nm, but only over small scales.
3C radio galaxies exhibit alignment in structures of order 15\,kpc and
50\,kpc; in the 7C sample, the effect seen at 15 kpc did not extend to
structure at 50\,kpc. Dunlop \& Peacock's \scite{Dunlop+93a} PSR sample
was tested in the $K$, $J$, $B$ and $R$-bands, with no evidence for
alignment being found in the red or infrared, and only a possible 
marginal effect in the
$B$-band. Wieringa \& Katgert \scite{Wieringa+92a} also found that the
optical morphology of less luminous radio galaxies was more rounded.
  
 Eales et al.\ \scite{Eales+97a} criticise Dunlop \& Peacock's 
\scite{Dunlop+93a} selection technique for the PSR galaxies: given that
nuclear light biases upwards the apparent brightness of 3CR galaxies,
the $K-z$ relation for 3CR galaxies cannot be used to estimate
redshifts for PSR galaxies unless a correction is made for the nuclear 
component of the brightness. Constructing their own sample of 6C/B2
galaxies at redshifts well matched to Dunlop \& Peacock's
\scite{Dunlop+93a} 3CR sample, and following the same position angle
analysis technique, Eales et al.\ \scite{Eales+97a} found no strong
evidence of an Alignment Effect, but a limited statistical analysis
showed a probability $< 20\%$ that the null hypothesis (no alignment
effect whatsoever) was true.

Best, Longair \& R\"{o}ttgering \scite{Best+96a} analyzed a complete
subsample of eight 3CR galaxies; all lay at $1.0 \la z \la 1.3$ and
emitted radio emission at $S_{\mathrm 178\,MHz} \sim 10$\,Jy, so the
set should be free of evolutionary or radio luminosity trends. The
sample showed a clear trend, such that those galaxies with small radio
structure had complex knotted structures, closely aligned with the
radio axis, which $K$-band imaging showed to be on the same scale as
the host galaxy. Galaxies with much larger radio structures showed
only one or two bright knots, and the alignment, if present, was not
so accurately matched with the radio hotspot axis.

 The observational evidence to date, therefore, shows a clear
infrared-radio alignment and a clear visible-radio alignment in 3CR
galaxies at $z \sim 1$, with the possibility that the
visible structures are slightly misaligned $(\Delta \phi \sim 10\degr)$ with
and/or more extended than their infrared counterparts. 
\label{misalign1} Galaxies at
lower radio luminosity show only marginal evidence for
radio-optical alignment in any optical band, except for the near
ultraviolet small-scale ($\la$ 15\,kpc)
alignments of Lacy et al.\ \scite{Lacy+98b}. Evidence
for an Alignment Effect in quasar host galaxies has also been reported
recently \cite{Ridgway+97a}.

 We will consider the possible explanations of the Alignment Effect
offered by the Unification Model and its alternatives in the following
sections after reviewing the evidence of polarized optical radiation.
It may be worth noting, however, the first glimpse of ordinary galaxies
at high redshift, as provided by the {\em Hubble Deep Field}. This
window on a younger universe has revealed many galaxies of disturbed
morphology \cite{Naim+97a}, including elongated objects which have
become known as `chain galaxies'. The evolutionary relationship between
these objects and the morphological classes of galaxies in today's
older universe remains to be resolved, but we cannot rule out the
possibility of some common factor at work in these chain galaxies and
the hosts of radio galaxies.

\subsection{Broadband polarization measurements}
\label{polevi}

 Motivated by the discovery of broad lines in the polarized spectra of
Seyfert 2s \cite{Antonucci+85a}, the late 1980s saw several NLRGs
analysed by spectropolarimetry in the hope of revealing broad lines in
their polarized flux. Bailey et al.\ \scite{Bailey+86a} and Hough et
al.\ \scite{Hough+87a} found that Centaurus A and IC 5063 respectively were
$\ga 10\%$ polarized perpendicular to their radio structures in light
at 2\,\micron . They suggested that this polarized light might be direct
emission from nuclei obscured at visible wavelengths, and that the high
polarization was indicative of blazar activity. Antonucci \& Barvainis
\scite{Antonucci+90a} agree that the nuclear light is more visible in
the infrared, partly because kpc-scale dust lanes optically thick in
the visible are more transparent to the near infrared; but they point
out that the lobe-dominated radio structure and the strong
perpendicular radio structure/infrared polarization alignment are not
characteristic of blazars. They review the discovery of polarized broad
lines in 3C 234, arguing that this object is an NLRG and very similar
to NGC 1068; and both Centaurus A and IC 5063 could be objects of the same
class. [3C 234 is a $z=0.185$ object now known to have a spectrum similar 
to that of Seyfert 2s, and is sufficiently luminous to be harbouring a 
quasar nucleus \cite{Tran+95a}.]

 More recently, detailed $K$-band imaging polarimetry studies have been
performed on Centaurus A \cite{Packham+96a}. It was found that in the
near-infrared, the polarization vectors mainly lie along the dust lane,
with the polarization being produced by dichroic absorption of the
radiation from stars embedded within it. But an additional larger
polarized component was detected in the nucleus at 2.2\,\micron , with
the position angle of polarization perpendicular to the inner radio jet
and the X-ray jet. Millimetre-wave observations at 0.8\,mm and 1.1\,mm
found no evidence of polarization at these wavelengths. Centaurus A can hence
be explained by the usual scattering model.

Other NLRG imaging polarization measurements rapidly followed 3C 234
in the literature. Polarization was found in regions distinct from the
nucleus in PKS 2152-69 $(z=0.028)$ \cite{Alighieri+88a}, and in the
first high-redshift $(z=1.132)$ radio galaxy successfully analysed, 3C
368 \cite{Scarrott+90a,Alighieri+89a}. Nuclear polarization was
detected in 3C 277.2 $(z=0.766)$ \cite{Alighieri+88a}. Results
published in other papers were as follows:

\begin{description}

\item{Antonucci \& Barvainis \scite{Antonucci+90a}} attempted to measure the
polarization of several other NLRGs but obtained only large upper
limits in most cases (they blame obscuring kpc-scale dust lanes for
their failure to detect nuclear light in these cases). They obtained a
significant result for 3C 223.1 $(z=0.108)$, with its 2.2\,\micron\
polarization
measured at $4.9 \pm 0.7 \%$ oriented at $116 \pm 4 \degr$, an offset
from the radio structure of $80 \pm 6 \degr$. The visible-light
polarization was found to be below 0.5\%.

\item{Impey, Lawrence \& Tapia \scite{Impey+91a}} took a complete sample of
radio sources covering both radio galaxies and radio-loud quasars, to
analyse their polarizations. Polarizations were successfully measured
or obtained from the literature for 20 of the 30 radio galaxies forming
the sample; those polarizations obtained by the authors themselves were
unfiltered, with a nominal wavelength of 570\,nm defined by the
properties of the GaAs phototubes used. Only two radio galaxies
consistently yielded polarizations higher than 3\%, {\em viz.} 3C 109
and 3C 234. Most had polarization values in the 1--2\% range.

\item{Jannuzi \& Elston \scite{Jannuzi+91a}} investigated the radio galaxy 3C
265, discovering that the orientation of its polarization in both $B$
and $R$ passbands is roughly perpendicular to the axis of the radio
emission and to the major axis of the structure seen in ultraviolet
emission. The data show no evidence of wavelength dependence in the
polarization between $B$ and $R$.

\item{Tadhunter et al.\ \scite{Tadhunter+92a}} compared the polarization of
medium and high redshift radio galaxies, imaging using either no
filter, or standard or broadened $V$-band filters. Seven objects at
$0.5<z<0.85$ were included in their sample, of which five exhibited raw
polarizations in the range 5--20\%, with generally perpendicular
alignments to the radio axes. None of the five intermediate-redshift
objects at $0.2<z<0.5$ showed polarizations over 5\%.

\item{di Serego Alighieri, Cimatti \& Fosbury \scite{Alighieri+93a}} 
measured six high-redshift radio galaxies in bands corresponding to
rest-frame wavelengths around 300\,nm. Four of these were found to have
polarizations of 4--18\%, all oriented perpendicular to their optical
structure.

\item{Shaw et al.\ \scite{Shaw+95a}} analysed four southern 
radio galaxies at $0.3<z<0.7$ in the $B$-band. Two, PKS 1602+01 and
PKS 2135-20, are BLRGs and have low polarization. PKS 1547-79 is also a
BLRG, and seems to be polarized but may be contaminated by dust.
PKS 2250-41 is a NLRG with high polarization.

\item{Cimatti \& di Serego Alighieri \scite{Cimatti+95a}} collected
data on eight 3C radio galaxies at $0.09 \leq z \leq 0.47$ in
Johnson-Cousins filters selected individually to reveal the rest frame
properties of each galaxy at around 300\,nm -- the most notable result
being findings of an ultraviolet alignment effect in the {\em
low}-redshift radio galaxy 3C 195. The same group also analyzed the
near ultraviolet properties of a $z=2.63$ object,
MRC 2025-218 \cite{Cimatti+94a}.

\end{description}

\subsection{Polarization trends}
\label{poltrend}

 Those various results listed above which had been published
by 1992, together with a few other individual objects in the
literature, were gathered together by Cimatti et
al.\ \scite{Cimatti+93a}. They took 42 radio galaxies at
$z\geq0.1$ from the literature and their own observations, and
looked for correlations between the optical polarization and
other properties. In cases where the object was extended, an integrated
polarization was taken. Trends were sought both with the observed
polarization (debiased), and with the corrected nuclear polarization 
which would be present if the light was being diluted by an
elliptical host galaxy. Because of the wide variety of filters used in
collecting the data, and the range in redshifts of the objects
observed, the polarizations represent rest-frame emissions at differing
wavelengths between 200\,nm and 700\,nm. Five trends were discerned for
observed polarizations, not all of which survive for the
underlying nuclear values.

\begin{description}

\item{{\em Redshift, $z$.}}
High polarization is observed preferentially at high redshift. The
observed values are almost perfectly bimodal: six out of the seven
objects at $z>0.6$ are polarized above 8\%, while all but one of the
lower redshift objects are polarized below 7\%. After dilution
correction, the nuclear polarization still appears to increase with
redshift, with $2\sigma$ significance.

\item{{\em Rest-frame wavelength, $\lambda_r$.}} High polarizations
are
preferentially
observed in wavebands corresponding to emission bluewards of 400\,nm.
Again, a $2\sigma$ significance correlation remains between the nuclear
polarizations and the wavelength of emission.

\item{{\em Radio power, $P_{\mathrm 178\,MHz}$.}} The total radio
power
emitted at 178\,MHz is highest for objects with the greatest observed
polarization; yet again, the nuclear polarizations retain this
correlation with $2\sigma$ significance.

\item{{\em Radio spectral index, $\alpha_r$.}} Most of the objects
surveyed possessed spectral indices between 0.5 and 1.0. The higher the
observed polarization, the closer to 1.0 the spectral index tended to
be. This correlation was weaker than the previous three, being
significant at the $3\sigma$ level for observed polarizations, and
dropping to $1.5\sigma$ for the corrected nuclear values.

\item{{\em Radio $Q$-structure.}} The $Q$-parameter
\cite{McCarthy+91a} is
a measure of the asymmetry of the radio structure, the ratio of the
longer radio arm to the shorter. No clear correlation could be
confirmed, but there was a noticeable absence of galaxies combining low
$Q$-values (i.e.\ symmetric radio structure) with high observed
polarization.

\end{description}

When the orientation of the polarization was assessed as well, the two
major observational results were (A) that radio galaxies at $z>0.6$ and
polarizations $p>5\%$ {\em always}\, show perpendicular polarization,
with a tendency to be aligned more closely with the structures observed
in the ultraviolet continuum than with the radio axis; and (B) that
those radio galaxies exhibiting parallel polarization/structure
alignments always had polarizations $p<5\%$.

\begin{small}
\begin{table*}
\caption{Polarimetry of radio galaxies $z \geq 0.2$ cited in the 
literature since 1993.}
\begin{tabular}{lclccrrrrr}
\hline
 Source & B/N & \multicolumn{1}{c}{$z$} & Band & $\lambda_r$ (nm) & $P
\pm \sigma_P$(\%) 
& $\theta \pm \sigma_{\theta}$(\degr) & $\delta_{o-r}$ & Ref\\

   PKS 2250-41 nuc & N &  0.310 &  $B$  & 336 & $\ell$\, 4.4 $\pm$ 0.8
& 152 $\pm$  5  &  58 & S+\\

   PKS 2250-41 W   & N &  0.310 &  $B$  & 336 & $\ell$\,$(<4.5)$
&  10 $\pm$ 8  &   93 & S+\\

  	  3C 313   & ? &  0.461 & $B$ & 303 & $(<6.0)$
& n/a &  n/a   & CA\\

   PKS 1602+01     & B &  0.462 &  $B$  & 301 &  $\ell$\,$(<4.8)$
& 119 $\pm$ 22  &  5 & S+\\

   PKS 1547-79     & B &  0.483 &  $B$  & 297 &  $\ell$\,2.9 $\pm$ 0.7
& 66  $\pm$ 2   &  42 & S+\\

   PKS 2135-20     & B &  0.635 &  $B$  & 269 &  $\ell$\,$(<2.7)$
& 180 $\pm$ 16  &  n/a & S+\\

    3C 277.2 & N &  0.766 & $i$ & 450 &  e\,9.9 $\pm$  1.4  
& 169 $\pm$   7 &  108 & A+\\

 3C 226   & N &  0.818 &  $i$  & 440 & e\,2.9 $\pm$ 1.4
&  84 $\pm$  13  &  120 & A+\\

  FSC 10214+4724& N &   2.286 & \O & 228 & 16.2 $\pm$ 1.8
& 75  $\pm$ 3   &  145 & L+\\

  MRC 2025-218 & ? & 2.627 & $R$ & 185  & w\,8.3 $\pm$ 2.3  
& 93 $\pm$ 8       & 63 & C+ \\

\hline
\end{tabular}

\medskip

Key: Source: Common name of object; B/N: BLRG / NLRG classification;
$z$: redshift; Band: waveband of observation, Johnson-Cousins or Gunn
designation [\O\ denotes no filter, effective passband 400--1000\,nm];
$\lambda_r$: central wavelength of observed frame transformed into
source's rest-frame; $P\pm\sigma_P$: percentage polarization
(debiased) with 1$\sigma$ error [e indicates corrected for suspected
emission line contamination;  $\ell$ indicates largest of cited
nucleocentric apertures taken; w indicates
whole galaxy, not nucleus]; $\theta\pm\sigma_\theta$: Electric vector
orientation E of N (\degr); $\delta_{o-r}$: Orientation offset of
optical polarization minus radio position angle (\degr); Ref: Data source,
as follows: A+ \cite{Alighieri+94a};  C+ \cite{Cimatti+94a}; CA
\cite{Cimatti+95a}; L+ \cite{Lawrence+93a}; S+ \cite{Shaw+95a}.
\label{litpol} \end{table*} \end{small}

Since Cimatti et al.\ \scite{Cimatti+93a} compiled their paper, further
broadband polarizations have appeared in the literature. Those for
radio galaxies at $z>0.2$ are given in Table \ref{litpol}. If the
observed polarizations are compared with the trends in redshift and
rest-frame wavelength seen in the 1993 data, most of the new data
points lie within the scatter of the existing points. Notable
exceptions are 3C 226, whose 3\% polarization is very low for a $z=0.8$
object, and PKS 2135-20, also at low polarization and high
redshift, whose upper limit of 2.7\% is much lower than
the polarization of any other object seen in rest-frame light emitted
below 280\,nm.

 Three sources show a distinct perpendicular alignment between
polarization orientation and radio structure: PKS 2250-41, 3C 277.2, 3C
226. It should be noted, however, that the two 3C objects both occur
already as perpendicular objects in Cimatti et al.'s\
\scite{Cimatti+93a} data, but are observed here in different wavebands.
PKS 1602+01 shows a distinct parallel alignment. Several objects (PKS 
2250-41 nuc, PKS 1547-49, FSC 10214+4724, and MRC 2025-219) have
polarization orientations which seem to be
about 20\degr\ offset from perpendicular alignment with the radio
structure --- it would be instructive to determine whether these
objects displayed a closer perpendicular alignment between
polarization and their optical structure. Results A and B still stand,
except in the case of the significant misalignment of FSC 10214+4724.
\label{misalign2}

\subsection{Radio galaxy spectropolarimetry and extended imaging}
\label{rgsspecpol}

Early data on nearby radio galaxies' polarized spectra was compiled by
Antonucci \scite{Antonucci-84a}. Although the initial sample included 
45 objects, several showed signs of variability in their optical
polarization, and high quality radio maps were not available for many
of the others. From the few objects for which it was possible to
compare the radio structure and the optical polarization orientation,
there was clear evidence for a class of galaxies polarized parallel to
the radio structure, and weak evidence that the non-parallel
galaxies might form a perpendicular class. The polarization of the {\em
radio\,} emission from the core was also analyzed, and was found to
have
a tendency to be aligned perpendicular to the radio structure. Too few
objects had both radio and optical polarization measurements to provide
meaningful data on any correlation between the two orientations.

The first spectropolarimetric analysis of high-redshift radio galaxies
was made on three targets for which the broadband polarization
orientations were already known \cite{Alighieri+94a} --- 3C 226, 3C 
277.2, and 3C 324. This enabled
spectra to be taken through a linear analyzer oriented to extract light
polarized parallel and perpendicular to the broadband polarization
angle, making the most efficient use of limited observing time, at the
price of the impeded detection of any lines polarised at intermediate
angles. All three galaxies yielded evidence of a polarized continuum, a
broad Mg\,{\sc ii} $\lambda$2798 line polarised to the same degree, and
narrow lines which were consistent with zero polarization, but not
compatible with polarization as high as that of the continuum.

The specific measurements suggested that 3C 226 has a
polarization of about 11\% in the range 210-370\,nm, possibly constant
but possibly declining to the red; its $i$-band polarization is
considerably lower (see Table \ref{litpol}). 3C 277.2 exhibited more
variation in its continuum polarization, with values between 11\% and
24\% seen between 200\,nm and 380\,nm. Again, a possible decreasing
polarization to the red is reinforced by a much lower $i$-band
measurement.
Finally, 3C 324 also seems to have an 11\% continuum. The built-in 
assumption about the polarization orientation means that these
values are only true polarizations if the assumption is correct;
and we cannot, of course derive polarization orientation angles from
such observational data.

Cimatti et al.\ \scite{Cimatti+96a} analyzed the same $z=1.206$ galaxy
3C 324 with the W. M. Keck telescope. The continuum
emission polarization
between 200\,nm and 400\,nm (rest-frame) remained fairly constant
at about 12\% $\pm$ 4\%, 17\degr\ $\pm$ 2\degr. The two most prominent
emission lines in the polarized flux were [O\,{\sc ii}] $\lambda$3727,
which bore a much lower polarization than the continuum, and at a
perpendicular angle; and Mg\,{\sc ii} $\lambda$2800, which is
more polarized than the continuum, but in the same orientation. 

 Keck observations of further 3C objects followed: firstly 3C 256
\cite{Dey+96a}; then 3C 13 $(z=1.351)$ and 3C 356 ($z=1.079$, one of the
subjects of this thesis) \cite{Cimatti+97a}, and later two powerful radio
galaxies at $z \sim 2.5$, 4C 00.54 and 4C 23.56 \cite[using the Keck
II]{Cimatti+98a}. In all these cases, linear polarization oriented
perpendicular to the radio structure was found. One important exception to
this trend was also discovered \cite{Dey+97a}: Keck spectropolarimetry of
the $z=3.798$ galaxy 4C 41.17 showed no evidence for polarization, with a
$2\sigma$ upper limit of 2.4\%, with a number of strong absorption
features. This anomalous galaxy is proposed as an example of a galaxy
caught in the act of star formation. 

 Tran et al.\ \scite{Tran+98a} used the Keck I to obtain both
spectropolarimetry and extended imaging polarimetry: their targets were 
3C 265 and further observations of 3C 277.2; 3C 324 (imaging only) and 
3C 343.1 (spectropolarimetry only). [The imaging data are not
reproduced in Table \ref{litpol} since there is no one overall figure;
the results are published in the form of polarization vector maps.]
The three galaxies with imaging maps all displayed a bipolar fan of
polarization vectors centred on the nucleus, perpendicular to the
optical structure and misaligned by tens of degrees with the radio
axis. The {\em nuclear}\, polarization of 3C 265 appears to be about
12\%; the diluted polarization of the near ultraviolet emission of 3C
277.2 is rated at 29\% $\pm$ 6\%. The third source analyzed in its
spectrum, 3C 343.1, was not found to be highly polarised but was
contaminated by an object lying at an intermediate redshift.

\section{Interpretations of Orientation Data}

\subsection{Disentangling the spectra of radio galaxies}

 How are we to interpret these polarization and Alignment Effect
observations? The most revealing findings, though historically the most
recent, are the spectropolarimetry results. Invoking Occam's Razor, it
is safe to assume that any polarization which displays a constant
position angle over a range of wavelengths is generated by a single
mechanism \cite{Alighieri+94a,Cimatti+96a}. Further, if two different
`features' (e.g.\ the continuum and emission lines) are polarized in
the
same direction and with comparable strength, the polarization mechanism
is probably one which polarizes light in transit rather than anything
intrinsic to the emission of light at the source. [Were we to find
polarization angles which systematically changed with wavelength, one
interpretation might be a relativistic effect violating the Einstein
Equivalence Principle; conversely, the observed constancy of the
orientation angle in galaxies at various different redshifts provides
further reinforcement for General Relativity, as noted by Cimatti et
al. \scite{Cimatti+94a}.]

 Two approaches can be used to disentangle a polarized spectrum ---
modelling the separate components thought to contribute to it, or
separating the 100\% polarized component from the unpolarized component
by producing the product spectrum $p(\lambda) \times S_{\nu}(\lambda)$.
Although the underlying polarised component will probably not be 100\%
polarized,
the polarized flux will isolate features unique to the partially
polarized component. Clearly if {\em all}\, the features in a spectrum
with several components are polarised, the most likely mechanism would
be external to the source object --- most likely to be transmission
through aligned dust grains, which might be near the source, in the
intergalactic medium, or in our own Galaxy. In such a case polarization
studies would tell us much about the extragalactic or Galactic dust,
but nothing about the source galaxy.

 The next assumption which can be made is to model the light from the
radio galaxy as two components: evolved stellar blackbody emission, and
a nuclear component to be explained. Since nearby radio galaxies are
clearly ellipticals, more distant radio galaxies must surely also
contain an evolved stellar population --- although we must remember
that part of the motivation for studying radio galaxies is to determine
the evolutionary changes in this population at high redshift.  
A third, weaker, component may also be included in modelling
\cite{Manzini+96a}: nebular continuum emission. Individual cases are
known where the nebular contribution may be significant
\cite[10\%]{Cimatti+98a} or even dominant \cite[3C 368]{Cimatti+97a}.

 Neither nebular emission nor stellar blackbody emission is
intrinsically polarized, so we expect these sources to contribute only
to the unpolarized spectrum, while the nuclear component may be
partially polarized. This nuclear contribution may itself consist of
several features with different polarization strengths, indicative of
the various mechanisms at work within the nucleus. The presence of an
unpolarized stellar (and nebular) component together with a
substantially polarized nuclear component is indicative of the fact
that the polarizing mechanism is contained inside the host galaxy
(otherwise the stellar emission would also become polarized) but does
not consist of many independent cells of polarized emission
\cite{Antonucci-84a} which, if
independently oriented, would tend to cancel out one another's
polarization and produce a low overall figure.

A key feature of evolved galactic spectra is the so-called 4000\,\AA\
spectral break \cite{Bruzual-83a}: the intensity of the blackbody
radiation from the stars in the galaxy drops substantially bluewards of
400\,nm, due to the scarcity of short-lived massive stars which would
be luminous in the near ultraviolet. Therefore, significant amounts of
ultraviolet radiation are diagnostic of star-forming activity or
non-blackbody processes at work.

The shape of the polarized component of the spectrum provides clues
about the likely polarization mechanism \cite{Cimatti+93a}. Synchrotron
radiation is most intense at long wavelengths and falls off to the blue
-- but will not result in polarized emission lines. Scattering by
electrons (Thomson scattering) will not change the spectral profile of
the light being
scattered. Dust scattering (Rayleigh scattering) is most effective at
short wavelengths, so the spectrum of light incident on dust clouds
will become blued as the light scatters. Elongated dust grains aligned
in a magnetic field can also polarize light passing through the dust
cloud. Of course, in dust transmission or scattering scenarios, we must
also bear in mind the possibility of the incident light being partially
absorbed and reddened by the dust \cite{Alighieri+94a}. Orientation
correlations between polarization angles, optical structure, and the
positions of the radio jets can be used together with the spectral
profile to identify the most likely candidate mechanism in each case.

 The three galaxies observed by di Serego Alighieri et al.\
\scite{Alighieri+94a}, 3C 226, 3C 277.2, and 3C 324, all displayed
constant polarization orientations, and Mg\,{\sc ii} lines polarized at
approximately the same level as the continuum. This, therefore, is
indicative that some scattering/transmission process is modifying the
light; and the decline to the red suggests scattering. The polarization
strength of the continuum could differ from that of the magnesium line
if the sources of emission were in slightly different positions and the
geometry of the scattering process made the polarization of light from
one source more efficient than that of the other. The narrow oxygen
line observed at low polarization in 3C 324 by Cimatti et al.\
\scite{Cimatti+96a} could be assumed to arise outside the nuclear
scattering region, and they suggest that this galaxy also possesses a
dusty region capable of producing polarization by transmission.

Similar findings were reported for 3C 256 \cite{Dey+96a} and 3C 356
\cite{Cimatti+97a}: broad
magnesium lines were visible in both unpolarized and polarized flux,
while narrow forbidden lines were only visible in unpolarized flux.
In 3C 356, the ultraviolet continuum appeared to contribute about 80\%
of the total light at 280\,nm, and the remainder could be modelled by 
an evolved stellar population aged $\sim$ 1.5--2.0 Gyr. Both of the 
$z \sim$ 2.5 radio galaxies, 4C 00.54 and 4C 23.56
\cite{Cimatti+98a} were found to be dominated by non-stellar emission
at 150\,nm, with young massive stars contributing no more than half the
total continuum flux. For all the sources, their polarization
orientations were approximately perpendicular to their major structural
axes, implying that the scattered light was originally travelling
parallel to the major axis.

 Recent spectropolarimetry shows, therefore, that without making any
assumptions about the nature of the central engine, some of the most
powerful high redshift radio galaxies have an evolved stellar
population, and contain continuum and broad line sources in a confined
region such that only their emission parallel to the radio/optical
structural axis is able to be scattered into our line of sight. It must
be noted, of course, that those radio galaxies chosen for
spectropolarimetry tend to be those known to have high polarization
{\em a priori}, and which are bright enough for spectra to be taken
in a reasonable time.

\subsection{Interpreting broadband polarizations}

 Although spectropolarimetry of high redshift radio galaxies has only
recently become available, a similar analysis can be carried out by
comparing broadband multiwavelength polarimetry with synthetic spectra
modelled from stellar, nebular and power law components. In such cases,
it is not possible to distinguish the presence or polarization states
of broad or narrow lines, but the wavelength dependence of the
continuum polarization will be apparent. 

We have already seen how Manzini \& di Serego Alighieri
\scite{Manzini+96a} used a synthesis technique to distinguish
components in unpolarised spectra (\S \ref{RGspectrum}); they were also
able to simulate the effects of dust scattering by different species of
dust grains and so synthesise polarized spectra which they then fitted
against the photopolarimetry available from the literature. By doing
so, they could estimate not only the fraction of light present in each
component, but also the age of the host galaxy and the most likely
properties of the scattering dust.
Cimatti et al. \scite{Cimatti+94a} note how in most cases, broadband
fitting shows that radio galaxies probably contain evolved stellar
populations --- it is also possible to interpret findings in
some galaxies, though, as due to young stellar populations born with
non-standard IMFs (initial mass functions) \cite{Bithell+90a}.

  In most cases, however, only one or two measurements of polarization
will be available, and then the best analysis which can be carried out
is that of orientation correlations with other properties, as in the
trends analysis of Cimatti et al.\ \scite{Cimatti+93a}, reviewed above 
(\S \ref{poltrend}). We saw that polarization tended to be highest at
the shortest rest-frame wavelengths and for the most distant radio
galaxies; it is not clear which of these correlations is primary, and
which is a consequence of the other. 

 One further finding which we must review is that the optical and radio
structures seem to be misaligned by $\sim 15\degr$\ in many cases; and
the optical polarization tends to be perpendicular to the extended
optical structure rather than the radio structure
\cite{Cimatti+94a}. Tran et al.\ \scite{Tran+98a} 
note how a similar effect has been observed in Seyfert galaxies.
This might be attributed to rotation in transit of light polarised
perpendicular to the radio structure; but it
seems most likely that the extended optical emission region is the
scattering zone for the polarised radiation, yielding the natural
perpendicular result. It must then be explained why the Alignment
Effect is not perfectly parallel to the radio structure; and why the
infrared Alignment Effect \cite{Rigler+92a,Dunlop+93a} traces the radio
structure more closely than the ultraviolet extended emission.

\subsection{Further interpretation of spectral features}

 One further property which is of note is the classification of radio
galaxies as BLRGs or NLRGs. Now that we know that some distant radio
galaxies contain broad emission lines visible in polarized light, we
must ask what makes a galaxy fall into the BLRG class. Broad lines will
become visible in the unpolarized spectrum if the scattered BLR
emission is sufficiently strong, or if the geometry permits a direct
view of part of the broad line region. [There are NLRGs known where
the broad lines are not totally obscured, while the narrow lines
appear to be partially extinguished; this may be explained by
carefully selecting the geometry so the obscuring torus covers some
narrow line clouds and not all of the broad line emission region
\cite{Alighieri+94a}.]

Since Cimatti et al.'s\ \scite{Cimatti+93a} sample contained only 8
BLRGs out of 42 objects, and the most distant lay at $z=0.306$, so it
would be dangerous to draw conclusions about differences in the
observed polarizations of BLRGs and NLRGs. Nevertheless, recalling that
the broad-lined Seyfert 1s tend to have low parallel polarizations, it
is noteworthy that three of the eight BLRGs have only upper limits to
their polarizations, and three more have low ($\la 6\%)$ parallel
polarizations. Only 3C 332 (3\%) and 3C 234 (6\%) are perpendicular.  
Three of the additional objects recorded in Table \ref{litpol} above
are BLRGs, all polarized at $<5\%$; and we note also that one of these
three objects is a definite parallel polarization, a second is closer
to parallel than perpendicular, and the radio axis of the third is
undetermined. Against this is the earlier observation
\cite{Antonucci-84a} that radio galaxies exhibiting parallel
polarizations tended to be NLRGs; or else that those polarised parallel
which do possess broad lines also have other anomalous features.

 Other spectral properties, in particular comparisons of radio galaxy
and Seyfert spectra, can be found in the literature: in particular,
Seyferts only exhibit magnesium lines in polarised spectra while this
emission line is sufficiently strong to appear in total spectra in
radio galaxies \cite{Alighieri+94a}. Seyfert 1s are compared to BLRGs
by Rudy et al.\ \scite{Rudy+83a} who find the BLRGs of similar
luminosity but with weaker Fe\,{\sc ii} lines, a steeper Balmer
decrement and a larger [O\,{\sc iii}]/H\,$\beta$ ratio.

\subsection{Unified models of radio galaxies}

 We have now gone as far as we can in interpreting our results in terms
of the properties of an abstract power-law source and broad line region
at the heart of an elliptical galaxy. Now we must ask: what physical
mechanism can explain the Alignment Effect, Cimatti et al.'s
\scite{Cimatti+93a} polarization trends, and the spectropolarimetric
evidence for scattered light from power-law and broad line components?

The obvious candidate is that radio galaxies contain the same kind of
central engine as is postulated to exist in Seyfert galaxies and
quasars. Since radio galaxies are radio-loud by definition, the mechanism 
must be closest to that at work in the radio-loud quasars. But radio 
galaxies are not quasars, and a plausible model must also explain the
differences.

 Let us assume that a radio galaxy consists of an evolved elliptical
galaxy containing a black hole, accretion disk, BLR clouds, obscuring
molecular torus, and narrow line emission clouds described in the
previous chapter (\S \ref{intunimod}); as a radio galaxy, the central
engine will possess a powerful bipolar outflow jet responsible for the
radio structure. There is evidence
\cite{Kormendy+95a,Willott+98a,Serjeant+98a} that the radio and
optical luminosity of the active nucleus may be correlated with the
mass of the galaxy, and hence its stellar luminosity; but we will
assume that the relative contributions of the stellar and nuclear
components can be varied freely with a wide dynamic range.

 We postulate that dust and/or electrons may be present in the outer
regions of the galaxy, and that these particles are capable of
scattering light into the line of sight to the Earth. Particles
illuminated by the central engine through the opening angle of the
molecular torus may therefore scatter nuclear light towards Earth; such
light will naturally become partially polarized orthogonal to the lines
joining the scattering region to the central engine and to the
observer. We assume that light is only scattered once, since multiple
scattering would randomize the polarization angle of emerging light;
and with such a scattering efficiency, much less than half the nuclear 
light travelling in a given direction can be scattered out of its original
path.

 In radio galaxies with relatively strong central engines, we will see
dust clouds illuminated in a broad cone, and to a considerable distance
from the nucleus. Where the central engine is weaker, or the dust
clouds more tenuous, only the region closest to the nucleus will be
discernibly illuminated and so 
at limited resolution, it will be harder to distinguish deviations from 
perfect alignment. This would 
be consistent with the small-scale (15\,kpc) Alignment Effect in 7C
objects \cite{Lacy+98b}, and also with the great radio strength/loose
optical structure correlation \cite{Best+96a} if powerful radio jets
tend to sweep space clear of scattering material in their path.  The
scattering efficiency of dust is much lower in the infrared than in the
ultraviolet, so again, tighter and closer aligned structures might be
expected before nuclear infrared light diverges below a detectable
intensity, while visible light from a nucleus of similar luminosity
might diverge further and create an impression of `misalignment' before
falling below the intensity threshold.

The tendency towards increasing observed polarizations at $z>0.6$ can
be largely explained by the rest-frame wavelength of images taken in
standard filters moving towards the blue; nuclear polarization is more
apparent as the diluting effects of the host galaxy fall off shortward
of the 4000\,\AA\ spectral break \cite{Bruzual-83a}. Any residual
tendency in the nuclear polarization \cite{Cimatti+93a} could be
attributed to the increasing efficiency of Rayleigh scattering at
shorter wavelengths. Similarly, the Alignment Effect, if due simply to
the structure of the scattering regions compared to the stellar
structure, will be most pronounced in sources with the strongest active
nuclei [i.e.\ visible even in the $K$-band for 3CR sources
\cite{Dunlop+93a}], and at rest frame wavelengths below 400\,nm where
the stellar emission falls off rapidly --- again explaining the
$z<0.6$ cut-off for visible images, and the indications of $U$-band
alignment in nearby radio galaxies. 

Hammer, LeF\`{e}vre \& Angonin \scite{Hammer+93a} find no evidence for
the 4000\,\AA\ spectral break in a composite spectrum of ten radio
galaxies at $0.75 \le z \le 1.1$; but it has been pointed out
\cite{Alighieri+94a,Cimatti+94a} that the composite spectrum was of
total light (stellar plus nuclear) in which the break can be masked by
the nuclear contribution, and can be hard to measure accurately with
several emission lines lying close to 400\,nm. In using the properties
of the spectral break to estimate the age of the stellar population of
the host galaxy, it must be remembered that other factors which can
affect the solution for the age include the timescale of formation and
the metallicity.

 In some cases it may be possible to identify the most likely
composition of the scattering medium, though in a real galaxy there are
likely to be regions containing dust and regions of free electrons \cite{Tran+98a}.
Rudy et al. \scite{Rudy+83a} attribute some of the features of their
BLRG spectra to dust extinction effects. Electron scattering produces
much stronger polarization [20--50\% as opposed to 10--20\% in dust
\cite{Alighieri+94a}] and so it may be necessary to invoke an
additional component of hot young stars to dilute the observed
ultraviolet polarization if electrons rather than dust are suspected in
particular cases --- Cimatti et al. \scite{Cimatti+94a} demonstrate how
different combinations of old and young stellar populations, direct
nuclear radiation and dust or electron scattered components may be
combined to form models consistent with observed polarizations. On the
other hand, certain distributions of dust grains may be able to produce
polarizations comparably high to those due to free electrons
\cite{Manzini+96a,Cimatti+96a}.  Cimatti et al. \scite{Cimatti+96a}
demonstrate how an upper limit can be set on the free electron
temperature by the width of observed lines, which would be broadened
beyond visibility \cite{Fabian-89a} by scattering in a too-hot plasma.

 So far, a Unification Model can be made to fit the observed facts. But
if the redshift dependence of polarization and the Alignment Effect is
merely an artifact of our standard filters and the 4000\,\AA\ spectral
break, is there any evidence for evolution in radio galaxies? We have
seen that the host galaxy may contain a naturally evolving stellar
population whose aging process is masked by evolution in the structure
of the whole galaxy \cite{Eales+96a,Best+98a}; high redshift radio
galaxies are akin to BCGs in this respect, and themselves tend to form
in clusters. Why, then, are some low redshift radio galaxies -- the
luminous FR IIs -- not in clusters? Have the FR IIs' powerful outputs
disrupted their clusters over time? Or does the formation of an active
nucleus take longer in an isolated galaxy, and ignite at FR II
luminosity -- in which case we are only now seeing the birth of the
first FR IIs outside clusters, and none have had time to decay into FR Is?

 Clearly one simple model cannot tell the whole story, however; while
the presence of perpendicular polarization lays out very strong
evidence that scattering must be an important mechanism, other
contributions are not ruled out. On the contrary, jet-induced star
formation models can provide an explanation for the presence of the
dust clouds needed to cause scattering \cite{Cimatti+98a}. 
In some cases this is clearly not the case \cite{Tran+98a}, for example
when imaging polarimetry shows polarization increasing with distance from
the nucleus: clearly stellar dilution from the host galaxy is decreasing,
and there are no young blue stars in the extended region to compensate. The
Alignment Effect has been observed in some galaxies which are not
strongly polarised, and alternative models are needed to account for
such cases --- especially where other observed factors do not
correspond to those commonly observed in radio galaxies which fit the
canonical scattering model.

Longair, Best \& R\"{o}ttgering \scite{Longair+95a}, for example,
invoke three very different models to account for three 3CR galaxies
for which they obtained {\em Hubble Space Telescope}\, images. 3C 265
$(z=0.81)$ displays optical structure poorly aligned (25\degr\ offset)
with the radio structure, and at one one tenth of the scale. The
observed structure is most likely to be attributable to the
interactions of two or more galaxies, they suggest, with a modicum of
alignment possibly due to scattering or jet-induced star formation.

 Their second example, 3C 324 $(z=1.21)$, appears as a giant
elliptical in the $K$-band but exhibits a very knotted structure in
their {\em Hubble} (690\,nm and 783\,nm) images. The structure is
aligned with the axis along which relativistic material is believed to
flow, although the line linking the radio hotspots is offset from this
axis by about 30\degr. If the optical knots are associated with
companion galaxies, some theory is needed to account for their close
alignment with the relativistic jet; possibly that of West
\scite{West-94a} which postulates the formation of a prolate galaxy
and central black hole rotating about the axis of the large scale
matter distribution. [Such a theory predicts the presence of structure
up to Mpc scales, however, while 3C 324 shows no aligned galaxies
beyond 100 kpc; in general there is no evidence for the radio axis to 
be aligned with a particular axis of a triaxially symmetry galaxy
\cite{Sansom+87a}.] Otherwise, this source could be a classic case with
knots of scattered light explicable by the Unification Model, as argued by 
Cimatti et al.\ \scite{Cimatti+96a}.

 Finally, 3C 368 $(z=1.13)$ exhibits optical structure of the same
scale as the radio structure and might best be explained by
jet-induced star formation; star-forming regions on the radio axes
would be rich in the dust needed to account for its high optical
polarization, apparently scattered light from an AGN.
The wavelength dependence of 3C 368's polarization 
\cite{Cimatti+93a} rules out both electron scattering and synchrotron
radiation as possible mechanisms, leaving dust scattering as the most
likely hypothesis.

 Similarly, Pentericci et al.'s \scite{Pentericci+98a} study of very
high redshift radio galaxies $(2.6<z<3.2)$ finds that the galaxies fall
into several classes: those where the ultraviolet emission closely
traces the radio structure (akin to 3C 368 above); those where there is
a clear triangular emission region (presumably a cone of scattered
light); those where there is a clear radio/ultraviolet alignment effect
but no close relationship between the structures; and a couple of
pathological misaligned cases which may be peculiar for other reasons.

\subsection{Alternatives to radio galaxy unification}

 The Unification Model with scattered light causing perpendicular
polarization and the Alignment Effect is far from universally accepted
as the explanation of the various observed radio galaxy phenomena.
Simulations of galaxy formation based on hierarchical clustering models
\cite[note fig. 3]{Baron+87a} produce knotted structures in young
galaxies similar to that observed in radio galaxies. McCarthy \scite[\S
5.1]{McCarthy-93a} notes that galactic objects with the morphology
characteristic of the most luminous distant radio galaxies must undergo
substantial orbit mixing in $\sim 100$ Myr \cite{Daly-90a}, so any
viable model must explain the radio and optical structures as a
short-lived phenomenon within that timescale. Longair, Best \&
R\"{o}ttgering \scite{Longair+95a} point out that aligned structure cannot
be explained by scattering alone, as the aligned regions often do not
exhibit the conical shape which scattered light would illuminate
\cite{Ridgway+97a}. 

Synchrotron emission -- which is polarized, although not necessarily
with the orientation alignments commonly observed in radio galaxy
polarization -- is another candidate mechanism. This can often be ruled
out in individual cases, however, by showing that the
extrapolation of the radio emission at the measured spectral index
would not produce optical synchrotron emission at the luminosity
required \cite{Cimatti+94a}. Among
other models suggested we have the following --- all of which are
wanting, since none provide for an optical polarization mechanism:

 The difference in the Hubble diagrams for 3CR and weaker radio galaxy
populations must logically be ascribed to the presence of two or more
components to their infrared light, a stellar component and one related
to the active nucleus. One possibility for the nuclear-related
component considered by Eales et al.\ \scite{Eales+97a} is that
emission lines --- known to be directly correlated with radio luminosity
\cite{Willott+99a} --- are polluting the $K$-band light;
but infrared spectroscopy of $z \sim 1$ galaxies \cite{Rawlings+91a}
shows that emission lines do not contribute more than a quarter of the
total light intensity. Nevertheless, warm emission nebulae excited by
the active nucleus are known to contribute a significant fraction of
the observed ultraviolet light in some cases \cite{Dickson+95a}.

 Another alternative \cite{Eales+97a} allows that quasars form the
central engines of radio galaxies, and posits that the dust obscuring
the visible light from the quasar nuclei is not thick enough to obscure
the near infrared emission -- as has already been proposed in the
specific case of 3C 22 \cite{Rawlings+95a}. But if this mechanism were
widespread, it would tend to concentrate light in a single nucleus (3C
22 is pointlike, showing only small, faint optical and infrared extended
structure -- see Chapter \ref{resulch}) and would not provide any explanation for
the knotted structures, aligned or otherwise. Quasars which do not
exhibit the blazar phenomenon have low polarization, so if
orientation-based theories are incorrect, radio galaxies with
polarizations $p > 3\%$ ought to show signs of blazar activity, which
would be distinctive. [The presence of broad lines \cite{Alighieri+94a}
is sufficient to show that even if a blazar component is present, the
quasar component is dominant over it.]

 McCarthy \scite[\S 5.1]{McCarthy-93a} and Dunlop \& Peacock
\scite{Dunlop+93a} review models which attempt to explain the
Alignment Effect as the result of a zone of star formation triggered
by the passage of a radio source. If such models are correct then the
newly-formed stars must be younger than the radio source, which would
give them an age of only 10--100 Myr (although the generally accepted
ages of radio sources are not indisputable). But such young ages are
hard to reconcile with the colours observed by Dunlop \& Peacock
\scite{Dunlop+93a} and with the low scatter \cite{Lilly+84a},
indicative of a settled population, in the Hubble diagram. Since the
alignments between radio and optical structure are good only to $\sim
10\degr$, it is also difficult to explain how an expanding radio
source can cause star formation so far off-beam. Neither is there any
evidence of star formation in more than one or two examples of the
tens of low-redshift radio galaxies which have been studied in detail
now \cite{Alighieri+89a}. Best, Longair \& R\"{o}ttgering's
\scite{Best+96a} findings, however, can best be interpreted in terms
of such a model: as radio hotspots pass through the intergalactic
medium of their host, they trigger bursts of star formation (hence the
complex knots associated with small radio structures). Later, they
have travelled well outside the visible region of the host galaxy and
stellar formation activity ceases -- explaining why the larger radio
structures do not exhibit so many optical knots.

 Several other models have been proposed to explain the Alignment
Effect, but these again do not account for the observed polarization
orientations. Suggested mechanisms include two-component blazar models
\cite{Brindle+86a}, thermal plasma emission \cite{Daly-92a} and the
illumination pattern of a Doppler-beamed continuum as seen in blazars
\cite{Tran+98a}. The consequences of selection effects following from
increased luminosity of radio sources in the plane of a flattened disk
of gas have also been suggested \cite{Eales-92a}, but McCarthy
\scite[\S 5.1]{McCarthy-93a} suggests that the timescale for this would
be too long for the 100 Myr transient phenomenon of radio galaxies.

For the purposes of this thesis, we need only consider models which are
relevant to the interpretation of broadband infrared
aperture polarimetry. We will not, therefore, review further these
other models which need to be invoked to explain non-polarimetric
features of radio galaxies, but turn instead to the matter of the
statistical techniques applicable to aperture polarimetry.

\section{Mathematical Glossary}

\begin{description}

\item[$h_0$] The Hubble constant in units of 100 km\,s$^{-1}$\,Mpc$^{-1}$.

\item[$P_\nu$] The radiated power per unit bandwidth of a source as
measured at frequency $\nu$.

\item[$p$] The degree of linear polarization.

\item[$S_\nu$] The flux density of a source as measured at frequency
$\nu$; measured in jansky, such that 1 Jy = $10^{-26}$
W\,m$^{-2}$\,Hz$^{-1}$. \cite{Illingworth-94a}

\item[$\alpha_r$] The spectral index of a radio spectrum, such that $S
\propto \nu^{-\alpha_r}$.

\item[$\lambda_r$] The rest-frame wavelength of light emitted
by a distant galaxy.

\item[$\nu$] The frequency of (radio) emission.

\item[$\phi$] The orientation (relative to celestial North) of the 
{\boldmath $E$}-vector of linearly polarized radiation.

\end{description}

\chapter{The Measurement and Publication of Polarization}
\label{stoch}

 \begin{quote}
 Although you can format an equation almost any way you want with \LaTeX,
 you have to work harder to do it wrong.
\attrib{Leslie Lamport, 
 {\em The \LaTeX\ Reference Manual}}
 \end{quote}

 At the start of this research project, the available literature seemed to
provide no coherent and unified account of the best way to reduce, analyse
and present data on the polarization of astronomical sources. Accordingly
the best method as described here was submitted for publication in {\em
Experimental Astronomy} \cite{Leyshon-98a}. The following chapter is based
on that paper, updated in the light of the new book by Tinbergen
\scite{Tinbergen-96a} and the work of S\'{a}nchez
Almeida~\scite{Almeida-95a}, and of Maronna, 
Feinstein \& Clocchiatti~\scite{Maronna+92a}.  In
an age when theses are becoming increasingly available via the World-Wide
Web, it seems most useful to retain the format of a `how-to' manual for
this chapter, in the hope that it will prove useful and instructive to
polarimetrists of the 21st Century. 

\section{The task of the polarimetrist}

 When performing optical polarimetry of astronomical objects, we wish to
answer several distinct, but related, physical questions.

 Firstly, is the object polarized at all? Secondly, if it is, what is the best estimate of the polarization?
And thirdly, what confidence can we give to this measure of polarization? It is also necessary to be able to
test whether the polarization has changed from one epoch to another, or differs in neighbouring spectral
bands. 

 In addition to these physical questions is a presentational one: in what 
format should the results be published, to be of most 
utility to the scientific community?

 The questions of quantifying and presenting data on linear polarization
have been discussed at length by Simmons \&
Stewart~\scite{Simmons+85a}, who note that the traditional method used
by optical astronomers, that of Serkowski~\scite{Serkowski-58a}, does
not give the best estimate of the true polarization under most
circumstances. More recently, S\'{a}nchez Almeida~\scite{Almeida-95a},
Maronna, Feinstein \& Clocchiatti~\scite{Maronna+92a}
and Clarke with colleagues~\cite{Clarke+86a,Clarke+93a,Clarke+94a}
have developed the statistical basis of how noise affects measurements of
polarization. Using their recommendations, I present here a recipe for
reducing polarimetric data. 

\section{Paradigm}
\label{paradigm}

 In this chapter, I will not consider the origin of the polarization of 
light. It may arise from intrinsic polarization of the source, from 
interaction with the interstellar medium, or within Earth's atmosphere. 
Each of these sources represents a genuine polarization, which must be taken 
into account in explaining the measured polarization values. Some possible
sources of such systematic polarization are discussed by Hsu \& 
Breger~\scite{Hsu+82a}.

 Most modern optical polarimetry systems employ a two-channel system, 
normally a Wollaston prism. Such a prism splits the incoming light 
into two parallel beams (`channels') with orthogonal polarizations -- it 
functions as a pair of co-located linear analyzers. The transmission 
axes of the analyzers can be 
changed either by placing a half-wave plate before the prism in the 
optical path, and rotating this, or by rotating the actual Wollaston 
prism. Such a system is incapable of distinguishing circularly 
polarized light from unpolarized light, and references to `unpolarized' 
light in the remainder of this chapter strictly refer to light which is 
not linearly polarized; such light may be totally unpolarized (i.e. randomly 
polarized), or may include a circularly polarized component.

 Where a half-wave plate is used, an anticlockwise rotation $\chi$ of the 
waveplate results in an anticlockwise rotation of $\eta = 2\chi$ of the 
transmission axes. [For the theory of Wollaston prisms and wave plates, see, 
for instance, Chapter 8 in Hecht~\scite{Hecht-87a}; for a general 
survey of the theory and practice of astronomical polarimetry, see 
Tinbergen~\scite{Tinbergen-96a} or the briefer accounts by 
Kitchin~\scite[\S 5.2]{Kitchin-84a} or McLean~\scite[\S \S 
3.4, 4.5]{McLean-97a}.\,]

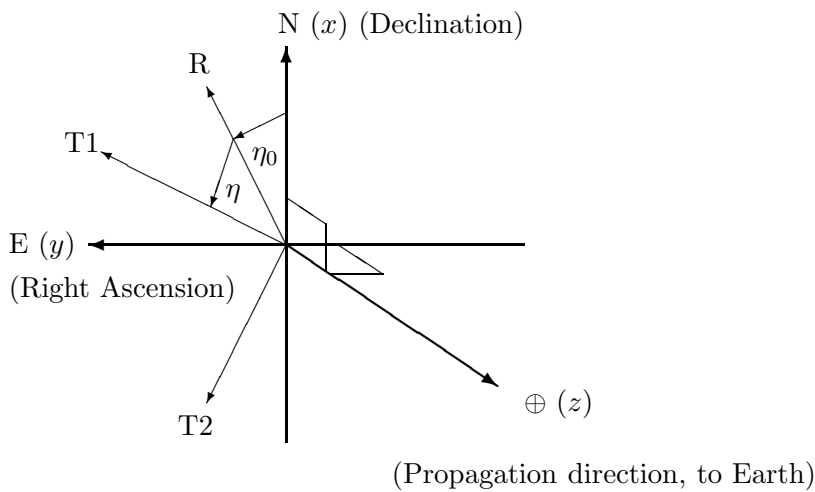
\begin{figure}
\begin{picture}(250,200)(-120,-100)
\thicklines
\put (0,-75){\vector(0,1){150}}
\put (-3,80){N $(x)$ (Declination)}
\put (90,0){\vector(-1,0){165}}
\put (-105,-3){E $(y)$}
\put (-105,-20){(Right Ascension)}
\put (0,0){\vector(3,-2){80}}
\put (90,-63){$\oplus\ (z)$}
\put (40,-90){(Propagation direction, to Earth)}
\thinlines
\put (20,0){\line(3,-2){16.64}}
\put (36.64,-11.09){\line(-1,0){20}}
\put (15,-10){\line(0,1){18.03}}
\put (15,8.03){\line(-3,2){15}}
\put (0,0){\vector(-1,2){30}}
\put (-37,65){R}
\put (-13,32){$\eta_0$}
\put (0,50){\vector(-2,-1){20}}
\put (-23,18){$\eta$}
\put (-20,40){\vector(-1,-3){8.57}}
\put (0,0){\vector(-2,1){70}}
\put (-84,35){T1}
\put (0,0){\vector(-1,-2){30}}
\put (-41,-72){T2}
\end{picture}
\caption{Reference axis, R, relative to celestial co-ordinates.}
\label{figeta}
\end{figure}

 We will suppose that Channel 1 of the 
detector has a transmission axis which can be rotated by some angle 
$\eta$ anticlockwise on the celestial sphere, relative to a reference 
position $\eta_0$ east of north. (See Figure~\ref{figeta}.)
The transmission axes T1, T2, of Channels 1 and 
2 are hence at $\eta_0 + \eta$ and $\eta_0 + 90\dg + \eta$ respectively.

 The reference angle $\eta_0$ will depend on the construction of the
polarizer, and will not, in general, be neatly due north. For mathematical
convenience in the rest of this chapter, we will take $\eta_0$ to define a
reference direction, `R', in our instrumental co-ordinate system and
relate all other angles to it. Such instrumental angles can then be mapped
on to the Celestial Sphere by the addition of $\eta_0$. 

Since the light emerging in the two 
beams has traversed identical paths until reaching the 
Wollaston prism, this method of polarimetry does not suffer from
the systematic errors due to sky fluctuation which affect single-channel 
polarimetry (where a single beam polarimeter alternately samples the 
two orthogonal polarizations).

 The two channels will each feed some sort of photometric array, e.g. a
{\sc ccd} or infrared array, which will record a photon count. Since such
images are often built up by a process of shifting the image position on
the array and combining the results, we will refer to a composite image
taken in one transmission axis orientation, $\eta$, as a {\em mosaic}. We
will denote the rate of arrival of photons recorded in Channel 1 and
Channel 2 by $n_{1}(\eta)$ and $n_{2}(\eta)$ respectively. From these
rates, we can calculate the total intensity ($I$) of the source, and the
difference ($S$) between the two channels: 

\begin{equation} 
I(\eta) = n_{1}(\eta) + n_{2}(\eta),
\label{Idef}
\end{equation}

\begin{equation} 
S(\eta) = n_{1}(\eta) - n_{2}(\eta).
\label{Sdef}
\end{equation}

We can also define a {\em normalized} difference:
\begin{equation}
\label{normdiff}
s(\eta) = \frac{S(\eta)}{I(\eta)}.
\end{equation}

The purpose of this chapter is to discuss how to interpret and present such 
data.

\section{Curve Fitting for $p$}

 Suppose we have a beam of light, which has a linearly polarized component
of intensity $I_p$, whose electric vector points at an angle $\phi$
anticlockwise of R. Its (linearly) unpolarized component 
is of intensity $I_u$. 
When such a beam enters our detector, we can use Malus' 
Law~\cite[\S8.2.1]{Hecht-87a} to deduce that
\[n_1(\eta) = \ha I_u + I_p.\cos^2(\phi - \eta)\]
and
\[n_2(\eta)  = \ha I_u + I_p.\sin^2(\phi - \eta),\]
from which we find
\begin{equation}
\label{Ipol}
I(\eta) = I_u + I_p,
\end{equation}
and, less trivially,
\begin{equation}
	S(\eta) = I_p.\cos[2(\phi - \eta)].
\label{Spol}
\end{equation}
The {\em degree of linear polarization}, $p$, is defined by
\begin{equation}
\label{ppol}
 p = \frac{I_p}{I_p + I_u}
\end{equation}
and so we can obtain the normalized difference by substituting Equations 
\ref{Ipol}, \ref{Spol} and \ref{ppol} into \ref{normdiff}:
 \[ s(\eta) = p.\cos[2(\phi - \eta)].\]

 Now, if observations have been made at a number of different angles,
$\eta_j$, of the transmission axis, then a series of values for $\eta_j$
and $s_j(\eta_j)$ will be known, and $p$ and $\phi$ may be determined by
fitting a sine curve to this data, weighted by errors $\sigma_{s_j}(\eta_j)$
as necessary. This method has been used, for example, by di Serego Alighieri
et al.~\scite[\S 2]{Alighieri+93a}. (Their refinement of the method 
allowed for 
the correction of the $s_j(\eta_j)$ for instrumental polarization at each 
$\eta_j$, which was necessary as they were rotating the entire camera, 
their system having no half-wave plate.)

We note that if there is any systematic bias of Channel 1 compared to 
Channel 2, this will show up as an $\eta$-independent ({\sc dc}) term 
added to the sinusoidal component when $s_j(\eta_j)$ is fitted to the data. 
Such bias could arise if an object appears close to the edge of the {\sc 
ccd} in one channel, for example.

\section{The Stokes Parameters}

\subsection{Basic definitions}

 Polarized light is normally quantified using Stokes' parameterisation. 
[For basic definitions see, for example, 
Clarke, in Gehrels (ed.)~\scite{Clarke-74a}.] Where interference properties
need not be treated, the intensity of polarized light can exhaustively be 
characterised by the four Stokes Parameters: one for the overall amplitude, 
two orthogonal linear components, and one circular component.

 Various conventions are known in the literature for the four 
Stokes Parameters; this thesis uses the most common, the $I,Q,U,V$ notation. 
The $V$ parameter will not be considered here, as it parameterises circular 
polarization, which a system involving only half-wave plates and linear 
analyzers cannot measure. The total intensity, $I$, of the light is an 
absolute Stokes Parameter. The other two parameters are defined relative 
to some reference axis, which in our case will be R, the $\eta_0$ 
direction. Thus we define:
\[Q = S\,(0\dg) = - S\,(90\dg),\] 
and
\[U = S\,(45\dg) = - S\,(135\dg).\]

{\em Normalized\,} Stokes Parameters are denoted by lower case letters 
($q$,$u$,$v$), and are found by dividing the raw parameters by $I$. We 
note that $S$ and the normalized $s$ can be thought of as a Stokes 
Parameter like $Q$ or $U$, generalised to an arbitrary angle -- and 
results which can be derived for $S$ (or $s$) will apply to $Q$ and $U$ 
(or $q$ and $u$) as special cases.

 If the Stokes Parameters are known, then the degree and angle of 
linear polarization can be found:
\begin{equation}
\label{defp}
 p = \sqrt{q^2 + u^2};
\end{equation}
\begin{equation}
\label{phidef}
\phi = \ha.\tan^{-1} \left( \frac{u}{q} \right) ,
\end{equation}
where the signs of $q$ and $u$ must be inspected to determine the correct
quadrant for the inverse tangent. Note that $S\,(\eta)$, $Q$ and $U$ {\em
must} be defined as above to be consistent with the choice of R as
Reference. 

 We must now distinguish between the true values of the Stokes Parameters 
for a source, and the values which we measure in the presence of noise. 
We will use the subscript $0$ to denote the underlying values, and 
the subscript $i$ for individual measured values.
 In the rest of this chapter, symbols such as $S_i$ and $\sigma_{S_i}$, 
where not followed by $(\eta)$, can be read as denoting `either $Q_i$ or 
$U_i$', `either $\sigma_{Q_i}$ or $\sigma_{U_i}$', etc.;
arithmetic means are 
denoted in the usual way, by an overbar, hence for $\nu_S$ measurements
$S_i$, $\bar{S}=\sum_{i=1}^{\nu_S} S_i / \nu_S$.

\subsection{The importance of Stokes Parameters}

 In particular, consider a source which is not polarized, so $q_0=u_0=0$, 
$p_0=0$, and $\phi_0$ is undefined. Since the $q_i$ and $u_i$ include noise, 
they will not, in general, be zero, and because of the quadrature form of Equation 
\ref{defp}, $p_i$ will be a definite-positive 
quantity. In short, $p_i$ is a {\em biased} estimator for $p_0$.

 There is no known {\em unbiased} estimator for $p_0$, and Simmons \&
Stewart~\scite{Simmons+85a} discuss at length the question of which
estimator should be used. They conclude that the Stokes Parameters
themselves are more useful than $p$ and $\phi$ in many applications. Since
$p$ inevitably suffers from bias while estimators of the Stokes Parameters
may, in principle, be unbiased, it is recommended that all published
polarimetric data should include the values of the normalized Stokes
Parameters. This would provide a standard format for further use by the
scientific community, whereas tabulated values of $p$ and $\phi$ would
always be sensitive to the debiasing scheme used to obtain them. 

 Given this preference for the Stokes Parameters it appears that one should
eschew the curve fitting method in favour of direct evaluation of the
parameters, at least when we only have data for the usual angles $\eta_j =
0\dg, 45\dg, 90\dg, 135\dg$. In practice, observers will take several
observations of an object at each transmission angle. This raises the
question of how best to combine all the measured values $q_i, u_i$ to
yield a single pair of `best estimators' for $q_0$ and $u_0$ -- a question
which is dealt with by Clarke et al.~\scite{Clarke+83b} and 
developed by Maronna, Feinstein \& Clocchiatti~\scite{Maronna+92a}.

A full discussion of the optimal method of estimating the true value of a 
normalised Stokes Parameter based on such individual measurements can be 
found in the
Appendix (sections \ref{optest} and \ref{optbin}). In accordance with the notation used there, 
$\tilde{s}=\bar{S}/\bar{I}$ represents the (approximate) optimal estimator of $s_0$. It is possible 
to determine $\bar{I}$ by pooling the $I_i$ values used to determine $Q$ and those used to determine $U$,
which would reduce the error on the mean; but to avoid systematic effects it is safer to calculate two 
separate $I$ values and produce $\tilde{q}=\bar{Q}/\bar{I_Q}, \tilde{u}=\bar{U}/\bar{I_U}$.

\section{Noise Affecting the Measurement of Stokes Parameters}
\label{realnoise}

The raw numbers which our photometric system produces will be a set of
count rates $n_{1i}(\eta)$ and $n_{2i}(\eta)$, together with their
errors, $\sigma_{n_{1i}}(\eta)$ and $\sigma_{n_{2i}}(\eta)$. These errors 
arise from three sources: photon counting noise; pixel-to-pixel variations in
the sky value superimposed on the target object;  and imperfect estimation
of the modal sky value to subtract from the image
\cite{Sterken+92a,IRAFphot}.

 The fundmental physical limitation on the measurement of any low
intensity of light is the quantum nature of light itself: low intensity
monochromatic light of frequency $f$ arrives in discrete photons of energy
$hf$. For a beam of light whose average intensity is $\cal R$ photons per
second, the probability of a given number of photons actually passing a
point in the beam during time $\tau$ is distributed according to a Poisson
distribution with mean ${\cal R}\tau$, and hence standard deviation
$\sqrt{{\cal R}\tau}$. Now ${\cal R}\tau$ will not necessarily be an integer,
but individual measurements must give integer results; and the fluctuation in the measured photon counts for
repeated integrations of time $\tau$ is termed {\em shot noise}. 
 
 As shown in Appendix \ref{photapp} (see Equation \ref{shotSN}), a photon counting system registers one
count for every $\Delta$ photons incident on it, and
the shot noise in one channel is related to the count rate as follows:
\begin{equation} \label{shotnSN}
\sigma_{\mathsf{shot}}^2 = \bar{n}_{\times i} /\tau \Delta.
\end{equation}

 The modal value of a sky pixel, $n_{\mathsf{sky}}$ can be found by 
considering, say, the pixel values in an annulus of dark sky around the 
object in question, an annulus which contains $\mD$ pixels 
altogether.
 The root-mean-square deviation of these pixels' values about the mode can 
also be found, and we will label this, $\sigma_{\mathsf{sky}}$. Hence we can
estimate the error on the mode, 
$\sigma_{\mathsf{sky}}/\sqrt{\mD}$.

 If we perform aperture-limited photometry on our target, with an 
aperture of area $\mA$ in pixels, we must subtract the modal sky 
level, $\mA.n_{\mathsf{sky}}$, which will 
introduce an 
error $\sigma_{\mathsf{skysub}} = 
\mA.\sigma_{\mathsf{sky}}/\sqrt{\mD}$.

 Each individual pixel in the aperture will be subject to a random sky 
fluctuation; adding these in quadrature for each of the $\mA$ 
pixels, we obtain an error $\sigma_{\mathsf{skyfluc}} =
\sqrt{\mA}.\sigma_{\mathsf{sky}}$.

 Ultimately, the error on the measured, normalized, intensity, is the sum 
in quadrature of the three quantities, $\sigma_{\mathsf{shot}}$, 
$\sigma_{\mathsf{skysub}}$, and $\sigma_{\mathsf{skyfluc}}$. If the areas 
of the aperture and annulus are comparable, then both the second and 
third terms will be significant; in practice, for long exposure times, 
the first (shot) noise term will be much smaller and can be neglected. 
This is important as, unlike the sky noise, the shot noise depends on the 
magnitude of the target object itself. If its contribution to the error 
terms is negligible, then sky-dominated error terms can be compared 
between objects of different brightness on the same frame.

\newtheorem{step}{Data Reduction Step}

\newtheorem{chek}[step]{Data Check}

\begin{chek}
\label{smallshot}

 For each object observed in each channel of each mosaic, the photometry
system will have produced a count rate $n_{\times i}$ with an error,
$\sigma_{n_{\times i}}$. For each such measurement, calculate
$\sqrt{n_{\times i}} / \Delta \tau$ and verify that it is much less than
$\sigma_{n_{\times i}}$. Then one can be certain that the noise terms are
dominated by sky noise rather than shot noise. 

\end{chek}

{\sl The treatment which follows in this chapter relies on sky noise being dominant. Other scenarios are
possible: in particular, an investigation into the case where shot noise is dominant is presented in
Appendix \ref{photapp}, and the effects of scintillation noise are considered by Clarke \& Stewart \scite[\S
3.2]{Clarke+86a}. These cases become more relevant for brighter sources but are not useful for the faint AGN
which form the subject of this thesis.}

\section{Testing for DC bias}

In practice, for each target object, we will have taken a number of 
mosaics at each angle $\eta_j$. We can immediately use each pair of 
intensities $n_{1i}, n_{2i}$ to find $I_{i}(\eta_j)$ 
and $S_{i}(\eta_j)$ using Equations \ref{Idef} and \ref{Sdef}. 

 Since the errors 
on the two channels are independent, we can trivially find the errors on 
both $I_{i}(\eta_j)$ and $S_{i}(\eta_j)$; the errors turn out to be 
identical, and are given by:
\begin{equation}
\label{erreq}
\sigma_{I_{i}} = \sigma_{S_{i}} = \sqrt{{\sigma_{n_{1i}}}^2 + 
{\sigma_{n_{2i}}}^2 }.
\end{equation}

\vspace{1ex}

\begin{chek} 
\label{cbias}
Take the mean value of all the $S_{i}(\eta_j)$ by summing over all the 
values $S_i$ at all angles $\eta_j$; and obtain an
error on this mean by combining in quadrature the error on each $S_i$. If
the mean value of $S_i(\eta_j)$, averaged over all the angles $\eta_j$, is
significantly greater than the propagated error, then there may be some
{\sc dc} bias.
\end{chek}

 Check \ref{cbias} uses $S_{i}(\eta_j)$ as a measure of excess intensity
in Channel 1 over Channel 2, and relies on the fact that there are 
similar numbers of observations at $\eta_j=\eta$ and 
$\eta_j=\eta+90\dg$ to average away effects due to polarization. If, as 
may happen in real data gathering exercises, there are not {\em 
identical} numbers of observations at $\eta_j=\eta$ and
$\eta_j=\eta+90\dg$, this could show up as apparent `{\sc dc} bias' in a 
highly polarized object. In practice, however, we are unlikely to 
encounter this combination of events; testing for bias by the above 
method will either reveal a bias much greater than the error (where the 
cause should be obvious when the original sky images are examined); or a 
bias consistent with the random sky noise, in which case we can assume 
that there is no significant bias. 

\section{Obtaining the Stokes Parameters}

 Once we are satisfied that our raw data are not biased, we can proceed.
At this stage in our data reduction, we will find it convenient to 
divide our set of $S_{i}(\eta_j)$ values, together with their associated 
$I_{i}(\eta_j)$ values, into the named Stokes Parameters,
\[Q_i = S_{i}(\eta_j=0\dg) = -S_{i}(\eta_j=90\dg)\]
and 
\[U_i = S_{i}(\eta_j=45\dg) = -S_{i}(\eta_j=135\dg).\]

\begin{step}
\label{getthei}

For each pair of data $n_{1i}(\eta_j), n_{2i}(\eta_j)$, produce the sum,
$I_{i}$, and the difference, $Q_{i}$ or $U_{i}$ as appropriate. 
Using Equation \ref{erreq}, produce the error common to the sum and
difference, $\sigma_{Q_i}$ or $\sigma_{U_i}$. Also find the normalized 
difference, $q_i$ or $u_i$.

\end{step} 

In practice, for a given target object, we will have taken a
small number of measurements of $Q_i$ and $U_i$ -- say $\nu_Q$ and 
$\nu_U$ respectively -- with individual
errors obtained for each measurement. If the errors on the individual
values are not comparable, but vary widely, we may need to consider taking a 
weighted mean.

\begin{chek} 
\label{maxbig} 
For a set of measurements of $(S_i,\sigma_{S_i})$, take all the measured 
errors,  $\sigma_{S_i}$; and so find the mean error (call this $\eps{phot}$) 
and the maximum deviation of any individual error
from $\eps{phot}$. If the maximum deviation is large 
compared to 
the actual error, consider whether you need to weight the data. 
\end{chek}

 If the deviations are large, we can weight each data point, $S_i$, by 
${\sigma_{S_i}}^{-2}$; but we will not pursue the subject of statistical 
tests on weighted means here. In practice, one normally finds that the
noise does not vary widely between measurements.

 We have already checked (see Check \ref{smallshot}) that the shot noise is 
negligible compared with the sky noise terms. Therefore, the main source 
of variation will be the sky noise. If the maximum deviation of the 
errors from $\eps{phot}$ is small, then we can infer that the 
fluctuation in the sky pixel values is similar in all the mosaics.

\begin{step}
\label{assumenorm}
In order to carry the statistical treatment further, we must assume 
that the sky noise is normally distributed. This is standard astronomical 
practice. 
\end{step}

\begin{step}
\label{getmean}

From the sample of Stokes Parameters $I_i$, $Q_i$ and $U_i$, obtained in 
Step \ref{getthei}, find 
the two means, $\bar{Q}$ and $\bar{U}$, with their corresponding intensities 
$\bar{I}_Q$ and $\bar{I}_U$; and find
the standard deviations of the two \bi{samples}, $\psi_Q$ and $\psi_U$.

\end{step}

\subsection{Photometric and statistical errors}

Since modern photometric systems can estimate the sky noise on each 
frame, we are faced throughout our data reduction sequence with a choice 
between two methods for handling errors. We can propagate the errors on 
individual measurements through our calculations; or we can use the 
standard deviation, $\psi_S$, of the set of sample values, $S_i$.

 In this chapter, I use the symbol $\sigma_{S_i}$ to denote the measured 
(sky-dominated) error on $S_i$, and $\sigma_{\bar{S}}$ for the standard 
error on the estimated mean, $\bar{S}$. The standard deviation of the 
population, which is the expected error on a single measurement $S_i$, 
could be denoted $\sigma_{S}$, but above I used $\eps{phot}$ to make its 
photometric derivation obvious.

 Using statistical estimators discards the data present in the 
photometric noise figures and uses only the spread in the data points to 
estimate the errors. We would expect the statistical estimator to be of 
similar magnitude to the photometric error in each case; and a cautious 
approach will embrace the greater of the two errors as the better error 
to quote in each case.

 Because we may be dealing with a small sample (size $\nu_S$) for some 
Stokes Parameter, $S$, the standard deviation 
of the sample, $\psi_S$, will not be the best estimator of the 
population standard deviation. The best estimator is~\cite[\S 10.5, for 
example]{Clarke+83a}:
\begin{equation}
\label{stateq}
\eps{stat} = \sqrt{\frac{\nu_S}{\nu_S - 1}}.\psi_S. 
\end{equation}

 In this special case of the \bi{population} standard deviation, I have used 
the notation $\eps{stat}$ for clarity. Conventionally, $s$ is used for the 
`best estimator' standard deviation, but this symbol is already in use here 
for a general normalized Stokes Parameter, so in this chapter I will use 
the variant form of sigma, $\varsigma$, for errors derived from the 
sample standard deviation, whence $\varsigma_S = \eps{stat}$, and the 
(statistical) standard error on the mean is
\[ \varsigma_{\bar{S}} = \frac{\psi_S}{\sqrt{\nu_S - 1}} = 
\frac{\eps{stat}}{\sqrt{\nu_S}}.\]

 The mean value of our Stokes Parameter, $\bar{S}$, is the best estimate 
of the true value $(S_0)$ regardless of the size of $\nu_S$. Given a 
choice of errors between $\sigma_{\bar{S}}$ and  $\varsigma_{\bar{S}}$, 
we will cautiously take the greater of the two to be the `best' error, 
which we shall denote $\sbb{\bar{S}}$.

\begin{chek}
\label{noiseOK}

We now have two ways of estimating the noise on a single 
measurement of a Stokes Parameter: 

\bllt $\eps{phot}$ is the 
mean sky noise level obtained from our photometry system: Check 
\ref{maxbig} obtains its value and verifies that the noise levels do not 
fluctuate greatly about this mean.

\bllt Statistical fluctuations in the 
actual values of the Stokes Parameter in question are quantified by 
$\eps{stat}$, obtained by applying Equation \ref{stateq} to the data 
from Step \ref{getmean}.

 We would expect the 
two noise figures to be comparable, and this can be checked in our data. 
We may also consider photometry of other objects on the same frame: 
Check \ref{smallshot} shows us that the errors are dominated by sky 
noise, and $\sigma_{\mathsf{sky}}$ should be comparable between objects, 
correcting for the different apertures used: 
\[ \sigma_{\mathsf{sky}} = \eps{X}/\sqrt{2\mA(1+\mA/\mD)}.\]

 We therefore take the best error, $\sbb{S}$, on a Stokes 
Parameter, $S$, to be the greater of $\eps{phot}$ and $\eps{stat}$.

\end{chek}

 If our data passes the above test, then we can be reasonably 
confident that the statistical tests we will outline in the next 
sections will not be invalidated by noise fluctuations.

\section{Testing for Polarization}

 The linear polarization of light can be thought of as a vector of length 
$p_0$ and
phase angle $\theta_0 = 2\phi_0$. There are two independent components to
the polarization. If either $Q_0$ or $U_0$ is non-zero, the light is said 
to be polarized. Conversely, if the light is to be described as 
unpolarized, both $Q_0$ and $U_0$ must be shown to be zero.

 The simplest way to test whether or not our target object emits polarized 
light is to test whether the measured Stokes Parameters, $\bar{Q}$ and 
$\bar{U}$, are consistent with zero. If either parameter is 
inconsistent with zero, then the source can be said to be polarized.

 To proceed, we must rely on our assumption (Step \ref{assumenorm}) that the 
sky-dominated noise
causes the raw Stokes Parameters, $Q_i, U_i,$ to be distributed normally. 
Then we can perform hypothesis testing~\cite[Chapters 12 and
16]{Clarke+83a} for the null hypotheses that $Q_0$ and $U_0$ are zero. 
Here, noting that the number of samples is typically small ($\nu_Q \simeq 
\nu_U < 30$) we face a choice: 

\vspace{1ex}

 \bllt {\sf Either:} assume that the sky fluctuations are normally 
distributed with standard deviation $\eps{phot}$, and perform 
hypothesis testing on the standard normal distribution with the statistic:
\[ z = \frac{\bar{S} - S_0}{\eps{phot}/\sqrt{\nu_S}}; \] 

 \bllt {\sf Or:} use the variation in the $S_i$ values to estimate the 
population standard deviation $\eps{stat}$, and perform 
hypothesis testing on the Student's $t$ distribution with $\nu_S - 1$ 
degrees of freedom, using the statistic:
\[ t = \frac{\bar{S} - S_0}{\eps{stat}/\sqrt{\nu_S}}.\] 

 In either case, we can perform the usual statistical test to determine
whether we can reject the null hypothesis that `$S_0 = 0$', at the
$C_S.100\%$ confidence level. The confidence intervals for retaining the
null hypothesis will be symmetrical, and will be of the forms $-z_0<z<z_0$
and $-t_0<t<t_0$.

 The values of $z_0$ and $t_0$ can be obtained from tables, and we define
$\cls{S}$ to be the greater of
$z_0.\eps{phot}/\sqrt{\nu_S}$ and
$t_0.\eps{stat}/\sqrt{\nu_S}$. Then the more conservative hypothesis test 
will reject that null hypothesis at the $C_S.100\%$ confidence level when
$\abs{\bar{S}}>\cls{S}$.

 In such a confidence test, the probability of making a `Type I Error',
i.e.\ of identifying an \bi{unpolarized} target as being polarized in {\em 
one} polarization sense, is simply $1-C_S$. The probability of correctly 
retaining the `unpolarized' hypothesis is $C_S$.

 The probability of making a `Type II Error'~\cite[\S 12.7]{Clarke+83a}
(i.e.\ not identifying a \bi{polarized} target as being polarized in one
polarization sense) is not trivial to calculate. 

Now because there are two independent senses of linear polarization, we 
must consider how to combine the results of tests on the two independent 
Stokes Parameters. Suppose we have a source which has no linear 
polarization. We test the two Stokes Parameters, $\bar{Q}$ and 
$\bar{U}$, for consistency with zero at confidence levels $C_Q$ and 
$C_U$ respectively. The combined probability of correctly retaining the 
null hypothesis for both channels is $C_Q.C_U$, and that of making the 
Type I Error of rejecting the null hypothesis in either or both channels 
is $1-C_Q.C_U$. Hence the overall confidence of the combined test is 
$C_Q.C_U.100\%$.

 Since the null hypothesis is that
$p_0=0$ and $\phi_0$ is undefined, there is no preferred direction in the 
null system, and therefore the confidence test should not prefer one 
channel over the other. Hence the test must always take place with 
$C_Q=C_U$.

 Even so, the test does not treat all angles equally; the 
probability of a Type II Error depends on the orientation of the 
polarization of the source. Clearly if its polarization is closely 
aligned with a transmission axis, there is a low chance of a polarization 
consistent with the null hypothesis being recorded on the aligned axis, 
but a much higher chance of this happening on the perpendicular axis. As 
the alignment worsens, changing $\phi_0$ while keeping $p_0$ constant, the 
probabilities for retaining the null hypothesis on the two measurement axes 
approach one another.

 Consider the case where we have taken equal numbers of measurements in
the two channels, so $\nu_Q = \nu_U = \nu$, and where the errors on the
measurements are all of order $\eps{phot}$. Hence we can calculate $z_0$
for the null hypothesis as above. Its value will be common to the $Q$ and
$U$ channels, as the noise level and the number of measurements are the
same in both channels. 

 Now suppose that the source has 
intensity $I_0$ and a true non-zero polarization $p_0$ oriented at position 
angle 
$\phi_0$. Then we can write $Q_0 = I_0 p_0 \cos(2\phi_0)$, and $U_0 = I_0 
p_0 \sin(2\phi_0)$. To generate a Type II error, a false null result 
must be recorded on both axes. The probability of a false null can be 
calculated for specified $p_0$ and $\phi_0$: defining $z_1 = 
\frac{I_0 p_0}{\eps{phot}/\sqrt{\nu}}$ then the probability of such a 
Type II error is
\begin{equation}
\label{IIprob}
P_{\rm II} = \frac{1}{2\pi}
\int_{x= z_1 \cos(2\phi_0) - z_0}^{x= z_1 \cos(2\phi_0) + z_0} 
\int_{y= z_1 \sin(2\phi_0) - z_0}^{y= z_1 \sin(2\phi_0) + z_0} 
\exp\left[ - \ha (x^2 + y^2) \right] dx\,dy.
\end{equation}
Clearly this probability is 
not independent of $\phi_0$.

\begin{step}
\label{findconfr}
Find the 90\%
confidence region limits, $\cln{Q}{90}$ and $\cln{U}{90}$, and inspect whether
$\abs{\bar{Q}}<\cln{Q}{90}$ and $\abs{\bar{U}}<\cln{U}{90}$.

\bllt If both Stokes 
Parameters fall within the limits, then the target is not shown to be 
polarized at 
the 81\% confidence level. In this case we can try to find polarization 
with some lower confidence, so repeat the test for 
$C_Q=C_U=85\%$. If 
the null hypothesis can be rejected in either channel, then 
we have a detection at the 72.25\% confidence level. There
is probably little merit in plumbing lower confidences than this.

\bllt If, however, polarization is detected in one or both of the Stokes
Parameters at the starting point of 90\%, test the polarized parameters to
see if the polarization remains at higher confidences, say 95\% and
97.5\%. The highest confidence with which we can reject the null
(unpolarized) hypothesis for either Stokes Parameter should be squared to
give the confidence with which we may claim to have detected an overall
polarization. 

\end{step}

 It is worth noting, {\em en passent}, that there is also a statistical test which is applicable to test
whether two polarization measurements taken at different epochs or in neighbouring spectral bands are likely
to indicate a common underlying value or not. Details of this, the Welch test, are given in the review by
Clarke \& Stewart \scite[\S 7]{Clarke+86a}. 

 In our hypothesis testing, we have made the {\em a priori}\ assumption 
that all targets are to be assumed unpolarized until proven otherwise. 
This is a useful question, as we must ask whether our data are worth 
processing further -- and we ask it using the raw Stokes 
Parameters, without resorting to complicated formulae. To publish useful 
results, however, we must produce the normalized Stokes Parameters, 
together with some sort of error estimate, and it is this matter which we 
will consider next.

\section{The Normalized Stokes Parameters}

 We have already derived an exact formula for the error on a normalised 
Stokes Parameter (Equation \ref{getsterr}).
 In order to simplify the calculation, we recall that in 
Check \ref{maxbig}, we 
checked that the errors on all the $S_i$ (and hence $I_i$) were 
similar. Thus the mean error on {\em one}\ rate in {\em one}\ channel is 
$\eps{phot}/\sqrt{2}$. Since the number of measurements 
made of $S$ is $\nu_S$, then
\[ \sigma_{\bar{n}_{1i}} \simeq \sigma_{\bar{n}_{2i}} \simeq 
\eps{phot}/\sqrt{2\nu_S}\]
and the error formula approximates to:
\begin{equation}
\label{normerr}
\sigma_{\tilde{s}} = 
\tilde{s}.\eps{phot}.\sqrt{(\bar{S}^{-2}+\bar{I}^{-2})/\nu_S}. 
\end{equation} 

In practice, we will be dealing with small polarizations, so $\bar{S} \ll 
\bar{I}$, and knowing $\tilde{s}$ from Equation \ref{stilde}, then Equation 
\ref{normerr} approximates to:
\begin{equation}
\label{simerrs}
 \sigma_{\tilde{s}} \simeq 
\frac{\tilde{s}.\eps{phot}}{\bar{S}.\sqrt{\nu_S}} = 
\frac{\eps{phot}}{\bar{I}.\sqrt{\nu_S}} 
\end{equation}

 As we had before with $\eps{stat}$ and $\eps{phot}$, so now we have a choice 
of using sky photometry or the statistics to estimate errors. The above 
method gives us the photometric error on a normalized Stokes' Parameter as 
$\vas{phot} = \eps{phot}/\bar{I} = \sigma_{\tilde{s}}.\sqrt{\nu_S}$; the 
statistical method would be to 
take the root-mean-square deviation of the measured $s_i$, obtained in 
Step \ref{getthei}, about Clarke et al.'s~\scite{Clarke+83b} best estimator 
value, $\tilde{s}$:
\begin{equation}
\label{Nstaterr}
\vas{stat} = \varsigma_{\tilde{s}}.\sqrt{\nu_S} =
 \frac{1}{\sqrt{\nu_S-1}}.\left[{\sum_{i=1}^{\nu_S}(s_i - 
\tilde{s})^2}\right]^{\ha} \end{equation}

\begin{step}
\label{hereNSPs}
Following the method outlined for finding $\tilde{s}$ and 
$\sigma_{\tilde{s}}$, apply Equations \ref{stilde} and \ref{simerrs} to 
the data obtained in Step \ref{getmean} to obtain $\tilde{q}$ with 
$\sigma_{\tilde{q}}$ 
and $\tilde{u}$ with $\sigma_{\tilde{u}}$. 
\end{step}

\begin{chek}
\label{stoeq}
Using $\tilde{q}$ and $\tilde{u}$, compute $\varsigma_{\tilde{q}}$ and 
$\varsigma_{\tilde{u}}$; find $\vas{stat}$ for both 
normalized Stokes Parameters, and compare it with $\vas{phot}$ in each case.
Verify also that the errors, $\vas{X}$, on the population standard 
deviations for the two Stokes Parameters are similar -- this should 
follow from the $S$-independence of Equation
\ref{simerrs} for small $\tilde{q}$ and $\tilde{u}$.
\end{chek}

So which error should one publish as the best estimate, 
$\sbb{\tilde{s}}$, on 
our final $\tilde{s}$ --- $\sigma_{\tilde{s}}$ or $\varsigma_{\tilde{s}}$\,? 
Again, a conservative 
approach would be to take the greater of the two in each case.

\begin{step}
\label{gotnorm}
Choose the more conservative error on each normalized Stokes Parameter, 
and record the results as $\tilde{q} \pm \sbb{\tilde{q}}$ and 
$\tilde{u} \pm \sbb{\tilde{u}}$. Record also the best population standard 
deviations, $\sbb{q}$ and $\sbb{u}$. 
\end{step}

\section{The Degree of Linear Polarization}

\subsection{The distribution of the normalised Stokes Parameters}

 Having obtained estimated values for $q$ and $u$, with conservative
errors, these values -- together with the reference angle $\eta_0$ -- can
and should be published as the most convenient form of data for colleagues
to work with. It is often desired, however, to express the polarization
not in terms of $q$ and $u$, but of $p$ and $\phi$. 

 Simmons \& Stewart \scite{Simmons+85a} discuss in detail the estimation 
of the degree of linear 
polarization. Their treatment makes a fundamental assumption that the {\em normalized}\ Stokes 
Parameters have a normal distribution \cite[\S4.2]{Clarke+86a}, and that 
the errors on $\tilde{q}$ and $\tilde{u}$ are similar. This latter 
condition is true for small polarizations (see Check 
\ref{stoeq}), but before we can proceed, we must test whether the 
former condition is satisfied. (Maronna, Feinstein \& 
Clocchiatti~\scite{Maronna+92a} outline cases where $\tilde{s}$
approximates to the normal distribution, but the criteria are vague: that $\nu_S$ and/or $I$ should be
`large'.)

If one assumes (Step \ref{assumenorm}) that $n_{1}$ and $n_{2}$ are 
normally distributed, 
one can construct, following Clarke et al.~\scite{Clarke+83b}, a joint 
distribution for $s$ whose parameters are 
the underlying {\em population}\ means $(n_{1_0}, n_{2_0})$ and standard 
deviations $(\sigma_1, \sigma_2)$ for the count rates $n_{1i}$ 
and $n_{2i}$. The algebra gets a little messy here, so we define three 
parameters, $A, B, C$:

\begin{equation}
\label{paralpha}
A = \ha \left[ \frac{1}{{\sigma_1}^2} + 
\frac{1}{{\sigma_2}^2} \left( \frac{1-s}{1+s} \right) ^2 \right],
\end{equation}

\begin{equation}
\label{parbeta}
B = \ha \left[ \frac{n_{1_0}}{{\sigma_1}^2} + 
\frac{n_{2_0}}{{\sigma_2}^2} \left( \frac{1-s}{1+s} \right) \right],
\end{equation}

\begin{equation}
\label{pargamma}
C = \ha \left[ \frac{{n_{1_0}}^2}{{\sigma_1}^2} + 
\frac{{n_{2_0}}^2}{{\sigma_2}^2} \right].
\end{equation}

 Using these three equations, we can write the probability distribution 
for $s$ as:
\begin{equation}
\label{sdist}
P(s) = \frac{B.\exp[\frac{B^2}{A} - 
C]}{\sigma_1.\sigma_2.\sqrt{\pi.A^3}.(1+s)^2}.
\end{equation}

This 
can be compared to the limiting case of the normal distribution whose mean 
$\tilde{s}_0$ and standard error $\sigma_0$ are obtained by 
propagating the underlying means $(n_{1_0}, n_{2_0})$ and standard 
deviations $(\sigma_1,\sigma_2)$ through Equations 
\ref{getstilde} and \ref{getsterr}:
\begin{equation} 
\label{snorm}
P_n(s) = 
\frac{\exp[\frac{-(s - 
\tilde{s}_0)^2}{2.{\sigma_0}^2}]}{\sigma_0.\sqrt{2\pi}}; 
\end{equation}

 We can derive an expression for the ratio $R(s) = P(s)/P_n(s)$, which 
should be close to unity if the normalized Stokes Parameter, $s$, is 
approximately normally distributed.

\begin{chek}
\label{nearnormal}

\bllt Estimate $n_{1_0}$ and $n_{2_0}$ using Equations \ref{Idef} and 
\ref{Sdef}, and the data from Step \ref{getmean}. Estimate $\sigma_1 
\simeq \sigma_2 \simeq \sbb{S}/\sqrt{2}$, where $\sbb{S}$ is obtained 
from Check \ref{noiseOK}.

\bllt Use the values of $\tilde{s}$ and $\sbb{s}$ obtained in 
Step \ref{gotnorm} as the best estimates of $\tilde{s}_0$ and 
$\sigma_{0}$.

\bllt Hence use a computer program to 
calculate and plot 
$R(s)$ in the domain $-3\sbb{s} < s < 
+3\sbb{s}$. If R(s) is close to unity throughout this domain, 
then we may treat the normalized Stokes Parameters as being normally 
distributed.
\end{chek} 

\subsection{Point estimation of $p$}

 If the data passes Checks \ref{stoeq} and \ref{nearnormal}, then we can 
follow 
the method of Simmons \& Stewart~\scite{Simmons+85a}. They `normalize' the
intensity-normalized Stokes Parameters, $q$ and $u$, by dividing them by 
their common population standard
deviation, $\sigma$. For clarity of notation, in a field where one 
can be discussing both probability and polarization, I will recast their
formulae, such that the {\em measured}\ degree of polarization, 
normalized as required, is here given in
the form $m = \tilde{p}/\sigma$; and the {\em actual}\ (underlying) degree of
polarization, also normalized, is
$a = p_0/\sigma$. It follows from the definition of $p$ (Equation 
\ref{defp}) that
\begin{equation}
\label{errp}
\sigma_p = \sqrt{\frac{q^2.{\sigma_q}^2 + u^2.{\sigma_u}^2}{q^2+u^2}}.
\end{equation}
If ${\sigma_q} = {\sigma_u} = \sigma$, then $\sigma_p = \sigma$.

 Now, Simmons \& Stewart~\scite{Simmons+85a} consider the case of a `single 
measurement' of each of $q$ and $u$, whereas we have found our best 
estimate of these parameters following the method of Clarke et 
al.\ \scite{Clarke+83b} However, we can consider the whole process 
described 
by Clarke et al.~\scite{Clarke+83b} as `a measurement', and so the 
treatment holds when applied to our best estimate of the normalized Stokes 
Parameters, together with the error on that estimate.

\begin{step}
\label{findperr}
Find $\sbb{p}$, and hence $\sigma=\sbb{p}$, by substituting our best 
estimates of $q$ and $u$ and their 
errors (Step \ref{gotnorm}) into Equation \ref{errp}. Hence calculate $m$:
\[ m = \sqrt{\tilde{q}^2+\tilde{u}^2}/\sigma.\]
\end{step}

 The probability distribution $F(m,a)$ of obtaining a measured value, 
$m$, for some underlying value, $a$, is given by
the Rice distribution~\cite{Simmons+85a,Wardle+74a}, which is cast in the 
current notation using the modified Bessel function, $I_0$~\cite[as defined in 
Ch.12, \S17]{Boas-83a}:
 \begin{equation} 
\label{rice}
 F(m,a) = m.\exp \left[ \frac{-(a^2+m^2)}{2} \right] .I_0(ma) \ldots 
(m\geq 0)
\end{equation}
\[F(m,a)=0 {\mathit{~otherwise}}.\]

 Simmons \& Stewart \scite{Simmons+85a} have tested various estimators 
$\ah{X}$ for bias. They find that when $\lapeq{a}{0.7}$, the best 
estimator is the `Maximum Likelihood Estimator', $\ah{ML}$, which 
maximises $F(m,a)$ with respect to $a$. So $\ah{ML}$ is the 
solution for $a$ of:
 \begin{equation}
\label{MLest}
a.I_0(ma) - m.I_1(ma) = 0.
\end{equation}
 
If $m<1.41$ then the solution of this equation is $\ah{ML} = 0$.
 
 When $\gapeq{a}{0.7}$, the best estimator is that traditionally used by
radio
as\-tron\-o\-mers, e.g.\ Wardle \& Kronberg~\scite{Wardle+74a}. In this case,
the best
estimator, $\ah{WK}$, is that which maximises $F(m,a)$ with respect to m, 
being the solution for $a$ of:
\begin{equation}
\label{WKest}
(1-m^2).I_0(ma) + ma.I_1(ma) = 0.
\end{equation}
If $m<1.00$ then the solution of this equation is $\ah{WK} = 0$.

 Simmons \& Stewart~\scite{Simmons+85a} graph $m(a)$ for both cases, and so 
show 
that $m$ is a monotonically increasing function of $a$, and that $\ah{ML} < 
\ah{WK} < m$\ $\forall m$. But which estimator should one use? Under their 
treatment, the selection of one of these estimators over the other depends 
on the underlying value of $a$; they point out 
that there may be good {\em a priori}\ reasons to assume greater or 
lesser polarizations depending upon the nature of the source.

 If we do not make any such assumptions, we can use monotonicity of $m$ and 
the inequality $\ah{ML} < \ah{WK}$\ $\forall m$, to find two limiting cases:

\bllt Let $\ml{WKmin}$ be the solution of the Wardle \& Kronberg Equation 
(\ref{WKest}) 
for $m$ with $a=0.6$. Hence if $m<\ml{WKmin}$, then $\ah{ML} < \ah{WK} 
< 0.7$ and the Maximum Likelihood estimator is certainly the most 
appropriate. 
Calculating, we find $\ml{WKmin} = 1.0982 \ll 1.41$ and so the Maximum 
Likelihood estimator will in fact be zero.

\bllt Let $\ml{MLmax}$ be the solution of Maximum Likelihood Equation 
(\ref{MLest}) for $m$ with $a=0.8$. We find $\ml{MLmax} = 1.5347$.
 Hence if $m>\ml{MLmax}$, then $0.7 < \ah{ML} < 
\ah{WK}$, and Wardle \& Kronberg's estimator will clearly be the 
most appropriate.

 Between these two extremes,
we have $\lapeq{\ah{ML}}{0}.\lapeq{7}{\ah{WK}}$.
This presents a problem, in that each estimator suggests 
that its estimate is more appropriate than that of the other estimator. 
If our measured value is $\ml{WKmin} < m < \ml{MLmax}$, what should we 
take as our best estimate? We could take the mean of the two estimators, 
but this would divide the codomain of $\hat{a}(m)$ into three discontinuous 
regions; there might be some possible polarization which this method 
could never predict! It would be better, then, to interpolate between the 
two extremes, such that in the range $\ml{WKmin} < m < \ml{MLmax}$,
\begin{equation}
\label{interpa}
\hat{a} = \frac{m-\ml{WKmin}}{\ml{MLmax}-\ml{WKmin}}.\ah{ML} +  
\frac{\ml{MLmax}-m}{\ml{MLmax}-\ml{WKmin}}.\ah{WK}.
\end{equation}

If we do not know, {\em a priori}, whether a source is likely to be 
unpolarized, polarized to less than 1\%, or with a greater polarization, 
then $\hat{a}$ would seem to be a reasonable estimator of the true 
noise-normalized polarization, and certainly better than the biased $m$.

\begin{step}
\label{esta}
Use the above criteria to find $\hat{a}$, and hence obtain the best 
estimate, $\hat{p} = \hat{a}.\sigma$, of 
the true polarization of the target.
\end{step}

\subsection{A confidence interval for $p$}

 As well as a point estimate for $p$, we would like error bars. The 
Rice distribution, Equation \ref{rice}, gives the probability of 
obtaining some $m$ given $a$, and can, therefore, be used to find a 
confidence 
interval for the likely values of $m$ given $a$. We can define two 
functions, $\mL{a}$ and $\mU{a}$, which give the 
lower and upper confidence limits for $m$, with some confidence $C_p$; 
integrating the Rice distribution, these will satisfy:
\begin{equation}
\label{Ldef}
\int_{m=-\infty}^{m=\mL{a}} F(m,a).dm = p_1
\end{equation}
and
\begin{equation}
\label{Udef}
\int_{m=\mU{a}}^{m=+\infty} F(m,a).dm = p_2
\end{equation}
such that 
\begin{equation}
\label{addprobs}
1 - C_p = p_1 + p_2.
\end{equation}

 Such confidence intervals are non-unique, and we need to impose an 
additional constraint. We could require that the tails outside the 
confidence region be equal, $p_1 = p_2$, but following Simmons \& 
Stewart~\scite{Simmons+85a}, we shall require that the confidence interval have 
the smallest possible width, in which case our additional constraint is:
\begin{equation}
\label{aconst}
F[\mU{a},a] = F[\mL{a},a].
\end{equation}

\begin{figure}
\psfig{file=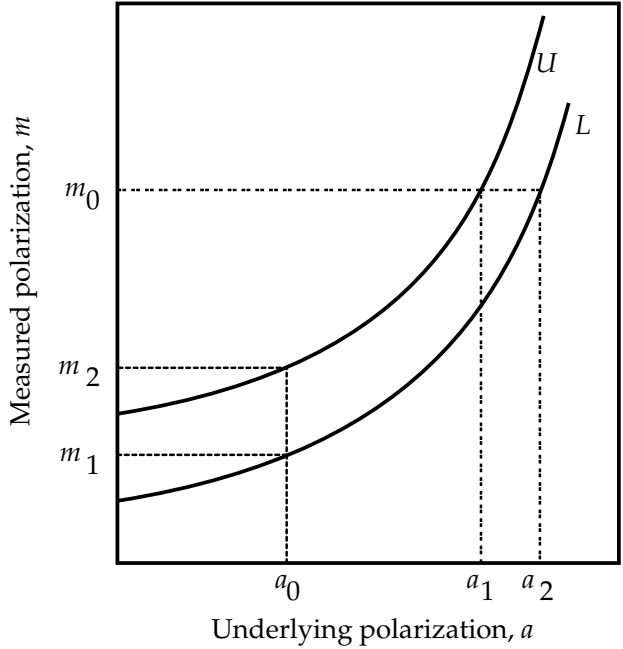,width=85mm} 
\caption{Confidence Intervals based on the Rice Distribution.}
Figure adapted from Leyshon \& Eales~\protect\scite{Leyshon+98a}.
\label{ricefig} 
\end{figure}

 From the form of the Rice distribution, $\mL{a}$ 
and $\mU{a}$ will be monotonically increasing functions of $a$, as shown 
in Figure \ref{ricefig}. Given a particular underlying polarization 
$a_0$, the $C_p$ confidence interval $(m_1,m_2)$ can be 
obtained by numerically solving Equations \ref{Ldef} thru \ref{aconst} to 
yield $m_1 = \mL{a_0}$ and $m_2 = \mU{a_0}$.

 Now, it can be shown~\cite[Ch.\
VIII, \S 4.2]{Mood+74a} that the process can also be inverted, i.e. if we 
have obtained some measured 
value $m_0$, then solving for $m_0 = \mU{a_1} = \mL{a_2}$ will yield a
confidence interval $(a_1,a_2)$, such that the confidence of $a$ lying 
within this interval is $C_p$.

 Since the contours for $\mU{a}$ and $\mL{a}$ cut the $m$-axis at 
non-zero values of $m$, we must
distinguish three cases, depending on whether or not $m_0$ lies above one or 
both of the intercepts. The values of $\mL{0}$ and $\mU{0}$ depend only 
on the confidence interval chosen; substituting $a = 0$ into Equations 
\ref{Ldef} thru \ref{aconst} results in the pair of equations
\begin{equation}
\label{cforz}
C_m = \exp \left[ -\frac{\mL{0}^2}{2} \right] - 
      \exp \left[ -\frac{\mU{0}^2}{2} \right]
\end{equation}
and
\begin{equation}
\label{transce}
 \mL{0}.\exp \left[ -\frac{\mL{0}^2}{2} \right] =
 \mU{0}.\exp \left[ -\frac{\mU{0}^2}{2} \right].
\end{equation}

A numerical solution of this pair of equations can be found for any given 
confidence interval, $C_m$; we find that, in 67\% 
$(1\sigma)$ interval, $\mL{0} = 0.4438,\: \mU{0} = 1.6968$, while in a 
95\% $(2\sigma)$ interval, $\mL{0} = 0.1094,\: \mU{0} = 2.5048$.
Hence, knowing $m_0$, and having chosen our desired confidence level, we 
can determine the interval $(a_1,a_2)$ by the following criteria:

\bllt {\boldmath $m_0 \geq \mU{0}$} \newline
\hspace*{1.2cm}\dotfill\ There are non-zero solutions for both 
$\mU{a_1}$ and $\mL{a_2}$.

\bllt {\boldmath $\mL{0} < m_0 < \mU{0}$} \newline
\hspace*{1.2cm}\dotfill\ In this case, $a_1=0$, and we must solve $m_0 = 
\mL{a_2}$.

\bllt {\boldmath $m_0 \leq \mL{0}$} \newline
\hspace*{1.2cm}\dotfill\ Here, $a_1=a_2=0$.

 Simmons \& Stewart~\scite{Simmons+85a} note that the third case is formally a
confidence interval of zero width, and suggest that this is
counter-intuitive; and they go on to suggest an {\em ad hoc}\ method of
obtaining a non-zero interval. However, it is
perfectly reasonable to find a finite probability that the degree of
polarization is identically zero: the source may, after all, be 
unpolarized. This can be used as the basis of estimating the probability 
that there is a non-zero underlying polarization, as will be shown in the 
next section.

\begin{step}
\label{getint}
Knowing $m$ from Step \ref{findperr},
 find the limits $(a_1,a_2)$ appropriate to confidence intervals of 67\% and
95\%. Hence, multiplying by $\sigma$, find the confidence 
intervals on the estimated degree of polarization. The 67\% limits may be 
quoted as the `error' on the best estimate.
\end{step}

\subsection{The probability of there being polarization}

Consider the contour $m=\mU{a}$ on Figure \ref{ricefig}. As defined by 
Equation \ref{Udef} and the inversion of Mood et 
al.~\scite{Mood+74a}, it divides the domain into two regions, such 
that there is a probability $p_2$ of the underlying polarization being 
greater than $a={\mathcal U}^{-1}(m_0)$. There is clearly a limiting case 
where the contour cuts the $m$-axis at $m_0$, hence dividing the domain 
into the polarized region $a>0$ with probability $p_P$, and the 
unpolarized region with probability $1-p_P$.

 Now we may substitute the Rice Distribution, Equation \ref{rice}, into 
Equation \ref{Udef} and evaluate it analytically for the limiting case, 
$a=0$:
\begin{equation}
\label{propol}
p_P = 1 - \exp(-{m_0}^2/2).
\end{equation}
  Equation \ref{propol} hence yields the probability that a measured
source actually has an underlying polarization.

\begin{step}
\label{estpolun}
Substitute $m$ from Step \ref{findperr} into Equation \ref{propol}. Hence 
quote the probability that the observed source is truly polarized.
\end{step}

 A more powerful method, applicable to cases where $\nu_Q \neq \nu_U$ and $\sigma_Q \neq \sigma_U$, is given
by Clarke \& Stewart~\scite[\S6.2]{Clarke+86a}: they define and tabulate values for a statistic
$Z_{\alpha-1}$ such that the $(\alpha-1).100\%$ confidence interval for $p$ is an ellipse in the $q,u$ plane
centred on $(\bar{q},\bar{u})$ and with semi-axes given by $\sqrt{{\varsigma_q}^2.Z_{\alpha-1}}$ and
$\sqrt{{\varsigma_u}^2.Z_{\alpha-1}}$. Values are only tabulated, however, for certain $\nu_S$, all
multiples of 5 or 10. In the current notation, the statistic is \begin{equation} \label{zstatdef}
Z_{\alpha-1} = \frac{(\bar{q}-q_0)^2}{{\varsigma_q}^2} + \frac{(\bar{u}-q_0)^2}{{\varsigma_u}^2}.
\end{equation}
 
\section{The Polarization Axis}

 It remains to determine the axis of polarization, for which an unbiased
estimate is given by Equation \ref{phidef}. Once again, we
have a choice of using the statistical or photometric errors --- and,
indeed, a choice of raw or normalized Stokes Parameters. Since 
\begin{equation}
\label{redf}
2\phi = \theta = \tan^{-1}(u/q) = \tan^{-1}(r),
\end{equation}
 our first problem is to obtain the 
best figure for $r = u/q$.

 Now, as we saw in our discussion of the best normalized Stokes 
Parameter, it is better to ratio a pair of means than to take the mean 
of a set of ratios. We could take
$r=\bar{U}/\bar{Q}$, but for a very small sample, there is the danger that 
the mean intensity of the $Q$ observations will differ from that of the 
$U$ values. Therefore, we should use the normalized Stokes Parameters, 
and the least error prone estimate of the required ratio will be 
$\tilde{r}=\tilde{u}/\tilde{q}$, yielding $\tilde{\phi}$.

 Knowing the errors on $\tilde{q}$ and $\tilde{u}$, we can find the 
propagated error in $\tilde{r}$:
\begin{equation}
\label{getrerr}
\sigma_{\tilde{r}} = 
\tilde{r}.\sqrt{\left(\frac{\tilde{q}}{\sigma_{\tilde{q}}}\right)^2 + 
\left(\frac{\tilde{u}}{\sigma_{\tilde{u}}}\right)^2};
\end{equation}
given the non-linear nature of the tan function, the error on 
$\tilde{\phi}$ should be found by separately calculating $\sigma_+ = \ha 
\tan^{-1}(\tilde{r}+\sigma_{\tilde{r}}) - \tilde{\phi}$ and $\sigma_- = \ha
\tan^{-1}(\tilde{r}-\sigma_{\tilde{r}}) - \tilde{\phi}$. Careful 
attention must be paid in the case where the error takes the phase 
angle across the boundary between the first and fourth quadrants, as the 
addition of $\pm \pi$ to the inverse tangent may be necessary to yield a 
sensible error in the phase angle.

\begin{step}
\label{propphi}
Obtain $\tilde{\phi}$, the best estimate of $\phi$, and the propagated 
error on it, $\sigma_{\tilde{\phi}} = \ha (|\sigma_+| + 
|\sigma_-|)$, using Equations \ref{redf} 
and \ref{getrerr}. Add $\eta_0$ to $\tilde{\phi}$ and hence quote the best 
estimate of the polarization orientation in true celestial co-ordinates.
\end{step}

 For the statistical error, we note that the probability distribution 
of observed {\em phase}\ angles, $\theta=2\phi$, calculated by 
Vinokur~\scite{Vinokur-65a}, and quoted elsewhere \cite{Wardle+74a,Clarke+86a,NaghizadehKhouei+93a}, is:
\[P(\theta) = \exp \left[ -\frac{a^2 \sin^2(\theta-\theta_0)}{2} \right] 
.\]
 \begin{equation}
\label{thetadis}
\left\{ \frac{1}{2\pi} \exp \left[ - \frac{a^2 \cos^2(\theta-\theta_0)}{2} 
\right] + \frac{a \cos(\theta-\theta_0)}{\sqrt{2\pi}}.\left\{ \ha + f[a 
\cos(\theta-\theta_0)] \right\} \right\}
\end{equation}
where
\begin{equation}
\label{thetasup}
f(x) = \frac{{\mathrm sign}(x)}{\sqrt{2\pi}} \int_0^x 
\exp \left(- \frac{z^2}{2} \right) \,dz 
={\mathrm sign}(x).{\mathrm erf}(x)/\sqrt{8}, \end{equation}
and ${\mathrm erf}(x)$ is the error function as defined by 
Boas~\scite[Ch.11, \S9]{Boas-83a}. We do not 
know $a=p_0/\sigma$, and will have to use our best estimate, $\hat{a}$, 
as obtained from Step \ref{esta}.
The $C_\phi.100\%$ confidence interval on the measured angle, 
$(\theta_1,\theta_2)$, is given by numerically solving
\begin{equation}
\label{thetaerr}
\int_{\theta_1}^{\theta_2}  P(\theta).d\theta = C_\phi;
\end{equation}
in this case we choose the symmetric interval, $\theta_2-\tilde{\theta} = 
\tilde{\theta}-\theta_1$.

\begin{step}
\label{findangle}
Obtain the limiting values of $\phi=\theta/2$ for confidence intervals of 
67\% $(1\sigma)$ and 95\% $(2\sigma)$. Quote the 67\% limits as 
$\varsigma_{\tilde{\phi}} = (\phi_2-\phi_1)/2$. Choose the more 
conservative error from $\varsigma_{\tilde{\phi}}$ and 
$\sigma_{\tilde{\phi}}$ as the best error, $\sbb{\tilde{\phi}}$.
\end{step}

\section{Comparison with Other Common Techniques}

 It may be instructive to note how the process of reducing polarimetric
data outlined in this chapter compares with the methods commonly used in the
existing literature. The paper by Simmons \& Stewart~\scite{Simmons+85a}
gives a thorough review of five possible point estimators for the degree
of polarisation. One of these methods is the trivial $m$ as an estimator 
of $a$. The other four methods all involve the calculation of thresholds 
$\mh{X}$: if $m < \mh{X}$ then $\ah{X}=0$. These four methods are 
the following:
\begin{enumerate}

\item Maximum Likelihood: as defined above, $\ah{ML}$ is the value of $a$
which maximises $F(m,a)$ with respect to $a$. Hence $\ah{ML}$ is the
solution for $a$ of Equation \ref{MLest}. The limit $\mh{ML}=1.41$ is 
found by a numerical method.

\item Median: $\ah{med}$ fixes the distribution of possible measured 
values such that the actual measured value is the {\em median}, hence 
$\int_{m'=0}^{m'=m} F(m',\ah{med}).dm' = 0.5$. The threshold is $\mh{med} 
= 1.18$, being the solution of $\int_{m'=0}^{m'=\mh{med}} F(m',0).dm'=0.5$.

\item Serkowski's estimator: $\ah{Serk}$ fixes the distribution of 
possible measured values such that the actual measured value is the {\em 
mean}, hence $\int_{m'=0}^{m'=\infty} m'.F(m',\ah{Serk}).dm' = m$. The 
threshold is $\mh{Serk} = 1.25 = \int_{m'=0}^{m'=\infty} m'.F(m',0).dm'$.

 \item Wardle \& Kronberg's method: as defined above, the estimator, 
$\ah{WK}$, is that which maximises $F(m,a)$ with respect to $m$ (see 
Equation \ref{WKest}), and $\mh{WK}=1.00$.

\end{enumerate}
Simmons \& Stewart~\scite{Simmons+85a} note that although widely used in the
optical astronomy literature, Serkowski's estimator is not the best for
either high or low polarizations; they find that the Wardle \& Kronberg
method commonly used by radio astronomers is best when $\gapeq{a}{0.7}$,
i.e.~when the underlying polarization is high and/or the measurement noise
is very low. The Maximum Likelihood method, superior when $\lapeq{a}{0.7}$
(i.e.~in `difficult' conditions of low polarization and/or high noise),
appears to be unknown in the earlier literature. [It seems to have been 
used independently shortly after
Simmons \& Stewart~\scite{Simmons+85a} in Appendix B of the paper by Killeen, Bicknell \&
Ekers~\scite{Killeen+86a}.]

 In this chapter, I have merely provided an interpolation scheme between the
point estimators which they have shown to be appropriate to the `easy' and
`difficult' measurement regimes. The construction of a confidence interval
to estimate the error is actually independent of
the choice of point estimator, although (as mentioned above) I believe 
that Simmons \& Stewart's \scite[\S 3]{Simmons+85a} unwillingness to `accept 
sets of zero interval as confidence intervals' is unfounded, since 
physical intuition allows for the possibility of truly unpolarised 
sources (i.e. with identically zero polarizations), and their arbitrary 
method of avoiding zero-width intervals can be dispensed with.

\section{Zero-Polarization Objects and a Residual Method}
\label{zpos}

 The data reduction process presented above has made no {\em a priori}
assumptions about whether the target object has a high or low polarization,
and is even general enough to cope with different numbers of observations of
the $q$ and $u$ Stokes parameters if difficult observing conditions limit
the data in this way. 

 Clarke et al.~\scite{Clarke+93a} suggest a method which can be
used to test whether the underlying polarization of a low polarization
object is actually zero. For a zero polarization $(a=0)$ object, the Rice 
distribution simplifies to the Rayleigh distribution:
 \begin{equation} 
\label{rayleigh}
 F(m,0) = m.\exp \left[ \frac{-m^2}{2} \right] \ldots 
(m\geq 0)
\end{equation}
\[F(m,0)=0 {\mathit{~otherwise}}.\]

 We can use the Rayleigh distribution to calculate the cumulative 
distribution function $\cdf{p}{m}$ for the probability of obtaining a 
measurement $0<m_{i}<m$, and compare this to the actual fraction of 
measurements which lie between 0 and $m$ -- the `empirical cumulative 
distribution', $\ecd{p}{m}$.

 Integrating the Rayleigh distribution, we find
\begin{equation}
\label{p-cdf}
\cdf{p}{m} = 1 - \exp \left[ \frac{-m^2}{2} \right].
\end{equation}
This equation gives us the probability that an {\em unpolarised} object 
might give a polarization measurement of $m$ or less, and is identical to 
Equation \ref{propol} for the probability that an object yielding a 
measurement $m_0$ is actually unpolarised. (This follows from the 
inversion argument illustrated in Figure \ref{ricefig}.)

To obtain the `empirical cumulative distribution', we must obtain and sort
a set of $m_i$ based on pairs of individual measurements $q_i$, $u_i$. To
calculate the $m_i$ we also need the relevant standard deviations. Now in
theory, as long as the noise level is constant (not necessarily true if
observations are pooled from different instruments) there should be a pair
of population standard deviations $\sigma_{q}$, $\sigma_{u}$, which
characterise the errors on any individual measurements of these normalised
Stokes parameters. But Clarke et al.~\scite{Clarke+93a} point out
that the true errors are not known and must be estimated. 

 We estimated $\hat{\sigma}_{q}$ and $\hat{\sigma}_{u}$ at Step
\ref{gotnorm} based on the whole sample; these figures should be good
estimates of the true value. Array-based photometry is also capable of
giving an error $\sigma_{s_{i}}$ for each individual measurement: if Data
Check \ref{maxbig} verified that individual errors did not vary widely
from the mean error, we can take $\sigma_{s_{i}} \simeq \hat{\sigma}_{s}$,
but if there is wide variation, then the individual errors should be 
used, being calculated in analogy with Equation \ref{getsterr} as:
\begin{equation}
\label{nspi-err}
 \sigma_{s_i} =
\frac{1}{n_{1}+n_{2}}.\sqrt{[(1-s_{i})\sigma_{{n}_{1}}]^2
+ [(1+s_{i})\sigma_{{n}_{2}}]^2}.
\end{equation}

 Ultimately, the $m_i$ can be calculated:
\begin{equation}
\label{m-i}
 m_{i} = \sqrt{ \left( \frac{q_i}{\sigma_{q_i}} \right)^{2} +
		\left( \frac{u_i}{\sigma_{u_i}} \right)^{2} }.
\end{equation}

 A similar exercise can be conducted for the direction of polarization. 
An unpolarised object subject to random noise should not display any 
preferred direction of polarization, so the probability distribution 
function for the measured angle will be uniform between $\phi=0$ and 
$\phi=\pi$; the cumulative distribution function for 
angles will be $\cdf{\theta}{\phi} = \phi/\pi$.

 Again the measured Stokes parameters must be paired, to give the
individual position angles $\phi_{i} = \tan^{-1}(u_{i}/q_{i})\div 2$
(compare Equation \ref{redf}). In this case it is not necessary to
consider the errors;  the empirical cumulative distribution
$\ecd{\theta}{\phi}$ is simply the fraction of $\phi_{i}$ values in the
range $0<\phi_{i}<\phi$. 

 Clarke et al.~\scite{Clarke+93a} explain how the 
Kolmogorov(-Smirnov) test can be used to compare the theoretical and 
empirical distributions, and discuss systematic effects which might cause 
the empirical distributions to deviate from those expected for an 
unpolarised source.

Clarke \& Naghizadeh-Khouei~\scite{Clarke+94a} point out that if a 
good estimate $\tilde{s}$ is available for a normalized Stokes 
parameter of a polarized source, then the residuals $\breve{s}_{i} = 
s_{i} - \tilde{s}$ should behave in the same way as the measured 
polarization of an unpolarised source. It would be possible, therefore, 
to proceed from Step \ref{gotnorm} for an equal number, $\nu$, of 
measurements of the two Stokes parameters, as follows:

 The normalised Stokes residuals $\breve{q}_{i}$, $\breve{u}_{i}$, may be
calculated, and may be treated in the same way as the Stokes parameters of
an unpolarised object; the empirical cumulative distributions of the
residuals $\ecd{p}{\breve{m}}$ and $\ecd{\theta}{\breve{\phi}}$ can be
obtained and tested for goodness-of-fit to the theoretical distributions
for an unpolarised object. It might also be possible to iteratively 
refine the values of $\tilde{q}$ and $\tilde{u}$ to improve the fit.

\section{Conclusion}

 The reduction of polarimetric data can seem a daunting task to the
neophyte in the field. In this chapter, I have attempted to bring together
in one place the many recommendations made for the reduction and 
presentation of polarimetry, especially those of Simmons \&
Stewart~\scite{Simmons+85a}, and of Clarke et al.~\scite{Clarke+83b}. 
In addition, I have suggested that it is possible to develop the
statistical technique used by Simmons \& Stewart~\scite{Simmons+85a} to
obtain a simple probability that a measured object has non-zero underlying
polarization.
 I have also suggested that there is a form of estimator for 
the overall degree of linear
polarization which is more generally applicable than either the Maximum
Likelihood or the Wardle \& Kronberg~\scite{Wardle+74a} estimators
traditionally used, and which is especially relevant in cases where the 
measured
data include degrees of polarization of order 0.7 times the estimated
error.

 Modern computer systems can estimate the noise on each individual
mosaic of a sequence of images; this is useful information, and is not to
be discarded in favour of a crude statistical analysis. A recurring theme
in this chapter has been the comparison of the errors estimated from
propagating the known sky noise, and from applying sampling theory to the
measured intensities. Bearing this in mind,
I have presented here a process for data reduction in
the form of \ref{findangle} rigorous steps and checks. The recipe might be 
used
as the basis of an automated data reduction process, and I hope that it
will be of particular use to the researcher -- automated or 
otherwise -- who is attempting polarimetry for the first time.

\newpage

\section{Mathematical Glossary}

Since this chapter uses a lot of mathematical terms common with Appendix
\ref{photapp}, and a few which differ in definition, I have given both this
chapter and that appendix a mathematical glossary defining the terms   
used. Latin symbols are listed in alphabetical order first, followed by
Greek terms according to the Greek alphabet -- except that terms of the
form $\sigma_\aleph$ are listed under the entry for $\aleph$.

\begin{description}

\item[$a$] The true error-normalised polarization, $a = p_0/\sigma$.
\begin{itemize}
\item[$\ah{X}$] A generic estimator of $a$.
\item[$a_0$] A specified value of $a$ used to estimate a confidence region for $m_1, m_2$.
\item[$a_1, a_2$] A confidence region corresponding to a measured polarization $m_0$.
\end{itemize}

\item[${\cal A}$] The number of pixels forming the aperture within which a source intensity is measured.

\item[ADU] Analogue-to-Digital Unit: another name for DN ({\em q.v.}).

\item[$A, B, C$] Variables used to parameterise the complicated 
expression for $P(s)$ ({\em q.v.}).

\item[$C_S$] Generic symbol for $C_Q$ and $C_U$, respectively the confidence levels 
for rejecting null hypothesis when testing $\bar{Q}$ and $\bar{U}$ for consistency with zero.

\item[$C_m, C_p$] The confidence specified for normalised polarization $m$ to lie within a given interval.

\item[$C_{\phi}$] The confidence specified for polarization axis orientation $\phi$ to lie within a given interval.

\item[${\cal D}$] The number of pixels forming the annulus surrounding a source,
 within which the dark sky intensity is measured.

\item[DN] Data Number: a photon counting system returns a count of 1 DN for every $\Delta$ 
photons incident. 

\item[$f$] The frequency of a beam of quasimonochromatic light.

\item[$F(m,a)$] The Rice distribution.

\item[$I_0(x), I_1(x)$] The modified Bessel Functions.

\item[$I$] The intensity of a source in DN counts per second, $I=n_1 + n_2$. It takes 
the same annotations as $S$ ({\em q.v.}), and also:
\begin{itemize}
\item[$I_u, I_p$] The linearly unpolarised and polarised components of a partially polarised beam of light.
\item[$I_Q, I_U$] Two estimates of $I$ made by taking separately the means of those $I_i$ values obtained
when determining $Q$ and those obtained in determining $U$.
\end{itemize}

\item[$\mL{a}$] Lower confidence limit for $m$.

\item[$m$] The measured error-normalised polarization, $m =
\tilde{p}/\sigma$. \begin{itemize} \item[$m_0$] A measured polarization
used to estimate a confidence region for the true polarization, $a_1,
a_2$. \item[$m_1, m_2$] A confidence region corresponding to likely
measured values for a true polarization $a_0$. \item[$\mh{X}$] A lower
limit: the lowest measurement likely to indicate that there is a true
underlying polarization. 
\item[$\breve{m}$] The residual measured polarization calculated from the
residual normalized Stokes Parameter measurements $\breve{s}_i$ ({\em
q.v.}). 
\end{itemize}

\item[$n_\times$] A count rate (in DN per unit time) measured in one channel of a 
two-channel polarimeter.
\begin{itemize}
\item[$n_{1}(\eta), n_{2}(\eta)$] The DN count rates measured in the two 
channels of a rotatable analyzer when set to orientation $\eta$.
\item[$n_{1i}, n_{2i}$] The individual DN count rates measured in the two 
channels of a two-channel polarimeter on the $i$\,th mosaic.
\item[$n_{1_0}, n_{2_0}$] The true values of $n_1, n_2$.
\item[$n_{\times i}$] Generic symbol for either of $n_{1i}, n_{2i}$.
\item[$\bar{n}_{1}, \bar{n}_{2}$] The mean values of a series of $\nu_S$ DN count rates.
\item[$\sigma_{n_{1i}}, \sigma_{n_{2i}}$] The errors on a pair of individual DN count rate measurements,
based on the sky errors returned by the photometry system.
\item[$\sigma_{\bar{n}_{1i}}, \sigma_{\bar{n}_{2i}}$] The errors on a pair of mean DN count rate measurements.
\end{itemize}

\item[$p$] This symbol is used both for probabilities, and for the
degree of polarization of partly linearly polarised light.
\begin{itemize}
\item[$p_0$] The true value of a polarization $p$.
\item[$\sbb{p}$] The best estimate of the error on a polarization $p$.
\item[$p_1, p_2$] Probabilities when estimating confidence intervals for polarization $p$.
\item[$p_P$] The probability that a source is not unpolarized.
\end{itemize}

\item[$P(s)$] The accurate distribution of a normalised Stokes Parameter, $s$, when
the two contributing channel intensities can be treated as Gaussian.

\item[$P_n(s)$] The Gaussian approximation to $P(s)$.

\item[$Q, q$] Absolute and normalised linear Stokes Parameters. See $S, s$.

\item[$r$]The ratio $U/Q$ used for finding the polarization axis.
\begin{itemize}
\item[$\tilde{r}$] The best estimate $\tilde{u}/\tilde{q}$.
\item[$\sigma_{\tilde{r}}$] The error on $\tilde{r}$.
\end{itemize}

\item[$R$] Symbol for the reference direction corresponding to $\eta_0$.

\item[$R(s)$] The ratio of the true distribution $P(s)$ to its approximation $P_n(s)$.

\item[${\cal R}$] The number of photons per second in a beam of light.

\item[$S$] A generalised absolute Stokes Parameter $S=n_1 - n_2$ illustrating the 
generic properties of $Q$ and $U$. 
\begin{itemize}
\item[$S_0$] The true value of $S$.
\item[$S_i$] The $i$\,th measurement of $S$.
\item[$\bar{S}_{C_S}$] The limiting value of $\bar{S}$ for accepting a null hypothesis at the
$C_S.100\%$ confidence level.
\item[$\bar{S}$] An estimate of $S_0$, the mean of the $S_i$ values.
\item[$\sigma_{S_{i}}$] The error on an individual source intensity measurement,
based on the sky errors returned by the photometry system.
\item[$\sigma_{\bar{S}}$] The standard error on the mean, indicating the accuracy with which
$\bar{S}$ has been determined, based on the sky errors returned by the photometry system.
\item[$\varsigma_{\bar{S}}$] The standard error on the mean, based on the spread of the $S_i$ values in the sample.
\item[$\sbb{\bar{S}}$] The `best estimate' of the standard error on the mean, taken as the greater of  
$\sigma_{\bar{S}}$ and  $\varsigma_{\bar{S}}$.
\item[$\sbb{S}$] The `best estimate' of the error on an individual measurement of $S$, 
taken as the greater of  $\eps{phot}$ and $\eps{stat}$.
\end{itemize}

\item[$s$] A general normalised Stokes Parameter $s=S/I$ illustrating the 
properties of $q$ and $u$. 
\begin{itemize}
\item[$s_0$] The true normalised Stokes Parameter of a source.
\item[$s_j$] The value of $s$ measured with a linear analyzer at the $j$th stepped
position angle orientation $\eta_j$.
\item[$s_i$] The ratio of individual measurements of the 
absolute Stokes Parameters, $s_i = S_i / I_i$.
\item[$\bar{s}$] The mean of the individual
$s_i$, such that $\bar{s} = \sum_{1}^{\nu_S} s_i / \nu_S$.
\item[$\tilde{s}$] The optimal estimator which is the ratio of the mean  
Stokes Parameters, $\tilde{s} = \bar{S} / \bar{I}$.
\item[$\breve{s}$] The residual Stokes Parameters when the optimal estimator is subtracted,
$\breve{s}_i = s_i - \tilde{s}$.
\item[$\sigma_{\tilde{s}}$] The error on $\tilde{s}$ estimated using 
$\eps{phot}$, based on the sky errors returned by the photometry system.
\item[$\varsigma_{\tilde{s}}$] The error on $\tilde{s}$ estimated using 
$\eps{stat}$, based on the spread of the $S_i$ values in the sample.
\item[$\sbb{\tilde{s}}$] The `best estimate' of the standard error on the mean, taken as the greater of  
$\sigma_{\tilde{s}}$ and  $\varsigma_{\tilde{s}}$.
\item[$\sbb{s}$] The `best estimate' of the error on an individual measurement of $s$, 
taken as the greater of  $\vas{phot}$ and $\vas{stat}$.
\end{itemize}

\item[$t$] The statistic of the Student $t$ distribution.

\item[$U, u$] Absolute and normalised linear Stokes Parameters. See $S, s$.

\item[$\mU{a}$] Upper confidence limit for $m$.

\item[$z$] The statistic of the standard normal distribution.

\item[$Z_{\alpha-1}$] A statistic listed here for compatibility with Clarke \& Stewart~\scite[\S6.2]{Clarke+86a}.

\item[$\Delta$] The {\em integer} number of photons which must be 
detected to give a count of 1 DN.

\item[$\eps{\times}$] A generic symbol for the error expected in making an individual measurement of $S_i$.
\begin{itemize}
\item[$\eps{phot}$] The mean of the errors on individual measurements of $S_i$, such that
$\eps{phot} = \sum_{i=1}^{\nu_S} \sigma_{S_i}/\nu_S$. {\sl Hence 
$\eps{phot}$ estimates
the error on an individual measurement based on the error data from the photometry array.}
\item[$\eps{stat}$] The estimated error on an individual measurement of $S_i$, based on the 
sample SD of all the $S_i$. {\sl Hence $\eps{stat}$ estimates
the error on an individual measurement without using the error data from the photometry array.}
\end{itemize}

\item[$\vas{\times}$] A generic symbol for the error expected in making an individual measurement of $s_i$.
\begin{itemize}
\item[$\vas{phot}$] The expected error on an individual measurement of $s_i$, based on $\eps{phot}$.
{\sl Hence $\vas{phot}$ makes use of the photometric error data.}
\item[$\vas{stat}$] The expected error on an individual measurement of $s_i$, based on the 
spread of all the $s_i$ about $\tilde{s}$. {\sl Hence $\vas{stat}$ estimates
the error on an individual measurement without using the error data from the photometry array.}
\end{itemize}

\item[$\eta$] The position angle measured East of North on the celestial sphere
of the transmission axis of a linear analyser.
\begin{itemize}
\item[$\eta_0$] The position of the reference direction relative to which others are measured 
by a particular polarimeter.
\item[$\eta_j$] One specific position angle setting of a stepped rotating analyzer.
\end{itemize}

\item[$\theta$] The phase angle of the polarization expressed as a vector in a phase space where
$\theta = 2\phi$. See $\phi$.

\item[$\nu_S$] The number of individual pairs of measurements made with a 
two-channel polarimeter in order to determine a set of $I_i$ and $S_i$.
Note that in practical cases, often $\nu_Q \neq \nu_U$.

\item{$\Xi$} An empirical cumulative probability (ECD): the empirical 
probability that a measured quantity does not exceed a stated value.
\begin{itemize}
\item[$\ecd{p}{m}$] The empirical probability that the 
measured polarization is not greater than $m$.
\item[$\ecd{\theta}{\phi}$] The empirical probability that the 
measured polarization axis orientation is not greater than $\phi$.
\end{itemize}

\item[$\sigma, \varsigma$] A standard deviation. Terms of the
form $\sigma_\aleph$ are listed under the entry for $\aleph$, but note:
\begin{itemize}
\item[$\sigma$] Without annotation, $\sigma$ is the best estimate of the common error on $q$ and $u$, 
in the case $\sbb{\tilde{q}} \simeq \sbb{\tilde{u}}$. 
\item[$\sigma_1, \sigma_2$] The expected errors on measurements of $n_1, n_2$ ({\em q.v.}).
\item[$\sigma_0$] The approximate SD of an approximately normally distributed $s$.
\end{itemize}

\item[$\tau$] The integration time for measuring light intensity. 

\item[$\phi$] The orientation position angle projected on the celestial plane
of the electric field vector of partially linearly polarised light.
\begin{itemize}
\item[$\phi_0$] The true values of $\phi$.
\item[$\tilde{\phi}$] The best estimate of $\phi$, derived from $\tilde{r}$.
\item[$\sigma_{\tilde{\phi}}$] The error on the best estimate of $\phi$, derived from photometric errors.
\item[$\varsigma_{\tilde{\phi}}$] The error on the best estimate of $\phi$, derived from 
its theoretical distribution.
\item[$\sbb{\tilde{\phi}}$] The best estimate of the error on $\phi$.
\item[$\breve{\phi}$] The residual measured polarization axis calculated 
from the residual normalized
Stokes Parameter measurements $\breve{s}_i$ ({\em q.v.}).
\end{itemize}

\item{$\Phi$} A cumulative distribution function (CDF): the theoretical 
probability that a measured quantity will not exceed a stated value.
\begin{itemize}
\item[$\cdf{p}{m}$] The CDF for the probability that the measured polarization
is not greater than $m$.
\item[$\cdf{\theta}{\phi}$] The CDF for the probability that the 
measured polarization axis orientation is not greater than $\phi$.
\end{itemize}

\item[$\chi$] The physical angle of rotation of a half-wave waveplate: $\eta = 2\chi$.

\item[$\psi_S$] The standard deviation of the sample of $S_i$ values about their mean.

\end{description}

\chapter{Observations, Reduction Procedure and Sample Selection}
\label{mediuch}
\label{obsch}

\begin{quote}
When you have eliminated the impossible, whatever remains,
however improbable, must be the truth. \attrib{Sherlock Holmes, 
{\em The Sign of Four}}
\end{quote}

 Observational data for this project were taken on two observing runs,
both using the same equipment in Hawaii, in August 1995 and August 1997. 
On the 1995 run, seven objects were studied, all radio galaxies featuring
in the 3C catalogue. The seven radio galaxies, at redshifts $0.7<z<1.2$,
were not selected according to any statistical criterion, but formed a
representative sample of the different morphologies present in this
redshift band. With the hindsight provided by Eales et al.\
\scite{Eales+97a}, one presumes that 3C galaxies, displaying the strongest
alignment effect, are also likely to display the strongest polarizations. 
 
 The second run, August 1997, looked at three objects. One, 3C 441, had
featured in the first run but had not yielded a conclusive polarization
value. Another, MRC 0156$-$252, possessed the brightest known absolute 
$V$-band
magnitude for a radio galaxy at $z \sim 2$, and had featured in Eales \&
Rawlings's \scite{Eales+96a} comparison of radio galaxies at redshifts $z
\sim 1$ and $z > 2$. Finally, LBDS 53W091 is a very red radio galaxy
visible at a very high ($z=1.552$) redshift --- especially interesting
since comparisons of its spectrum with synthetic and real elliptical
galaxies suggest that it must be at least 3.5 Gyr old
\cite{Dunlop+96a,Spinrad+97a} -- which is only consistent with its high
redshift in certain cosmologies.
 
 The instruments and procedure followed are summarised below.
Consideration is also given to possible contaminating polarization due to
Galactic dust.  Full details of the nine sources investigated and other
data concerning them can be found together with our results in Chapter
\ref{resulch}.

 As Mark Neeser reminds us in the cautionary note to his
recent thesis \cite{Neeser-96a}, such small samples of powerful objects
are not likely to be typical of the Universe at large. But {\em this}\,
author hopes that the results and discussion presented here in the final
chapters will add a little to our understanding of the powerful objects
that are high redshift radio galaxies; and especially that the methodology
presented in Chapter \ref{stoch} will be a useful guide to those who
follow this work and take polarizations of statistically meaningful
samples of these objects in future.

\section{Instrumentation}

 All the $K$-band polarised images taken for this project were obtained
using the IRPOL2 instrument, designed by the University of
Hertfordshire and installed at UKIRT (the United Kingdom InfraRed
Telescope, Hawaii).  Our August 1995 run was the first common-user
project undertaken by this instrument after its commissioning run.

The IRPOL2 polarimeter consists of a rotatable half-wave plate and a
Wollaston prism, working in conjunction with the IRCAM3 InSb array
detector. We used IRCAM3 at the default pixel scale, 0.286 arcsec/pixel
\cite{Aspin-94a}, with a $K$-band filter. The IRPOL2 system has
negligible instrumental polarization \cite{hpc,Chrysostomou-96a}.  The
Wollaston prism -- a two-channel polarimeter following the paradigm
of \S \ref{paradigm} -- causes each source in its field of view to
appear as a
pair of superimposed images with orthogonal polarizations, separated by
$-0.93$ pixels in right ascension and $+69.08$ pixels in declination
\cite{Aspin-95a}. A focal plane mask is available: if used, it divides
the array plane into four horizontal strips imaging light of alternate
orthogonal polarizations. If not used, one polarised image is displaced
and superposed on its orthogonal complement.

 The design of the instrument is such that when the waveplate is set to
its 0\degr\ reference position, an object totally linearly polarised with
its electric vector at 83\degr\ (i.e.  celestially East of North) would
appear only in the upper (Northern) image, and an object totally linearly
polarised at -7\degr\ would only appear in the lower image. Hence
the reference axis `R' is oriented at $\eta_0 = 83\degr$. There are
four
standard offset positions for the waveplate: 0\degr, 22.5\degr, 45\degr\
and 67.5\degr. 

\section{Observing Procedure}

 We used slightly different techniques on the two observing runs to
build up our images. On the first run, the focal plane mask was not
used: a slight error in pointing could have caused the extended target
objects to lie partially behind the mask. The absence of the mask meant
that more field objects (useful controls for instrumental and local
Galactic polarization) would also be imaged. The array was shifted
equally in right ascension and declination to build up final image.  
On the second run, to reduce the background noise around our target
objects, the mask was used; and so the array was shifted principally in
right ascension to build up a final image.

\subsection{First run procedure}

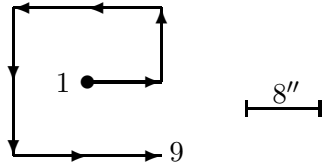
\begin{figure}
\centering
\begin{picture}(126,60)(-30,-30)
\thicklines
\put (-12,-4){1}
\put (0,0){\circle*{5}}
\put (0,0){\vector(1,0){28}}
\put (28,0){\vector(0,1){28}}
\put (28,28){\vector(-1,0){28}}
\put (0,28){\vector(-1,0){28}}
\put (-28,28){\vector(0,-1){28}}
\put (-28,0){\vector(0,-1){28}}
\put (-28,-28){\vector(1,0){28}}
\put (0,-28){\vector(1,0){28}}
\put (31,-30){9}
\put (60,-10){\line(1,0){28}}
\put (60,-13){\line(0,1){6}}
\put (88,-13){\line(0,1){6}}
\put (70,-8){8\arcsec }
\end{picture}
\caption{Mosaicing pattern for observing run 1.}
\label{mospat1}
\end{figure}

 For each target object, we took a `mosaic' of nine images with the
waveplate at the 0\degr\ offset. One image consisted of a 60 second
exposure (the sum of six ten-second co-adds), and the mosaic was built
up by taking one image with the target close to the centre, and eight
images with the frame systematically offset from the first by $\pm 28$
pixels (8 arcsec) horizontally and/or vertically, as illustrated in Figure
\ref{mospat1}. Since IRCAM3 is a square array of 256 pixels each side,
each individual image had side 73\arcsec, and the final mosaics had
side 92\arcsec, with the greatest sensitivity being achieved in the
central square of side 54\arcsec .

\begin{figure}
\centerline{\psfig{file=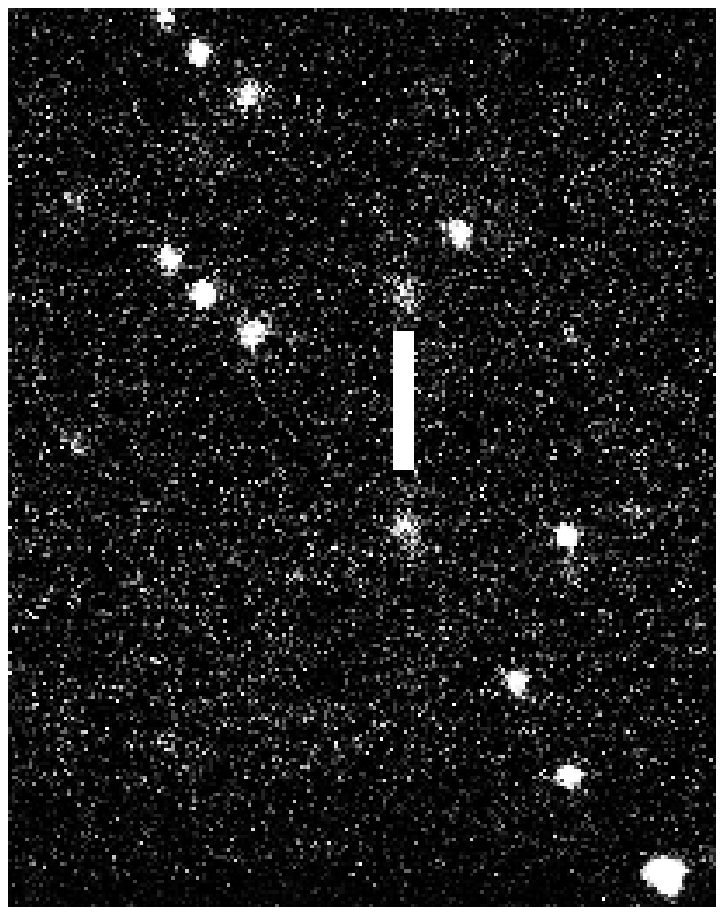,width=85mm}}
\caption{K-band image of 3C 54 (at either end of the bar) and surrounding 
field.} 
North is at the 
top, East at the left. The image was composed by mosaicing a series of 
nine 60-second exposures taken with the waveplate offset at 22.5\degr.)
\label{54image} \end{figure}

The same source was then similarly
observed with the waveplate at the 22.5\degr, 45\degr\ and 67.5\degr\
offsets, completing one cycle of observations;  hence one such cycle took
36 minutes of integration time. Between two and five observation cycles
were performed over the three nights for each target; the total integration
time for each target is given with the observational data in  
Table \ref{extab}. Not all of the times quoted are
exact multiples of 36 minutes, as in some cases, mosaics were
corrupted by
problems with the UKIRT windblind, and excluded from our analysis. An 
example of a mosaic, 3C 54 and its
surrounding field, observed with the waveplate at 22.5\degr, is shown in
Figure \ref{54image}. This image shows clearly the effect of using the 
Wollaston prism without the focal plane mask: note the double images of 
most of the objects, and the partnerless objects on the right hand side 
whose upper or lower channels fell outside the detector array.

\subsection{Second run procedure}

\begin{figure}
\centerline{\psfig{file=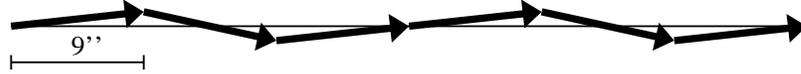,width=120mm}}
\caption{Mosaicing pattern for observing run 2.}
\label{mospat2}
\end{figure}

For our 1997 run, making use of the instrument's focal plane mask, we
created mosaics of each source by combining seven 60-s exposures at
horizontal spacings of $9 \arcsec$ and vertical spacings of $\pm 1
\arcsec$ -- see Figure \ref{mospat2}. This procedure creates
rectangular strip images measuring
$127\arcsec \times 20\arcsec$, of which the central $19\arcsec \times
16\arcsec$ displays the maximum sensitivity.  The total exposure time, 
summed over all four waveplate settings, is again listed in
Table \ref{extab}.

\section{Data Reduction}

\subsection{First run procedure}
\label{firstrunDR}

 The raw data from IRCAM3 were stored as {\em Starlink}\, {\sc ndf} ({\tt
foo.sdf}) images. These were converted to {\sc iraf} ({\tt foo.imh})
format for subsequent analysis by one of two methods: the earliest data to
be analysed was handled by conversion to intermediate {\sc fits} files,
which were then read into {\sc iraf} by its {\tt rfits} 
routine\footnote{Doctoral thesis declaration: Thesis supervisor Dr Stephen
Eales had already converted these images to {\sc iraf} format before the
author began his work; all subsequent data reduction was performed by
the author.}; later,
when the {\em Starlink}\, {\sc figaro} package was enhanced, its new
one-step {\tt ndf2iraf} routine was employed.  [Documentation for {\sc
iraf} and {\em Starlink} packages can be found at their respective
websites, {\tt iraf.noao.edu} and {\tt star-www.rl.ac.uk}.]

 We reduced the data by marking bad pixels, subtracting dark frames,
and flat-fielding. Dr Stephen Eales at Cardiff University (private
communication) made available software containing a list of known bad
pixels on IRCAM3, which was used to mark the hot and dead pixels; he
also provided mosaicing software which ignored such pixels when
combining shifted frames to produce a final composite mosaic. The dark
frames used were the means of 10-s dark exposures made at the start,
middle and end of the night on which the corresponding target images
were taken. Flat-fielding frames were obtained by median-combining the
nine images of each mosaic without aligning them, and normalizing the
resulting image by its mean pixel value. In order to align each set of
nine images, we chose a star present on each frame, and measured its
position with the {\sc apphot.center} routine in the {\sc iraf}
package. Using these positions, the nine images were melded into a
single mosaic image.

 Photometry was performed on the pairs of images of field stars and of
radio galaxies in each mosaic using {\sc apphot.phot} from the {\sc iraf}
package. For each target, we chose one mosaic arbitrarily, and tested this
to determine the best photometry aperture as follows: Using the arbitrary
mosaic, we performed photometry on the two images of the source at a
series of radii increasing in unit pixel steps. The measured magnitude in
each aperture decreases as the light included in the aperture increases;
we noted the first pair $(r,r+1)$ of radii where the change in magnitude
was less than half the measured error on the magnitude. We then earmarked
the next radius, $(r+2)$, for use in determining the polarization. In this
way, we hope to include most of the source light but as little as possible of
the surrounding sky. Where the $(r+2)$ aperture sizes differed for the two
channels' images of the source, we earmarked the larger of the two. This
chosen aperture was then used as the photometric aperture on all the
mosaics containing the target (i.e.\ on every waveplate setting from every
observing cycle), with {\sc iraf} output yielding a set of flux 
counts, magnitudes and errors. 

 The {\sc apphot.phot} routine corrects for the sky brightness by
measuring the modal light intensity in an annulus around the target; the
position of the annulus was chosen in each case such that the outer radius
did not extend to the nearest neighbouring object, and the inner radius
was normally set one pixel greater than the photometry aperture. (Where we
attempted to perform photometry on a knot within a larger structure, the
inner radius of the annulus was set sufficiently large to exclude all the 
knots comprising the object.) The output file of {\sc apphot.phot}
contains data on the sky brightness and consequent errors on each
photometric measurement, as well as `magnitudes' (relative to an
arbitrary zeropoint) for each target. These values were extracted and
analyzed on a PC spreadsheet package ({\em Microsoft Works}). The 
spreadsheet programming follows the scheme described in Chapter 
\ref{stoch} and is documented in Appendix \ref{codeapp}.

 The spreadsheet analysis calculated both absolute $(Q,U)$ and 
normalized $(q,u)$ Stokes parameters for the $\eta_0 = 83\degr$ reference 
frames, together with the best estimates of the errors in each case.
 Finally, the optimal normalised
Stokes parameters were used as input to {\sc fortran} routines (also 
documented in Appendix \ref{codeapp})
designed to follow Steps \ref{findperr} thru
\ref{findangle} of Chapter \ref{stoch}'s data reduction
scheme. The program's resultant output values consist of point
estimates and confidence
intervals of the debiased polarization and orientation measures, and
an estimate of the probability that the source is actually polarised.

\subsection{Second run procedure}

 Data reduction for the August 1997 run was performed using the
standard UKIRT {\sc ircamdr} software to flat field images, and {\sc
iraf} to align and combine the mosaics. Since spreadsheet analysis of
the August 1995 data had shown no source of systematic error, it was
not felt necessary to repeat the spreadsheet system of rigorous checks
before producing algebraic optimal estimates of the normalised Stokes
parameters. Rather, in this case, the images were then averaged to form
a single image for each waveplate, as this is equivalent to the
algebraic method which gives the best signal-to-noise in the final
polarimetry \cite{Leyshon-98a,Maronna+92a,Clarke+83b}. The 3C 441 data
was then combined, at each waveplate position, with our stacked 1995
images. Results are discussed in the next chapter, and summarised in
Table \ref{restab}.

\section{Our Samples}
\label{GalPolims}
 
 Our first observing run took place on the nights of 1995 August 25,
26, and 27, and covered a sample of seven high-redshift 3C radio
galaxies at redshifts $0.7<z<1.2$. These objects were not selected to
form any kind of statistically complete sample; rather, they form a
representative sample of the different morphologies present in this
redshift band. Since this run represented the first known attempt to
obtain infrared polarimetry of high redshift radio galaxies, objects
known to be bright in the $K$-band were preferred.
 
The seven objects surveyed included 3C 22 and 3C 41, which are bright
and appear pointlike; 3C 114 and 3C 356, which display complex knotted
morphologies with large scale alignments between the $K$-band structure
and the radio axis;  and 3C 54, 3C 65 and 3C 441, which are faint
sources with some indication of $K$-band structure. Of these three
faint sources, 3C 54 displays an alignment between the $K$-band
morphology and the radio axis \cite{Dunlop+93a}, 3C 65 shows no
preferred direction in its $H$-band structure \cite{Rigler+94a}, and 3C
441 has a broad-band optical polarization which is roughly
perpendicular to its radio structure \cite{Tadhunter+92a}. Five of the
sources (not 3C 22 or 3C 356) were early radio sources observed by   
Longair \scite{Longair-75a}.
 
 The second observing run, conducted on the nights of 1997 August 18
and 19, was awarded primarily to study the controversial radio galaxy
LBDS 53W091. This galaxy (see \S\ref{D53W091}) appears to be at least
3.5 Gyr old \cite{Dunlop+96a,Spinrad+97a} --- which is only consistent
with its $z=1.552$ redshift in certain cosmologies. The timing of
the run also made it possible to re-observe 3C 441, for which our
1995 data were inconclusive, and to observe MRC 0156$-$252, which has the
brightest known absolute $V$-band magnitude for a
radio galaxy at $z \sim 2$.
 
 Table \ref{extab} lists all the sources observed together with their
redshifts and the rest-frame wavelength of the observed light. As
discussed below in \S\ref{ULGalPol}, it is possible to estimate an upper
limit on the interstellar polarization imposed on $K$-band light during
its passage through our Galaxy. Hence the {\sc ned} values for the
Galactic extinction $A_B$ at each observed source and the
corresponding upper limits on $p_{\rm K}$ are given in this Table, too.
 
\begin{small}
\begin{table*}
\caption{Redshifts, integration times, total $B$-band extinctions and  
upper limits on $K$-band interstellar
polarization for our sample of radio
galaxies.}
\label{extab}
\begin{tabular}{|llrrrrrrr|} 
\hline
Source & IAU form & \ccen{$z$} & $\lambda_r$($\mu$m) & $t_{\rm
int}$(min) & \ccen{$l(\degr)$} & \ccen{$b(\degr)$} & $A_B$ & $p_K$ \\
\hline
3C 22        & 0048+509 & 0.936 & 1.14 & 72   & 122.9 & -11.7 & 1.09 & 
0.76 \\
3C 41        & 0123+329 & 0.794 & 1.23 & 108  & 131.4 & -29.1 & 0.17 &
0.12 \\
3C 54        & 0152+435 & 0.827 & 1.20 & 135  & 135.0 & -17.6 & 0.37 &
0.26 \\
3C 65        & 0220+397 & 1.176 & 0.92 & 72   & 141.5 & -19.5& 0.16 &
0.11 \\
3C 114       & 0417+177 & 0.815 & 1.21 & 189  & 177.3 & -22.2 & 1.26 &
0.88 \\
3C 356       & 1723+510 & 1.079 & 1.06 & 99   & 77.9& 34.2& 0.10 & 0.07 \\
3C 441 (1995)& 2203+292 & 0.707 & 1.29 & $\dagger$144  & 84.9 & -20.9& 
0.34 &
0.24 \\
\hline
3C 441 (1997)& 2203+292 & 0.707 & 1.29 & $\dagger$112 &  84.9 & -20.9 &
0.34 & 0.24\\
LBDS 53W091  & 1721+501 & 1.552 & 0.86 & 364 &  76.9 & +34.5 & 0.08 &
0.06\\
MRC 0156$-$252 & 0156$-$252 & 2.016 & 0.73 & 196 & 208.6 & -74.8 & 0.00 &
0.00\\
\hline 
\end{tabular}
 
\medskip

Key: $z$: redshift; $\lambda_r$ (\micron): rest-frame equivalent of   
observed-frame 2.2\,\micron;
$t_{\rm int}$ (min) : total integration time (min) summed over all
waveplate
settings ($\dagger$: 3C 441 data from the two runs was pooled
giving a composite image
of 256 minutes integration time in total) ; $l$ (\degr):
Galactic longitude ({\sc ned}); $b$ (\degr): Galactic latitude ({\sc
ned}); $A_B$: Blue-band extinction (mag), from {\sc ned},
derived from Burstein \& Heiles \scite{Burstein+82a}; $p_K$: maximum
Galactic interstellar contribution to $K$-band polarization (per cent).
\end{table*}
\end{small}

\section{Calibration}

 The purpose of the two observing runs made for this project was to obtain
polarimetry of faint objects rather than to obtain absolute photometry.
Two-channel polarimetry depends on measuring the difference in signal
between two channels simultaneously; this eliminates systematic errors
which might arise due to imperfect calibration if the channels were
instead measured consecutively. Such measurements can be carried out in
atmospheric conditions less stringent than those required for accurate
photometry. Normalised Stokes parameters, by definition, do not require 
calibration as they depend on the ratio of two intensities measured by 
the same system. The source intensities and absolute Stokes Parameters 
calculated in the course of the data reduction are not reproduced in this 
thesis; they were expressed in IRCAM3 data number rates throughout the 
reduction process.

The {\sc iraf apphot.phot} routine produces a set of output magnitudes for
the target objects, relative to a calibrated zero point. We did not use
the resultant magnitudes, but the raw data number counts from the
photometry aperture, as explained in Appendix \ref{codeapp} -- so it was
not necessary to calibrate the zero-point for this purpose. 
We did not, therefore, observe photometric standard stars as part of the 
observations for this project.

 A {\em polarimetric}\, standard was measured as part of the August 1997 
run to verify the accuracy of the hardware and software forming our 
polarization reduction chain.  The standard star HD 215806, recorded by 
Whittet et al.\ \scite{Whittet+92a} to have polarization 
0.55\% $\pm$ 0.06\% at 77\degr, 
was measured twice by our data reduction chain.
The first measurement yielded 0.36\% $\pm$ 0.11\% (debiased to 0.35\%) at 74\degr\ $\pm$ 6\degr: clearly consistent within a $2\sigma$ error box.
The second measurement similarly yielded  
0.30\% $\pm$ 0.11\% (debiased to 0.29\%) at 73\degr.
If instrumental polarization were present, it would increase the measured
value, and probably skew it to a different orientation; the
agreement between the published and measured orientations confirms that
instrumental polarization is negligible in the IRPOL2 chain.
The fact that both measurements are lower than the published value is not
statistically significant, but is interesting enough that future workers 
measuring HD 215806 might want to check for variability in its polarization.

In general, it should be noted that polarized standards must be carefully
chosen as polarization angle can only be measured to 0.1\degr\ and varies
with wavelength \cite{Dolan+86a} --- but with such inaccurate measurements
being returned for polarization orientations on our radio galaxies, this
is hardly relevant in this particular case. Neither would attempting to 
refine the reference axis zero point $\eta_0$ using the standard star add 
any meaningful accuracy to the measured orientations, given their large 
error figures.

\section{Galactic Interstellar Polarization}
\label{ULGalPol}

 In the previous chapter, we studied at length the best way to
recover a measurement of the true polarization of light from a noisy
system. One step remains, however, before the figure obtained can be
said to be that of the active galaxy: the interstellar medium of both
our own galaxy and the active nucleus's host galaxy may modify the
linear polarization of light passing through it.

 Ordinary stars in our own galaxy are not expected to emit
intrinsically polarised light; and measured polarization of starlight
is presumed to be due to transmission through dust grains. Gehrels
\scite{Gehrels-60a} was the first to note that the
interstellar polarization varied
over visible wavelengths, and subsequent work by Serkowski and
colleagues \cite{Coyne+74a,Serkowski+75a} found an empirical
relationship applicable throughout the visible spectrum:
\begin{equation}
\label{skemper}
p/p_{\rm max} = \exp [-K \ln^{2}(\lums{max}/\lambda)]
\end{equation}
where $\lums{max}$ is the wavelength at which
the polarization
peaks, usually around 0.5\,\micron, and empirically in the range
0.3--0.8\,\micron\ \cite{Serkowski+75a,Wilking+80a}.

 The parameter $K$ was fitted as a constant by Serkowski, Mathewson \&
Ford \scite{Serkowski+75a}, who found the best value to be $K = 1.15$.
Wilking et al.\ \scite{Wilking+80a}, however, investigated whether
Serkowski's empirical formula remained valid at infrared wavelengths,
and found that a better fit was obtained by taking $K$ to be linearly
dependent on $\lums{max}$; an adequate approximation for our purposes
is $K = 1.7\,$\micron$^{-1}\, \lums{max}$. That the empirical formula, so
modified, holds up to around 2\,\micron, was confirmed by Martin \&
Whittet \scite{Martin+90a}. The best value of the constant coefficient was 
refined slightly by
Whittet et al.\ \scite{Whittet+92a} but the value 1.7 remains an adequate
approximation for our purposes.

 In general, $p_{\rm max}$ for a given set of galactic co-ordinates is
not known. But suppose we take the ratio of polarizations in two
wavebands, $V$ and $K$, and rearrange:
\begin{equation}
\label{sratund}
p_{K} = p_{V} \exp \left\{ -3.4 \lums{max} \left[ \ln 
\left( \frac{\lums{max}}{\sqrt{\lambda_{K}\lambda_{V}}}\right) \ln
\left(\frac{\lambda_{V}}{\lambda_{K}}\right) \right] \right\}.
\end{equation}
Hence $p_{K} = c.p_{V}$ where $c$ depends on $\lums{max}$ but for
$0.3<\lums{max}<0.8$ we find $0.15<c<0.30$.
 
 An empirical upper limit for $p_{V}$ (expressed as a percentage) is
given by Schmidt--Kaler \scite{SchmidtKaler-58a} as $p_{V} \leq
9E_{B-V}$ and typically, $p_{V} = 4.5E_{B-V}$. [Clarke \& Stewart \scite[\S4.2]{Clarke+86a} point out that determinations of an 
empirical upper limiting polarization also tend to find an empirical lower
limit, suggesting imperfect debiasing, and that the true empirical
upper limit is in fact lower than the one determined.] It is well established
\cite{Savage+79a,Koorneef-83a,Rieke+85a} that the ratio of total to
selective extinction is $A_{V}/E_{B-V} \sim 3$; and so we can use the
extinctions $A_{B}$ \cite[figures can be obtained from the {\sc ned}
database]{Burstein+82a} to obtain $E_{B-V} = A_{B} - A_{V} = A_{B}/4$.
 
 Taking the maximum values, $c=0.3$ and $p_{V} \leq 9E_{B-V}$, we find
an upper limit for infrared polarization $p_{K} \leq 0.7A_{B}$. We have
seen (Table \ref{extab}) that none of the objects surveyed for
this thesis lie beyond regions of our Galaxy with $A_B > 1.3$, and all
but two have $A_B < 0.4$, so the Galactic medium cannot contribute more
than 1.2\% to the $K$-band polarization of the two high-extinction
sources, or more than 0.3\% to $K$-band polarization of any of the
others. A further check can be made by measuring the polarizations of
sources (presumably intrinsically unpolarized stars) which lie in the
same fields as our target objects (this will be commented on during
discussions of individual targets in the next chapter).

 Like our own Galaxy, the host galaxies of the active nuclei may
contain dust regions capable of polarizing light passing through them.
Goodrich \& Cohen \scite{Goodrich+92a} argue that 3C 109 is polarised
in this way; 3C 234 could be a similar example \cite{Tran+98a}.
Such polarization effects affect the observed $K$-band
light at its rest-frame wavelength and will be considered in the
context of the discussion of individual sources -- since the source of 
such polarization is, by definition, the host galaxy of an active
galactic nucleus. The possibility of polarizing effects being produced at
intermediate wavelengths by any intergalactic medium cannot be ruled out 
{\em a priori}, but we invoke Occam's Razor to assume the absence of 
significant amounts of any intergalactic medium without evidence to the 
contrary (e.g.\ from the observed colours of the target galaxies).

\chapter{General Observations and Individual Objects} 
\label{resulch}

\begin{quote}
Look up at the heavens and count the stars -- if indeed you can count 
them.\attrib{Genesis 15:5 (NIV)}
\end{quote}

 The observational datasets obtained were reduced and analysed using the
procedure described in the previous chapter. As is recommended for
polarimetry, the normalised Stokes parameters $q$ and $u$ were obtained;
the reference axis for IRPOL2 is $\eta_0 = 83\degr$, i.e.\ that $q>0, u=0$
corresponds to a polarization orientation of 83\degr\ E of N and that for
$q=0, u>0$, the polarization orientation is 128\degr.  Table \ref{obstab}
gives the polarizations of all target objects and associated objects, but
not of the field objects also analyzed. 

Some target objects displayed extended structure:  Reference is made in
the text of this chapter to the `moment analysis' of Dunlop \& Peacock
\scite{Dunlop+93a}, who devised an automated routine to evaluate a
position angle for the extended structure they saw in $K$-band sources.
For those sources in which there are clear distinct components to the
structure, the identification of these components is given on the
labelled images which follow. In most of our images, the target object
is labelled {\tt T}; other objects have been labelled following
earlier maps in the literature, where available.  Composite images are
shown in each case with the images from all waveplate positions stacked
together; in some cases, edited images are also provided where one set
of images from the Wollaston prism's `double image' have been removed
using the image patching facilities of the {\sc starlink gaia} package.

\begin{small}
\begin{landscape}
\begin{table*}
\centering
\caption{Observational results of $K$-band polarimetry of our nine
target radio galaxies and associated objects.}
\label{obstab}
\begin{tabular}{lrrrrrlrrrrrrr} 
\hline \small
Source & $r$(\arcsec) & $q (\sigma_q) $(\%) & $u (\sigma_u)$(\%)
& $P \pm \sigma_P$(\%) & $2\sigma$.UL & Prob &
$\theta$(\degr) & \cduo{$\sigma_\theta$(\degr)} \\

\hline
3C 22 & 2.6 & -1.3 (1.4) & -3.2 (1.4) &  3.3 $\pm$ 1.4 & - & 0.95 &
27 & -71 & +17 \\
\hline
 3C 41 & 2.3 & +0.8 (1.1) & -3.1 (1.1) & 3.1 $\pm$ 1.1 & - & 0.98 & 
45 & -14 & +79 \\
\hline
3C 54  & 4.0 & -1.4 (2.5) & -6.0 (2.5) & 5.9 $\pm$ 2.6 & - & 0.94 &
32 & -79 & +17 \\
\hline
3C 65 & 2.9  & -4.3 (4.2) & -1.2 (4.0) & 2.2 $\pm$ \tbn{L}4.5 & 10 & 
0.42 & 1 & \cduo{$\pm$71} \\
\hline
3C 114 (Whole) & 3.6 &+11 (3) & -4.1 (2.7) & 11 $\pm$ 3 & - &
0.99 & 73 & -21 & +34 \\
3C 114 (Knee)& 1.7 & +3.0 (1.6) & +4.3 (1.8) & 5 $\pm$ 1.7 & - &
0.99 & 111 & -63 & +14 \\
\hline
3C 356 (Whole)& 4.0 & -10 (5) & +3.5 (6) &9 $\pm$ 5 & 16 &
0.85 & 172 & 
\cduo{$\pm$22} \\

3C 356 $a$ (North)& 2.3 &  -13 (8) & +4.6 (8) & 13 $\pm$ 8 & 41
& 0.62 & 164 & \cduo{$\pm$25} \\ 

3C 356 $b$ (SE)     & 2.6 &  -10 (9) &-19 (17) &  19 $\pm$
15& 24& 0.78 & 24&-54&+33 \\
\hline
3C 441 {\bf a}  & 3.1 & +4 (5) &-0.7 (5) & 1 $\pm$ \tbn{L}6 &
        10 & 0.46 &  78 & \cduo{$\pm$51} \\
3C 441 B$\ddagger$ & 3.1 & +0.5 (5)  &-1.7 (5)   & 0.1 $\pm$ 
\tbn{L}2.8 & 7  & 0.14 &  47 & \cduo{$\pm$59}  \\
3C 441 {\bf c}$\ddagger$ & 3.1 & +1.3 (12) & +10 (11) & 3.5 $\pm$
\tbn{L}16 & 24 & 0.54 & 124 & \cduo{$\pm$45} \\
3C 441 E  & 2.6 & +5 (12) & +19 (13) & 18
$^{+9}_{-8}$ &
        36 & 0.91 & 120 & \cduo{$\pm$15} \\
3C 441 F & 2.0 & +6 (19) & +16 (22) & 6 $\pm$ \tbn{L}28
&
        43 & 0.48 & 118 & \cduo{$\pm$48}  \\
3C 441 G  & 2.6 &-0.1 (13) &-6 (13)  & 0.3 $\pm$ \tbn{L}10
&
        21 & 0.18 &  38 & \cduo{$\pm$59} \\
3C 441 H  & 2.3 &-12 (14)&-9 (15)  & 7 $\pm$
\tbn{L}22 &
        33 & 0.67 & 11 & \cduo{$\pm$38}  \\
\hline
LBDS 53W091 $\natural$  & 1.1 & 0 (17) & -7.5 (22) & 0.4
 $\pm$ \tbn{L}8 &  31 & 0.11 & 38 & \cduo{$\pm$60} \\
LBDS 53W091 $\flat$ & 1.1 & +0.6 (16) &  -3.4 (18) & 0.17
 \tbn{L,U} & 22 & 0.20 & 43 & \cduo{$\pm$60}  \\
Object 3a $\natural$   & 1.1 & -12 (18) &  +17 (18) & 16
 $\pm$ \tbn{L}14 & 43 & 0.70 & 146 & \cduo{$\pm$24}  \\
Object 3a $\flat$   & 1.1 & -4.5 (21) &  +21 (20) & 10
 $\pm$ \tbn{L}21 & 46 & 0.67 & 134 & \cduo{$\pm$36}  \\
\hline
MRC 0156$-$252 & 2.3 & -2.5 (7) &  +1 (7) & 0.14 $\pm$
\tbn{L}4.3 &
 10.5 & 0.14 & 161 & \cduo{$\pm$59} \\
MRC 0156$-$252 & 2.7 & -0.5 (8) & -0.5 (8) & 0.04 \tbn{L,U} &
 4 & 0.01 & 60 & \cduo{$\pm$60}  \\
MRC 0156$-$252 & 3.4 &  +0.4 (8) & +0.8 (9) & 0.05 \tbn{L,U} &
 5.5 & 0.01 & 114 & \cduo{$\pm$60} \\
\hline
\end{tabular}

\medskip

\parbox{185mm}{Key: Source: Source name and component ($\natural$: natural 
image; $\flat$: `despiked' image; $\ddagger$: data is based on
1995 observations only); r: radius of
photometry aperture (arcseconds);
 $q \pm \sigma_q, u \pm \sigma_u$: normalized Stokes
Parameters (per cent) with respect to 83\degr\ E of N;
 $P\pm\sigma_P$: percentage polarization (debiased) with 1$\sigma$
error
(\tbn{L} --- the $1\sigma$ lower limit for polarization is zero;
\tbn{L,U}
--- the $1\sigma$ `confidence interval' is identically zero even though
the best point estimate polarization is non-zero); Prob: the
probability
that there is underlying polarization, given by Equation \ref{propol};
$2\sigma$.UL: $2\sigma$ upper
limit
(in per cent) for polarization in objects unlikely to be polarised;
$\theta\pm\sigma_\theta$: Electric vector orientation E of N (\degr).}
\label{polvals}
\label{restab}
\end{table*}
\end{landscape} 
\end{small}

\section{Have We Detected Polarization?}
\label{havewe}

 Equation \ref{propol} allows us to quantify the probability that a given
object is polarised. The probabilities of each object being polarised are 
listed in Table \ref{polvals}. Three of our nine sources have a 95\%
 or better probability of being polarised; and of these, 3C 22 and
3C 41 are polarised at the 3 per cent level, and 3C 114 at the 12 per cent
level. 

The number of prominent starlike objects (in addition to the target)
featuring on the 1995 set of 3C object mosaics varies between one and
seven, depending on the target. (We will refer to these as `stars' but 
have no spectroscopic evidence to confirm their identity as such.)
Where possible, we have performed
polarimetry on these stars; out of the 21 stars so observed, only one
has
a greater than 95\% probability of being polarised. This object
was a bright starlike object on the mosaic containing 3C 114, but is
only polarised at the 0.7 per cent level, which is explicable by the
interstellar medium (see Table \ref{extab}). Even without such special
pleading, it would not be surprising for random noise to cause one star out 
of 21 to appear to be polarised at such a level.

 Within the bin of sources having a probability 80-95\% of being
polarized, fall three further stars; of these, one is extremely faint,
and another appears to be polarised at only the 0.3 per cent level. The
third falls on the same mosaic as 3C 54, and appears to be polarised at
the 5.6 $\pm$ 2.6 per cent level, with a 94\% chance of the
polarization being genuine. This star, however, straddles the edge of
three of the nine component frames of the mosaics, so the validity of
the result is called into question.  Two of our sources also fall in
the 80-95\% probability bin: 3C 54 itself, polarised at the 6
per cent level, and 3C 356, at the 9 per cent level. (Object E of the
3C 441 complex may also fall in this bin, on the basis of the pooled
1995 and 1997 observations.)

 For the 1995 observations, given that 17 out of 21 stars, but only 2
out of 7 sources, have a probability of less than 80\% of being
polarised at all, we feel confident of having detected polarization in
three 3C sources, and possibly in a further two. The three targets for
the 1997 run were faint objects in comparison to most of those observed
in the earlier run: one component of the 3C 441 complex displays
marginal evidence for polarization, but there is no
strong evidence for polarization in 53W091 or MRC 0156$-$252.

\section{Individual Objects}

In the following object-by-object discussion, we will examine each target
object in the context of other observational data about the same object
from the literature. Evidence for parallel or perpendicular alignments
will be noted, but discussion of the implications of our data for the
properties of the central engines and host galaxies will be deferred to
the next chapter. In this context, the {\em Hubble Space Telescope}\, is
abbreviated {\em HST}.\,\ Certain papers will be cited very
frequently, and
will be abbreviated in this chapter: Dunlop \& Peacock \scite{Dunlop+93a}
will be denoted D\&P, Leyshon \& Eales \scite{Leyshon+98a} is
abbreviated 
L\&E, and a series of papers by Best, Longair \&
R\"{o}ttgering will be denoted BLR-I \scite{Best+96a}, BLR-II
\scite{Best+97a}, and BLR-III \scite{Best+98a}. 

 The polarizations of other objects on the target frames will be noted
here, using normalised Stokes parameters of the form $\tilde{q} \pm
\sigma_{\tilde{q}} , \tilde{u} \pm \sigma_{\tilde{u}}$, as obtained through 
Data Reduction
Step \ref{hereNSPs}; estimates of $P$ (debiased) and orientation will also
be quoted. Since the probability that these objects are polarized is low,
formal errors on the nominal degree and angle of polarization are not
quoted; these can easily be calculated from the normalised Stokes
parameters and their errors if required. These field objects serve as
useful controls which would immediately indicate regions of high $K$-band
Galactic polarization -- though they cannot, by themselves, rule out the
presence of polarizing material beyond their locations. 

\subsection{3C 22}
\label{D3C22}

\subsubsection*{Structure}

\begin{figure}
\centerline{\psfig{file=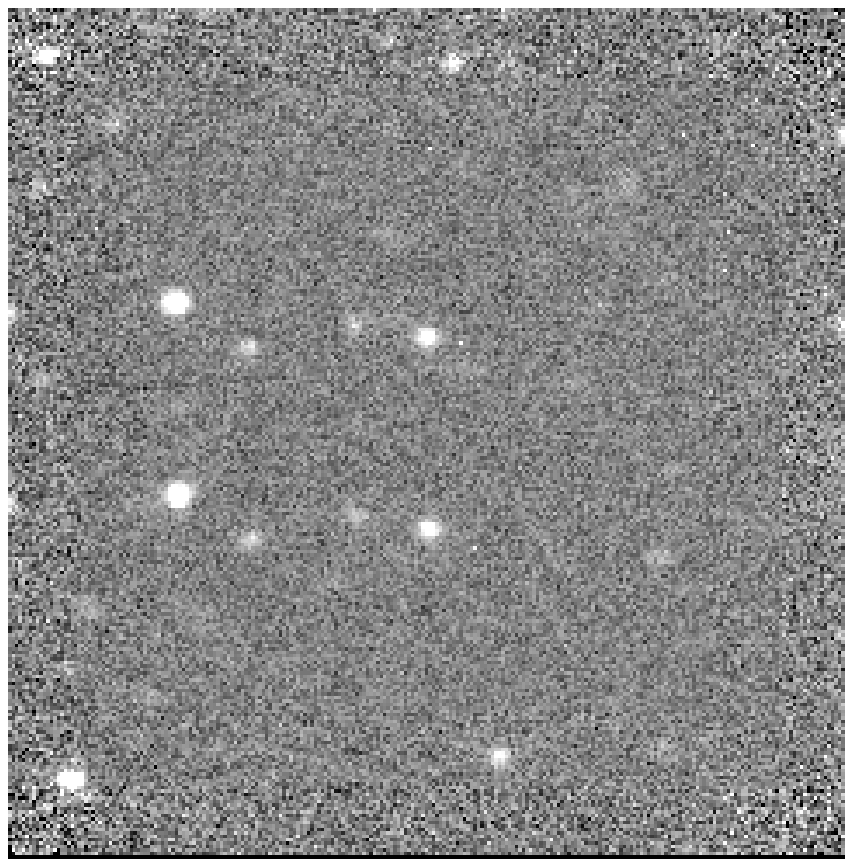,width=85mm}}
\centerline{\psfig{file=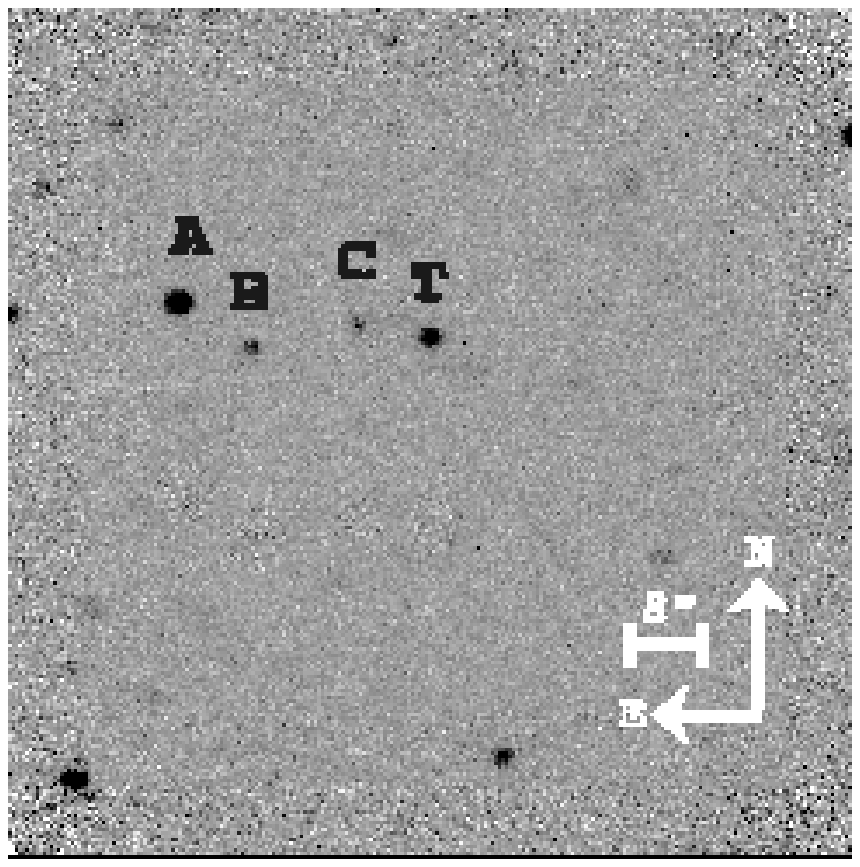,width=85mm}}
\caption{Images of 3C 22.}
Raw (above) and annotated negative (below) $K$-band structure 
of 3C 22 -- lower 
channel objects have been edited out of the negative image.
\label{22figs} \end{figure}

 Radio galaxy 3C 22 appears close to three other starlike objects which
we have designated A, B and C (see Figure \ref{22figs}); our star A is
object C in the
notation of Riley, Longair \& Gunn \scite{Riley+80a}. UKIRT $K$-band
imaging shows a red companion about 4\arcsec\ to the south-west. This
is placed at bearing 237\degr\ from the core by D\&P's
moment analysis algorithm; they note that it was not
apparent in the optical image of McCarthy \scite{McCarthy-88a}. Our
$K$-band image shows no evidence for extended structure, though this
companion is clearly visible in the UKIRT $K$-band image of BLR-II,
and also (clearly resolved
as a separate object) in their {\em HST}\, image. At the resolution of
the {\em HST}, the true bearing of this companion is seen to be
208\degr\ $\pm$ 1\degr.

The most recent review of the radio, visible and infrared properties of
3C 22 (BLR-II \S3, and references therein) finds two slight
extensions in {HST}\, images of 3C 22
itself. They find that the nucleus can be fitted as a combination of a
point source and a de Vaucouleurs law, with the point source
contributing 37\% of the total $K$-band intensity (BLR-III).  
The $J-K$ colour of 3C 22 is typical of a radio galaxy at its redshift,
but the $R-K$ colour is one of the reddest of the 3CR subsample of
D\&P. The $\sim 0.5\arcsec$\ extensions
are interpreted (BLR-II) as a possible close companion,
marginally redder than the core, due south (bearing $\sim 180\degr$), and 
as a blue extension slightly south of west (bearing $\sim 250\degr$).

The radio position angle of 3C 22 is given as 102\degr\ (equivalently
282\degr) by Schilizzi, Kapahi \& Neff \scite{Schilizzi+82a} and
Jenkins, Pooley \& Riley \scite{Jenkins+77a}; the
8.4\,GHz VLA radio map of BLR-II confirms this to $\pm 1\degr$. 

\subsubsection*{Polarimetry}

 Polarimetry results for 3C 22, designated T (for Target), and the
three field objects, are reported in Table \ref{3C22objtab}. All three
companions produce normalised Stokes parameters around $1\sigma$. 3C 22
itself has a 95 per cent chance of truly being polarised, and its
debiased polarization is $3.3\pm1.4$\%. We expect that no more than
0.8\% is due to the interstellar medium; most of the polarization is
therefore intrinsic to the source. In the optimal 2.6\arcsec\ radius
aperture, the two extended features observed by the {\em HST}\, will be
included, but light from the red SW companion at 4\arcsec\ will not.

\begin{small}
\begin{table}
\centering
\caption{Normalised Stokes Parameters: Objects in 3C 22 field.}
\label{3C22objtab}
\begin{tabular}{|c|rr|rr|rrr|}
\hline  
Source & $q$ & $\sigma_q$   & $u$ & $\sigma_u$ 
&	Prob   &	$P$(\%)	&	$\theta(\degr)$ \\
A	&	-3.05	&	3.92	&	-3.84	&	4.03	
&	53.2	&	4.22	&	18.8	\\
B	&	-3.13	&	2.23	&	2.62	&	2.23
&	81.3	&	3.77	&	153.0	\\
C	&	-0.61	&	0.60	&	-0.72	&	0.58	
&	72.0	&	0.84	&	17.9	\\
T	&	-1.27	&	1.38	&	-3.16	&	1.38	
&	95.3	&	3.26	&	27.0	\\

\hline
\end{tabular}

\medskip

\parbox{153mm}{Key: Source objects are as identified in Figure 
\ref{22figs}, T is 3C
22; $q$ and $u$ are `best estimator' normalised Stokes parameters (\%)
with R-axis at $\eta_0$=83\degr. Prob (\%) is the probability of a 
nonzero polarization being present; $P$(\%) and $\theta(\degr)$ 
are the nominal debiased degree of polarization and its 
orientation.}
\end{table}
\end{small}

The measured orientation of the {\boldmath $E$}-vector is
${+27}^{+17}_{-71} \degr$ East of North. The error is
large, but at a nominal 27\degr\
$(= 207\degr)$ our measurement suggests that the true direction is more
likely to be perpendicular to the radio axis, than parallel to it. 
Comparing the polarization orientation with extended structure, we find
that the blue western optical extension is not remarkably close to
being perpendicular or parallel to the polarization orientation; the
red southern extension/companion might be in parallel alignment for a
plausible error in the orientation angle. The red companion to the SW
at bearing 208\degr\ effectively lies on the nominal polarization
axis. 

Jannuzi \scite{jpc} \cite{Elston+97a} has performed imaging polarimetry
on 3C22 at shorter optical wavelengths, and reports $3\sigma$ upper
limits in $V$ and $H$ of 5 per cent and 3 per cent respectively.

\subsection{3C 41}
\label{D3C41}

\subsubsection*{Structure}

\begin{figure}
\centerline{\psfig{file=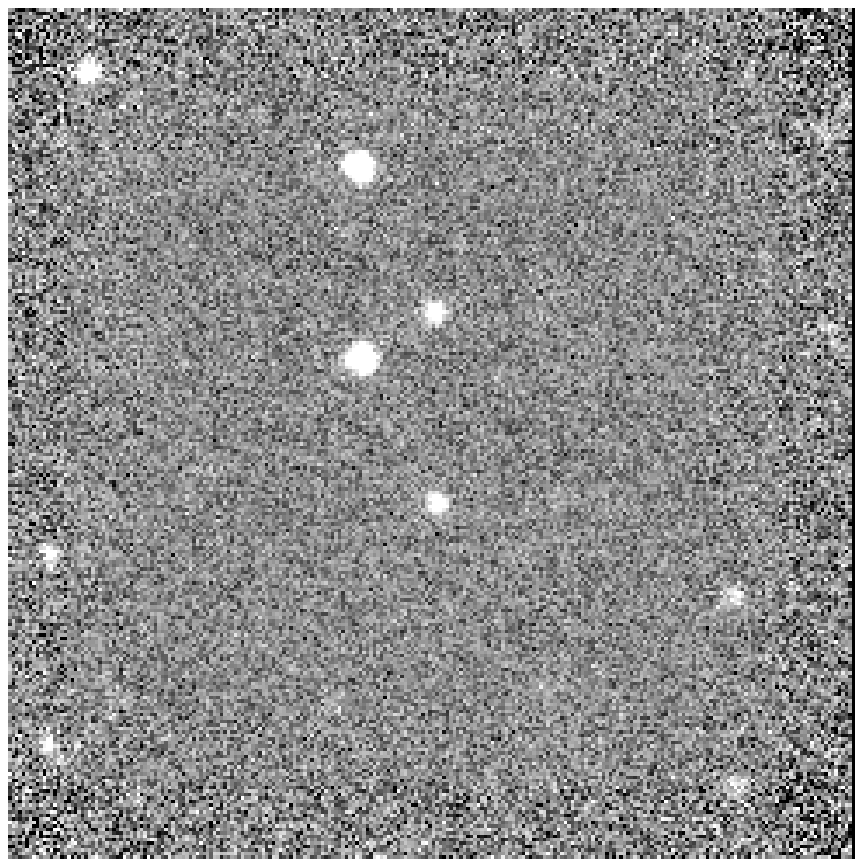,width=85mm}}
\centerline{\psfig{file=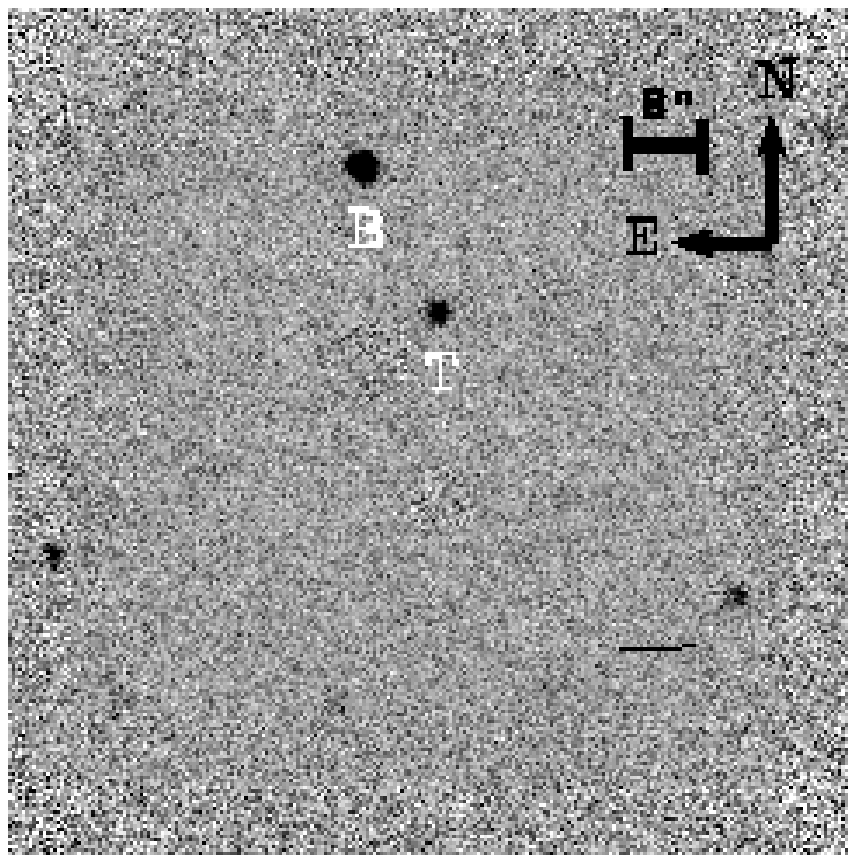,width=85mm}}
\caption{Images of 3C 41.}
Raw (above) and annotated negative (below) images
of 3C 41 -- lower channel objects have been edited out of the negative 
image.
 \label{41figs} \end{figure}

3C 41 appears with a field object labelled B (see Figure \ref{41figs})
following the notation of Riley, Longair \& Gunn \scite{Riley+80a}.  
The radio position angle of 3C 41 is 147\degr\ \cite{Longair-75a},
confirmed to $\pm 2\degr$ by the 8.4\,GHz VLA radio map of BLR-II.

Our $K$-band image shows no evidence for extended structure, but BLR-II
detect two distinct companions in their {\em HST}\, image which can
also be discerned as `extensions' in their UKIRT $K$-band image; the
WSW extension can also be distinguished in the $K$-band contour map of
Eisenhardt \& Chokshi \scite{Eisenhardt+90a}, who note that $K$-band
emission from the source extends for at least 12\arcsec. Both
companions are more than a magnitude bluer than the core: positioned
ESE and WSW of the core they lie on a line oriented at a position angle
of $127\degr \pm 2\degr$. They are hence misaligned with the radio axis
by about 20\degr\ (BLR-II).

\subsubsection*{Polarimetry}

Normalised Stokes parameter measurements for 3C 41 (Target T) and field
object B are given in Table \ref{3C41objtab}.
3C 41 has a 98 per cent chance of having a nonzero underlying
polarization,
which we measure to be 3.1 $\pm$ 1.1 per cent. Our upper limit for
extinction-induced polarization is only 0.1 per cent, so we are
confident of having detected intrinsic polarization in this object. The
orientation of the {\boldmath $E$}-vector is ${+45}^{+79}_{-14} \degr$
East of North.

\begin{small}
\begin{table}
\centering
\caption{Normalised Stokes Parameters: Objects in 3C 41 field.}
\label{3C41objtab}
\begin{tabular}{|c|rr|rr|rrr|}
\hline  
Source & $q$ & $\sigma_q$   & $u$ & $\sigma_u$ 
&       Prob   &    $P$(\%) &       $\theta(\degr)$ \\
B	&	-0.27	&	0.16	&	0.18	&	0.16  
&	86.7	&	0.30	&	156.6	\\
T	&	0.82	&	1.11	&	-3.08	&	1.11	
&	98.4	&	3.08	&	45.4	\\
\hline
\end{tabular}

\medskip

\parbox{153mm}{Key: Source objects are as identified in Figure 
\ref{41figs}, T is 3C
41; $q$ and $u$ are `best estimator' normalised Stokes parameters (\%)
with R-axis at $\eta_0$=83\degr. Prob (\%) is the probability of a 
nonzero polarization being present; $P$(\%) and $\theta(\degr)$
are the nominal debiased degree of polarization and its
orientation.}
\end{table}
\end{small}

 Jannuzi \scite{jpc} \cite{Elston+97a} have firm $V$ and $H$ band
polarizations for this source: at $V$, the polarization is 9.3 $\pm$
2.3\% at 58\degr $\pm$ 7\degr; at $H$, the polarization is 6.6
$\pm$ 1.6\% at 57\degr $\pm$ 7\degr. The {\boldmath $E$}-vector
orientations in the three wavebands are consistent with one another.  
We therefore find a very good perpendicular alignment between the radio
structure and optical polarization axes; the small errors on the $V$
and $H$ polarizations show that their alignment is perpendicular to the
radio structure rather than the {\em HST}\, visible structure. Our
$K$-band polarization orientation error is large enough to permit it to
be perpendicular to the optical structure rather than the radio; but
Occam's razor invites us to assume that the true orientation in $K$
should correspond to that in $V$ and $H.$

\subsection{3C 54}
\label{D3C54}

\subsubsection*{Structure}

\begin{figure}
\centerline{\psfig{file=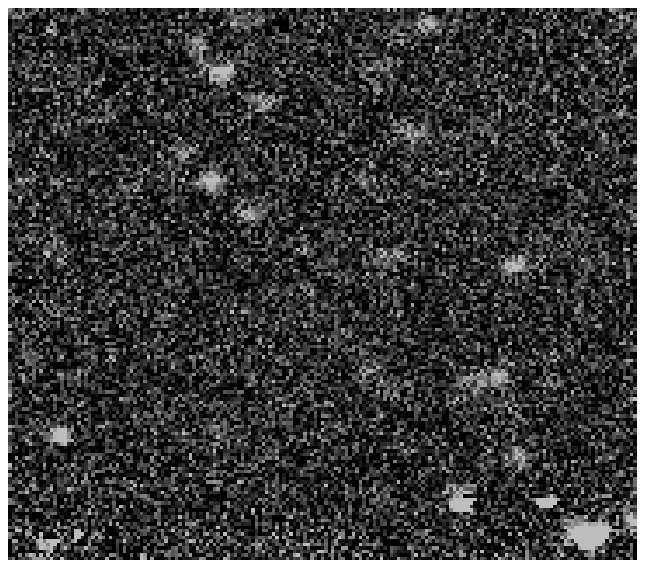,width=85mm}}
\centerline{\psfig{file=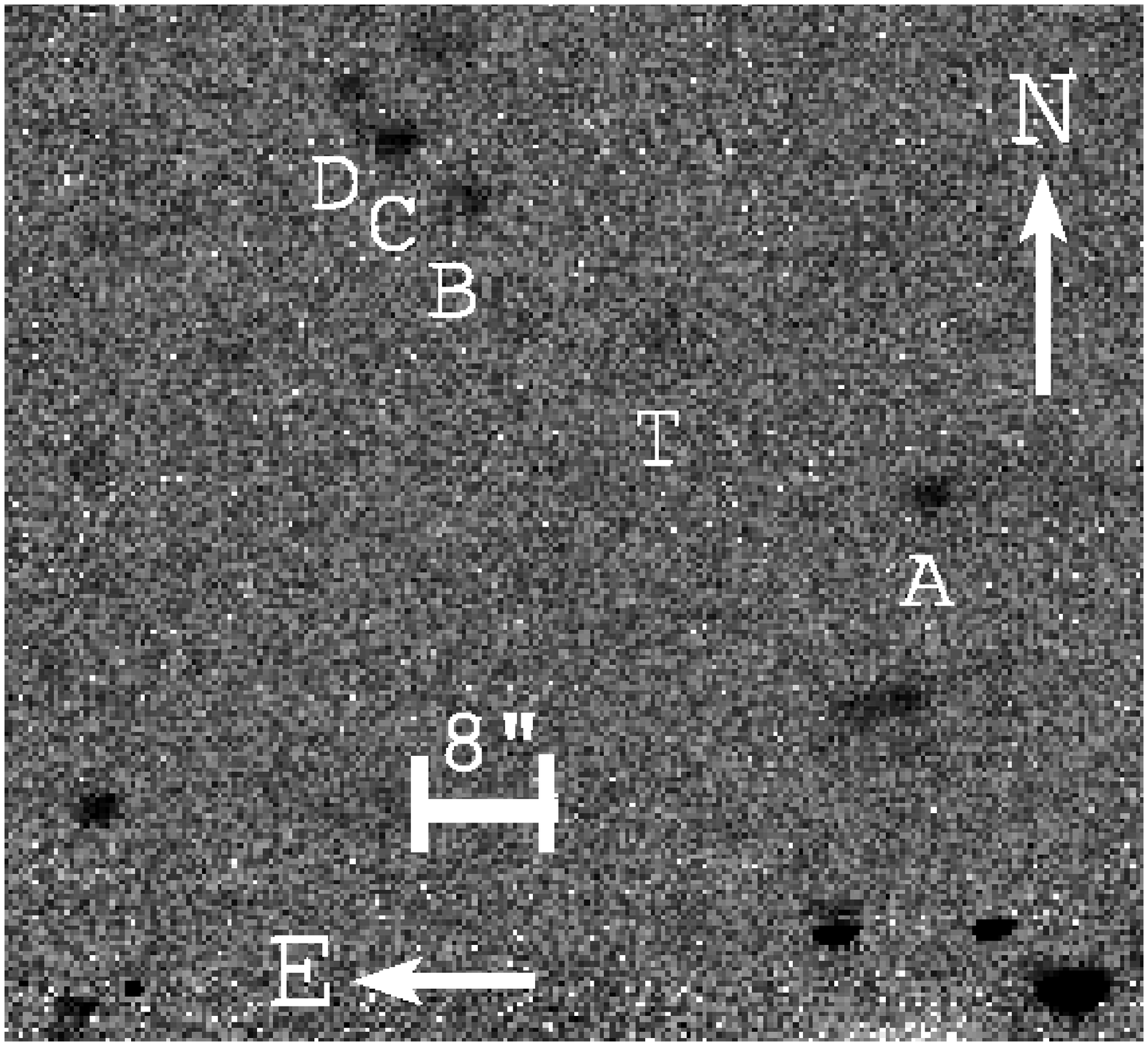,width=85mm}}
\caption{Images of 3C 54.}
Raw (above) and annotated negative (below) images 
of 3C
54 -- lower channel objects have been edited out of the negative image.
\label{54figs}
\end{figure}

No wide-aperture image of 3C 54 and its surrounding field could be
found in the literature; the target was identified on the grounds that
it lay close to the nominal position at the centre of our UKIRT images
and displayed a slight southern extension corresponding to that seen in
D\&P's $K$-band contour map. This extension is known to be more
prominent in $K$ than in $J$ \cite{Eisenhardt+90a}, and is also known
as structure $b$ at a bearing of 200\degr\ in the $R$-band
\cite{McCarthy+87a}.
Bright field objects have been designated A to D as indicated in Figure
\ref{54figs}; the target object itself may be seen more clearly in
Figure \ref{54image}.

The position angle of 3C 54's radio structure is 24\degr\
\cite{Longair-75a}. A moment analysis gives the $K$-band structure's
major axis orientation as 27\degr\ (D\&P), essentially
parallel to the radio axis; the visible continuum structure is reported
as very similar to the infrared, while the [O\,{\sc ii}] emission is
less similar, although elongated in the same sense
\cite{McCarthy+87a,McCarthy-88a}. The $J-K$ colour is typical of 
similar objects at the same redshift \cite{Eisenhardt+90a}.

\subsubsection*{Polarimetry}

The target and those field objects (A, B) bright enough to be analysed
have their normalised Stokes parameters reported in Table
\ref{3C54objtab}. The probability that 3C 54 is polarised is 94 per
cent; our measured value of polarization is 5.9 $\pm$ 2.6\% at
${+32}^{+17}_{-79} \degr$ East of North. Dust is not expected to
contribute more than 0.3\%. We therefore appear to have a
genuine polarization oriented parallel to both the radio and extended
optical structures of this source.

\begin{small}
\begin{table}
\centering
\caption{Normalised Stokes Parameters: Objects in 3C 54 field.}
\label{3C54objtab}
\begin{tabular}{|c|rr|rr|rrr|}
\hline  
Source & $q$ & $\sigma_q$   & $u$ & $\sigma_u$ 
&       Prob   &    $P$(\%) &       $\theta(\degr)$ \\
A	&	1.82	&	2.75	&	2.41	&	3.17	
&	39.3	&	1.08	&	109.5	\\
B	&	5.82	&	2.50	&	-0.35	&	2.66	
&	93.4	&	5.54	&	81.3	\\
T	&	-1.36	&	2.52	&	-6.01	&	2.55	
&	94.6	&	5.88	&	31.6	\\
\hline
\end{tabular}

\medskip

\parbox{153mm}{Key: Source objects are as identified in Figure 
\ref{54figs}, T is 3C 54; $q$ and $u$ 
are `best estimator' normalised Stokes parameters (\%) with 
R-axis at $\eta_0$=83\degr. Prob (\%) is the probability of a   
nonzero polarization being present; $P$(\%) and $\theta(\degr)$
are the nominal debiased degree of polarization and its
orientation.}
\end{table}  
\end{small}

\subsection{3C 65}
\label{D3C65}

\subsubsection*{Structure}

\begin{figure}
\centerline{\psfig{file=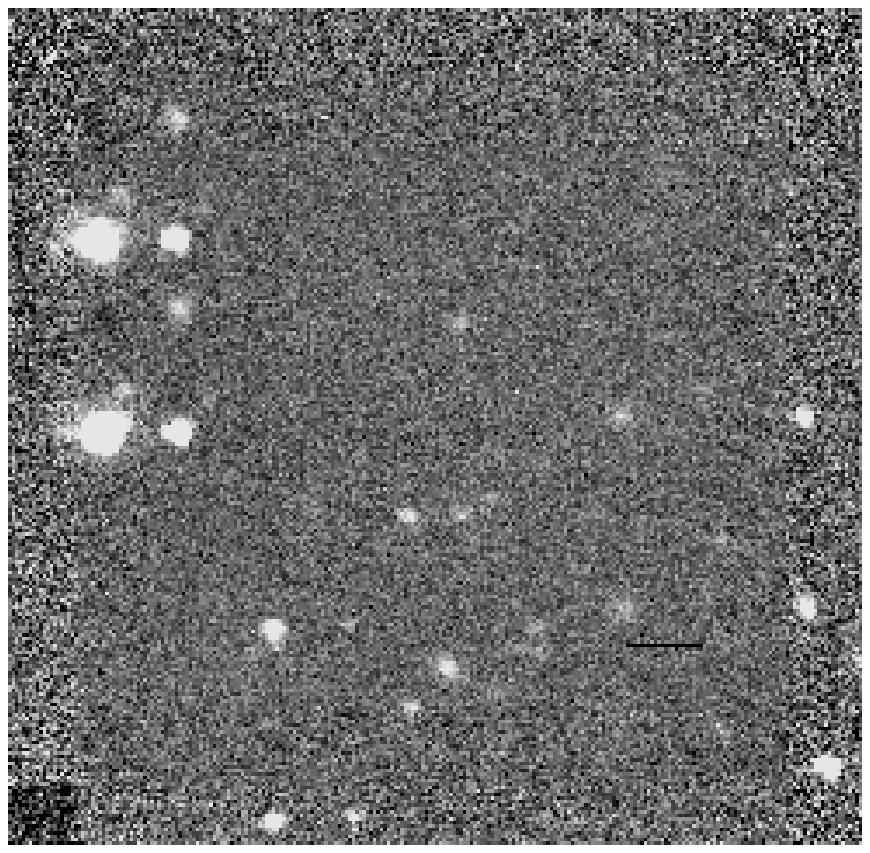,width=85mm}}
\centerline{\psfig{file=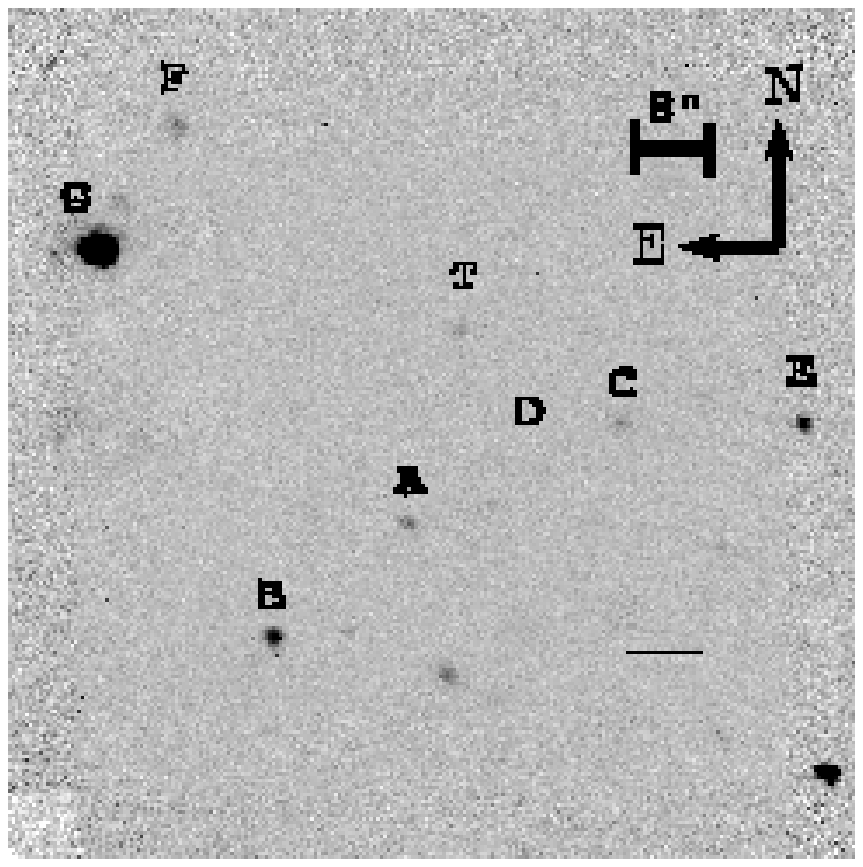,width=85mm}}
\caption{Images of 3C 65.} Raw (above) and annotated negative (below) 
images of 
3C 65 -- lower channel objects have been edited out of the negative 
image.
\label{65figs} \end{figure}

 Source 3C 65 was identified using the chart provided by Gunn et al.\
\scite{Gunn+81a}, and its field objects have been designated A--G as
indicated in Figure \ref{65figs}.

BLR-I describe 3C 65 as a fairly round central object, and one of the
reddest in the 3CR sample with $V-K\sim 6$.  A 4000\,\AA\ break in the
off-nuclear spectrum \cite{Lacy+95a,Stockton+95a} indicates the
presence of an old stellar population, $\sim 3-4$ Gyr. Their {\em
HST}\, visible image shows that 3C 65 is slightly elongated NE--SW.
Both their {\em HST}\, and UKIRT images show a blue companion galaxy
3\arcsec\ to the west, lying approximately on the radio axis, but this
was too faint to be distinguished from the noise on our $K$-band
images. Their 8.4\,GHz VLA radio map shows a radio structure at
position angle $100\degr \pm 3\degr$.

 Lacy et al.\ \scite{Lacy+95a} have claimed evidence for an infrared 
point source, possibly an obscured quasar nucleus, in the infrared core
of 3C 65; Rigler \& Lilly \scite{Rigler+94a}, however, found that the
infrared profile could be satisfactorily fitted by a de Vaucouleurs
law. BLR-II\,\&\,III's {\em
HST}\, image of 3C 65 also yields an adequate fit for a cD galaxy (de
Vaucouleurs law plus faint halo) model.

\subsubsection*{Polarimetry}

 Field object D proved too faint for accurate photometry; normalized
Stokes parameters for 3C 65 (target T) and the other field objects are
reported in Table \ref{3C65objtab}. Our polarimetry indicates a 57\%
probability that 3C 65 is an {\em unpolarized}\, source; our nominal
polarization orientation angle (from Table \ref{obstab}) is
perpendicular to the radio axis, but the statistical significance of
our $K$-band polarization measurement is at best dubious.

\begin{small}
\begin{table}
\centering
\caption{Normalised Stokes Parameters: Objects in 3C 65 field.}
\label{3C65objtab}
\begin{tabular}{|c|rr|rr|rrr|}
\hline  
Source & $q$ & $\sigma_q$   & $u$ & $\sigma_u$ 
&       Prob   &    $P$(\%) &       $\theta(\degr)$ \\
A	&	-0.99	&	4.00	&	-1.51	&	6.58	
&	4.6	&	0.09	&	21.3	\\
B	&	4.21	&	15.12	&	1.97	&	14.73	
&	4.6	&	0.23	&	95.5	\\
C	&	2.12	&	4.32	&	2.95	&	4.20	
&	30.6	&	1.27	&	110.2	\\
E	&	-0.07	&	2.29	&	2.41	&	2.21	
&	44.9	&	1.87	&	128.8	\\
F	&	-0.71	&	5.16	&	-2.81	&	3.07	
&	33.0	&	1.00	&	30.9	\\
G	&	-0.05	&	0.41	&	0.14	&	0.44	
&	5.4	&	0.01	&	137.2	\\
T	&	-4.33	&	4.25	&	-1.17	&	4.03	
&	42.9	&	2.12	&	0.6	\\

\hline
\end{tabular}

\medskip

\parbox{153mm}{Key: Source objects are as identified in Figure 
\ref{65figs}, T is 3C
65; $q$ and $u$ are `best estimator' normalised Stokes parameters (\%)
with R-axis at $\eta_0$=83\degr.  Prob (\%) is the probability of a   
nonzero polarization being present; $P$(\%) and $\theta(\degr)$
are the nominal debiased degree of polarization and its
orientation.}
\end{table}
\end{small}

\subsection{3C 114}
\label{D3C114}

\subsubsection*{Structure}

\begin{figure}
\centerline{\psfig{file=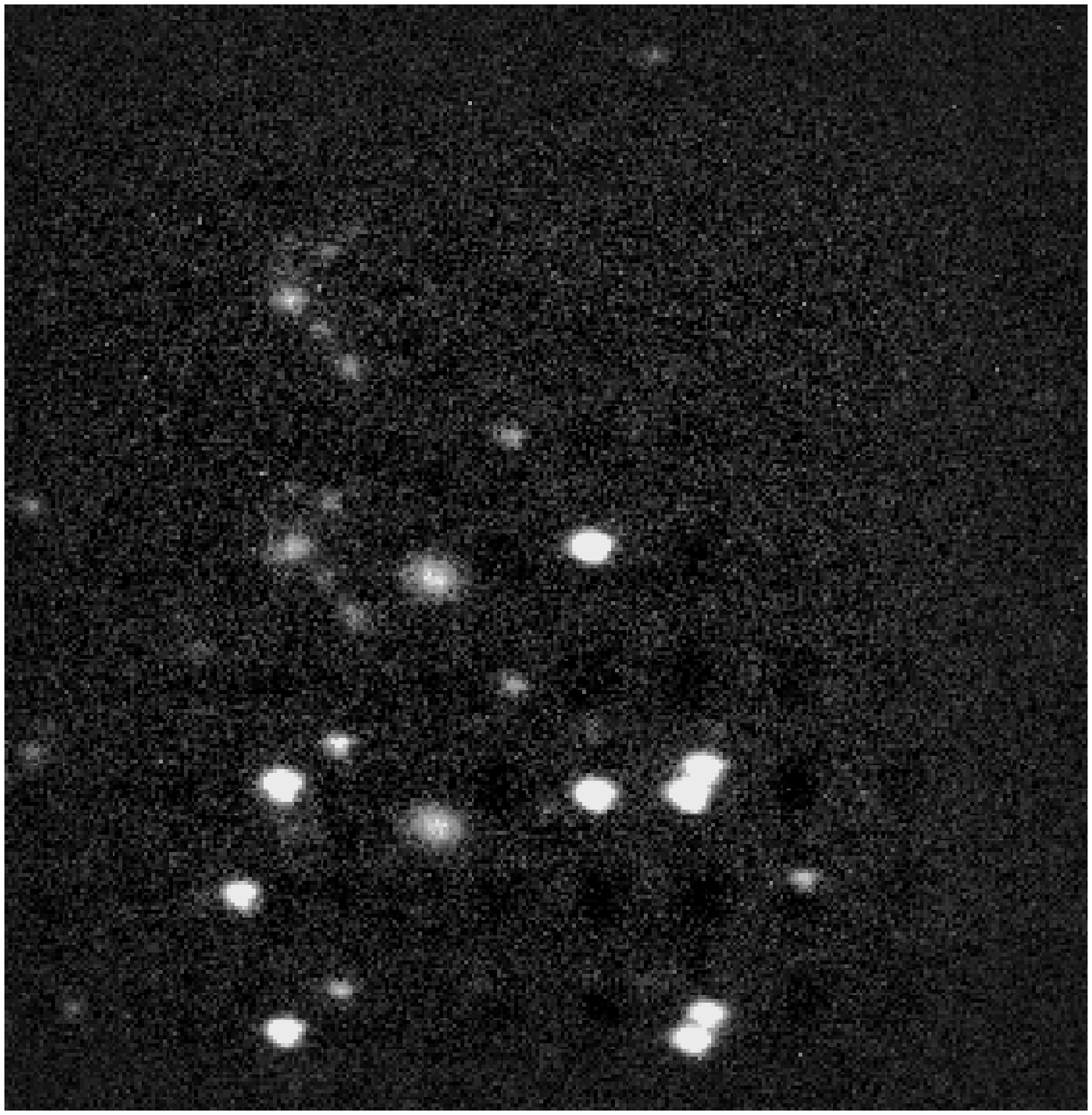,width=85mm}}
\centerline{\psfig{file=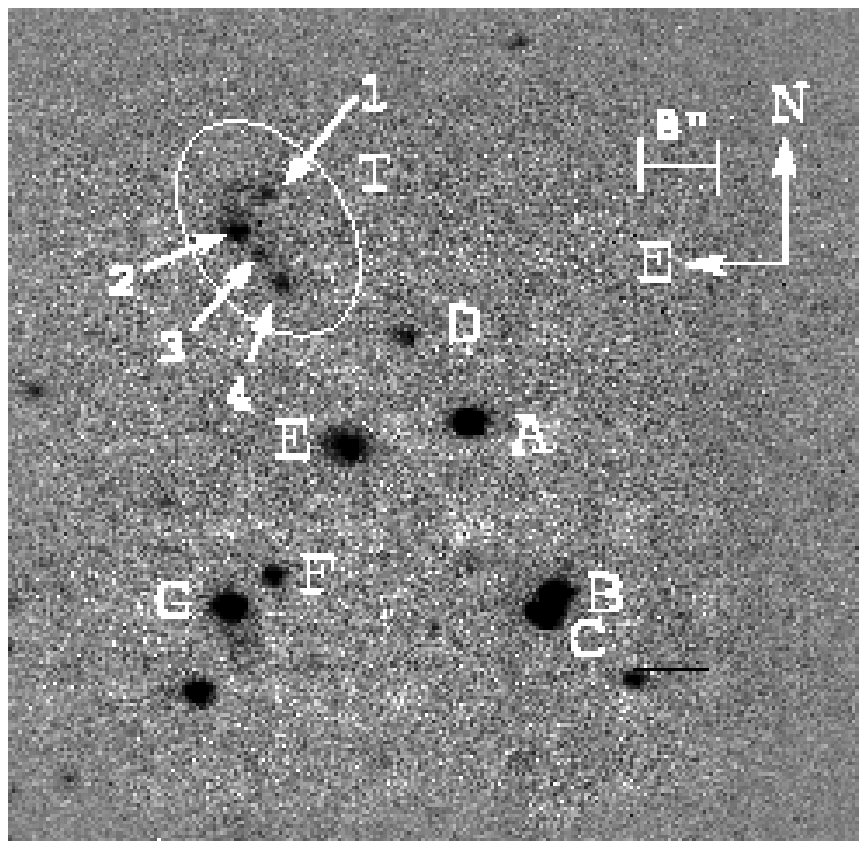,width=85mm}}
\caption{Images of 3C 114.}
Raw (above) and annotated negative (below) images
of 3C 114
-- lower channel objects have been edited out of the negative 
image.
\label{114figs} \end{figure}

\begin{figure}
\centerline{\psfig{file=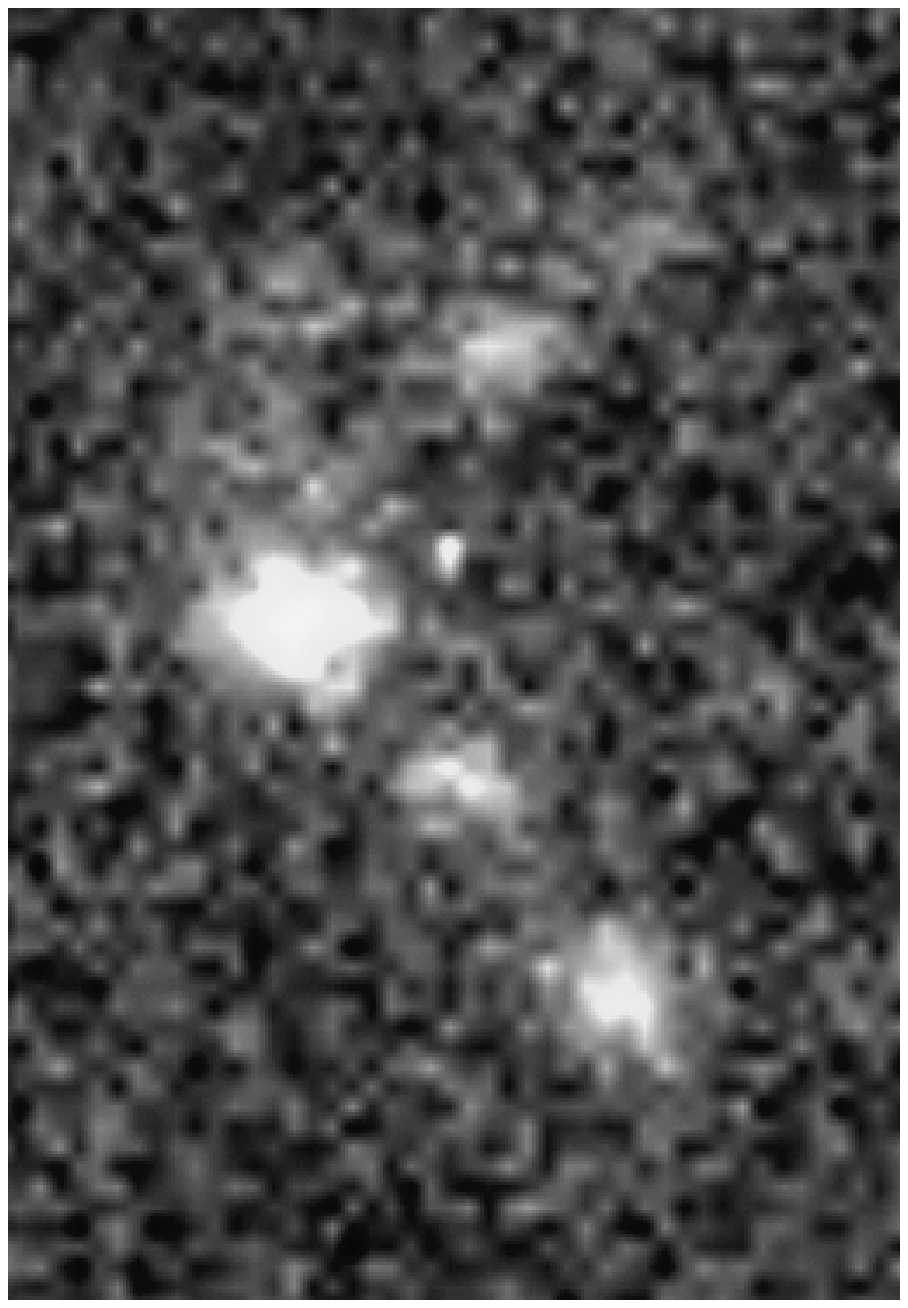,width=85mm}}
\caption{Close-up of the $K$-band structure of 3C 114.} (North is at the 
top, East is on the left.)
\label{114image}
\end{figure}

The field containing 3C 114 is illustrated in Figure \ref{114figs}; a
close-up image detailing the structure of knots is shown as Figure
\ref{114image}. Its distinctive shape (D\&P) makes it easily identifiable;
it consists of at least four knots, where the brightest knot forms the
knee of a $\Gamma$-shaped structure. The radio position angle of 3C 114 is
44\degr\ \cite{Strom+90a}, and the large size of the radio structure
(54\arcsec) is noteworthy. 

D\&P's $K$-band contour map reveals
more detail in the structure of 3C 114 with additional knots of lower
intensity close to the four obvious ones in our image; the three major 
knots forming the NE--SW line seem to be joined by an underlying
luminous structure. This structure dominates a moment analysis of the
whole source, yielding a $K$-band optical structure angle of 52\degr,
close to the 44\degr\ for radio structure.

\subsubsection*{Polarimetry}

Only the knee knot proved bright enough to
analyze on its own; Table \ref{polvals} includes results for both that
knot and the structure as whole. There is a probability in excess of 99
per cent that there is genuine polarization in both the knee knot and
the overall structure.

\begin{small}
\begin{table}
\centering
\caption{Normalised Stokes Parameters: Objects in 3C 114 field.}
\label{3C114objtab}
\begin{tabular}{|c|rr|rr|rrr|}
\hline  
Source & $q$ & $\sigma_q$   & $u$ & $\sigma_u$ 
&       Prob   &    $P$(\%) &       $\theta(\degr)$ \\
A	&	0.77	&	0.23	&	0.12	&	0.29	
&	99.7	&	0.77	&	87.3	\\
B	&	5.56	&	3.20	&	0.78	&	3.21	
&	78.5	&	5.14	&	87.0	\\
C	&	5.55	&	3.20	&	0.67	&	3.21	
&	78.3	&	5.12	&	86.5	\\
D	&	0.12	&	2.86	&	0.43	&	2.85	
&	1.2	&	$\ddag$	&	$\ddag$	\\
E	&	-0.80	&	1.12	&	-1.39	&	1.13	
&	63.4	&	1.41	&	23.1	\\
F	&	1.90	&	2.81	&	3.29	&	2.74	
&	61.2	&	3.32	&	113.0	\\
G	&	5.28	&	3.31	&	-0.18	&	3.14	
&	72.0	&	4.75	&	82.0	\\
T0	&	11.19	&	3.07	&	-4.14	&	2.68
&	100.0	&	11.73	&	72.9	\\
T2	&	3.00	&	1.63	&	4.31	&	1.77	
&	99.0	&	5.10	&	110.6	\\
\hline
\end{tabular}

\medskip

\parbox{153mm}{Key: Source objects are as identified in Figure 
\ref{114image}, T0 is the whole of 3C
114 and T2 its `knee' knot; $q$ and $u$ are `best estimator' normalised 
Stokes parameters 
(\%) with R-axis at $\eta_0$=83\degr.  Prob (\%) is the probability of 
a nonzero polarization being present; $P$(\%) and $\theta(\degr)$
are the nominal debiased degree of polarization and its
orientation. $\ddag$: These data are not given for object D; it seems 
unpolarized and our debiasing algorithm did not converge to a solution.}
\end{table}
\end{small}

 We measured polarization in the knee of 5.1 $\pm$ 1.7 per cent, at
${+111}^{+14}_{-63} \degr$ East of North. Overall, the whole object has
a polarization of 11.7 $\pm$ 3.0 per cent at ${+73}^{+34}_{-20} \degr$.
The extinction contribution could be as high as 0.9 per cent, but our
detections of polarization are much higher than this, so the
polarization appears to be intrinsic. With the radio and optical
structure axes at $\sim 48\degr$, there is no clear alignment of
polarization either parallel or perpendicular to the structure. 

\subsection{3C 356}
\label{D3C356}

\subsubsection*{Structure}

 Source 3C 356 has provoked much discussion in the literature. The
radio structure is a large (72\arcsec) double with position angle
161\degr\ \cite{Leahy+89a}. D\&P's $K$-band contour map displays three
knots: two brighter knots lying NW--SE along the radio axis, and a much
fainter knot off-axis to the south-west. Both bright knots are
associated with radio cores \cite{Fernini+93a}, and it is not clear
which hosts the radio source; both lie at $z=1.079$ (BLR-II). In the
convention established by LeF\`{e}vre, Hammer \& Jones
\scite{LeFevre+88b}, and used by others
\cite{Eales+90a,Eisenhardt+90a,Lacy+94a,Cimatti+97a}, the NW component is 
denoted $a$
and the SE component, $b$. Fainter components are labelled following 
BLR-II.

The SE component, $b$, is elongated roughly perpendicularly to the
radio source \cite{Rigler+92a}. Lacy \& Rawlings \scite{Lacy+94a} 
note how this SE radio core
has a flatter spectral index and may be the host to the radio source,
with galaxy $a$ interacting with the jet; $b$'s spectral index is
$\alpha \approx 0.1$ between 8.4\,GHz and 5\,GHz (BLR-II), which is
typical of a compact core in an extended radio source.
Eales \& Rawlings \scite{Eales+90a} also favoured $b$ as the radio core
due to its colour, magnitude and shape being typical of radio galaxies.

The more recent {\em HST}\, observations of BLR-I\,\&\,II, however,
reveal the NW
component $(a)$ to have the same dumbbell morphology they observe in
other radio galaxy hosts, while the SE object $(b)$ seems much more
diffuse than their other 3CR sources. Component $a$ is also favoured
as the radio core by Eisenhardt \& Chokshi \scite{Eisenhardt+90a} and
McCarthy \scite{McCarthy-88a}: it has bluer infrared--optical colours
and dominates the visible continuum and [O\,{\sc ii}] images. D\&P
dispute LeF\`{e}vre et al.'s \scite{LeFevre+88a} claim that $b$ has the
bluer colours; {\em HST}\, observations show that $b$ is redder, but
some of the diffuse emission is as blue as $a$
\cite[BLR-II]{Eisenhardt+90a}.
Component
$a$'s spectral index is that of a compact steep spectrum source,
$\alpha \approx 1.1$, and the 8.4\,GHz radio flux is only a quarter of
that of component $b$ (BLR-II).

BLR-III find that 3C 356 (presumably meaning the NW component, measured
in a 5\arcsec\ aperture), can be well modelled purely by a de
Vaucouleurs profile. A 4000\,\AA\ spectral break has been detected in
both $a$ and $b$ \cite{Lacy+94a}, indicating that both components
contain stars and are aged at least 10 Myr. D\&P's $K$-band image
moment analysis algorithm gives a position angle of 159\degr\ for the
overall structure, but Cimatti et al.\ \scite{Cimatti+97a} state that
the two $K$-band knots taken together as a single structure lie at a
position angle of 145\degr, with the two dumbbell components of $a$ 
separated along a line at 152\degr.

\begin{figure}
\centerline{\psfig{file=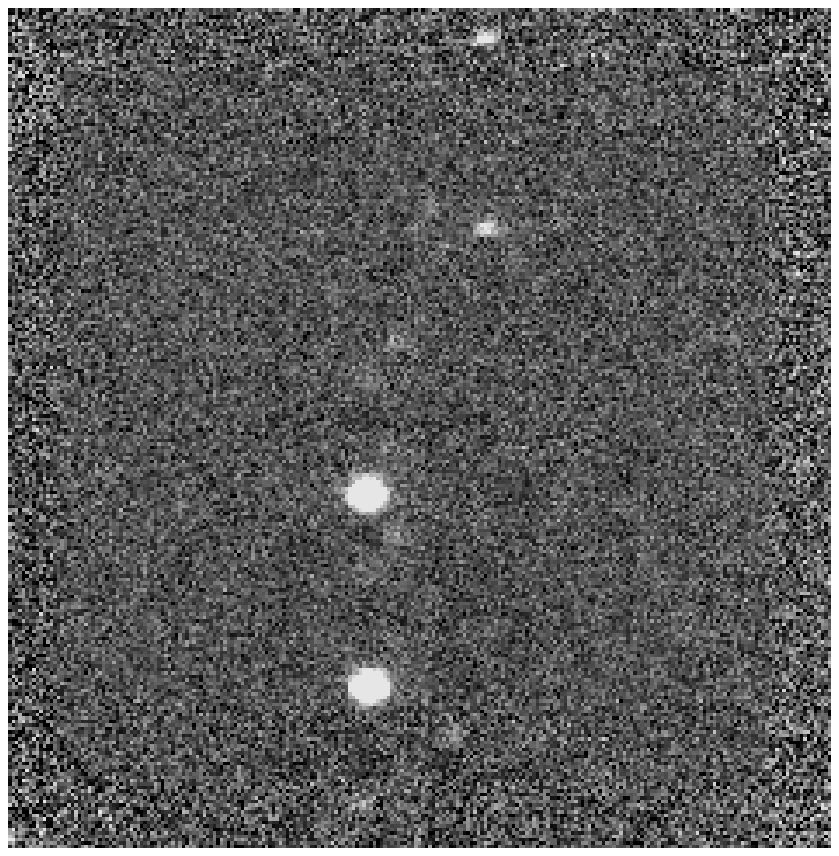,width=85mm}}
\centerline{\psfig{file=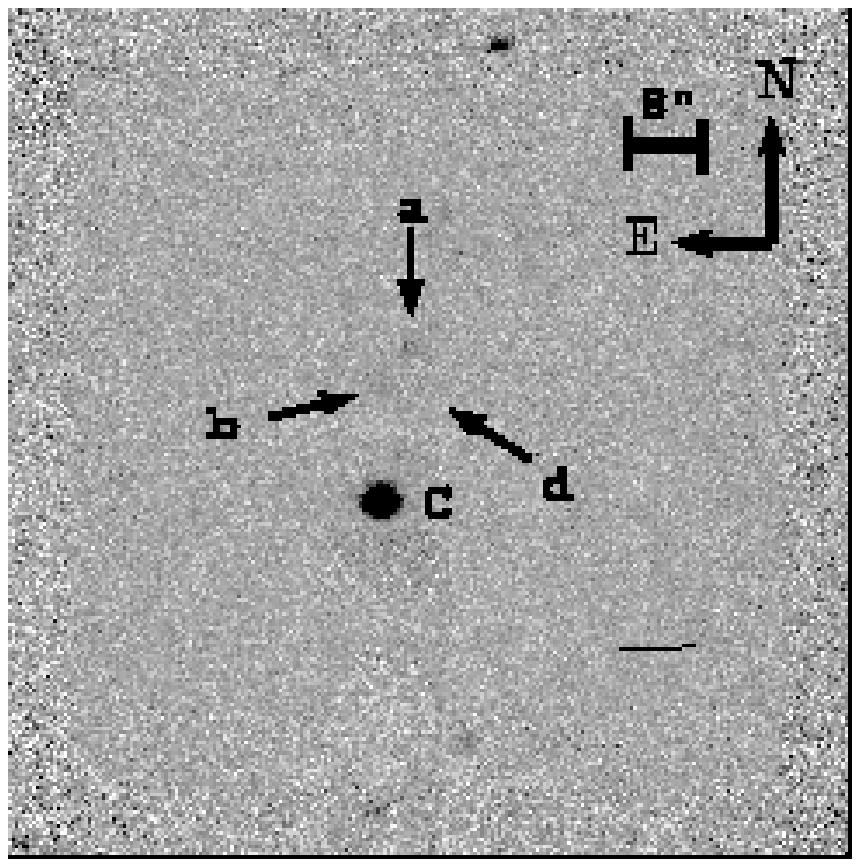,width=85mm}}
\caption{Images of 3C 356.} Raw (above) and annotated negative (below) 
images of 
3C 356 -- lower channel objects have been edited out of the negative 
image.
\label{356figs} \end{figure}

Our $K$-band image of 3C 356 (Figure \ref{356figs}) reveals the three
knots indicated: $b$ at the south-east, $a$ at the north-west,
and the very faint south-west component denoted $d$ in
the {\em HST}\, image of BLR-II. A prominent field star, object C of
Riley, Longair
\& Gunn \scite{Riley+80a}, is also labelled.

\subsubsection*{Polarimetry}

Cimatti et al.\ \scite{Cimatti+97a} have obtained Keck I
spectropolarimetry of 3C 356's two main components for light emitted
between 200\,nm and 420\,nm. They find polarization which rises towards
the ultraviolet: source $b$'s polarization was low, reaching $4.0 \pm
1.2 \%$ at 200\,nm, while $a$'s polarization rose from 3\% at 420\,nm
to about 15\% at 200\,nm. A distinct Mg\,{\sc ii}\,$\lambda$2800 line
was also detected in component $a$'s light, polarized to the same
degree as the continuum, and with the same orientation of 64\degr.

 The perpendicular axis to the polarization vector lies at 154\degr,
and is therefore within two degrees of the dumbbell separation observed
in object $a$, nine degrees anticlockwise of the radio structure, and
seven degrees clockwise of the $a$-$b$ axis.

 Our $K$-band images show the galaxy in light emitted in a band centred
on 1060\,nm. The SW knot (object $d$) was not bright enough to permit
polarimetry. Normalised Stokes parameters for the two major components
($a$, $b$), the 3C 356 complex as a whole (T0) and field object C are
recorded in Table \ref{3C356objtab}. We find no strong evidence for
$K$-band polarization in 3C 356; the probability of non-zero $K$-band
polarization being present in $a$ and $b$ is 62\% and 79\%
respectively, and the polarization orientations derived from our noisy 
measurements do not include 64\degr\ in their $\pm 1\sigma$ error
boxes.

\begin{small}
\begin{table}
\centering
\caption{Normalised Stokes Parameters: Objects in 3C 356 field.}
\label{3C356objtab}
\begin{tabular}{|c|rr|rr|rrr|}
\hline  
Source & $q$ & $\sigma_q$   & $u$ & $\sigma_u$ 
&       Prob   &    $P$(\%) &       $\theta(\degr)$ \\
C	&	0.15	&	0.15	&	0.27	&	0.21	
&	71.0	&	0.27	&	113.4	\\
T0	&	-9.64	&	4.94	&	0.35	&	5.63	
&	85.1	&	8.97	&	172.0	\\
$a$	&	-13.32	&	7.95	&	4.60	&	8.38	
&	62.4	&	18.76	&	23.9	\\
$b$	&	-10.10	&	9.39	&	-18.88	&	16.62	
&	78.9	&	12.90	&	163.5	\\
\hline
\end{tabular}

\medskip

\parbox{153mm}{Key: Source objects are as identified in Figure 
\ref{356figs}, T0 is the whole structure of 3C 356; $q$ and $u$ 
are `best estimator' normalised Stokes parameters (\%) with 
R-axis at $\eta_0$=83\degr.  Prob (\%) is the probability of a   
nonzero polarization being present; $P$(\%) and $\theta(\degr)$
are the nominal debiased degree of polarization and its
orientation.}
\end{table}  
\end{small}

\subsection{3C 441}
\label{D3C441}

\subsubsection*{Structure}

\begin{figure}
\centerline{\psfig{file=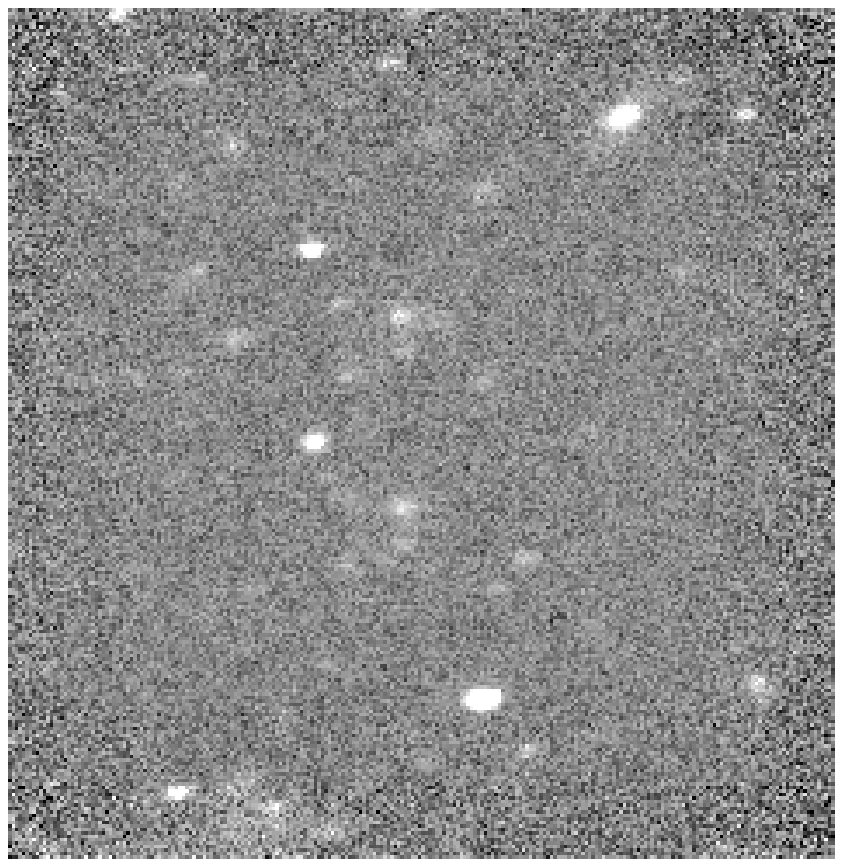,width=85mm}}
\centerline{\psfig{file=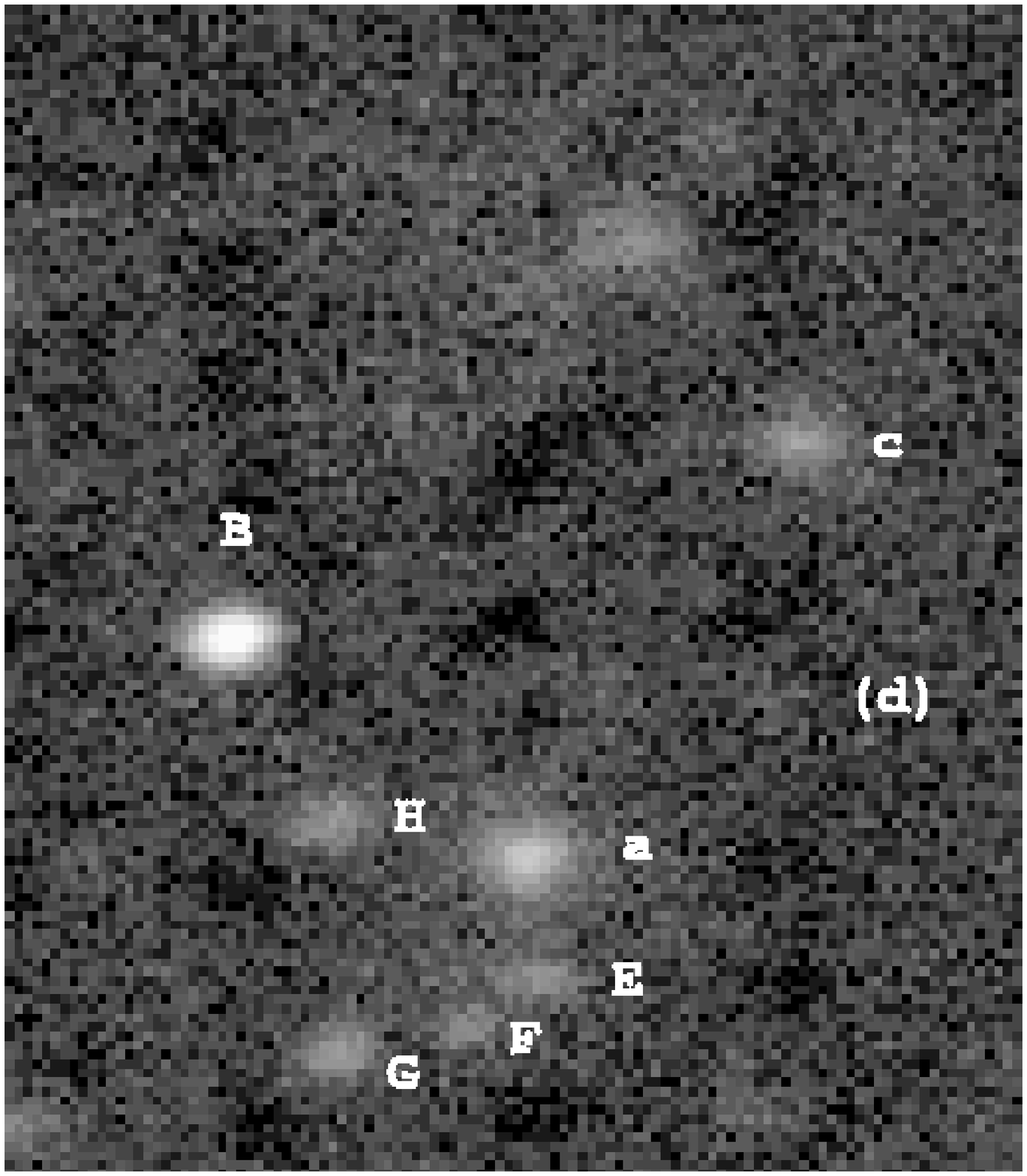,width=85mm}}
\caption{Images of 3C 441.} Raw (above) and annotated negative (below) 
images of 3C 441 --
 lower channel objects have been edited out of the negative image.
North is at 
the top and East to the left; 
the scalebar is given in Figure \ref{Cpic}. \label{441figs} \end{figure}

\begin{figure*}
\psfig{file=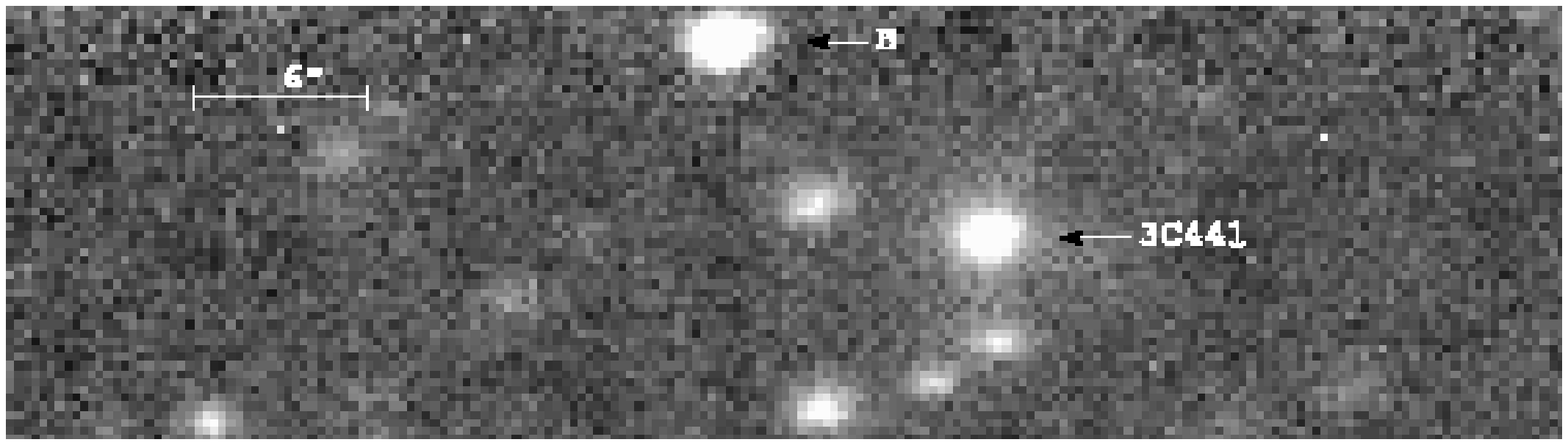,width=153mm}
\caption{3C 441: stacked \Kb-band image of August 1995 and August 1997 
data.}
Total integration time 4 hours 16 minutes. [1 hour 52 minutes (1997) plus 2
hours 24 minutes (1995).]  North is at the
top, East at the left. The star B is labelled as in Riley, Longair \&
Gunn \protect\scite{Riley+80a}. \label{Cpic} 
\end{figure*}

3C 441 appears in a rich field (Figure \ref{441figs}) with five
neighbours;  identification of the radio core is based on the
observations of Riley, Longair \& Gunn \scite{Riley+80a} and is
apparently confirmed by the work of McCarthy \scite{McCarthy-88a} and
of Eisenhardt \& Chokshi \scite{Eisenhardt+90a}. Figure \ref{441figs}
is based on our August 1995 data (L\&E) which was taken without a focal
plane mask.  The objects labelled {\bf a} and {\bf c}, and the position of 
unseen object {\bf d}, follow the notation of Lacy et al.\ 
\scite{Lacy+98a}; the
star B is labelled as in Riley et al. \scite{Riley+80a}, and the
remaining objects are labelled E thru H. An unedited image with scale
bar is given later as Figure \ref{Cpic}. Object {\bf a} itself is shown to
have 0.5 mag bluer extension protruding to the south-west in its {\em
HST}\, image (BLR-II). Object {\bf c} is more compact in the infrared than in
$R$-band imaging, while object {\bf a} appears more extended east-west in the
infrared than in $R$  \cite[BLR-II]{Eisenhardt+90a}.

 Fabry-Perot imaging by Neeser \scite{Neeser-96a} shows that none of
the other objects in the field lie within a velocity range $(-1460,
+1180)$ km\,s$^{-1}$ of 3C 441 itself, but since the $J-K$ colours of
most of these neighbouring objects lie between 1.6 and 1.85 (BLR-II),
3C 441 could be part of a cluster. Neeser \scite{Neeser-96a} questions
whether the identification of 3C 441 is correct -- arguing that it may
in fact be our object F. Recent imaging by Lacy et al.\
\scite{Lacy+98a} suggests that 3C 441's jet is impacting object {\bf c} to
its north-west -- this object has the same infrared colour as object
{\bf a} and is the only other area of strong [O\,{\sc ii}] emission in the
field \cite[BLR-II]{Eisenhardt+90a}.
 
The 5\,GHz radio map of 3C 441 \cite{Longair-75a} shows a double radio
source with the separation between sources running East-West, and
extended structure trailing off to the South-East. Lacy et al.\ 
\scite{Lacy+98a} show that the North-West jet (at 8\,GHz) is deflected to
the south where it would otherwise have encompassed object {\bf c}.

{\sc nb}  Any future worker planning to observe 3C441 should note that 
there is
a $z=4.4$ quasar which falls in the same field \cite{McCarthy+88a}. 
Judicious planning could enable studies of this object to be conducted in
parallel with the radio galaxy. 

\subsubsection*{Polarimetry}
\label{objE}

It is known that 3C 441 has a broad-band optical polarization of 1.5 
$\pm$ 0.7\% at 70\degr\ $\pm$ 13\degr\ in a 2\arcsec\ diameter 
aperture about the core: this orientation is
roughly perpendicular to the radio structure \cite{Tadhunter+92a}. Our
1995 and 1997 images were stacked together to obtain polarization
measurements of objects {\bf a} (the putative core of 3C 441) and E thru 
H; the 1995 data alone was used to obtain polarization data on B and {\bf 
c}. Given the uncertainty posited by Neeser \scite{Neeser-96a} over the
identification of 3C 441, and the interest in object {\bf c} of Lacy et al.\ 
\scite{Lacy+98a}, we performed polarimetry on all the objects on the
field, with the full results presented in Table \ref{obstab}.
 
 The only object with a strong indication (90 per cent chance
genuine) of polarization is E.
The orientation is 120\degr, which would be roughly parallel with
the radio jet --- but the position angle which E makes with the presumed
core {\bf a} is close to 0\degr, which means that a model of E scattering 
light from {\bf a} is possible. It would not be necessary for 
light from {\bf a}
to be beamed into E; if E subtends only a small solid angle of the light
emitted by {\bf a}, any light from {\bf a} scattered by E would be 
quasi-unidirectional.
 
 There is a weak indication that object {\bf c} might be polarised. If 
so, the best estimate is 3.5\% polarization at 124\degr\ -- an
orientation roughly parallel to any jet from {\bf a} which might be 
scattered into
our line of sight. Similarly, object H might possibly be polarised at 6.7\%,
11\degr, which is roughly perpendicular to its line of sight   
with {\bf a}, and parallel to that with B. We have, however, no redshift 
data  on any source other than {\bf a}, and therefore cannot eliminate chance
alignments should any of these sources be located at other
redshifts.
 
 For the presumed radio galaxy at {\bf a}, the best estimate polarization
is 1\% at 78\degr --- consistent with both the magnitude and orientation of
Tadhunter et al.'s \scite{Tadhunter+92a} broad-band visible measurement ---
but there is a 54 per cent chance that {\bf a} is unpolarised with this
result being merely an artifact of the noise. Even our measurement for E
has a ten percent chance of being a noise-induced spurious result.

\subsection{LBDS 53W091}
\label{D53W091}

\subsubsection*{Structure}

 The galaxy LBDS 53W091 has aroused great excitement in recent   
years. First investigated by Dunlop et al.\ \scite{Dunlop+96a} as an
extremely red radio source, it was found to be a very red radio galaxy
visible at a very high ($z=1.552$) redshift. Its spectrum exhibits
late-type absorption features, and no prominent
emission features. Comparisons of its spectrum with synthetic and real   
elliptical galaxies suggest that it must be at least 3.5 Gyr old
\cite{Dunlop+96a,Spinrad+97a} -- which is only consistent with its high
redshift in certain cosmologies, requiring a low density Universe 
$(\Omega \sim 0.2)$, or
else an unacceptably low Hubble constant $(H_0 \la
50\,$km\,s$^{-1}$\,Mpc$^{-1})$ in an $\Omega=1$ cosmology. [See Leyshon, 
Dunlop \& Eales \scite{Leyshon+99a} for further discussion of 53W091's age 
and internal chemistry, considering issues which do not affect the 
interpretation of its polarization.]
 
The lack of emission features suggests that the active nucleus
responsible for its $\sim 25$ mJy 1.4\,GHz radio emission contributes
very little light to the optical/ultraviolet;  Dunlop et al.\ 
\scite{Dunlop+96a} argue that the galaxy is unlikely to be an obscured
quasar. Yet unpublished evidence (Chambers, private communication)
suggests that 53W091 has a high infrared polarization --- of order 40\%
--- which would be extremely difficult to account for in an
object with such a weak active nucleus.

 Spinrad et al.\ \scite{Spinrad+97a} note that radio galaxies with
weak active nuclei ($S_{\rm 1.4GHz}<$ 50 mJy) generally are not expected to
be dominated by optical nuclear emission, and do not display the   
alignment effect. Their 4.86\,GHz radio map of 53W091 reveals a
double-lobed FR-II steep spectrum radio source, where the radio lobes are 
separated by approximately 4\farcs3 at position angle 131\degr.

\begin{figure*}
\centerline{\psfig{file=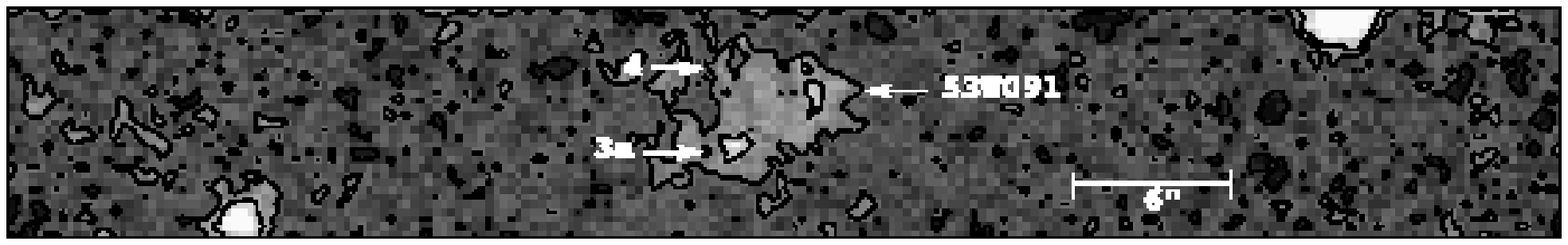,width=153mm}}
\caption{53W091: stacked \Kb-band image of August 1997 and July 1997 data.}
Total integration time 5 hours 56 minutes. [4 hours 4 minutes (August) plus
1 hour 52 minutes (July).]  (North is at the top, East at the left. The 
image has been overlaid with contours fitted for 8\%, 16\% and 24\% of the
peak intensity present.)
\label{Wpic}  \end{figure*}

 Our data of 53W091 were stacked together with earlier observations made
by Dr James Dunlop (private communication) in July 1997, and the total
image is seen in Figure \ref{Wpic}. The companion object to the south-east
of 53W091 is known to be at the same redshift, and is labelled `3a' in
accordance with the labelling of Spinrad et al.\ \scite{Spinrad+97a}. The
position of their third component at the same redshift is also marked
(labelled `4') although there is not a distinct source on our image. 
 
Two stars on our image (the star on the top right of Figure \ref{Wpic}
and a brighter one in the lower slot of the focal plane mask, not
shown) were identified with stars whose B1950 co-ordinates were
obtained from the {\em Digitized Sky Survey} \cite{Lasker+90a}.
Offsetting from these stars, the B1950 co-ordinates of the \Kb-band
sources were obtained and are given in Table \ref{positab}.

\begin{table}
\centering
\caption{B1950 co-ordinates of \Kb-band sources in 53W091 field.}
\label{positab}
\begin{tabular}{lcc}
\hline  
Source &  $\alpha$ & $\delta$\\
53W091 &  17h 21m 17\fs898 $\pm$ 0\fs057
& $+50\degr 08\arcmin 48\farcs34 \pm 0\farcs29$ \\
3a & 17h 21m 18\fs156 $\pm$ 0\fs029 &
 $+50\degr 08\arcmin 46\farcs34 \pm 0\farcs57$\\

\hline
\end{tabular}

\medskip

Key: $\alpha$: B1950 Right Ascension; $\delta$: B1950 Declination.
\end{table}  
 
Spinrad et al.\ \scite{Spinrad+97a} ask whether 3a and 53W091 together
might form a system displaying the alignment effect, but note that both
sources' colours suggest they are composed of old stars, for which
there is no plausible alignment mechanism. (They allow that some
interaction of jets from the active nucleus with the material
surrounding the galaxy may cause some appearance of filamentary
structure.) The axis connecting the two objects is at a position angle
of 126\degr, comparable to the radio axis at 131\degr.
 
 It is noteworthy that the diagonal distance between 53W091 and 3a is
4\arcsec, the same distance as between the radio lobes in the 4.86\,GHz
map of Spinrad et al.\ \scite{Spinrad+97a}. The radio positions
suggest
that the south-east radio lobe lies due south of 53W091 (which is the
north-west partner of the $K$-band pair): we cannot definitively claim
that the radio and infrared pairs are congruent, nor that the infrared
source of 53W091 lies between the two radio lobes. Systematic error in
registering the astrometry between the two wavebands might allow either
eventuality.

\subsubsection*{Polarimetry}
\label{53Wgeom}
 
 Polarimetry was performed on both 53W091 and on Object 3a; results for
both are given in Table \ref{restab}. Since the objects were very faint
and close together, the photometry aperture was not chosen according to
the method in \S\ref{firstrunDR}, but was set to a radius of 4 pixels
($1\farcs1$). We also attempted to prune the frames with the greatest
noise from our data\footnote{Doctoral thesis declaration: this
despiking process and photometry of the despiked frames was carried out
by Dr James Dunlop of the University of Edinburgh. Conversion of the
photometry to polarimetry and subsequent debiasing was performed by the
author. There is no qualitative difference between the natural and
despiked results and so the details of the despiking process are not
recorded here.} and repeated the polarimetric analysis. Results for
both the natural ($\natural$) and `despiked' ($\flat$) data are given
in Table \ref{restab}.
 
 There is a weak indication that Object 3a is polarised, with a 70 per 
cent chance of the polarisation being genuine. If there truly is
polarisation at a level of 10--15\%, then 30--45 per cent of the  
light from object 3a could be scattered, and the source could consist
entirely of scattered light within the error bars. (Dust scattering and   
non-perpendicular electron scattering will not result in total linear
polarization of the scattered light.)
 
 Is it possible that a beam from 53W091 is being scattered by a cloud
at the position of 3a? The geometry suggests that this cannot be the
case, since the polarization orientation is around 140\degr, which is
nearly parallel to the line connecting 3a to 53W091. If 3a were a
hotspot induced by a beam emerging from 53W091, a polarization
orientation nearer 30\degr\ would have been expected.
 
The core of 53W091 itself provides no evidence for polarization, and it
would be difficult to obtain results as high as the 40\% which Chambers
(private communication) has suggested; nevertheless, short integration
times on faint objects are subject to large errors in their
polarimetry, so such a result is not impossible. [James Dunlop (private 
communication) reports that Chambers' integration time was not greater 
than three hours in total, compared to our six hours.] Our polarization
orientation, interestingly, is 38\degr, nearly perpendicular to the
radio axis and line to object 3a; but this is unlikely to be
significant with errors of $\pm60\degr$\ on our formal measure of the
polarization angle.

\subsection{MRC 0156$-$252}
\label{D0156-252}

\subsubsection*{Structure}

 Eales \& Rawlings \scite{Eales+96a} have compared radio galaxies at
redshifts $z \sim 1$ and $z > 2$, and find that those radio galaxies at
$z>2$ have brighter absolute $V$-band magnitudes, very low
Ly$\alpha/$H$\alpha$ ratios, and may be subject to strong reddening by
dust. Such results might be attributed to evolution in radio galaxies, or
to a selection bias for more powerful active nuclei at high redshift.   
 
\begin{figure*}
\psfig{file=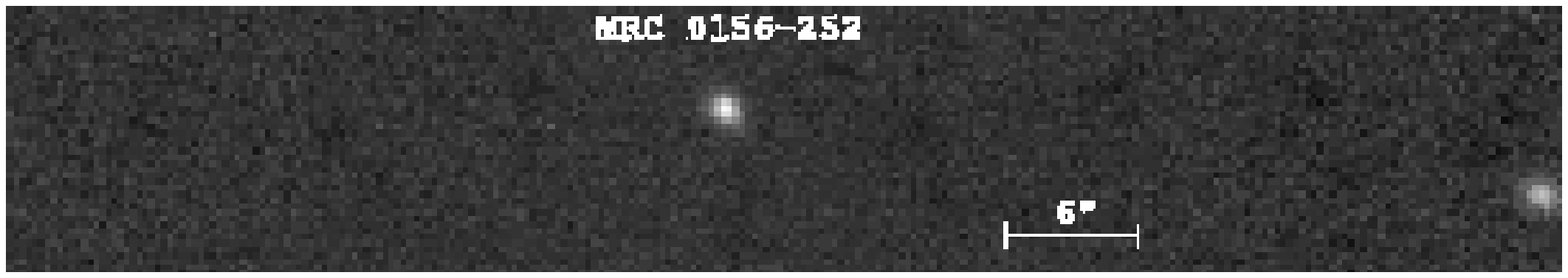,width=153mm}
\caption{MRC 0156$-$252: stacked \Kb-band image of August 1997 data.}
Total integration time 3 hours 44 minutes -- including one cycle of
observations which was not used for our subsequent polarimetry. North
is at
the top, East at the left. \label{Mpic} \end{figure*}

 MRC 0156$-$252 has the brightest known absolute $V$-band magnitude for a
radio galaxy at $z \sim 2$. The cause of its high luminosity is uncertain:
it may be being viewed during an epoch of star formation, or Eales \&
Rawlings \scite{Eales+96a} have suggested that the source is actually a
quasar obscured by dust. Broad H$\alpha$ lines suggest that some of its 
light is originating in an active nucleus. McCarthy et al.
\scite{McCarthy+90a} earlier classified it as a radio galaxy and
suggested \cite{McCarthy-93a}, on the basis of its red spectral energy
distribution, that it was a galaxy at an advanced stage of evolution.
The galaxy appears unresolved in our $K$-band image (Figure \ref{Mpic}),
verifying the findings of McCarthy, Persson \& West
\scite{McCarthy+92a}, who did, however, find extended structure in
their visible-band images. 

\subsubsection*{Polarimetry}

 The criterion used by L\&E to select the 
aperture for polarimetry did
not yield a unique result for this object, so photometry is given in Table
\ref{restab} for apertures of radius 8, 10 and 12 pixels. In all cases
the best point estimate of the polarization is less than 0.15 per cent;   
and for the 10 and 12 pixel apertures, the formal $1\sigma$ confidence
interval indicates that the source is totally unpolarised.

\chapter{Discussion}

\begin{quote}
$\star$ There is a theory which states that if ever anyone discovers 
exactly what 
the Universe is for and why it is here, it will instantly disappear and 
be replaced by something even more bizarre and inexplicable.\newline
$\star$ There is another which states that this had already happened.
\attrib{Douglas Adams, {\em The Restaurant at the End of the 
Universe}} 
\end{quote}
  
The nine objects studied for this thesis project were selected for their
diverse properties, and do not form any kind of statistically complete
sample. No one single approach, therefore, can interpret all the new data
presented in this thesis. In some cases, the published data available in
the literature complements our polarimetry and enables a more detailed
analysis to be made of the properties of certain targets. 

One factor which can be calculated for all radio galaxies is some
constraint on the contribution of the active nucleus to the total light
intensity observed in the $K$-band. This is performed using some basic
assumptions about the undiluted polarization of the active nucleus. Best,
Longair \& R\"{o}ttgering \scite{Best+98a} have fitted radial profiles to
UKIRT $K$-band images of some of our targets, and in each of these cases
their figure can be used in place of our assumptions, allowing an estimate
of the undiluted polarization to be made. 

 Spatial and spectral modelling can be performed for some interesting 
sources. 3C 22 and 3C 41 have had polarizations determined at other 
wavebands, and taken with our data the properties of the scattering 
medium can be modelled. The complicated morphology of sources like 3C 
114 and 3C 356 invites consideration of what spatial scattering processes 
may be at work, and this is modelled for a simple axisymmetric scattering 
geometry.

 We have already noted how Cimatti et al.\ \scite{Cimatti+93a} reviewed 
the properties of high-redshift radio galaxies whose optical polarizations 
had been measured by 1993. The 
results of this thesis, and of the recent literature (Table \ref{litpol}) 
can be used to extend the parameter space in which Cimatti et al.\ 
\scite{Cimatti+93a} looked for trends, and this will form the final part 
of our discussion.

As in the previous chapter, a series of papers by Best, Longair \&
R\"{o}ttgering will be denoted BLR-I \scite{Best+96a}, BLR-II
\scite{Best+97a}, and BLR-III \scite{Best+98a}. The dust scattering model 
paper of Manzini \& di Serego Alighieri \scite{Manzini+96a} is 
abbreviated MdSA.

\section{The Fractional Contribution of Quasar Light}
 
 As we have reviewed in Chapter \ref{reviewRG}, radio galaxies are
thought to have quasar nuclei at their cores, but to be oriented such
that no direct radiation from the core can reach us. Nearby radio
galaxies are known to have the morphology of giant ellipticals;
spectral modelling of sources at higher redshifts suggests that more
distant radio galaxies, too, are dominated by old, red stars. If our
target objects are typical radio galaxies, the total light received
from our sources will be a combination of starlight, light from the
active nucleus scattered into our line of sight, and nebular continuum
emission. Direct optical power-law emission from the active nuclei of
radio galaxies is normally considered to be totally obscured, but in
this analysis we will also need to consider the potential contribution
of such light: in some radio galaxies the obscuring material may be
less efficient owing to its orientation or optical depth.

\subsection{The dilution law for polarization}
 
 Let us denote by $\Phi_{\mathrm W}$, the fraction of the total flux density,
$\Itt{W}$, in a given waveband, W, which originates in the active
nucleus and is scattered into our line of sight. We expect that in the
visible wavebands, $\Phi_{U,B,V}$ will be a
significant fraction of unity. From our observations, we wish to
determine whether $\Phi_{K}$ is small, or whether a significant component
of the stellar-dominated infrared also arises in the active nucleus.
If we denote the flux density scattered into our line of sight from the
quasar core by $\Iq{W}$, then $\Phi_{\mathrm W} = \Iq{W}/\Itt{W}$
 --- at this stage we make the assumption that there is no
contribution by direct radiation from the active nucleus.

 Following MdSA,  we assume that only the scattered component of the light
from radio galaxies is polarised. Recalling from Equation \ref{ppol} that 
the degree of linear polarization is defined by $P = I_p/(I_p + I_u)$, we 
define the unpolarized component to be $I_c$ for the scattered core light 
only, and $I_c + I_h$ for the scattered core light and host galaxy together.
Hence defining
$\Pq{W}$ as the intrinsic polarization produced by the scattering process,
and $\Ptt{W}$ as the observed polarization after dilution, it follows that
\begin{equation}
\label{fracflux}
\frac{\Ptt{W}}{\Pq{W}} = \frac{I_p/(I_p + I_c + I_h)}{I_p/(I_p + I_c)} = 
\frac{I_p + I_c}{I_p + I_c + I_h} = \frac{\Iq{W}}{\Itt{W}} = 
\Phi_{\mathrm W}. \end{equation}
Therefore the fraction of the total light which is the scattered 
nuclear component, is 
equivalent to the ratio of the diluted and undiluted polarizations.

\subsection{The dilution law in a Unification Model scattering
geometry}
 
 If we know the restrictions on possible values of the intrinsic
polarization $\Pq{W}$, we can use our corresponding measurements of
$\Ptt{W}$ to put limits on $\Phi_{\mathrm W}$ for the measured sources. 

The 
appropriate restrictions depend on the nature of the scattering centres.
If the scattering centres are electrons \cite{Fabian-89a}, then Thomson
scattering will take place: the effects of the geometry and of the
wavelength can be treated independently. The spectral energy distribution
of the light scattered in a given direction is independent of the
scattering angle:  $\Pq{W}$ will be the same constant $\Ptt{Q}$ at 
all wavelengths. 
The degree of polarization of the scattered light is given simply by
\begin{equation}
\label{elecpol} 
\Ptt{Q} = \frac{1-\cos^2\chi}{1+\cos^2\chi},
\end{equation}
where $\chi$ is the scattering angle. For an AGN observed as a radio
galaxy, we assume \cite{Barthel-89a} an orientation 45\degr\ $\leq \chi 
\leq$ 90\degr, whence $1/3 \leq \Ptt{Q} \leq 1\ \forall\ $W.
 
 The case where the scattering centres are dust grains has been modelled
recently (MdSA); the fraction of the light scattered by the dust, $f_{\rm
W}$, and the polarization of the scattered light, $\Pq{W}$, both depend  
strongly on wavelength. The exact relationship depends critically on the 
size distribution of the dust grains, and the amount of extinction they  
introduce; MdSA provide a series of graphs for the variation of $f_{\rm W}$
and $\Pq{W}$ with rest-frame wavelength
0.1\,\micron\ $< \lambda_r <$ 1.0\,\micron, corresponding to many different 
dust grain compositions and size distributions. At the redshifts of the 
objects 
in our sample, light observed in the $H$ and $J$ bands originates at
rest-frame wavelengths below 1.0\,\micron, but $K$-band light originates in
the region
1.0\,\micron\ $< \lambda_r <$ 1.15\,\micron. To accommodate the $K$-band 
light within our models, we linearly
extrapolated MdSA's curves to $\lambda_r = 1.15$\,\micron.
 
In MdSA models where the smallest dust grains have a radius not less
than 40\,nm, $\Pq{W}$ approaches zero twice: once at a (rest frame)
wavelength around 0.2\,\micron, and again at some wavelength between 
0.1\,\micron\ and 0.7\,\micron\ which depends critically on the dust grain 
size distribution. But in all cases, $\Pq{W}$ extrapolated into the 
1.0\,\micron\
 $< \lambda_r < 1.15$\,\micron\ region gives 0.3 $< \Pqi{K} <$ 0.5;
the MdSA graphs show that the intrinsic polarizations $\Pqi{V}$ and
$\Pqi{H}$ should be lower than $\Pqi{K}$ for the objects where we have 
$V$- and $H$-band polarimetry.
 
\subsection{Constraints on the nuclear component intensity of our
sources}
\label{constraincalc}

 For the sources in which we have evidence of $K$-band polarization
(here we will consider those with a $\geq 80\%$ probability of genuine
polarization), we can hence estimate $\Phi_{K}$ under both electron
and dust models. The values and upper limits are given in Table
\ref{quasfrac}. For the dust models, we take $ 1/\Pq{W} = 2.5 \pm 0.5$;
the error takes into account all dust models, and the different
redshift corrections for the different galaxies, but assumes that the
scattering angle is 90\degr. If the scattering angle is less, we assume
that less polarization occurs (see MdSA, Figure 20), and hence
$\Phi_{K}$ will be greater than our estimate. For the electron models,
we multiply the observed polarization by $1/\Pq{W} = 2 \pm 1$; this
takes into account all possible $\chi \geq 45\degr$\ orientation
effects.

 Similarly, in all sources we can at least estimate an upper limit for
the nuclear contribution contingent on our assumption of a quasar core
and a $\chi \ge 45\degr$\ scattering geometry. In sources which we
evaluate as having a less than 80\% probability of genuine
polarization, we will multiply the $2\sigma$ upper limit on their
measured polarizations by the $1\sigma$ upper limit on the
reciprocal of the modelled intrinsic polarizations --- which for both
dust and electrons under the
above assumptions requires a multiplication by three. These
limiting values, too, are listed in Table \ref{quasfrac}.

\begin{small}
\begin{table}
\centering
\caption{Percentage of $K$-band light estimated to be arising from the
postulated active nucleus in our sample of radio galaxies.}
 
\medskip
 
\label{quasfrac}
\begin{tabular}{|l|c|cc|c|c|}
\hline 
Source & $\Ptt{K}$ & ${\Phi_{K}}_{\it e}$ & ${\Phi_{K}}_{\it d}$ &
${\Phi_{K}}_{\it r}$ & ${\Phi_{K}}_{\it s}$ \\
\hline
	3C 22 & 3.3 \tpm 1.4 & 7 \tpm 4 & 8 \tpm 4 & 37 &-- \\
\hline
	3C 41 & 3.1 \tpm 1.1 & 6 \tpm 4 & 8 \tpm 3 & 24 &-- \\
\hline
	3C 54 & 5.9 \tpm 2.6 & 12 \tpm 8 & 15 \tpm 7 &-- &-- \\
\hline
	3C 65 & $<9.7$  & \cduol{$<29$}  & $<8$ & 1 \\
\hline
3C 114 (T0)   & 11.7 \tpm 3.0 & 23 \tpm 13 & 29 \tpm 10 &-- &-- \\
3C 114 (T2)   &  5.1 \tpm 1.7 & 10 \tpm 6 & 13 \tpm 5 &-- &-- \\
\hline
3C 356 (T0)   & 9 \tpm 5 & 18 \tpm 13 & 23 \tpm 13 & $<14$ & 16 \\
3C 356 $a$    & $<41$ & \cduol{$\ominus$} & $<14$ &-- \\
3C 356 $b$    & $<24$ & \cduol{$<72$} & $<14$ & --\\
\hline
3C 441 {\bf a} & $<10$ &  \cduol{$<30$}  & $<5$ & 4 \\
3C 441 {\bf c} & $<24$ & \cduol{$<72$} &--&-- \\
      3C 441 E & 18 \tpm 9  & 35 \tpm 26 & 44 \tpm 25 &-- &-- \\
\hline
 LBDS 53W091 $\flat$ &  $<22$ &  \cduol{$<65$}  & --&-- \\
 53W091-3a $\flat$   &  $<46$ &  \cduol{$\ominus$} &-- &-- \\
\hline
MRC 0156$-$252 (2\farcs3) & $<10.4$ &  \cduol{$<31$} & -- &-- \\
\hline
\end{tabular}
 
\medskip
 
\parbox{153mm}{Key: $\Ptt{K}$: measured $K$-band polarization; 
${\Phi_{K}}_{\it e}$:
percentage of light from quasar according to electron model; ${\Phi_{K}}_{\it
d}$: percentage of light from quasar according to dust model.
Depolarization from multiple scattering (both models), and shallower
scattering angles (dust model only) will tend to increase $\Phi_{K}$. 
 The $\ominus$ symbol denotes that the
polarization is unconstrained since the formal upper limit exceeds
100\%. ${\Phi_{K}}_{\it r}$: estimate of the percentage
of light from a quasar core according to BLR-III radial profile fitting;
${\Phi_{K}}_{\it s}$: estimate of the percentage
of light from a quasar core according to BLR-III flat spectrum fitting.
 A dash (--) indicates where no data is 
available from the literature. If $\Ptt{K}<80\%$ then an upper limit
has been calculated as detailed in the text (\S\ref{constraincalc}). The 
last two columns may include a contribution from nuclear light arising 
from processes other than scattering --- see text for details.} \end{table} 
\end{small}

 The effects of multiple scattering have been ignored for both models;
multiple scattering tends to depolarise light, and so the true value of
$\Phi_{K}$ under multiple scattering will again be greater than our
estimate. The only physical mechanism which could cause the true
$\Phi_{K}$ to be {\em lower}\, than our estimate, is polarization of
light in transit by selective extinction; and as we have already seen
(Table \ref{extab}), any such contribution to the polarization of our
targets will be small.

\subsection{Constraints on the nuclear polarization based on BLR-III data}
\label{blr3constr}

 Another approach to estimating the possible contribution of a quasar
component is to fit the radial intensity profile of the radio galaxy with
a combination of a de Vaucouleurs law and a point source. This has been
done with {\em HST}\, and UKIRT imaging for five of our 3C sources
(BLR-III), and the fitted values or upper limits (${\Phi_{K}}_{\it r}$)
are also reproduced in Table \ref{quasfrac}. The same paper made a further
estimate of the fraction of nuclear light present by fitting the spectrum
as a combination of a (nuclear) flat spectrum and an old stellar
population with the spectral energy distribution of Bruzual \& Charlot
\scite{Bruzual+93a}; again, these fractions (${\Phi_{K}}_{\it s}$) are
reproduced in Table \ref{quasfrac}. All of the estimates of $\Phi_{K}$ are
consistent with one another; the significance in individual objects will
be considered below. 

It should be noted, however, 
that BLR-III's `nuclear light' need not include only
scattered light -- their simple spectral model distinguishes light from an
evolved stellar population but lumps everything else (scattered central
engine light, the spectral profile of newly formed stars, and nebular
emission) into the `nucleus'. Similarly the radial profile fit 
distinguishes only the light sources which contribute to the
de Vaucouleurs structure of a normal galaxy. Our earlier definition makes 
$\Phi_K$ the fraction of the total light which is scattered nuclear 
light. In particular, if the active nucleus is not perfectly shielded, 
the nuclear light fractions derived from BLR-III's analysis will include 
direct nuclear light and may therefore be higher than the $\Phi_K$ values 
derived from our $K$-band data.

 In two cases (3C 65 and 3C 441 {\bf a}), the BLR-III radial fitting
yields upper limits which are in fact much lower than (but obviously
consistent with) the upper limits estimated on the basis of our
$K$-band polarimetry. Dividing the nominal $K$-band polarization by
${\Phi_{K}}_{{\it r\!,}{\mathrm max}}$ yields a nominal lower limit to
the intrinsic polarization of the nucleus, namely 27\% for 3C 65 and
21\% for 3C 441 {\bf a}; since our $K$-band polarizations are
consistent with zero within their error bars, however, these nominal
nuclear polarizations are of limited usefulness.

Similarly, the spectral fitting approach yielded definite values of
${\Phi_{K}}_{{\it s}}$ in these two objects; dividing the measured
polarization by this light-fraction gave nominal nuclear polarizations
of 215\% {\em (sic)}\, in 3C 65 and 26\% in 3C 441 {\bf a}. But the
errors on the measured polarizations make any intrinsic polarization
between zero and 100\% possible. If the spectral fitting figure
${\Phi_{K}}_{{\it s}} = 1\%$ is accurate for 3C 65, then our nominal
2.15\% diluted polarization measurement for this object is clearly too
high, assuming the polarization occurs only in the nuclear component of
the light.

 In 3C 356 (T0), we find a better-constrained case: given a 
spectral fit light-fraction ${\Phi_{K}}_{{\it s}} = 16$\%, the intrinsic 
nuclear polarization is $\Pqi{K} = 57 \pm 31 \%$ (or for radial
fitting, 
the {\em lower limit}\, nuclear polarization is 65 $\pm$ 35\%). Again, 
the large error on our polarization measurement gives us a constraint of 
limited usefulness, but we might cautiously conclude (with 1$\sigma$ 
confidence) that the nuclear source in 3C 356 is at least 25\% polarized.
 
 The recent finding \cite{Eales+97a} that 3C galaxies
at $z \sim 1$ are 1.7 times as bright as the radio-weaker 6C/B2 galaxies
in a similar sample requires that $\Phi_{K} \ga 40\% (= 7/17)$
for 3C galaxies, if the scattering of nuclear nonstellar light is
responsible for the brighter $K$-band magnitudes of 3C galaxies. Most of 
the results presented in Table \ref{quasfrac}, whether based on the 
polarimetry of this thesis or the BLR-III data, produce $\Phi_{K}$ values 
which are somewhat lower. This may be indicative of some correlation 
between the strength of the active nucleus and the number of passively 
evolving stars in the galaxy, allowing the polarizations and spectral 
and spatial fits to produce lower $\Phi_{K}$ values, while still producing 
the enhanced $K$-band brightnesses measured in the most powerful (3C) 
radio galaxies.

\section{Optically Compact Sources}

 The most obvious division which can be made in our sample of nine
sources is between those whose $K$-band image is dominated by a clear
source object, and those where there is a complex structure of knots or
components of comparable brightness. First we consider as a group the
`optically compact sources', dominated by one bright object: 3C 22, 3C
41, 3C 54, 3C 65 and MRC 0156$-$252. 

 Among the optically compact sources, the radio galaxies 3C 22 and 3C
41 are particularly noteworthy for the compact rounded morphologies in
their observed $K$-band structure. They are also prominent for having
the brightest $K$-band excess over the mean locus of the $K$-$z$ Hubble
plot for 3C galaxies. When BLR-III fitted radial profiles to {\em
HST}\, visible and
UKIRT $K$-band images of eight high redshift radio galaxies, six could
be modelled by a simple elliptical galaxy de Vaucouleurs profile; but
in these two galaxies alone, an additional point source was required to
give a good fit in the central region. These two sources also have
polarization figures available in visible wavebands, data which enable
simple spectral modelling to be performed for these two sources.

\subsection{Determination of the scattering angle from the BLR-III 
light-fraction}

 The BLR-III radial profile fitting suggests that in 3C 22, the point
source contributes 37\% of the total $K$-band intensity, and in 3C 41,
24\%; but their fitting method is known to be biased low for sources with
a high point component. By simulating the effects of an additional point
source, a revised estimate could be made, suggesting that 3C 22 actually
had a $50^{+20}_{-10}\%$ nuclear contribution, and 3C 41's light included
$31^{+10}_{-8}\%$ from the nucleus. Flat-spectrum component fitting was
not attempted in these cases. 

 Clearly these very high nuclear contributions are not easily reconciled
with the low polarizations of order 3\% which we have measured in the
$K$-band, provided our model assumptions are correct. If we assume the
nuclear source is obscured from direct view and has no diluting effect on
the scattered light, then the intrinsic nuclear polarizations implied by
our polarization and BLR-III's fits are of the order of 10\%. Such a value
is too low to be consistent with a $\chi > 45\degr$\ scattering geometry. 

If we assume a shallower scattering angle, we should also allow for the
diluting effect of nuclear light which may have become visible along a
direct line of sight to the nucleus. We can put a lower limit on the 
scattering angle by neglecting this component; if electron
scattering is taking place, we can identify the limiting scattering angle.
Substituting Equation \ref{elecpol} into Equation \ref{fracflux} and
rearranging yields
\begin{equation}
\label{getchi} 
\cos^{2}(\chi) = \frac{\Phi_{K}-P_{K}}{\Phi_{K}+P_{K}}.
\end{equation}

 We find that a 3.3\% polarised 3C 22 with 50\% of its $K$-band
emission originating in a nuclear source could be scattering light at
21\degr\ $\pm$ 5\degr\ if the scattering medium were electrons, or
could contain dust scattering light at a slightly higher angle.
Similarly the figures for 3C 41 imply electron scattering at 25\degr
$\pm$ 6\degr, or a correspondingly higher angle for dust. 

 These models suggest that these two objects are not oriented at the
$\chi \geq 45\degr$\ positions of radio galaxies but are being viewed
`down the jet'. Either the direct contribution of nuclear light is weak 
owing to some other factor (e.g.\ dust obscuration) and these angles are 
correct; or the direct contribution is stronger, in which case the true 
scattering angles are somewhat higher.

Both objects are also atypical in other properties: 3C 22
appears to emit a broad line component
\cite{Rawlings+95a,Economou+95a}, and both 3C 22 \cite{Fernini+93a} and
3C 41 (BLR-II) exhibit radio jets, a feature rare in radio
galaxies at high redshift. For these reasons, independently of the
evidence of the $K$-band light fraction and polarization, it has been
already suggested that these two objects may be oriented close to the
threshold between radio galaxy and radio quasar properties. This is 
entirely consistent with the polarization which is suggesting that these 
objects are oriented at $\chi \sim 30\degr$\ with some direct nuclear 
light contribution.

 A further line of enquiry is open to us, since polarization
measurements in other optical wavebands are available in the literature
or by private communication for these two sources. These additional
wavebands give a broad baseline against which models of the scattering
medium can be tested.

\subsection{Modelling the scattering medium given
multiwaveband polarimetry}
 
 Where the magnitude, W, has been measured in a given waveband for
which the zero-magnitude flux density is $\Itt{W0}$, the total
flux density can be calculated:
\begin{equation}
\label{totint}  
\Itt{W}=\Itt{W0}.10^{-0.4{\rm W}}.
\end{equation}

 The optical flux density of quasars can be modelled well by a power law
of the form $C.\nu^{-\alpha}$ where $\alpha$, the `spectral index', is of
order unity \cite[\S 1.3]{Peterson-97a}. Since the efficiency with
which a given species of scattering
centre
scatters light may depend on wavelength, we denote that efficiency by
$f_{\rm W}$, and the scattered quasar component can be expressed
\begin{equation}
\label{scatcomp}
\Iq{W}= C.f_{\rm W}.\nu^{-\alpha}.
\end{equation}
 
It will be computationally convenient to define an `unscaled model
light ratio', $\Phm{W}$, as
\begin{equation}
\label{phiunsc}
\Phm{W} =
\frac{f_{\rm W}.\nu^{-\alpha}}{\Itt{0W}.10^{-0.4{\rm W}}};
\end{equation}  
then the actual model light ratio will be $\PHI{{\mathrm W}}=C.\Phm{W}$.
There is clearly an upper limit set on $C$ by the fact that 
$\PHI{{\mathrm W}}$
may not exceed unity in any waveband; hence $C \leq 1/\Phm{W}$. 
Allowing for errors in the observed magnitudes, $\Delta$W, the maximum 
permissible value of $C$ in a model can be constrained by inspecting
every relevant waveband, since the inequality must hold in all bands: 
\begin{equation}
\label{conmax}  
C_{\rm max} = {\rm min} \left[ \frac{\Itt{0W}.10^{-0.4({\rm
W}-\Delta{\rm W})}}{f_{\rm W}.\nu^{-\alpha}}\right], \forall\ {\rm W}.
\end{equation}
Hence, for a given choice of scattering model, which determines the value
for $\alpha$ and the form of $f_{\rm W}$, $C_{\rm max}$ is 
the minimum value obtained by substituting W by each wavelength
observed, in turn.

Given a measurement of the magnitude of a radio galaxy, we can predict
its polarization as a function of
wavelength, up to a multiplicative constant. Rearranging Equation
\ref{fracflux}, and employing our `unscaled model light ratio',  
we first define an `unscaled model polarization' $\Ptm{W} = \Pq{W}.\Phm{W}$,
and so express our modelled polarization as:
\begin{equation}
\label{polmod}  
\Ptt{W,modelled} = \Pq{W}.\PHI{W} =
C.\Pq{{\mathrm W}}.\Phm{W} = C.\Ptm{W}.
\end{equation}
 
To fit a dust scattering model, we can calculate $\Ptm{W}$ by obtaining
$f_{\rm W}$ and $\Pq{W}$ from suitable
curves in MdSA. For electron models, the wavelength-independent term
$f_{\rm W}$ can be considered to have been
absorbed into the multiplicative constant, $C$, while $\Ptt{Q}$, also
wavelength-independent, can be assumed to be its minimum value, 1/3. 
We cannot separately identify $C$ and $\Ptt{Q}$; the physical
constraints on these constants are $1/3 \leq \Ptt{Q} \leq 1$ and $0 \leq C
\leq C_{\rm max}$. If $\Ptt{Q}$ is greater than the assumed 1/3, then $C$ 
will be correspondingly smaller.
 
 Given a set of $N$ polarization measurements 
$\Ptt{W} \pm \Stt{W}$,
and a
corresponding set of unscaled model polarizations, 
$\Ptm{W} \pm \Stm{W}$, 
based on measured magnitudes, we can calculate the deviation of the fit:  
\begin{equation}
\label{fitdev}  
\delta = \sqrt{\frac{1}{N}.\sum_{{\mathrm W}_{1}\ldots
{\mathrm W}_{N}}{\frac{(\Ptt{W}-C.\Ptm{W})^2}{\Stt{W}^2+(C.\Stm{W})^2}}}.
\end{equation}
The best fit is that with the value of $C$ which minimizes $\delta$,
subject to the physical constraint $0\leq C\leq C_{\rm max}$.
 
\subsection{A model for 3C 41}
 
 The source for which we had the most data was 3C 41, with polarimetry in 3
bands: $P_V = 9.3$ \tpm 2.3\%;
$P_H = 6.6$ \tpm 1.6\%; $P_K=3.1$ \tpm 1.1 \%.
We attempted to fit two models; an electron model and
 a typical dust model with a minimum grain radius of 80\,nm.
To fit the models to the observed polarizations, we tested a discrete series
of possible spectral indices, $-0.5 \leq \alpha \leq 2$, with a
step size of $1/3$.
For each value of $\alpha$ we calculated the `unscaled polarizations'
$\Pi_V \pm \epsilon_V$, 
$\Pi_H \pm \epsilon_H$, 
$\Pi_K \pm \epsilon_K$. 
We then
iteratively determined the best fit value of $C$ for each $\alpha$,
and took as our overall best fit that
combination of $\alpha$ and $C$ which gave the lowest $\delta$.
 
Lilly \& Longair's \scite{Lilly+84a} data shows
that 3C 41, at $K = 15.95$ \tpm 0.10, is significantly
brighter than the
mean $K$-$z$ relationship, by about 0.6 mag.
Magnitudes for 3C 41
were available in 5 bands: $J$, $H$ and $K$
\cite{Lilly+84a} and the narrow filters g and r$_S$ \cite{dpc}. The $H$ and
$K$ values yielded direct estimates of the corresponding `unscaled
polarizations' $\Pi_H$ and $\Pi_K$; $\Pi_V$ was 
estimated by linear interpolation between $\Ptm{g}$ and $\Ptm{\rS}$.
 
\begin{figure}
\centerline{\psfig{file=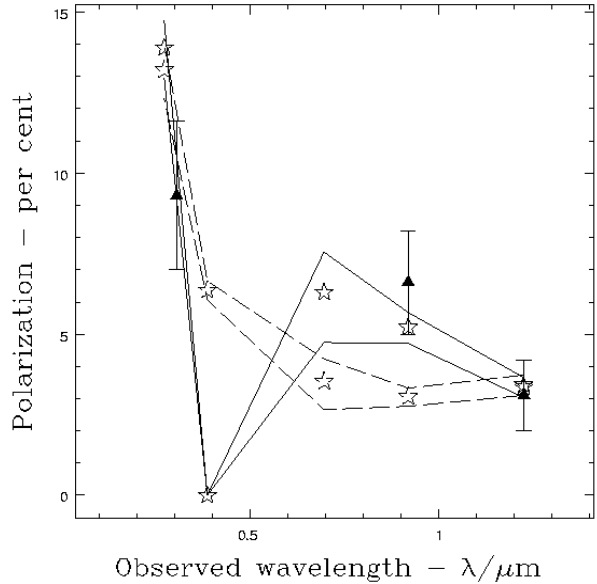,width=85mm}}
\caption{Measured and best-fit model polarizations of 3C 41 as a function of
rest-frame wavelength.} Solid line: dust
model, $\alpha = 1.167$; dashed line: electron model, $\alpha = 1.733$.
\label{pmodel41} \end{figure}
 Figure \ref{pmodel41} shows the measured polarizations for 3C 41 as
triangles (\psfig{file=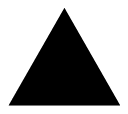,height=0.9em}) and the magnitude-based
polarization estimates, after best-fit scaling, as stars
(\psfig{file=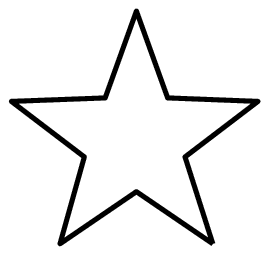,height=0.9em}). The lines give the error envelope on
the modelled polarization (based only on the errors on the magnitudes).
The solid lines correspond to the dust model, and the dashed lines to the
electron model.
 
 As can be seen from Figure \ref{pmodel41}, the model curves lie below
the data point at $H$, but above those at $K$ and $V$. The shape of the
curve depends more strongly on the measured magnitudes (and on $f_{\rm
W}$ for dust) than on the spectral index, and the models for all
reasonable values of $\alpha$ will have a broadly similar shape; the
best fit will necessarily pass below the measured point at $H$, and
above that at $V$.
 
 Consider the electron model. Figure \ref{pmodel41} shows us that the
theoretical polarization curve for electron scattering is concave with
respect to the origin, whereas a curve through the three data points
would be convex; clearly it will not be possible to obtain a close fit
for the central ($H$-band) point. Fitting the electron model curve, we
found that the best fit occurred for $\alpha = 1.733$, with a deviation
$\delta = 0.430$. From Table \ref{quasfrac}, we have $\Phi_{K} = 6 \pm
4$ per cent for 3C 41. Multiplying the polarizations observed at $H$
and $V$ by $1/\Pq{W} = 2 \pm 1$, we predict $\PHI{H}=13 \pm 7$ per
cent, and $\PHI{V}=19 \pm 10$ per cent in these bands.
 
 The dust model chosen as typical from MdSA was that for a cloud of
spherical dust grains, with radii 250\,nm $> a >$ 80\,nm, with the
number density per unit dust mass following an $a^{-3.5}$ law. This
model produced a curve which fitted the data points very well. The best
fit indicated that the optimum spectral index was $\alpha = 1.167$, for
which $\delta = 0.177$.
 
 This dust model was also used to estimate the proportion of scattered
light at shorter wavelengths: $\Phi_{K} = 8 \pm 4$\%, $\Phi_{H}
= 24 \pm 6$\%, and $\Phi_{V} = 155 \pm 38$\% {\em (sic)}.
MdSA's
curve for polarization as a function of rest-frame wavelength predicts
a 6 per cent polarization in the observed $V$-band, lower than the 9
per cent {\it after dilution} measured by Jannuzi \scite{jpc}
\cite{Elston+97a}. This is still consistent, within error bars, as long
as the true value of $\Phi_{V}$ for 3C 41 is less than, but very close
to, unity;  the observed $V$-band corresponds to the near-ultraviolet
in the rest frame of 3C 41, and it is reasonable (MdSA) to suppose that
the scattered quasar light in that band could form in excess of eighty
per cent of the total light.

We noted earlier that the shape of the dust model polarization curve
between 0.2\,\micron\ and 0.7\,\micron\ (rest frame) is very sensitive to
the choice of dust grain distribution. The particular dust model chosen
approaches zero polarization at a wavelength corresponding to the r$_S$
band when redshifted into our frame. This causes the `well' visible in
the model polarization curve (Figure \ref{pmodel41}), whose presence is
essential for the dust model curve to fit the data points closely.
 
 We also considered other dust models from the selection given by MdSA.
Of those which differed significantly from the `typical' one considered so
far, many of them will not predict polarizations in the observed $V$-band   
which are sufficiently high to be reconciled with the observed 9 per cent;
and those which do, do not possess the deep well needed to fit the
polarimetry across the spectrum. We conclude, therefore, that
the best model for 3C 41 is that of an obscured quasar core with $\alpha
\sim 1.2$ beaming its optical radiation into a dust cloud, although an  
electron model cannot be ruled out within the error bars.
 
\subsection{A model for 3C 22}

 As we have already seen, 3C 22 is suspected of being an obscured
quasar \cite{Dunlop+93a,Rawlings+95a}; according to data in Lilly \&
Longair \scite{Lilly+84a}, its $K$-band magnitude ($15.67 \pm 0.10$) is
brighter than the mean $K$-$z$ relationship by about 0.9 mag.
 
\begin{figure}
\centerline{\psfig{file=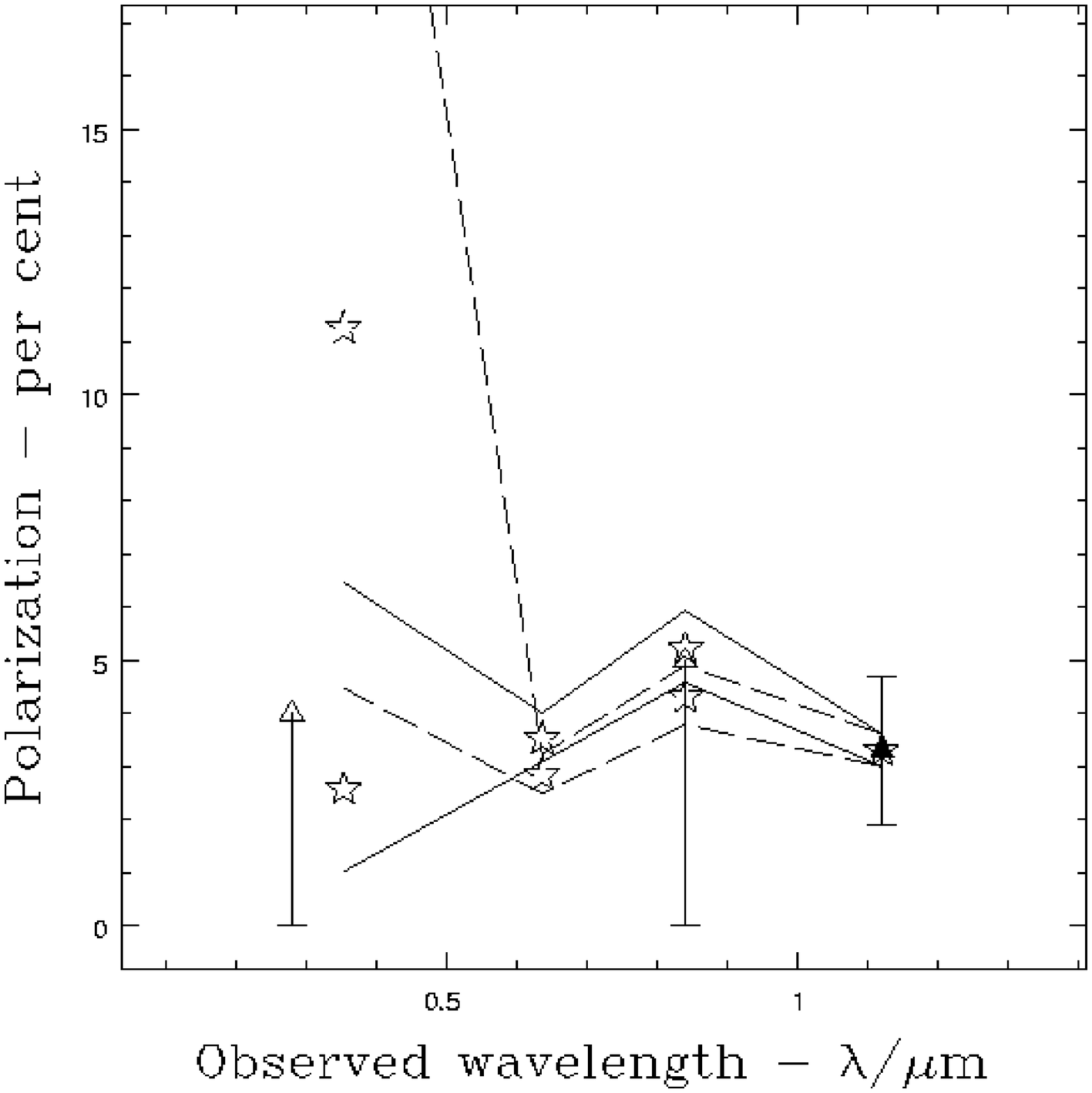,width=85mm}}
\caption{Modelled ($\alpha = 1$) and measured polarizations of 3C 22 as a
function of
rest-frame wavelength.} Solid line: dust model; dashed line: electron model.
\label{pmodel22}
\end{figure}

For 3C 22, we have one firm polarimetry point (this thesis) and two upper
limits in $V$ and $H$ \cite{jpc} \cite{Elston+97a}; magnitudes were
available
in 4 bands including $J$, $H$ and $K$ \cite{Lilly+84a}, and a crude eye
estimate in r \cite{Riley+80a}. The measurements and models are shown in
Figure \ref{pmodel22}, with the same symbols as Figure \ref{pmodel41};  
open triangles represent upper limits. We
have taken $\alpha = 1$, and normalised the theoretical curves to the
$K$-band data point.
 
Here the models are inconclusive. In the observed near-infrared, both
models can easily fit, with some slight scaling, within the $K$-band   
measurement error bars; both models suggest that about 8 per cent of the
$K$-band light arises in the active nucleus (Table \ref{quasfrac}). The   
error on the observed r-band magnitude is so
large that both models are consistent with the observed $V$-band upper
limit of polarization.
 
 \subsection{Other compact sources}

Manzini \& di Serego Alighieri (MdSA) comment that given the range of
possible dust models, `the wavelength dependence of polarization is not
necessarily a discriminant between electron and dust scattering.' The  
data available to us are insufficient to indicate whether the scattering
centres in these objects are electrons or dust; it is not possible to give
an unambiguous fit of the polarization curves with so few data points,
although 3C 41 does seem to fit a model (MdSA) with a minimum dust radius
of 80\,nm particularly well, and we suggest that it does indeed consist  
of a quasar obscured by dust.

 Since our sample of sources was selected for the variety rather
than homogeneity of sources, it is difficult to come to any general
conclusions on the properties of radio galaxies as a whole. There is no 
obvious trend of alignments, either with
elongated structure of the compact optical source or the faint optical
companions. $K$-band polarization may be present at any level from at
least 6\% down to zero. Table \ref{compalign} 
indicates the different alignments (with respect to the radio jets) 
observed in the faint companions, optical cores, and polarizations of 
these objects; alignments are identified as parallel or perpendicular 
rather than skew if the $2\sigma$ error bars allow an aligned
interpretation.

\begin{small}
\begin{table}
\centering
\caption{Alignments in compact objects.}
\label{compalign}
\begin{tabular}{rccc}
\hline
Source & Companions & Optical Core & Polarization \\
\hline  
3C 22 & $\perp$ & $\angle$ & $\perp$ \\
3C 41 & $\angle$ & $\odot$ & $\perp$ \\
3C 54 & $\cdot$ & $\parallel$ & $\parallel$ \\
3C 65 & $\parallel$ & $\perp$ & $\perp$ \\
MRC 0156$-$252 & $\cdot$ & $\parallel$ & $\circ$ \\
\hline
\end{tabular}
 
\medskip
 
\parbox{153mm}{Key: 
$\perp$: perpendicular to radio structure; $\parallel$: parallel to radio 
structure;  $\angle$: skew to radio structure; $\cdot$\,: no faint 
companions detected; $\odot$: optical core is round, no extended structure; 
$\circ$: no polarization detected.}
\end{table}
\end{small}

 Two of these compact sources, MRC 0156$-$252 and 3C 54, appear to be
truly isolated. Both of these are extremely faint sources in the
exposures available to us (Figures \ref{54image} and \ref{Mpic}) but
there is no indication of close companions of comparable brightness.
BLR-II's {\em HST}\, images of the other three reveal much fainter
objects in close proximity (within 5\arcsec), but there is no clear
trend of alignment with the radio structure. 3C 22's one companion is
offset almost perpendicular to the radio jet; 3C 41 has two companions
on opposite sides along an axis offset 20\degr\ from the radio
structure position angle; and 3C 65 has one companion lying
between the optical source and its northwestern radio lobe.

 There is some evidence for structure within the source galaxies. 3C 65
is slightly elongated NE-SW (BLR-II), roughly perpendicular to its
radio structure; 3C 54 is extended along its radio axis
\cite{Dunlop+93a}; MRC 0156$-$252 is extended over 8\arcsec\ with three 
knots in the $r$-band visible (rest frame ultraviolet)
corresponding to its radio core and lobes, 
though it appears compact in $J$, $H$ 
and $K$ (rest frame visible bands) \cite{McCarthy+92a}. Two tiny 
(subarcsecond) extensions are known 
south and west of 3C 22's core, neither clearly related to the radio jet
orientation (BLR-II).

 There is no clear trend, therefore, for alignment in host galaxy
structure or the positioning of companions. One skew companion is most
likely a chance association of objects; two skew companions on the same
axis, as seen in 3C 41, might be evidence for precession of the radio jets
from the skew axis to their present position. Parallel core structure or 
companions suggest that the radio jet may be responsible in some way for 
the formation or excitation of structure along its path. Perpendicular 
structure could be indicative that the radio jet is orthogonal to the 
plane of rotation of the galaxy and its satellites. Any 
comprehensive model of radio galaxies clearly needs to allow, therefore, 
for the possibility of core structure and companions, both aligned with 
and orthogonal to, the large-scale radio structure.

 Evidence that 3C 22 and 3C 41 may be close to the $\chi = 45\degr$
boundary between radio galaxy and quasar profiles encourages us to
consider the other sources in this light. The most recent radio maps of
MRC 0156$-$252 \cite{Carilli+97a} do not display a marked head-tail
asymmetry in radio lobe intensity at 4710\,MHz or 8210\,MHz, although
the eastern lobe has a much greater intensity in linearly polarized
4710\,MHz radiation than the core or western lobe.  The formal
$1\sigma$ confidence interval indicates that the source is totally
unpolarised in $K$. At most, if we assume that it has radio jets
perpendicular to our line of sight, 16 per cent of its \Kb-band light
arises in the active nucleus.

 Nevertheless, it is possible that MRC 0156$-$252 is an obscured quasar
\cite{Eales+96a}, and McCarthy, Persson \& West \scite{McCarthy+92a} note
that its properties are comparable to the red quasars observed by Walsh et
al.\ \scite{Walsh+85a}. If this is the case, we must be looking close to
`straight down the jet' with a shallower scattering angle for infrared
light; and hence our upper limit for $\Phi_{K}$ would be weaker. The fact
that McCarthy, Persson \& West \scite{McCarthy+92a} detect $r$-band
structure on the same 8\arcsec\ scale as the radio structure could support
this, since at radio galaxy orientation, optical structure is usually less
extended than radio structure, especially in the most powerful radio
galaxies (BLR-I). Alternatively, if this galaxy's high luminosity is due
to star formation or direct emission from an active nucleus which is not
well shielded, any scattered component might easily be diluted below a
detectable level. 

 The five optically compact objects in our sample have been selected
for their radio strength and should be presumed to share the radio
emission mechanism common to all radio galaxies, unless proven
otherwise. Their compact appearance gives no indication of a history of
merging or recent star formation, unless the rest-frame ultraviolet 
elongated 
structure of MRC 0156$-$252 is interpreted as such. The detection of 
$K$-band polarization requires both the presence of a sufficiently dense
scattering medium (probably situated within the host galaxy itself),
and a well-shielded nucleus strong enough to yield sufficient scattered
infrared light to be detectable despite dilution by the host galaxy.

 The perpendicular polarizations of 3C 22, 3C 41 and (if genuine) the
marginal 3C 65 are all, therefore, consistent with the Unification
Hypothesis for radio galaxies. Failure to detect polarization in MRC
0156$-$252 is consistent with Unification given dilution, the absence
of a suitable scattering medium, or an obscured quasar scenario. Finally, 
the parallel-polarized 
3C 54 is the most difficult candidate to reconcile with the Unification
model, requiring a special scattering geometry or possibly the presence
of aligned dust grains causing polarization by transmission in the
outer structure of what seems to be a quite diffuse source.

\section{Spatial Modelling of Knotted Sources}

 Our four remaining sources clearly display several distinct or joined
knots in their $K$-band structure: 3C 114, 3C 356, 3C 441 and
53W091. The scattering models of the Alignment Effect invite us to
investigate scenarios where one knot contains an active nucleus and
other knots are regions of scattering material which have intercepted a
particle jet or radiation cone emerging from the central engine. 

\subsection{Polarized companions?}
\label{polcom}

We have already discussed (\S\ref{objE}) companion E to 3C 441 as a 
possible case of nuclear light scattered and polarized by a cloud 
illuminated by the central engine. Other cases giving evidence for 
polarization by scattering have been published recently, as the new 
generation of telescopes and instruments begins to make possible
high-resolution imaging polarimetry of high redshift radio galaxies.

 Tran
et al.\ \scite{Tran+98a} used the Keck I to obtain extended imaging
polarimetry of 3C 265, 3C 277.2 and 3C 324. In all three cases, the  
polarization maps displayed bipolar fans of polarization vectors centred
on the nucleus, perpendicular to the optical structure and misaligned by  
tens of degrees with the radio axis. Earlier structural information on one
radio galaxy was obtained by di Serego Alighieri, Cimatti \& Fosbury   
\scite{Alighieri+93a}; their $V$-band polarimetry of the
$z=0.567$ object $1336+020$ showed perpendicular polarization in
a northern knot and in extended emission, higher than in the core.
 
Contour maps of the three Tran et al.\ \scite{Tran+98a} sources are
provided, at levels relative to the peak intensity of the central
knot, and all three sources include companion objects. Little or no
polarization is seen in the bright ($\sim 20$ per cent of peak) companions
of 3C
277.2 and 3C 324. But in 3C 265, a faint companion object also exhibits
the polarization seen in the fan -- the object is a knot less than 8 per 
cent of peak intensity and lies beyond the extension of the $V$-band optical
structure, in the same direction but unconnected with the optical core in
contours down to 2 per cent of peak. Such a faint polarized knot could  
readily be
identified with light redirected by a cloud of scattering particles.
 
 Assuming an $\Omega_0 = 1.0, \Lambda=0$ cosmology with $H_0 = h_0$
km\,s$^{-1}$\,Mpc$^{-1}$, $h_0 = 100$, the knot in 3C 265 which lies about
9\arcsec\ from
the core, is separated from the core by about 36$h_0$ kpc. The extended
structure of $1336+020$ of about 3\arcsec\ corresponds to 12$h_0$ kpc. In
comparison the 53W091 to $3a$ separation and the distance
between 3C 441 {\bf a} and E both correspond to approximately 16$h_0$ kpc.
So the structure of these companion
objects is of comparable scale to those in the literature.
 
\subsection{Alignments in the knotted sources: an overview}

 The most striking feature of our small sample of four knotted sources 
(Table \ref{knotalign})
is that in all four cases, the core and at least one other knot lie
along the line of the radio structure axis. In 3C 114, three prominent
knots lie on this axis and the fourth is offset perpendicular; in 3C
356, the displacement vector between the major components $a$ and $b$
lies within a few degrees of the radio axis; 3C 441 has component {\bf
c} closely associated with the NW radio lobe and component F fairly
close to the radio axis; and 53W091 is displaced from companion 3a,
again within a few degrees of the radio axis. Only 3C 441 has
relatively bright companions in skew positions; and since this source
seems to be in a rich field, these companions can easily be accounted
for as cluster members rather than effects of the active nucleus.

\begin{small}
\begin{table}
\centering
\caption{Alignments in knotted objects.}
\label{knotalign}
\begin{tabular}{rccc}
\hline
Source & Companions & Optical Core & Polarization \\
\hline  
3C 114 & $\parallel,\perp$ & $\odot$ & $\perp$\ (T2), $\parallel$\ (T0) \\
3C 356 & $\parallel$ & $\parallel$ & $\parallel\ (a); \angle\ (b)$ \\
3C 441 & $\parallel,\perp$ & $\parallel$ & $\parallel$ (E); $\perp$\ ({\bf 
a}) \\
53W091 & $\parallel$ & $\angle$ & $\parallel\ (3a); \perp$\ (core) \\
\hline
\end{tabular}
 
\medskip
 
\parbox{153mm}{Key: 
$\perp$: perpendicular to radio structure; $\parallel$: parallel to radio 
structure;  $\angle$: skew to radio structure; $\odot$\,: primary knot 
has no extended structure.}
\end{table}
\end{small}

\subsection{A conical sector model}

 We shall create a `toy model' to help us investigate the properties of
light scattered from a cloud of dust or electrons. Brown \& McLean
\scite{Brown+77a} have modelled the case of axisymmetric Thomson
scattering in a stellar envelope, and we can easily adapt this model to
the case of scattering by an axisymmetric electron cloud in the conical
region illuminated by a quasar nucleus embedded in an obscuring torus. We 
take an $(r,\theta,\phi)$ spherical co-ordinate system with the polar 
axis as the axis of the obscuring torus, and define $\mu = \cos(\theta)$. 
For convenience of integration we shall consider the electron cloud to 
have a constant number density, $n_0$, and to have the shape of a conical 
sector with boundaries $R_1 \le r \le R_2$ and $0 \le \theta \le 
\Theta_1$ [hence $1 \ge \mu \ge \mu_1 = \cos(\Theta_1)$].

Brown \& McLean's \scite{Brown+77a} treatment of the problem 
gives 
expressions for the light intensity (their Equation 5) and polarization 
(their Equation 16) of light scattered from an axisymmetric cloud of any 
number density $n(r,\mu)$ integrated over $r: 0 \rightarrow +\infty$ and 
$\mu: -1 \rightarrow +1$. They define the axial inclination, $i$, as the 
angle between the equatorial plane of the scattering cloud (in our case, 
the plane of the obscuring torus) and the sky plane (the perpendicular to 
the line of sight from the source to Earth). In this convention, our 
scattering angle, $\chi$, is such that $\chi = 180\degr - i$. 

Substituting our special case of a constant density 
cloud with conical sector boundaries, we obtain:
\begin{equation}
\label{scatintens}
I_1 = \frac{3 I_0 \sigma_T n_0}{16} (R_2 - R_1) \left[ (1-\mu_1)(2+\sin^2 
i) + \frac{(1-{\mu_1}^3)(2-3\sin^2 i)}{3} \right],
\end{equation}
and
\begin{equation}
\label{scatpol}
\frac{1}{P} = 1 + 2 ({\mathrm cosec}^2 i) \left[ 
\frac{3(1-\mu_1)+(1-{\mu_1}^3)}{3(1-\mu_1)-3(1-{\mu_1}^3)} \right];
\end{equation}
if $I_0$ is the intensity (power radiated into unit solid angle) of the 
central source, then $I_1$ is the intensity of the scattered radiation; 
$P$ is the polarization of this radiation, and $\sigma_T$ is the 
wavelength-independent cross-section for Thomson scattering. In this 
formalism, a negative $P$ corresponds to polarization perpendicular to 
the symmetry axis.

 The most striking feature of Equation \ref{scatpol} is that it is 
independent of the radial boundaries $R_1,R_2$. It follows that for small 
opening angles, where a spherical cap can be approximated as a disc of 
constant (Cartesian) $z$, the
(undiluted) polarization of {\bf any} axisymmetric distribution of electrons 
$n(r)$ depends only on the opening angle of the illuminating aperture;
we can model 
any axisymmetric distribution of dust $n(r)$ as the sum of scattered 
light from infinitessimal slices of constant density.

 We will now use our toy model to investigate two scenarios: scattering 
from a conical region of dust stretching from the origin $(R_1=0)$ to 
some finite distance $R_2$; and scattering from comparable clouds at 
different distances from the origin.

\subsection{A patchy cloud model}

 In order to use our toy model to investigate galaxy structures like
that of 3C 114, we will consider what happens when similar small
clouds are placed at different distances from a point source. We
will consider all clouds to have the same volume, $V$, the same
diameter, $w$, and the same fixed particle density, $n$.
Further, for ease of integration, we will consider all clouds to be
placed on the axis, and bounded by radial lines forming conical
surfaces, and by spherical caps.

 Let us consider a cloud which subtends some half-angle $\Theta_1$ and
extends from $R_1$ to $R_1 + \Delta R$. Now the width of this cloud we
can take as the linear distance subtended by the spherical cap at $R_1
+ \Delta R/2$, {\em viz.}
\begin{equation}
w = 2 \sin (\Theta_1).(R_1 + \Delta R/2);
\end{equation}
the volume is obtained by the trivial integration of $r^2 \sin \theta
\,dr\,d\theta\,d\phi$ over $(R_1,R_1+\Delta R)$ in $r$, 0 to
$\Theta_1$ in $\theta$, and 0 to $2\pi$ in $\phi$, whence
\begin{equation}
V = (2\pi/3).(1-\cos \Theta_1).[(R_1+\Delta R)^3 - {R_1}^3].
\end{equation}
For given $w$, $V$ and distance from the illuminating source, $R_1$,
the appropriate opening angle $\Theta_1$ and radial thickness $\Delta
R$ can easily be calculated. Trial and error revealed that with $V=6.6$
and $w=2$, the calculated $\Delta R$ values were also of order 2 at
various radii $R_1$, yielding a model cloud about as broad as it is
deep. Such a quasi-symmetric cloud seems the most appropriate for a toy
model mimicking fairly symmetrical knots in galaxies. The different
dimensions of this conical sector cloud at differing radii are given in
Table \ref{cloudsize}.

\begin{small}
\begin{table}
\centering
\caption{Dimensions of a conical sector of fixed volume and width.}
 
\medskip
 
\label{cloudsize}
\begin{tabular}{rrrr}
\hline 
$R_1$ & $\Delta R$ & $\Theta_1$ & $\mu_1$ \\
\hline
1 & 1.81 & 31.68 & 0.851 \\	
2 & 1.97 & 19.58 & 0.942 \\	
3 & 2.02 & 14.43 & 0.968 \\	
4 & 2.05 & 11.48 & 0.980 \\	
5 & 2.07 &  9.54 & 0.986 \\
6 & 2.08 &  8.17 & 0.990 \\
7 & 2.08 &  7.14 & 0.992 \\
\hline
\end{tabular}

\medskip
 
\parbox{153mm}{Key: The dimensions above are for a conical sector
bounded by
spherical caps at $R_1$ and $R_1 + \Delta R$, and a conical
surface subtending a semi-vertical angle of $\Theta_1$, such that
the volume enclosed is 6.6 units and the diameter of the disc bounding
the spherical cap at $R_1 + \Delta R/2$ is 2 units. We define $\mu_1 = 
\cos (\Theta_1)$.}
\end{table}
\end{small}

By substituting the boundaries for such a cloud into Equations
\ref{scatintens} and \ref{scatpol}, we can immediately obtain the
polarization and relative intensity of the light scattered by similar
clouds at different densities. The {\em undiluted}\, polarization is,
in fact, independent of both the cloud density and the intensity of the
illuminating source. The polarizations and intensities (relative to the
$R=1$ case) are given in Table \ref{patchpol}. It is also useful to
calculate the surface brightness, $B=I_1/A$: approximating the side-on
profile of
the cloud as a sector subtending half-angle $\Theta_1$, the area
bounded between $R_1$ and $R_1 + \Delta R$ is $A = 2\Theta_1 [(R_1 +
\Delta R)^2 - {R_1}^2]$. 

\begin{small}
\begin{table}
\centering
\caption{Polarization and intensity of light scattered by a small cloud
at various distances from an illuminating nucleus.}
 
\medskip
 
\label{patchpol}
\begin{tabular}{rrrrr}
\hline 
$R_1$ & $P$(\%) & $I_r$(\%) & $B_r$(\%) & $P_r$(\%)\\
\hline
1 & 73.6 & 100.0 & 100.0 & 100.0 \\	
2 & 89.0 &  40.6 &  38.5 &  55.7 \\	
3 & 93.9 &  22.5 &  20.9 &  34.0 \\	
4 & 96.0 &  14.4 &  13.2 &  22.6 \\	
5 & 97.3 &  10.0 &   9.1 &  16.1 \\
6 & 98.0 &   7.3 &   6.7 &  12.0 \\
7 & 98.5 &   5.6 &   5.1 &   9.3 \\
\hline
\end{tabular}

\medskip
 
\parbox{153mm}{Key: $R_1$: Distance of inner edge of cloud from
illuminating nucleus; $P$: polarization of light scattered by cloud
(always perpendicular to the nucleus-cloud axis in the cases above)
(\%); $I_r$: relative intensity $I_1(R_1)/I_1(1)$\ 
(\%); $B_r$: relative surface intensity $B(R_1)/B(1)$\ 
(\%); $P_r$: relative diluted polarization $P_d(R_1)/P_d(1)$\ (\%).}
\end{table}
\end{small}

It is evident from Table \ref{patchpol} that clouds of a given size
lying further from their source of illumination are more strongly
polarized, but the intensity of the polarized light is diminished.
Obviously at greater distances, the angle (and hence fraction of the
source emission) subtended by the cloud is smaller, but the beam
entering the cloud is more collimated and hence there is less
cancellation of polarization from light being scattered in opposite
senses.

 The toy model does not require that the illuminating source be
shielded, only that it be pointlike; so a scattering cloud illuminated
by an external small non-AGN source (e.g. a star cluster) would also
produce polarised light in the manner of this model. If the scattering
cloud's only source of light is the external illumination, it is
clearly not possible to obtain a structure of knots of similar
brightness at different distances from the nucleus, since the relative
intensity (and surface intensity) drop off so rapidly with distance.
Where the intensity of the intrinsic emission of the knot, $I_k$, is
much greater than that of the scattered light, the polarization of the
diluted light can be approximated as $P_d = P.I_1/I_k$, and the values
of $P_d$ relative to that at $R_1=1$ (since the absolute value depends
on the knot's intrinsic brightness) are also tabulated in Table
\ref{patchpol}.

 Is a scattering model adequate, therefore, to account for the knotted
structures seen in many high-redshift radio galaxies? The central
engine, by definition, must be many times brighter than any scattering
clouds shining purely by scattering some small portion of its output;
but if the nucleus is well-shielded then the residual starlight of the
host galaxy might be of comparable brightness to a knot of scattered
light. If there are multiple knots, those at greater distances from the
nucleus should be much fainter, following Table \ref{patchpol}, unless
they contain denser clouds of scattering material than the nearer
clouds. Knots offset more than 45\degr\ from the radio axis ought to be
shielded from the central engine by its obscuring torus; the light of
such knots cannot easily be attributed to scattering.

\subsection{A continuous cloud model}

 Another useful model to examine is that of a scattering cloud bounded
by $0 \le r \le R_1$ and $0 \le \theta \le \Theta_1$. This case can
model an AGN whose conical region of illumination is filled with
scattering particles close to the active nucleus. In a canonical model
with $\Theta_1 = 45\degr$ where the particles fill the opening angle of
the central engine, then $P = 53.3\%$. The undiluted polarizations
owing to cones filled at other opening angles are given in Table
\ref{conepol}. Again, the undiluted polarizations are perpendicular to
the conical axis and independent of the radial extent of the scattering
region. One important consequence of this model is that any measurement
implying an undiluted polarization much greater than 50\% is indicative
of a scattering medium subtending an angle smaller than the whole
45\degr\ illumination zone.

\begin{small}
\begin{table}
\centering
\caption{Polarization of scattering cones filled to various opening
angles.}
 
\medskip
 
\label{conepol}
\begin{tabular}{lrrrrrrrrrrr}
\hline
$\Theta_1$ & 5 & 10 & 15 & 20 & 25 & 30 & 35 & 40 & 45 & 50 & 55\\
\hline
$P$ & 99.2 & 97.0 & 93.4 & 88.5 & 82.6 & 75.9 & 68.7 & 61.1 & 53.3 &
45.6 & 38.2\\
\hline
\end{tabular}

\medskip
 
\parbox{153mm}{Key: $\Theta_1$: opening angle of scattering cone
(\degr);
$P$: polarization due to scattering in this cone (\%).}
\end{table}
\end{small}

\subsection{3C 114}
\label{M3C114}

We turn first to 3C 114, with three knots lying along the radio jet and
one offset perpendicularly (Figure \ref{114figs}). In this source, the
`knee' knot (T2) is presumably part of the parallel aligned structure,
and its polarization is roughly perpendicular to the radio jet, as
expected if some scattering process is at work inside the knot, perhaps
a region of dust in the scattering cone as discussed immediately above.
The rest of the structure, however, is difficult to interpret in terms
of a scattering model. We have no polarimetry on the knots T3 or T4,
but if T2 contains the central engine, T4 is brighter than T3
though it lies beyond T3. Clearly the light from T4 cannot primarily be
scattered light and some other mechanism must be at work (e.g. jet
induced star formation) to produce the Alignment Effect in this object.

 When polarimetry is performed over the whole aperture, including the
perpendicular knot T1, the overall polarization seems {\em parallel}\, to the
radio jet --- i.e.\ perpendicular to the axis connecting this anomalous
knot to the main structure. Is this indicative of some scattering
mechanism perpendicular to the main radio structure? It is difficult to
conceive of a mechanism which allows a brilliant central engine to emit
light in two perpendicular directions without also leaking such light
towards Earth; yet the morphology (and indeed the high polarization
which implies that any dilution is limited) do not suggest that we are
seeing a central engine directly in T2. Could T1 be a chance alignment
or a satellite galaxy? This would explain its presence but not the 
overall parallel polarization. The brightness of the knots in 3C
114 make it a prime candidate for spectroscopy or spectropolarimetry,
which would shed more light on the chemistry and light emission
mechanisms at work in each of the knots.

\subsection{3C 356}
\label{M3C356}

We have already reviewed the structure of 3C 356 (\S\ref{D3C356}) and
noted how the SE radio core $b$ has been proposed as the central engine
with component $a$ interacting with its jet \cite{Lacy+94a,Eales+90a}.
More recent imaging (BLR-III) and spectropolarimetry
\cite{Cimatti+97a}, however, now point towards $a$ as being the more
likely host of the active nucleus. Broad lines are clearly visible in
the polarised spectrum of $a$, which itself is now known to have two
components (BLR-II); two components are aligned along the $a-b$
axis with the component nearer $b$ providing 60\% of the total emission
from $a$. The Keck spectropolarimetry unequivocally demonstrates the
presence of a polarized component strong in the near-ultraviolet and
declining into the green; our $K$-band polarimetry cannot identify this
polarized component in the near-infrared, suggesting that nebular
emission and starlight are dominating this waveband.

 The weight of evidence in the more recent literature, though not
conclusive, indicates $a$ as the more likely host of the 3C radio
source, though in this case $b$ seems to be a flat-spectrum radio
galaxy in its own right. Comparison of the spectra of $a$ and $b$
presented by Cimatti et al.\ \scite{Cimatti+97a} does not suggest that
$b$'s main source of light is scattering from $a$'s emission, nor is
the polarization in $b$ high; they also show that the polarization in
$a$ can be modelled both by dust and by electron scattering. The high
$(\sim 15\%)$ polarization present in the ultraviolet shows that
scattered light must form a large proportion of the light from $a$ at
these frequencies, and Cimatti et al.\ \scite{Cimatti+97a} estimate
that at 280\,nm, 50 \tpm 15\% of the total flux from $a$ is scattered
light, and the intrinsic undiluted polarization is 21 \tpm 7\%. Our
$K$-band data can only add to this the knowledge that the diluting
component is much stronger in the infrared.

\subsection{3C 441}
\label{M3C441}

Analysis of 3C 441 is complicated by the richness of the field in which 
it lies; without knowledge of the redshifts of all the objects, and hence 
of which could be true neighbours to the radio core, any interpretation 
must be tentative. Object E is known to be much fainter than the presumed 
active nucleus in $K$ (this thesis) and in $I$ and 
[O\,{\sc ii}] \cite{Neeser-96a}. This is consistent with a scattered 
light hypothesis and it is highly plausible
that companion E to 3C 441 is an
illuminated object scattering light in the manner of the extended
structure seen in 3C 265 and $1336+020$, pending confirmation of its 
redshift --- though it must be noted that the result for E does have a 
ten percent chance of being a noise-induced spurious result.

\subsection{LBDS 53W091}
\label{M53W091}

 Unlike the other objects in this thesis, LBDS 53W091 was chosen on the 
basis of the {\em weakness}\, of its radio emission. Its apparent age and 
redshift are hard to reconcile and the reported 40\% polarization did not 
help to clarify the picture. We have ruled out a $K$-band polarization of 
that order; most of the discussion of this galaxy's peculiar properties 
can be found in \S\ref{D53W091} and in more detail\footnote{Doctoral thesis
declaration: that discussion is not reproduced in detail in this thesis
as the major part of it was developed by Dr James Dunlop of the University of
Edinburgh.} in Leyshon,
Dunlop \& Eales \scite{Leyshon+99a}.

 We have seen how radio-weak galaxies only display the Alignment Effect 
over small scales, up to about 15\,kpc \cite{Lacy+98b}; the separation 
between 53W091 and companion 3a is of this order (16$h_0$\,kpc) and so 
its good alignment is in keeping with what is known for 7C galaxies. We 
have seen (\S\ref{53Wgeom}), however, that the polarization in 3a cannot be 
reconciled with a scattering model; and the result for 3a is too 
tentative to warrant developing alternative models.

\section{Radio Galaxy Trends: The Big Picture}

 The data presented in this thesis represents the first $K$-band 
polarimetry of high redshift radio galaxies; that is, the measurements 
are the first on record for the rest frame near-infrared (0.7--1.3\,\micron) 
polarizations of radio galaxies significantly less evolved than the local 
universe. The results can be compared with the infrared properties of nearby 
radio galaxies and with the visible-light properties of 
high redshift objects (\S\ref{polevi}) --- always bearing in mind that 
our sample of nine diverse objects is of limited statistical significance.

 Some nearby radio galaxies (Cen A, IC 5063, 3C 234) have been observed in
the $K$-band (\S\ref{polevi}) and found to have polarizations of order
10\%, oriented perpendicular to the radio jet. In these cases, it seems
that the standard scattering hypothesis is the best explanation. 3C 233.1
is 5\% polarized in $K$ but less than 0.5\% in the visible, suggesting
that dust extinction may reduce the contribution of visible scattered
light.  Our sources are being observed at rest-frame wavelengths somewhat
shorter than the 2.2\,\micron\ $K$-band, so if 3C 233.1 were used as a
benchmark, we might expect perpendicular polarizations somewhat lower 
than 5\%. In fact our measurements range between zero and 20\%, all in 
the `ballpark' defined by these earlier $K$-band observations.

 We have noted (\S\ref{poltrend}) how Cimatti et al.\ \scite{Cimatti+93a}
analyzed the properties of 42 radio galaxies at $z\geq0.1$ and searched 
for trends with both the observed polarization and their estimate of the 
underlying nuclear polarization. All their conclusions were based on light 
emitted at rest frame wavelengths between 0.2\,\micron\ and 0.7\,\micron, 
so the observations of this thesis do not overlap in $\lambda_r$ with 
theirs. Nevertheless, it is valuable to try to interpret our results in 
the context of their trends analysis.

  Cimatti et al.\ \scite{Cimatti+93a} found that a good rule of thumb was
that radio galaxies at $z>0.6$ were polarized above 8\% and those at lower
redshift, less than 7\%. All of our sources lie at $z>0.6$, but some are
certainly polarized below 7\% in $K$-band, the firmest results being the
unpolarized MRC 0156$-$252 and the 3\% polarized 3C 22 and 3C 41. Jannuzi
\scite{jpc} reports that 3C 22 is no more than 5\% polarized in the
$V$-band; 3C 41 fulfills the rule of thumb in $V$ but not in $H$. The rule,
however, is a rule for radio galaxies, so if these objects are actually
obscured quasars the rule is not applicable to them. Spectropolarimetry
for 3C 356 $a$ \cite{Cimatti+97a} shows the rule satisfied in this source
at 0.2\,\micron\ but not at 0.4\,\micron, while our results indicate that
it probably obeys the rule at 1.1\,\micron. The handy rule of thumb,
therefore, must be used with the caveat that it applies only to light
emitted at visible/ultraviolet rest wavelengths. 

 There is also a rule of thumb based on general results about alignments.
Radio galaxies with $z>0.6$ and $P>5\%$ were always found to have
perpendicular polarizations. Again, this is not found to carry into the
$K$-band, since 3C 54's polarization is parallel, and 3C 114 is ambiguous
depending on which knots are included. Neither can we sustain the rule
that if a radio galaxy polarization is in parallel alignment, the
polarization is always lower than 5\%: the same two objects and 3C 356 may 
all have higher parallel polarizations.

 Except for 53W091, all the objects in our sample are very radio loud; all
lie at high redshift. They would therefore be expected to display a clear
Alignment Effect in the visible (\S\ref{aligneff}) and at least a marginal
effect in the infrared. In fact the knotted sources (Table
\ref{knotalign}) do display parallel aligned structures, though some (most
strikingly 3C 114) also have knots in perpendicular alignment --- a
feature not unknown in nearby radio galaxies \cite{Crane+97a}. 
Discerning alignments present in compact objects (Table \ref{compalign}) 
is, by definition, more difficult, and no clear trend is apparent in our 
five compact sources. Clearly bright knots, where present, dominate any 
analysis of alignments and structure, whereas the slight extensions or 
faint companions of radio galaxies may not be related to the mechanism 
which sometimes causes aligned or perpendicular knots.

\chapter{Conclusions}

\begin{quote}
It seemed to me that in one of my innumerable essays, here and elsewhere,
I had expressed a certain gladness at living in a century in which we
finally got the basis of the Universe straight. \attrib{Isaac 
Asimov, {\em The Relativity of Wrong}}
\end{quote}

 The work contained in this thesis has pushed forward scientific 
knowledge on two fronts: the practice of astronomical polarimetry, and 
our knowledge of the $K$-band polarization properties of radio galaxies. 
As always in science, new results reveal to us how little we know about 
the Universe at large and suggest future avenues of exploration. It is 
clear that there is great room for improvement in the polarimetric 
analysis software available to the astronomical community, and some 
recommendations are made here on functions which should be included in 
any comprehensive analysis package of the future.

 Our sample of targets was selected for its diversity and provides a
snapshot of some interesting objects; even so, some of these provide
marginal results over the maximum realistic integration time on a
world-class infrared telescope. Today, we can only dream of taking a
sample large enough to yield good statistics: the number of objects and
the integration times which would be required conspire to place such a
project in the realms of spaceborne infrared telescopes, Keck-size
telescopes, or weeks of dedicated observing time. In the meantime,
individual $K$-band objects will surely be subjected to polarimetric 
analysis,
and it is possible to give some pointers for properties to look out for. 

\section{Requirements for a Comprehensive Polarimetric Software Package}

The software currently available for polarimetric analysis ({\sc
aaopol} from the Anglo-Australian Observatory and {\sc polpack} from
{\em Starlink}\,) concentrates on the generation of `vector maps'
illustrating the polarization of different parts of an image by means
of arrows of appropriate length and orientation. Such software is fine
for imaging polarimetry where there is a high-quality signal, but is
inadequate to deal with pioneering research where the signal-to-noise
ratio is low. The experience gained in performing the analysis for this
thesis suggests that a future comprehensive polarimetry package should
include the following features for two-channel Wollaston prism systems
with a waveplate rotating in 22.5\degr\ steps.

\subsection{Generation of the Stokes Parameters}

 An imaging polarimetry package would normally function as an accessory
to an imaging photometry system.  It would be
necessary to define one
or more polarimetry apertures (a whole object, a series of knots, etc.)
and define the binning resolution (the whole aperture, individual
pixels, or some intermediate level). It should be possible to define a
list of input images (possibly several for each waveplate) and tag each
of them with the orientation of the
waveplate used for that exposure; the orientation, $\eta_0$, of the
reference axis of the waveplate system should also be noted. It would
be desirable to provide automatic and manual facilities for registering
the waveplate images rather than assuming perfect alignment.

 Having defined the sampling apertures and resolution, the software
should be capable of calling the photometry package, accepting the
returned photometry data, and converting the results to absolute Stokes
Parameters (with errors)  relative to the waveplate reference axis. The
reference angle and Stokes parameters for each pixel bin of each source
on each image should be stored in a file for further analysis. This
part of the software should implement Steps \ref{smallshot} to
\ref{noiseOK} of Chapter \ref{stoch}.

\subsection{Stokes Parameter analysis routines}

 The nucleus of a polarimetry package should be a versatile system for
performing analysis on sets of Stokes Parameters, absolute or
normalized, derived from the photometry procedure detailed above, or
directly entered from the literature. Among the analysis routines
available should be the following:

\begin{enumerate}

\item \label{optestfunc} Derive the optimal estimates of the
normalized Stokes Parameters from the absolute Stokes Parameters,
following (i) Steps \ref{hereNSPs} to \ref{nearnormal}, or (ii) the
iterative method of \S\ref{zpos}.

\item Estimate the probability that the true polarization of a bin is
zero/non-zero using any of three methods: (i) the absolute Stokes
Parameter confidence interval test (Step \ref{findconfr}), (ii) the
polarization debiasing test (Step \ref{estpolun}) and (iii) the
residual method (\S\ref{zpos}).

\item Convert normalized Stokes parameters to percentage polarization
(Steps \ref{findperr} to \ref{getint}) and orientation (Steps
\ref{propphi} and \ref{findangle}) format, providing point estimates
and/or confidence intervals.

\item Convert data in the form of percentage polarization and
orientation into Stokes Parameter form, both for `debiased'
polarizations and non-debiased crude estimates of the percentage
polarization. (Such a function allows rapid conversion of data from the
literature into a form comparable with other data.)

\item Test two sets of Stokes parameters for consistency with one
another (for studying temporal variability etc.).

\item
Convert Stokes Parameters from a given reference frame to a standard
frame where the axis points North. (Working in the instrumental frame
is best as the errors on the two channels are independent of one
another; but standard orientation may be needed if data from two
telescopes with different reference axes are to be combined.)

\item Combine two sets of normalized Stokes parameters. (Combining two
sets of absolute Stokes Parameters is a trivial extension of function
(\ref{optestfunc}) unless the two reference axes are misaligned.)

\end{enumerate}

 A software package capable of performing all these analyses would be a
powerful tool enabling the rigorous analysis of new polarimetric data
and efficient comparison with the existing literature. Ideally the
software should be able to output the debiased data in a form
compatible with existing software for displaying polarization vector
maps. Limited by the shot noise inherent in photometry of faint
sources, and providing optimal estimates, such software would
yield the most accurate estimates of true polarizations
theoretically possible, and these recommendations are commended to
astronomical programmers for their consideration.

\section{A Summary of the Properties of Our Sample}

We have taken $K$-band polarimetry for seven 3CR radio galaxies, and found
a diverse range of results. Out of our seven sources,
two (3C 65 and 3C 441) display no evidence for polarization (though a 
companion to 3C 441 may be polarised).
For the sources which do display some evidence of polarization, we
have estimated the fraction of observed $K$-band light which originated in
the active nucleus (Table \ref{quasfrac}). Most of our findings are lower 
than the $\Phi_{K} \ga 
40\%$ suggested by the recent findings of Eales et al.\ \scite{Eales+97a} 
but are consistent with the 
hypothesis that radio galaxies consist of quasar
nuclei embedded in giant elliptical galaxies.
 
 All of the galaxies which appear to be polarised have large errors on
the orientation of their {\boldmath $E$}-vectors; hence any apparent
alignment effects are suggestive rather than definitive. But with
this caveat, we note that two sources (3C 54 and 3C 114) 
have high polarizations oriented in roughly parallel
alignment with the radio axis and extension of the optical structure ---
i.e.\, in the opposite sense to the perpendicular alignment expected under a
simple scattering model.
 
 The compact galaxies 3C 22 and 3C 41 display significant polarizations of
around 3\% with a polarization alignment perpendicular to their radio
axes; both appear in the $K$-band as pointlike objects. We suggest,
therefore, that in these objects, infrared light from a quasar core is
being scattered into our line of sight, and forms a significant part of
the total $K$-band flux received from these sources; both objects may be 
inclined close to the $\chi \sim 45\degr$ `boundary' between quasar and 
radio galaxy properties. 
 
 In the case of MRC 0156$-$252,
which lies beyond a virtually dust-free part of our own Galaxy, we can be
reasonably certain that this radio galaxy is not polarised, and the
$K$-band light has not been scattered before reaching us. If some of
the $K$-band light has originated in the active nucleus, its
contribution should be smaller than at visible wavelengths
\cite{Manzini+96a}; this being the case,
subtraction of our image or a synthetic symmetrical galaxy could well  
reveal the structure of the active component at visible wavelengths,
given the visible structure observed by McCarthy et al.\ 
\scite{McCarthy+92a}. It is possible that this galaxy, like 3C 22 and 3C 41, 
is an obscured quasar at an intermediate orientation.

 In LBDS 53W091, we can rule out the contribution of an active nucleus to
providing more than $\sim 25$ per cent of the observed light. The
majority of its $K$-band light, therefore, must be
presumed to be due to its stellar population, and its $R-K$ colour
remains consistent with an age in the range $2.5-3$ Gyr.
The nature of its companion object $3a$, possibly polarised and of
unclear physical relationship with 53W091, warrants further investigation.

 In 3C 441, the polarization from object E may indicate that E is
scattering light from {\bf a} (whose identification as the central engine
would thus be vindicated); the orientation of E's polarization would not
be consistent with the source being located within E or F and emitting
jets at $\sim 145\degr$. Therefore, we favour the traditional
identification of the central engine with {\bf a}. Although this object was 
observed at two epochs, the observational errors cannot rule in or out 
temporal variability in polarization over a two year period.

 It is noteworthy that when radial profile or spectral fitting
estimates \cite{Best+98a} are combined with our $K$-band data
(\S\ref{blr3constr}), there are hints that the true $K$-band nuclear
polarizations of several sources (3C 65, 3C 356, 3C 441 {\bf a}) are of
the order of 25\%. The measurement errors and uncertainties in the
derivation mean that these figures are no more than indicative; but
following Enrico Fermi's rule of thumb that in a sufficiently
complicated problem, uncertain contributions tend to cancel out each
other, this can be taken as a very tentative indication for $K$-band
nuclear polarizations of order 25\% in radio galaxies.

  Polarimetry of faint objects requires long integration times. The 
observing time
available has permitted us to rule out the existence of very high
polarizations in many of the objects studied, at least for light emitted
along the line of
sight to Earth. It would still be possible for light emitted in other 
directions from these objects to be polarised. `Polarization' mentioned
in these conclusions should be understood in the restricted sense
of light leaving the source in the direction of Earth. Under the Unification
Model, radio galaxies (a class of AGN assumed to be oriented with their
jets perpendicular to that line of sight) would be more likely to display
polarization originating in scattering or synchrotron radiation in the 
light travelling Earthwards than in directions closer to the jet.

\section{Recommendations for Future $K$-band Radio Galaxy Polarimetry}

 Interpreting polarization measurements is intimately linked with
understanding the morphology of the knots or extensions which accompany a
radio source. It is a distinct advantage to study sources in which the
redshifts of the companion objects are known, so that chance alignments
can be ruled out, and true near-neighbours can be identified as such. 
Where redshifts are not available and the source lies in a rich field,
multifibre spectroscopy of the sources should be scheduled. Then, if an
imaging polarization measurement suggests that a knot may be scattering
nuclear light, the AGN spectrum could be scaled and subtracted from that 
of the knot.

 The spectral and spatial profile fitting methods of Best,
Longair \& R\"{o}ttgering \scite{Best+98a} greatly complement the data 
available from polarimetry, since stronger limits can be placed on the 
nuclear polarizations if the fraction of the light due to stars can 
be independently identified. There would be a distinct advantage in 
performing polarimetry on the other objects already analysed by them, or 
applying their analysis (de Vaucouleurs fitting could be done on the 
imaging polarimetry images in good seeing) to stacked polarimetric images.

 Polarimetry is a cheap `overhead' which could be seriously considered 
whenever imaging or photometry is being carried out on a radio galaxy 
whose linear extent is such that it can easily be viewed through a focal 
plane mask. Polarimetry complements mosaicing by spreading the image over 
the pixels available, and local effects are cancelled out when the 
waveplate is rotated through 45\degr. Combining the two slits for imaging 
purposes can be integrated into the mosaicing process. The photon 
rate in each channel is 50\% of that which would be achieved with no 
Wollaston prism (for both source and sky noise), so doubling the time per 
exposure would produce noise at the same level per integration as using the 
system without the Wollaston prism. In the same spirit, applying a 
Wollaston prism before a grating or multifibre spectroscopy system allows 
spectropolarimetry AND spectroscopy of a given quality to be done in just 
twice the time of spectroscopy alone.

 Sources in which there is evidence for scattered light from a distinct 
knot -- such as 3C 441 -- are prime candidates for spectropolarimetry to 
be applied to their distinct knots. Tracing the nuclear spectrum in the 
polarized spectrum of the knot would confirm the scattered light hypothesis.

Spectropolarimetry should also be applied to sources in which there are 
indications of parallel polarization, such as 3C 114, 3C 54 and 53W091 
3a. Only low parallel polarizations can be produced by flattened scattering 
discs, and 
untangling the spectrum of the parallel polarized component would give a 
better hint at what is taking place in such sources.

 Sources in which their are indications of a skew Alignment Effect with
infrared structure 10\degr--20\degr\ out of line with visible structures
(\S \ref{misalign1}) are also particularly interesting targets for further
study, as are those where optical polarization is somewhat misaligned with
radio structure (\S \ref{misalign2}). It would be a valuable exercise to
perform a full literature review and observing campaign to compare the
polarization orientation with the position angles determined for structure
in the radio, near-infrared, visible and near-ultraviolet bands for every
AGN with a published optical polarization.  Understanding the
wavelength-dependence of the skew alignment would probe the mechanism at
work in these objects. 

 Finally, all polarimetric studies must be realistic about the time 
required to get a useful result. Low polarizations are harder to isolate 
than high ones; we spent more than an hour on 3C 22 and nearly two on 3C 
41, our brightest sources; and the errors on these are still quite large. 
Six hours on the faint 53W091 has refuted any suggestion of 40\% levels 
of polarization but cannot give a definitive answer on the presence or 
absence of lower levels. The author looks forward to the days when larger 
telescope mirrors, parallel use of Wollaston prisms and imaging 
spectroscopy allow the radio galaxy trends survey of Cimatti et al.\ 
\scite{Cimatti+93a} to be extended to $K$-band observations, and the true 
contribution of scattered $K$-band light to the properties of 
high-redshift radio galaxies will be known.

\appendix

\chapter{Minimum Theoretical Errors in Stokes Parameters}

\label{photapp}

\newtheorem{theorem}{Theorem}
\newtheorem{approximation}[theorem]{Approximation}

\newtheorem{conq}{Consequence}

 The recent work of S\'{a}nchez Almeida~\scite{Almeida-95a}, and of
Maronna, Feinstein \& Clocchiatti~\scite{Maronna+92a}, considers the ideal
case of polarimetry limited only by the shot noise intrinsic to quantized
light. For completeness their results are presented here in the notation
of this thesis, and extended slightly in the case of normalised Stokes
Parameters. (See Chapter \ref{stoch} for definitions of the Stokes
Parameters.) This treatment also considers which estimators are optimal 
for estimating Stokes
Parameters, and the 
consequences of binning the photon counts.
\newpage

\section{Light Intensity as a Poissonian Quantity}

 Consider a quasimonochromatic beam of light of intensity $I$, where the 
units of $I$ are photons per second. In a time interval $\tau$, the 
number of photons expected to arrive is $\lambda = I\tau$. As bosons, 
there will be some correlation between the arrival of individual photons, 
but this effect is negligible at optical wavelengths, and we can assume that 
the arrival of photons 
can be characterised by a Poisson distribution \cite[Ch. 2]{Walker-87a}. 
The 
number of photons actually arriving is hence a Poissonian random variable 
$X$, such that \begin{equation} \label{poissdef}
P(X=x) = \frac{e^{-\lambda} \lambda^{x}}{x!}.
\end{equation}
In the remainder of this appendix we will speak of such a Poisson 
distribution as having an intensity $I$, indicating that the mean of 
the distribution is $\lambda = I\tau$ for some arbitrary integration time 
$\tau$.

\subsection{Combining beams}

\begin{theorem}
 Combining two beams of light of intensity $I_1$ and $I_2$ produces a 
Poissonian beam of intensity $I = I_1 + I_2$.
\end{theorem}

If the beams
yield $\lambda_1\tau$ and $\lambda_2\tau$ photons respectively in the 
obvious notation, then
\begin{equation} \label{addbeam}
P(X=x) = \sum_{i=0}^{x} P(X_1=i).P(X_2=x-i) = 
 \sum_{i=0}^{x} \frac{e^{-\lambda_1} 
{\lambda_1}^{i}}{i!}.\frac{e^{-\lambda_2} {\lambda_2}^{x-i}}{(x-i)!}
\end{equation}
and factorizing out the exponential term,
\begin{equation} \label{addbeamb}
P(X=x) = e^{-(\lambda_1+\lambda_2)} 
\sum_{i=0}^{x}\frac{{\lambda_1}^{i}{\lambda_2}^{x-i}}{i!(x-i)!}. 
\end{equation}
 
Anticipating the result, we substitute $\lambda=\lambda_1+\lambda_2$ into 
Equation \ref{poissdef} and use the binomial expansion to obtain
\begin{equation} \label{poissex}
P(X=x) = \frac{e^{-(\lambda_1+\lambda_2)}}{x!} (\lambda_1+\lambda_2)^{x} = 
\frac{e^{-(\lambda_1+\lambda_2)}}{x!} 
\sum_{j=0}^{x} {\lambda_1}^{j}{\lambda_2}^{x-j} \frac{x!}{j!(x-j)!}.
\end{equation}

 Since the factors of $x!$ cancel and the indices $i$ and $j$ are summed 
over, then Equation \ref{addbeamb} is identical to Equation \ref{poissex} 
and the combined beams produce a Poisson distribution of mean 
intensity $I_1 + I_2$. {\em QED.}

\begin{conq}
 {\sl It follows that when Poissonian light from two sources is combined --
e.g.  light from a pair of stars, or from a host galaxy and an active
nucleus -- the resultant beam is also a Poissonian.}
\end{conq}

\subsection{Attenuating beams}

\begin{theorem}
\label{fracbeam}
 Passing light of intensity $I$ through a filter which passes a fraction 
$f$ of the photons results in a
Poissonian beam of intensity $fI$.
\end{theorem}

 The number of photons arriving at the filter follows a Poisson 
distribution for intensity $I$, so $X$ photons arrive at the filter. 
There is a binomial distribution such that $W$ photons penetrate the filter 
given that $X$ arrive, where
\begin{equation} \label{binpen}
P(W=w\,|\,X) = \frac{X!}{w!(X-w)!} f^w (1-f)^{(X-w)}.
\end{equation}
Overall the probability that $w$ arrive and penetrate is given by
\begin{equation} \label{binover} 
P(W=w) = \sum_{x=w}^{\infty} P(X=x).\frac{x!}{w!(x-w)!} f^w (1-f)^{(x-w)}.
\end{equation}

Making the substitution $k=x-w$ and expanding $P(X=x)$ from Equation 
\ref{poissdef}, we obtain \begin{equation} \label{binoverb}
P(W=w) = \sum_{k=0}^{\infty}
 \frac{e^{-\lambda}\lambda^{k+w}}{x!}.\frac{x!}{w!k!} f^w 
(1-f)^k \end{equation}
and the $x!$ terms cancel. We take out the terms in $w$ and obtain
\begin{equation} \label{binoverc}
P(W=w) = \frac{(f\lambda)^w}{w!}.\left[e^{-\lambda} \sum_{k=0}^{\infty}
 \frac{\lambda^k}{k!}(1-f)^k\right]. \end{equation} 

To complete the proof we must show
that the term in square brackets is equivalent to $e^{-f\lambda}$.
So, let the term in square brackets be denoted $G$. Expressing 
$e^{-\lambda}$ as a series, we have
\begin{equation} \label{binoverd}
G = \sum_{j=0}^{\infty} \frac{(-\lambda)^j}{j!} \sum_{k=0}^{\infty}
 \frac{\lambda^k}{k!}(1-f)^k. \end{equation} We can express the 
double series as a single series if we group together terms of the same 
power of $\lambda$:
\begin{equation} \label{binoverg}
G = \sum_{i=0}^{\infty} \lambda^i \left[  
\sum_{j=0}^{i} \frac{(-1)^j}{j!}\frac{(1-f)^{i-j}}{(i-j)!}\right].
\end{equation}
We now let $h=1-f$, and  recast this as
\begin{equation} \label{binoverh}
G = \sum_{i=0}^{\infty} \lambda^i \left[
\sum_{j=0}^{i} \frac{(-1)^j}{j!}\frac{h^{i-j}}{(i-j)!}\right],
\end{equation}
where we recognise the square bracket as $1/i!$ times the binomial expansion
\begin{equation} \label{binexh}
(h-1)^i = \sum_{j=0}^{i} \frac{(-1)^j h^{i-j} i!}{j!(i-j)!}.  
\end{equation} Thus Equation \ref{binoverg} can be simplified to 
\begin{equation} \label{proof2}
G = \sum_{i=0}^{\infty} \frac{\lambda^i (h-1)^i}{i!} = \sum_{i=0}^{\infty} 
\frac{\lambda^i (-f)^i}{i!} = \sum_{i=0}^{\infty} \frac{(-f\lambda)^i}{i!}.
\end{equation}
Thus $G$ is shown to be the series expansion of $e^{-f\lambda}$ and so
Equation \ref{binoverc} is the Poisson probability for a distribution of 
mean $f\lambda$. This completes the proof. 

\begin{conq}
{\sl Any optical filter which passes a fraction $f$ of the incident light 
in practice passes a fraction $f$ of the incident photons and removes 
the remainder. Theorem \ref{fracbeam} shows that any filters employed in 
an astronomical experiment will not affect the Poissonian properties 
of a quasimonochromatic beam. (Naturally the wavelength dependence of the 
filter would change the spectrum of a polychromatic beam.) It also 
follows that a photon detector of quantum efficiency $f$ also produces a 
Poissonian output.} \end{conq}

\subsection{Binning Poisson distributions}
\label{binsec}

The work in the following sections is based on the ideal case of a detector 
which records photon counts limited only by
the shot noise intrinsic to photons. In such a case, the arrival of detections is Poissonian. In practice,
however, real astronomical detectors \cite{McLean-97a} first allow incoming photons to excite electrons
which can be trapped, and then amplify and digitize the voltage due to 
these electrons. 

 Some incoming photons will fail to excite electrons, because the system will never have 100 per cent
quantum efficiency; but Theorem \ref{fracbeam} above shows that if the success or failure of a photon to do
so is random (i.e. does not depend on the photon's energy) then the photons which succeed in exciting
electrons will also follow a Poisson distribution. 

 If we assume an idealized system where a fraction $f$ of the photons excite exactly one electron and the
remainder go undetected, the next source of error is quantization error. The analogue-to-digital converter
of the detector will measure the number of electrons with a conversion factor of $\Delta$ electrons per data
number (DN). It can be shown \cite{Scarrott+83a,McLean-97a} that 
quantization contributes a noise of
$0.289\Delta$. More importantly, the output in DN no longer follows a Poisson distribution: the distribution
has now been binned in units of width $\Delta$. 

 It is possible to give an exact formula for a DN distribution simply by adding up the Poisson
probabilities for each number of photons which would yield a given quantized output: If $D$ is the random
variable `count in DN units' then (assuming $\Delta$ is an integer) we have
 \begin{equation}
\label{D_prob}
P(D=d) = \sum_{k=d\Delta}^{d\Delta+\Delta-1} e^{-\lambda}\lambda^{k}/k!,
\end{equation}
and the mean of the distribution is hence
 \begin{equation}
\label{D_mean}
E(D) = \sum_{d=0}^{\infty} d.e^{-\lambda} \sum_{k=d\Delta}^{d\Delta+\Delta-1}
 \lambda^{k}/k!.
\end{equation}
 There is no obvious analytic simplification of $E(D)$, but clearly 
binning the Poisson photon distribution in bins of width $\Delta$, since 
the photon distribution has $E(X) = \lambda,$~${\mathrm{SD}}(X) = 
\sqrt{\lambda}$, then the
DN readout must have a mean of approximately $E(D) \simeq \lambda/\Delta$ 
and ${\mathrm{SD}}(D) \simeq (\sqrt{\lambda})/\Delta$.

 The square root of $E(D)$ is hence $\sqrt{\lambda/\Delta}$, 
which is equivalent to $\sqrt{\Delta}$
times ${\mathrm{SD}}(D)$. Thus if a count, $d$, has been measured 
in DN units, it follows that ${\mathrm{SD}}(D) =
\sqrt{E(D)/\Delta}$ or equivalently ${{\mathrm{SD}}({D})}^2 = 
E(D)/\Delta$. This approximation will be good for typical bin sizes
($\Delta \sim 6$) as long as the integration time is such that  
$\lambda > \Delta$. (This was verified empirically on a spreadsheet.)

 We will not consider rigorously here the case of a detector where an incident photon is likely to excite
more than one electron. Again, however, if the number of incident photons is significant over the
integration time, it will be possible to define an overall gain $\Delta$ encompassing the photon-to-electron
and analogue-to-digital conversions, and the crucial  
${{\mathrm{SD}}(D)}^2 = E(D)/\Delta$ relationship will be retained
to a first approximation. 

 Finally, in a realistic application (this thesis, \S \ref{realnoise}),
the units of choice will often be `DN per unit time'.
In this case, the system's output will be a count rate of $O_D = E(D) / \tau$ 
with shot noise of $\sigma_{\mathsf{shot}} = {{\mathrm{SD}}(D)} / \tau$, whence the relationship between
noise and signal becomes
\begin{equation} \label{shotSN}
\sigma_{\mathsf{shot}}^2 = [SD(D)]^2/\tau^2 = E(D)/\tau^2\Delta =
O_D/\tau \Delta.
\end{equation}

\newpage

\section{Noise in a Generalised Polarimeter}

\subsection{Absolute Stokes Parameters}

S\'{a}nchez Almeida~\scite{Almeida-95a} considers the most general case of
a polarimeter which splits the light from a source into $m$ different
optical trains, which each produce a photo-count $n_i$. At least four
distinct optical trains are needed to determine all four Stokes Parameters
but this treatment also applies to systems with $m<4$ which can only
determine $m$ Stokes Parameters. 

 We presume that this general polarimeter produces the Stokes Parameters
$I, Q, U, V$ (or a subset if $m<4$) with their respective errors
$\sigma_I, \sigma_Q, \sigma_U, \sigma_V$. (S\'{a}nchez Almeida takes these
Stokes Parameters to be actual numbers of photons, but the treatment
remains valid when normalized for unit time.) Using $S$ to denote any of
$Q, U, V$, and where $\cal N$ is the total number of photons received
summed over all $m$ optical trains, S\'{a}nchez
Almeida~\scite{Almeida-95a} proves the following (his Equation numbers
denoted SA): 

\begin{theorem}[SA 11a] \label{ithm}
The signal-to-noise on the intensity cannot be better than $\sqrt{\cal 
N}$ : \begin{equation}
\label{sa_inoise}
\sigma_{I} \geq I / \sqrt{\cal N}.
\end{equation}
\end{theorem}
It is possible to build `polarizers of minimum $I$ error'~\cite[\S 
4.2]{Almeida-95a}~which have $\sigma_I =I / \sqrt{\cal N}$.

\begin{theorem}[SA 11a] \label{sthm}
The signal-to-noise on the other Stokes Parameters cannot be better than 
$\sqrt{\cal N}$: \begin{equation} \label{sa_snoise}
\sigma_S \geq \abs{S} / \sqrt{\cal N}.
\end{equation}
\end{theorem}

\begin{theorem}[SA 13]
{\bf For unpolarized light,} the noise on the other Stokes 
Parameters is limited by the intensity, and
cannot be better than $I/\sqrt{\cal N}$: \begin{equation} \label{sa_unoise}
\forall S=0: \sigma_S \geq I / \sqrt{\cal N};
\end{equation}
this does not necessarily hold true for polarised light.
\end{theorem}

\begin{theorem}[SA 17,18] 
{\bf For polarizers with $m=4$ and polarizers of minimum $I$ error,} the 
errors on the Stokes Parameters are correlated such that
\begin{equation} \label{sa_anoise}
{\sigma_Q}^2 + {\sigma_U}^2 + {\sigma_V}^2 \geq {\sigma_I}^2 \geq I^2/ 
{\cal N}.
\end{equation}
\end{theorem}

\subsection{The case of binned absolute Stokes Parameters}

S\'{a}nchez Almeida's~\scite{Almeida-95a} logic can also be applied to Stokes Parameters expressed in DN
units from a real detector, if the Poissonian substitution $n_i 
\rightarrow {\sigma_i}^2$ is replaced by the
binned substitution $n_i \rightarrow \Delta {\sigma_i}^2$. 

Whence for Stokes Parameters $I_{D}$, $S_{D}$ and total count ${\cal N}_{D}$ expressed in DN units with
$\Delta$ photons per DN: 

\begin{theorem}[SA 11a] \label{DNithm}
The signal-to-noise on the intensity is restricted to:
\begin{equation}
\label{DNsa_inoise}
\sigma_{I_{D}} \geq I_{D} / \sqrt{\Delta {\cal N}_{D}}.
\end{equation}
\end{theorem}

\begin{theorem}[SA 11a] \label{DNsthm}
The signal-to-noise on the other Stokes Parameters cannot be better than 
$\sqrt{\cal N}$: \begin{equation} \label{DNsa_snoise}
\sigma_{S_{D}} \geq \abs{S_{D}} / \sqrt{\Delta{\cal N}_{D}}.
\end{equation}
\end{theorem}

\begin{theorem}[SA 13] \label{DNsa_last}
{\bf For unpolarized light,} the noise on the other Stokes 
Parameters is limited by the intensity: \begin{equation} \label{DNsa_unoise}
\forall S_{D}=0: \sigma_{S_{D}} \geq I_{D} / \sqrt{\Delta {\cal N}_{D}};
\end{equation}
this does not necessarily hold true for polarised light.
\end{theorem}

\begin{conq}
{\sl It follows that Theorems \ref{ithm} to \ref{DNsa_last} allow us to 
estimate, {\em a priori}, the minimum errors obtainable when absolute 
Stokes Parameters are measured for an object of known magnitude, both 
for detectors registering raw photon counts and for the binned case.}
\end{conq}

\subsection{Extension to normalised Stokes Parameters}

S\'{a}nchez Almeida's~\scite{Almeida-95a} method can be extended to 
give the minimum possible error on a normalised Stokes Parameter. He 
defines a calibration matrix $M_{ji}$ such that the measured Stokes 
Parameters (including $I=S_1$) are
\begin{equation} \label{meas_matx}
S_j = \sum_{i=1}^{m} M_{ji} n_i;
\end{equation}
we could include division by the exposure time in the matrix $M_{ji}$ if
we wish. We see that in the most general case the $j$th Stokes Parameter
$S_j$ could depend on all $m$ optical trains, and hence the error on 
$S_j$ could depend on errors on all the $n_i$.

Now consider a normalized Stokes Parameter $s_j = S_j/I = S_j/S_1$. By 
the rule of adding errors in quadrature, the noise on $s_j$ must be given by
\begin{equation} \label{nsp_noise}
{\sigma_{s_j}}^2 = \sum_{k=1}^{m} \left( \frac{\partial s_j}{\partial n_k} 
\right)^2 {\sigma_{n_k}}^2.
\end{equation}
But because each photon-count is assumed to be affected by 
independent Poissonian noise, ${\sigma_{n_k}}^2 = n_k$ and so
\begin{equation} \label{nsp_noiser}
{\sigma_{s_j}}^2 = \sum_{k=1}^{m} \left( \frac{\partial s_j}{\partial n_k}
\right)^2 n_k.
\end{equation}
Using $s_j = S_j/I$, this becomes
\begin{equation} \label{nsp_noises}
{\sigma_{s_j}}^2 = \sum_{k=1}^{m} 
\left( \frac{\partial S_j}{\partial n_k} - s_j\frac{\partial S_1}{\partial 
n_k} \right)^2 \frac{n_k}{I^2}.
\end{equation}
Substituting the matrix form Equation \ref{meas_matx} into the 
partial derivatives yields
\begin{equation} \label{nsp_noisep}
{\sigma_{s_j}}^2 = \sum_{k=1}^{m}
(M_{jk} - s_j M_{1k})^2 {n_k}/{I^2}.
\end{equation}
Expanding the brackets gives
\begin{equation} \label{nsp_noisee}
{\sigma_{s_j}}^2 = I^{-2} \sum_{k=1}^{m}
{M_{jk}}^2 n_k + {s_j}^2 n_k {M_{1k}}^2 - 2s_j n_k M_{1k} M_{jk}.
\end{equation}
S\'{a}nchez Almeida~\scite{Almeida-95a} shows (Equation SA 9a) that
\begin{equation} \label{sasum}
{\sigma_{S_j}}^2 = \sum_{i=1}^{m} {M_{ji}}^2 n_i
\end{equation}
which allows us to substitute terms in Equation \ref{nsp_noisee} yielding
\begin{equation} \label{nsp_noisex}
{\sigma_{s_j}}^2 = I^{-2} \left[
{\sigma_{S_j}}^2 + {s_j}^2 {\sigma_{I}}^2 - 2s_j {\sigma_{\times}}^2 
\right] \end{equation}
 where we define the (not necessarily positive) quantity
\begin{equation} \label{xdef}
{\sigma_{\times}}^2 = \sum_{i=1}^{m}
n_i M_{1i} M_{ji}.
\end{equation}

We already know the limits on $\sigma_I$ and $\sigma_{S_j}$ from Theorems 
\ref{ithm} and \ref{sthm}; to obtain a limit on 
$\sigma_{\times}$ we follow
S\'{a}nchez Almeida's~\scite{Almeida-95a} use
of the Cauchy-Schwarz inequality~\cite[\S 2.1]{Froberg-85a} for series.
Consider
\begin{equation} \label{cs_pow4}
{\sigma_{\times}}^4 = 
\left[ \sum_{i=1}^{m} M_{ji}M_{1i}n_i \right]^2 \leq
\left[ \sum_{i=1}^{m} {M_{ji}}^2 n_i \right]
\left[ \sum_{k=1}^{m} {M_{1k}}^2 n_k \right]
\end{equation}
by the Cauchy-Schwarz inequality. But the two bracketed terms on the 
right, by  Equation \ref{sasum}, are errors on $I$ and $S_j$, whence
\begin{equation} \label{cs_ineq}
{\sigma_{\times}}^4 \leq {\sigma_{I}}^2 {\sigma_{S_j}}^2.
\end{equation}
Taking square roots and not assuming the positive root,
\begin{equation} \label{cs_ineqr}
\abs{{\sigma_{\times}}^2} \leq {\sigma_{I}} {\sigma_{S_j}}. 
\end{equation}
Returning to Equation \ref{nsp_noisex}, we see that the first two terms 
in the square bracket must be positive, and the third term is in the 
range $\pm 2 s_j {\sigma_{I}} {\sigma_{S_j}}$. Recognising that both signs 
enable the bracket to be written as a square, and noting that $s_j$ is 
itself a signed quantity, we obtain:

\begin{theorem} \label{gathm}
The error on the determination of a normalised 
Stokes Parameter $s_j$ satisfies:
 \begin{equation} \label{cs_lims}
(\sigma_{S_j} - \abs{s_j} \sigma_{I})^2 \leq I^2 {\sigma_{s_j}}^2
\leq (\sigma_{S_j} + \abs{s_j} \sigma_{I})^2.
\end{equation} \end{theorem}

 Now this theorem is not particularly useful for the general case; we can rewrite the lower bound as
\begin{equation} \label{cs_lowl} \sigma_{s_j} \geq \left| {\frac{\sigma_{S_j}}{I} - \frac{\abs{s_j}
\sigma_{I}}{I}}\right| , \end{equation} and rearranging Theorems \ref{ithm} and \ref{sthm} (with $\abs{s_j}
= \abs{S_j}/I$) compare with $\sigma_I/I \geq 1/\sqrt{\cal N}$ and $\sigma_{S_j}/\abs{s_j}I \geq
1/\sqrt{\cal N}$. It becomes apparent that both terms in the difference must be greater than or equal to
$1/\sqrt{\cal N}$. This merely tells us that the error on the normalized Stokes Parameter must be
non-negative; hardly a surprising result. (In the case of binned Stokes Parameters the same result is
obtained, since the $\Delta$ terms cancel out by the time Equation \ref{cs_lims} is obtained.) But this
result is presented here because Equation \ref{cs_lims} produces a useful result in the special case when
$\sigma_I = \sigma_{S_j}$ (see Theorem \ref{gathmeq} below). 

\newpage

\section{Noise in a Two-Channel Polarimeter}

 While S\'{a}nchez Almeida~\scite{Almeida-95a} treated the general case 
of a polarimeter with an arbitrary number of optical trains, which could 
combine the data in every train to estimate $I$, Maronna, Feinstein \& 
Clocchiatti~\scite{Maronna+92a} consider the case of a two-channel 
polarimeter simultaneously measuring $I$ and {\bf one other} Stokes 
Parameter $S$ to obtain $s=S/I$. They produce a number of theorems 
(denoted here by MFC) deduced by assuming that the polarised light 
arrives at the detector according to a Poissonian distribution.

\subsection{Optimal estimation}
\label{optest}

As in Chapter \ref{stoch}, we must distinguish between the true values of 
the Stokes Parameters
for a source, and the values which we measure in the presence of noise.  
We will use the subscript $0$ to denote the underlying values, and
the subscript $i$ for individual measured values. We assume that $\nu_S$ 
individual sets of photon-count measurements have been made.
 
 Consider a general normalized Stokes Parameter:
\[ s_i = \frac{S_i}{I_i} =
\frac{n_{1i} - n_{2i}}{n_{1i} + n_{2i}}.\]
Clarke et al.~\scite{Clarke+83b} point out that the signal/noise
ratio obtained by calculating
\begin{equation}
\label{stilde}
 \tilde{s} = \frac{\bar{S}}{\bar{I}} =
\frac{\sum_{i=1}^{\nu_S} S_i}{\sum_{i=1}^{\nu_S} I_i}
\end{equation}
is much better than that obtained by simply taking the mean,
\begin{equation}
\label{sbar}
\bar{s} = \frac{1}{\nu_s} \sum_{i=1}^{\nu_S} s_i
= \frac{1}{\nu_s} \sum_{i=1}^{\nu_S} \frac{S_i}{I_i},
\end{equation}
 since the Equation \ref{stilde} involves the taking of only one
ratio, where the two terms $\bar{S}$ and $\bar{I}$ have better  
signal/noise ratios than the individual $S_i$ and $I_i$ which are ratioed
in Equation \ref{sbar}. Maronna, Feinstein \& 
Clocchiatti~\scite{Maronna+92a} prove the following results:
\begin{theorem}[MFC 1] \label{MFC1}
$\tilde{s}$ is the maximum 
likelihood estimator of $s_0$;
\end{theorem}
\begin{theorem}[MFC 2]Both $\tilde{s}$ and 
$\bar{s}$ are unbiased estimators of $s_0$.
\end{theorem}

\subsection{The maximum likelihood estimator of binned data}

 Consider again the case of taking polarimetric measurements using a device which produces 1 DN count for
every $\Delta$ incoming photons, and where the population means for the number of photons arriving in the
two channels of our detector are $\lambda_1$ and $\lambda_2$ respectively. Ultimately our interest is in
estimating the normalised Stokes Parameter characteristing that population, $s_0 = (\lambda_1 -
\lambda_2)/(\lambda_1 + \lambda_2)$. The proof of Theorem \ref{MFC1} hinges on the fact that the Maximum
Likelihood Estimator (MLE) of a function is given by applying the 
function to the MLEs of its parameters (the so-called {\em substitution 
principle} of MLEs):
since $\bar{S}$ and $\bar{I}$ are shown to be the MLEs of $S_0=
(\lambda_1 - \lambda_2)/\tau$ and
$I_0=\lambda/\tau=(\lambda_1 + \lambda_2)/\tau$, the proof follows. 

 The probability $P(D=d)$ is given by Equation \ref{D_prob} as a function of $\lambda$. The MLE of $D$ is
obtained by maximizing $P(D=d)$ with respect to $\lambda$, whence
 \begin{equation}
\label{D_MLEcalc}
 0 = \frac{\mathrm{d}}{\mathrm{d}\lambda} P(D=d) = 
\sum_{k=d\Delta}^{d\Delta+\Delta-1} 
\frac{1}{k!} [k.e^{-\lambda}\lambda^{k-1} - e^{-\lambda}\lambda^{k}]
= e^{-\lambda} \sum_{k=d\Delta}^{d\Delta+\Delta-1} \frac{\lambda^{k-1}.k}{k!} - \frac{\lambda^{k}}{k!}.
\end{equation}
Defining $T_k = \lambda^{k-1}.k/k!$, it follows that $T_{k+1} =
\lambda^{k}.(k+1)/(k+1)!  = \lambda^{k}/k!$, allowing us to cast Equation
\ref{D_MLEcalc} as

 \begin{equation} 0 = \sum_{k=d\Delta}^{d\Delta+\Delta-1} T_k - T_{k+1}. \end{equation} 
All the terms in the power series cancel out apart from the first and 
last, and substituting the limits of the sum gives $0= T_{d\Delta} - 
T_{(d+1)\Delta}$. Using the definition of $T_k$ and 
rearranging terms yields
\begin{equation}
\label{simD_MLE}
[(d_{\mathsf ML}+1)\Delta]!d_{\mathsf ML}\Delta = (d_{\mathsf 
ML}\Delta)!\lambda^{\Delta}(d_{\mathsf 
ML}+1)\Delta \end{equation} and hence \begin{equation}
\label{D_MLE}
(d_{\mathsf ML}\Delta -1)!\lambda^{\Delta} = [(d_{\mathsf ML}+1)\Delta -1]!.
\end{equation}

Now we would like Equation \ref{D_MLE} to provide $d_{\mathsf ML}$ as a 
function of
$\lambda$ and $\Delta$ to see how the MLE, $d_{\mathsf ML}$, compares to the 
intuitive
approximation $\lambda/\Delta$. The factorials allow no obvious analytic
solution, but useful upper and lower limits may be obtained as follows:
Equation \ref{D_MLE} can be recast as
 \begin{equation} \lambda^{\Delta} = 
(d_{\mathsf ML}\Delta)(d_{\mathsf ML}\Delta+1)(d_{\mathsf 
ML}\Delta+2) \cdots (d_{\mathsf ML}\Delta+\{\Delta-1\}), 
\end{equation}
 where the right hand side is
a product of $\Delta$ distinct terms, none smaller than $d_{\mathsf 
ML}\Delta$ and none
larger than $d_{\mathsf ML}\Delta+(\Delta-1)$. The RHS $(= 
\lambda^{\Delta} )$ 
is hence clearly larger than $(d_{\mathsf ML}\Delta)^\Delta$ and smaller than
$[d_{\mathsf ML}\Delta+(\Delta-1)] ^\Delta$, whence $d_{\mathsf ML}\Delta < 
\lambda < 
d_{\mathsf ML}\Delta + (\Delta-1)$. Rearranging the inequalities yields
 \begin{equation}
\label{dlims_MLE}
\frac{\lambda+1}{\Delta} - 1< d_{\mathsf ML} < \frac{\lambda}{\Delta}.
\end{equation}

The MLE of $D$ is hence slightly smaller than the simplistic $\lambda/\Delta$: this is not unexpected as a
few photons failing to fill the highest bin will not be measured, and the binned measurement will be biased
to slightly underestimate the photon count. But the MLE will never be lower than $1-\frac{1}{\Delta} < 1$
DN units (i.e.\ $<$ 1 DN unit) below the simplistic estimate.

\subsection{Normalized Stokes Parameters under binning} \label{optbin} We noted above that the MLE of $s_0$
is obtained by substituting the MLEs of $\lambda_1,\lambda_2$ into $s = 
(\lambda_1 - \lambda_2)/(\lambda_1 +
\lambda_2)$. Now we know that MLE$(D) = \frac{\lambda}{\Delta} - 
\epsilon$, where $\epsilon =
\frac{1}{2} . \left( 1-\frac{1}{\Delta} \right) \pm \frac{1}{2} . \left( 1-\frac{1}{\Delta} \right)$. If we
assume the Absolute Stokes Parameters $\bar{I_D}$ and $\bar{S_D}$ have been 
measured in DN units, then
MLE$(\tau \bar{I_D}) = (\lambda/\Delta) - 2\epsilon$ and 
MLE$(\tau \bar{S_D}) = (\lambda_1-\lambda_2)/\Delta$. Taking
their ratio, \begin{equation} \label{MLEofs} \mathrm{MLE} (s) = 
\frac{\bar{S_D}}{\bar{I_D}-2\epsilon} \simeq
\frac{\bar{S_D}}{\bar{I_D}} . \frac{1}{\left(1- 
\frac{1-1\Delta}{\bar{I_D}}\right)}. \end{equation}

 Equation \ref{MLEofs} is not an exact formula for the MLE of $s$ since $\epsilon$ is an approximation
half way between the known limits. But it is clear that use of the 
formula $\tilde{s} = \bar{S_D}/\bar{I_D}$
will give us within a factor $1/(1-2\epsilon/\bar{I_D})$ of the MLE, and 
this error factor may easily be calculated. 

\subsection{Minimum errors on the normalized parameters}

Returning to the case where $I$ and $S$ are measured in photons rather than DN, we note that errors on
$\bar{S}$ and on $\bar{I}$ are not independent of one another. We can write: \begin{equation}
\label{getstilde} \tilde{s} = \frac{\bar{n}_{1} - \bar{n}_{2}}{\bar{n}_{1} + \bar{n}_{2}}. \end{equation}
 
If we propagate through the errors on the intensities, we find:
\begin{equation}
\label{getsterr}
\sigma_{\tilde{s}} =
\frac{1}{\bar{n}_{1}+\bar{n}_{2}}.\sqrt{[(1-\tilde{s})\sigma_{\bar{n}_{1}}]^2
+ [(1+\tilde{s})\sigma_{\bar{n}_{2}}]^2}.
\end{equation}

\begin{theorem} \label{gathmeq} The error on a normalised Stokes Parameter $s_j$ determined with a
two-channel polarimeter cannot be better than $(1-\abs{s_j})/{\sqrt{\cal N}}.$ \end{theorem}

We can put a lower limit on the error on $s$ using a special case of Theorem \ref{gathm} defined above. In
this case where the system is a two channel polarimeter taking a sum and difference of counts, then the
errors on $S$ and $I$ are identical: the two channels have independent errors, and so
\begin{equation}
\label{equiverr}
\sigma_{\bar{I}} = \sigma_{\bar{S}} = \sqrt{{\sigma_{\bar{n_{1}}}}^2 + 
{\sigma_{\bar{n_{2}}}}^2 }.
\end{equation}
Equation \ref{cs_lowl} hence simplifies to 
\begin{equation} \label{cs_lowleq}
\sigma_{s_j} \geq \frac{\sigma_{I}}{I}  \abs{1-\abs{s_j}}.\end{equation}
The lower limit for $\sigma_{I}/I$ may be substituted from Theorem \ref{ithm} and hence Theorem
\ref{gathmeq} above is proven. Maronna, Feinstein \& 
Clocchiatti~\scite{Maronna+92a} follow an alternative treatment, as follows: 
 
\begin{theorem}[MFC 3]
The errors associated with the two estimators of $s_0$ satisfy
\begin{equation} \label{best_err} 
{\sigma_{\tilde{s}}}^2 = \frac{1}{\nu_S I_0}(1-{s_0}^2)\left[ 1 
+\frac{1}{\nu_S I_0} + \frac{\tilde{b}}{(\nu_S I_0)^2} \right]
\end{equation}
and
\begin{equation} \label{bar_err} 
{\sigma_{\bar{s}}}^2 = \frac{1}{\nu_S I_0}(1-{s_0}^2)\left[ 1 
+\frac{1}{I_0} + \frac{\bar{b}}{{I_0}^2} \right]
\end{equation}
where $\tilde{b}$ and $\bar{b}$ are non-negative constants dependent on 
$\nu_S$ and $I_0$.
\end{theorem}

 Equation \ref{best_err} should be consistent with the lower limit set by Theorem \ref{gathmeq}: squaring
the latter, we have a lower limit
\begin{equation} \label{sigmin} 
{\sigma_{\mathsf{min}}}^2 = \frac{1}{\cal N} (1-\abs{s_0})^2.
\end{equation}
Substituting ${\cal N} = \nu_S I_0$ and $\iota = \left[ 1 
+\frac{1}{\nu_S I_0} + \frac{\tilde{b}}{(\nu_S I_0)^2} \right]$,  we can 
write Equation \ref{best_err} as \begin{equation} \label{tilde_subbed} 
{\sigma_{\tilde{s}}}^2 =  \frac{1}{\cal N} (1-\abs{s_0})(1+\abs{s_0})(1+\iota)
= {\sigma_{\mathsf{min}}}^2  \frac{1+\abs{s_0}}{1-\abs{s_0}}(1+\iota).
\end{equation}
Since $\abs{s_0}$ lies between 0 and 1, and $\iota>0$, then clearly
$\frac{1+\abs{s_0}}{1-\abs{s_0}}(1+\iota) > 1$ and $ {\sigma_{\tilde{s}}}$ will never be lower than
$\sigma_{\mathsf{min}}$: {\em QED}.

\subsection{Estimating errors on the normalized parameters}

 We define a variance ${\sigma_0}^2$:
\begin{equation} \label{sig0def}
{\sigma_0}^2 = \frac{1-{s_0}^2}{\nu \tau I_0} =  \frac{1-{s_0}^2}{\cal N}. 
\end{equation} 
Equations \ref{best_err} and \ref{bar_err} show that both 
$\sigma_{\tilde{s}}$ and $\sigma_{\bar{s}}$ tend
towards $\sigma_0$ for large $I$, as does $\sigma_{\tilde{s}}$ (but {\em 
not}\, necessarily $\sigma_{\bar{s}}$) for large $\nu_S$. 

Now define $\bar{\sigma}^2 = ({1-\bar{s}^2})/{\cal N}$ and 
$\tilde{\sigma}^2 =
({1-\tilde{s}^2})/{\cal N}$, in which case in can be shown 
\cite{Maronna+92a}: 

\begin{theorem}[MFC 5]  \label{MFC5} $\tilde{\sigma}^2$ is the maximum likelihood 
estimator of $\sigma_0$;
\end{theorem}

\begin{theorem}[MFC 6]  \label{MFC6} $\tilde{\sigma}^2$ has the lowest 
variance of any 
possible estimator of the variance of $s$, and is hence the optimal error estimator. \end{theorem}

 The $\sigma_0$ variance can also be expressed ${\sigma_0}^2 = 
\frac{1-s^2}{\nu_S \lambda}$, and substituting
the MLE for $\lambda$ and (approximately) for $s$ in the case where $I_D$ 
is measured in binned units, we
obtain the MLE variance in the binned case. $\lambda = 
$MLE$(D)=\Delta [\tau.$MLE$(\bar{I_D}) + 2\epsilon]$, hence   
\begin{equation} \label{bin5} 
\mathrm{MLE}({\sigma_s}^2) = \frac{1-\tilde{s}^2}{\nu_S \Delta 
(\tau\bar{I_D} + 
2\epsilon)} \simeq \frac{1-\tilde{s}^2}{\nu_S \tau \Delta \bar{I_D}}.
\end{equation}

N.B. The proof of Theorem \ref{MFC6} depends on the proof that $\tilde{s}$ is an unbiased estimator for
$s_0$. Since we have not given formal proof of an unbiased estimator of $s_0$ in the case of binned counts
in DN units, we cannot extend this result to the binned case.

\subsection{The distribution of the normalized parameters}

 The Central Limit Theorem \cite[Ch. 16]{Boas-83a} suggests that even a
normalised Stokes Parameter $s$ must be distributed approximately normally
for a sufficiently large sample.  Now if $I \rightarrow \infty$ then we
have both \begin{theorem}[MFC 4]\label{til_tend} \begin{equation} 
\sigma(\tilde{s}) \rightarrow \sigma_0, \frac{\tilde{s}-s}{\sigma_0} \sim
N(0,1) \end{equation} \end{theorem} and \begin{theorem}[MFC 4]
 \label{bar_tend} \begin{equation} \sigma(\bar{s}) 
\rightarrow \sigma_0,
\frac{\bar{s}-s}{\sigma_0} \sim N(0,1). \end{equation} \end{theorem}
Further, for many measurements of a low intensity source, $\nu_S
\rightarrow \infty$, and Theorem \ref{til_tend} still holds -- but in this
case, Theorem \ref{bar_tend} no longer holds. Furthermore, (MFC 5)
both Theorems continue to hold under the same conditions if $\sigma_0$ is
replaced by the optimal estimator $\tilde{\sigma}$. 

\subsection{The effect of sky noise}

 The night sky is not totally dark, and contributes errors twice: from the sky superimposed on the target
object, and from the measurement of adjacent sky used to make a sky correction. If the sky has a constant
brightness per unit area, its intensity is subject to Poisson fluctuation like any other light source. But
the sky brightness itself may vary from point to point, too. 

 Maronna, Feinstein \& Clocchiatti~\scite{Maronna+92a} consider the 
effects of sky noise, and show that (MFC 8 -- for raw
photon counts), subtracting the MLE of the sky noise from the total MLE of the light in the target aperture
yields the MLE of the light from the source alone. Modern aperture photometry systems such as IRAF's {\sc
apphot} automatically subtract the estimated sky noise from the total signal and provide the correct output
for obtaining the MLE of the source.  Maronna, Feinstein \& Clocchiatti 
\scite{Maronna+92a} do not, however, evaluate whether
any estimators of $s$ are biased by the presence of sky noise. 

\begin{theorem}[MFC 9]
When counting photons in the presence of sky noise, the error on the resulting normalised Stokes Parameter
is $\sigma_{\dag}$ such that, with $\nu_\dag$ measurements of the 
background sky and an expected photon count $2\phi$ from the sky,
\begin{equation} \label{MFC9}
{\sigma_{\dag}}^2 = \frac{1}{\nu_S \lambda} \left[ (1-s^2) + 
\frac{2(1+s^2)\phi}{\lambda} +\frac{2s^2 \nu_S \phi}{\nu_{\dag} \lambda} 
\right]. \end{equation} \end{theorem}

 We can easily obtain MLE$(\sigma_{\dag})$ by substituting $\tau \bar{I}=$
MLE$(\lambda)$, $\tilde{s} =$ MLE$(s)$ and the sky half-intensity
MLE$(\phi)$. Similarly, in a system binning counts we may substitute
MLE$(\lambda)\simeq \tau \bar{I_D}\Delta$, MLE$(s) \simeq \tilde{s}$ and 
the sky half-intensity MLE$(\phi) \simeq \tau \bar{I}_{\mathsf
annulus}\Delta$. It is also shown (MFC 9) that $\tilde{s}$
calculated from noise-corrected values is normally distributed provided
$\nu_S$ and $\nu_\dag$ both tend to infinity, and also if $\lambda$ tends to
infinity with $\phi/\lambda$ bounded. 

\newpage

\section{Mathematical glossary}

Since this Appendix uses a lot of mathematical terms common with Chapter 
\ref{stoch}, and a few which differ in definition, I have given both this 
Appendix and that Chapter a mathematical glossary defining the terms 
used. Latin symbols are listed in alphabetical order first, followed by 
Greek terms according to the Greek alphabet -- except that terms of the 
form $\sigma_\aleph$ are listed under the entry for $\aleph$.

\begin{description}

\item[$D, d$] The random variable $D$ and its particular value $d$ 
expressing the output of a photon detector in DN units.

\item[$E(X)$] The expected value (arithmetic mean) of the random variable 
$X$.

\item[$f$] The fraction of photons transmitted by an attenuating filter.

\item[$I$] The intensity of a beam of light in photons per second.
\begin{itemize}
\item[$I_0$] The true value of $I$.
\item[$I_1, I_2$] The intensities of two component beams in a two-beam case. 
\item[$\bar{I}$] An estimate of $I_0$ such that $\bar{I} = \bar{X}/\tau$.
\item[$\sigma_{I_0}$] The true SD of $I$.
\item[$\sigma_{I}$] The measured SD of $I$. 
\item[$\sigma_{\bar{I}}$] The standard error on $\bar{I}$. 
\item[$I_i$] In a two-channel polarimeter, one of $\nu_S$ individual
measurements of the light intensity: for photon counts $n_1, n_2$ in the
two channels, $I_i = (n_{i1} + n_{2i})/\tau$. 
\item[$I_D$] The intensity of a beam of light in DN per second.
\end{itemize}

\item[$m$] The number of independent optical trains in a general polarimeter.

\item[MLE$(\bar{S})$] The Maximum Likelihood Estimator of $\bar{S}$.

\item[$n_\times$] A photon count measured in one channel of a 
multi-channel polarimeter.
\begin{itemize}
\item[$n_i$] The photon count measured in the $i$\,th optical train of a 
general polarimeter.
\item[$n_{1i}, n_{2i}$] The individual photon counts measured in the two 
channels of a two-channel polarimeter.
\item[$\bar{n}_{1}, \bar{n}_{2}$] The mean values of a series of $\nu_S$
photon 
counts measured in the two channels of a two-channel polarimeter.
\item[$\sigma_{n_i}$] The noise (error) on an individual photon 
count measurement $n_i$ of the $i$\,th channel of a generalised polarimeter.
\end{itemize}

\item[${\cal N}$] The total photon count summed over the $m$ optical 
trains in a general polarimeter, ${\cal N} = \sum_{1}^{m} n_i$.

\item[$P(X=x)$] The probability that random variable $X$ is some given 
value $x$.

\item[$Q, q$] Absolute and normalised linear Stokes Parameters. See $S, s$.

\item[$S$] A generalised absolute Stokes Parameter illustrating the 
properties of $Q$ and $U$ (and, where applicable, $V$). It takes the same annotations 
as $I$.
\begin{itemize}
\item[$S_j$] A generalised absolute Stokes Parameter in the sense that 
$S_1 \equiv I$, and $S_{2,3,4} \equiv Q,U,V$.
\item[$S_i$] In a two-channel polarimeter, one of $\nu_S$ individual
measurements of the absolute Stokes Parameter $Q$ or $U$: for photon 
counts $n_1, n_2$ in the two channels, $S_i = (n_{1i} - n_{2i})/\tau$. 
\end{itemize}

\item[$s$] A generalised normalised Stokes Parameter illustrating the 
properties of $q$ and $u$. It takes some of the same 
annotations as $I$.
\begin{itemize}
\item[$s_0$] The true normalised Stokes Parameter of a source: $s_0 = 
(\lambda_1 - \lambda_2)/(\lambda_1 + \lambda_2)$.
\item[$s_j$] A generalised absolute Stokes Parameter in the sense that 
$s_{2,3,4} \equiv q,u,v$, and $j \neq 1$.
\item[$s_i$] For a two-channel polarimeter, the ratio of individual 
absolute Stokes Parameters, $s_i = S_i / I_i$.
\item[$\bar{s}$] For a two-channel polarimeter, the mean of the individual
$s_i$, such that $\bar{s} = \sum_{1}^{\nu_S} s_i$.
\item[$\tilde{s}$] For a two-channel polarimeter, the ratio of the mean  
Stokes Parameters, $\tilde{s} = \bar{S} / \bar{I}$.
\end{itemize}

\item[SD$(X)$] The Standard Deviation of random variable $X$.

\item[$U, u$] Absolute and normalised linear Stokes Parameters. See $S, s$.

\item[$V, v$] Absolute and normalised circular Stokes Parameters. See $S$.

\item[$W, w$] The random variable $W$ and its particular value $w$ for 
photons counted from a beam of intensity $I$ attenuated by a factor $f$.

\item[$X$] A random variable: the number of photons which might arrive 
from a beam of intensity $I$ in time $\tau$.
\begin{itemize}
\item[$X_i$] The $i$\,th measurement of a set of $\nu$ measurements of 
the random variable $X$.
\item[$\bar{X}$] The arithmetic mean of a set of $X_i$, such that 
$\bar{X} = \sum_{1}^{\nu} X_i / \nu$. 
\item[$\sigma_{\bar{X}}$] The standard deviation of a set of $X_i$,  such 
that ${\sigma_{\bar{X}}}^2 = \left[ \sum_{1}^{\nu} {X_i}^2 / \nu \right] 
- {\bar{X}}^2$.
\item[$x$] A possible value of the random variable $X$.
\end{itemize}

\item[$\Delta$] The {\em integer} number of photons which must be 
detected to give a count of 1 DN.

\item[$\epsilon$] The approximate amount $\pm \epsilon$ by which the 
intuitive $\lambda/\Delta$ overestimates the MLE of a binned measurement 
of $\lambda$.

\item[$\lambda$] The parameter characterising the Poisson 
distribution of the number of photons expected to be received in time 
interval $\tau$, such that $\lambda = I\tau$.

\item[$\nu_S$] The number of individual pairs of measurements made with a 
two-channel polarimeter in order to determine a set of $I_i$ and $S_i$.

\item[$\nu_\dag$] The number of individual pairs of measurements of 
empty sky made with a 
two-channel polarimeter in order to determine the sky noise.

\item[$\sigma$] A standard deviation. Terms of the
form $\sigma_\aleph$ are listed under the entry for $\aleph$; note also:
\begin{itemize}
\item[$\sigma_0$] An idealised SD, such that ${\sigma_0}^2 = 
(1-{s_0}^2)/{\cal N}$
\item[$\tilde{\sigma}$] The SD corresponding to the MLE estimator of $s_0$, 
such that ${\tilde{\sigma}}^2 = (1-{\tilde{s}}^2)/{\cal N}$
\item[$\bar{\sigma}$] The SD corresponding to the mean estimator of $s_0$, 
such that ${\bar{\sigma}}^2 = (1-{\bar{s}}^2)/{\cal N}$
\item[$\sigma_\dag$] The SD corresponding to the MLE estimator of $s_0$ 
in the presence of sky noise.
\end{itemize}

\item[$\tau$] The integration time for measuring light intensity. 

\item[$\phi$] The parameter giving half the expected number of photons 
from the sky background which would be received in integration time $\tau$.

\end{description}

\chapter{Computer Codes}
\label{codeapp}

The data reduction for this thesis was accomplished with the use of 
several home-made {\sc fortran} routines and {\em Microsoft Works} 
spreadsheets. The programming of the most important routines and 
spreadsheets is recorded here for reference.

\section{Spreadsheet Analysis of Two-Channel Photometry} 

Any system for reducing and analyzing polarimetry begins with photometry.
Polarized images were presented to {\sc iraf}'s {\sc apphot.phot} routine 
as described in Chapter \ref{obsch}; photometry was performed on both images 
(i.e.\ channels) on a given mosaic. The {\sc phot} output consisted of 
ASCII files rich in detail, including a calculation of the magnitude of 
the source in each specified aperture; the zeropoint of the magnitude 
scale was not calibrated, however, and this portion of the output was not 
used in the current data reduction scheme. The {\sc phot} output also 
returned values, in data number count rate units, for the flux attributed 
to the source object (corrected for sky values using an annulus) and the 
error on this quantity. A {\sc fortran} routine by the author 
(not recorded here) stripped these data fields from the output of 
{\sc phot} into tabbed ASCII files which could be pasted into the 
analysis spreadsheet, {\em Microsoft Works}. 

 The spreadsheet was hand-coded with other data to accompany the 
aperture count rate and error: a normalization factor of either 1 or 10 
was included because some images (those with initial reduction performed 
by Dr Stephen Eales) had been normalized for the 10 second exposure time 
of each co-added component. The number of components in the mosaic was 
also coded, normally 9 but less for those images where some frames of the 
mosaic had been corrupted and therefore rejected. The shot noise is 
calculated from the square root of the total flux, and the sky noise is 
obtained by subtracting the shot noise in quadrature from the overall 
error. The spreadsheet tests each individual count rate to ensure that 
the sky noise is much greater than the shot noise, and an error flag is 
set to indicate a warning if any of the individual count rates in the 
dataset fail this test, thus fulfilling Check \ref{smallshot}.

 For each pair of photometry values, the spreadsheet next calculates the 
sum, difference, and common error on the sum and difference. The 
set of differences is summed, and if it exceeds three times the 
quadrature sum of the common errors, an error condition is flagged, thus 
implementing Check \ref{cbias}. The sums, differences, and errors are 
repeated in the next group of columns, but rearranged to group together 
all the $Q_i$ values and then all the $U_i$ values. The normalised Stokes 
Parameters $q_i$ and $u_i$ are also calculated in these blocks, 
completing Step \ref{getthei}.
All normalised Stokes parameters throughout the reduction process were
calculated in the instrumental $(\eta_0 = 83\degr)$ reference frame.

 The standard errors $\eps{phot}$ are calculated for the
$Q_i$ and $U_i$, following Step \ref{getmean}; the sample means and 
statistical errors $\eps{stat}$ are also calculated. The difference between 
each individual measured error and $\eps{phot}$ is calculated, and the
spreadsheet extracts the maximum deviation. If this is more than 30\% of
$\eps{phot}$, an error condition is flagged, satisfying Check
\ref{maxbig}. A normal distribution of sky noise (Step \ref{assumenorm})
is automatically assumed. Check \ref{noiseOK} is left to human inspection 
where the values of $\eps{phot}$ and $\eps{stat}$ are presented together.

The statistics for the $Q_i$ and $U_i$ samples are used to calculate the 
Student $t$ and normalized Gaussian $z$ statistics which can be used for 
hypothesis testing. The spreadsheet requires manual entry of the limits 
on $t$ or $z$ for a given confidence level in order to test the 
no-polarization hypothesis with that confidence (Step \ref{findconfr}).

The errors $\vas{stat}$ on the best-estimator Stokes Parameters are 
trivially 
calculated (Step \ref{hereNSPs}) using the statistics obtained above; and 
having obtained the statistical value the spreadsheet can compare with 
the entire dataset to obtain the photometric errors $\vas{phot}$. As 
before, the spreadsheet presents these errors for manual comparison, 
fulfilling Check \ref{stoeq}. The more conservative (larger) errors are 
duly obtained and recorded (Step \ref{gotnorm}).

Finally, the noise-normalised polarization $(m)$ and the nominal 
polarization orientation $(\phi)$ are calculated (Steps \ref{findperr} 
and \ref{propphi}) and returned in the results row, from which they can 
be manually passed on to the debiasing software documented below.

\section{Debiasing Software}

 A number of {\sc fortran} routines were developed in the course of the 
data analysis for this thesis. Most of these are not documented in 
detail here since 
they are trivial implementations of the formulae of Chapter \ref{stoch}, 
or else specific codes to convert {\sc iraf} output into ASCII files 
suitable for {\em Microsoft Works}. The exceptions are the routines used 
for debiasing, since these require an iterative solution using Bessel 
functions. Double precision arithmetic is used, and the Bessel functions 
{\tt dbesi0(y)} and {\tt dbesi1(y)} are drawn from standard double-precision 
reference libraries.

\subsection{Program {\tt debpol}}

{\flushleft
Program {\tt debpol}: takes in a noise-normalised measured polarization 
$m_0$ and returns output of the form $a_\ell , \hat{a}, a_u$ where 
$(a_\ell ,a_u)$ are the $1\sigma$ confidence limits and $\hat{a}$ the 
best point estimate of the true noise-normalised polarization. Note that 
the output must be multiplied by the normalizing error $\sigma$ used to 
obtain $m_0$ from $p$ in the first place, in order to obtain a meaningful 
value in percentage units.}

{\tt

\begin{tabbing}

\label{debpol}

\hspace{7ex} \= \hspace{7ex} \= \hspace{7ex} \= 
\\
\>\texttt{program debpol}\\
\\
\>\>implicit none\\
\\
\>double precision m0\\
\\
$*$ m0 is the measured normalized polarization P$/$sigmaP\\
\\
\>integer unit\_no\\
\\
\>character$*$80 my\_filename\\
\\
\>unit\_no = 1\\
\\
\>print $*$, "Welcome to DebPol: Debiaser for polarimetry"\\
\\
\>print $*$, "(c) Gareth Leyshon, 1997"\\
\\
\>print $*$, " "\\
\\
\>print $*$, "Output filename?"\\
\\
\>read $*$, my\_filename\\
\\
\>open (unit=unit\_no, file=(my\_filename), form="formatted")\\
\\
\>print $*$, " "\\
\\
\>print $*$, "N.B. input 0 for a blank output line, -1 to quit."\\
\\
\>print $*$, " "\\
\\
 145\>continue\\
\\
\>print $*$, "Enter the measured normed polarization:"\\
\\
\>read $*$, m0\\
\\
\\
$*$ this version gets m0 from keyboard input\\
\\
$*$ can escape here to end of programme or output a blank and repeat\\
\\
\\
\>if (m0.eq.-1.0d0) then\\
\>\>go to 149\\
\>else\\
\>\>if (m0.eq.0.0d0) then\\
\>\>\>write (unit\_no,$*$), " "\\
\>\>\>go to 145\\
\>\>end if\\
\>endif\\
\\
\>call debcalc (unit\_no, m0)\\
\\
\>goto 145\\
\\
 149 \>continue\\
\\
\>close(UNIT=unit\_no, STATUS="KEEP")\\
\\
\>print $*$, " "\\
\\
\>print $*$,"Routine concludes."\\
\\
\>print $*$, " "\\
\\
\>end\\
\\
\\
\\
\>\\
\>subroutine debcalc (unit\_no, m0)\\
\\
\>integer unit\_no\\
\\
\>double precision m0, aWK, aML, tol, ahat\\
\\
$*$ aWK and aML are the estimates of the Wardle \& Kronberg \& Maximum Likelihood\\
\\
\>double precision estML, estWK, mMLmax, mWKmin, den\\
\\
\>mMLmax = 1.5347d0\\
\\
\>mWKmin = 1.0982d0\\
\\
$*$ These are the fixed thresholds for applying different methods\\
\\
\>den =  (mMLmax - mWKmin)\\
\\
\>tol = 1.0d-7\\
\\
$*$ allows the tolerance for convergence to be hard-wired into software\\
\\
\>aML = estML (m0,tol)\\
\>\\
$*$ makes a maximum likelihood estimate\\
\\
\>aWK = estWK (m0,tol)\\
\\
$*$ makes a Wardle and Kronberg estimate\\
\\
\>if (m0.lt.mWKmin) then\\
\\
\>\>ahat = aML\\
\\
\>else\\
\>\>if (m0.gt.mMLmax) then\\
\>\>\\
\>\>\>ahat = aWK\\
\>\>else\\
\\
\>\>ahat = (((m0-mWKmin)$*$aML$/$den) + ((mMLmax-m0)$*$aWK$/$den))\\
\\
\>\>end if\\
\\
\>end if\\
\\
$*$ has set the best estimator ahat according to most appropriate function\\
\>\\
\\
\>write (unit\_no,$*$), m0,aML,aWK, ahat\\
\\
\>print $*$, " "\\
\\
\>print $*$, "Results for ", m0\\
\\
\>print $*$, aML, ahat, aWK\\
\\
\>end\\
\\
\\
$*$ outputs all the estimators (Maximum Likelihood, my best, Wardle \& Kronberg)\\
\\
\\
\>double precision function estML (m0,tol)\\
\\
\>double precision m0, y, yp, tol, dbesi0, dbesi1\\
\\
$*$ This works out the Maximum Likelihood Estimator of the true\\
$*$ noise-normalised polarization\\
 \\
$*$ y is the current value of the best estimate \\
$*$ yp is its previous value during iteration\\
\\
$*$ m0 is the measured value\\
$*$ tol the given tolerance\\
\\
$*$ debesi0 and dbesi1 are Bessel functions from double-precision libraries\\
\\
\>y = m0\\
\\
\>if (m0.lt.(1.4)) then\\
\\
\>\>y = 0.1d0\\
\\
\>else\\
\>\>if (m0.lt.2.5d0) then\\
\>\>\>y = m0$/$3\\
\>\>end if\\
\\
\>end if\\
\\
$*$ the start value of y for iteration is a constant if m0 is small,\\
$*$ otherwise a third of m0\\
\\
\\
 7827\>continue\\
\\
\>yp = y \\
\\
$*$ yp is the past value of y\\
\\
\>y = m0 $*$ m0 $*$ dbesi1(yp) $/$ dbesi0(yp)\\
\\
\>if ((y-yp).gt.tol) then\\
\\
\>\>go to 7827\\
\\
$*$ keep iterating until the change produced is lower than the given tolerance\\
\\
\>end if\\
\\
\>estML = y$/$m0\\
\\
\>end\\
\\
\\
\\
\\
\>double precision function estWK (m0,tol)\\
\\
\>double precision m0, y, yp, tol, dbesi0, dbesi1\\
\\
$*$ This works out the Wardle \& Kronberg estimate, notation as in ML case\\
\\
\>y = m0+0.8D0\\
\\
$*$ a fixed starting value is appropriate here\\
\\
\\
 7829\>continue\\
\\
\>yp = y \\
\\
\>y = ((m0 $*$ m0)-1.0d0) $*$ dbesi0(yp) $/$ dbesi1(yp)\\
\\
\>if ((y-yp).gt.tol) then\\
\\
\>\>go to 7829\\
\\
\>end if\\
\\
$*$ iterate until change is within specified tolerance\\
\\
\>estWK = y$/$m0\\
\\
\>end
\\

\end{tabbing}

}

\subsection{Program {\tt thcl}}

{\flushleft
Program {\tt thcl}: obtains the confidence limits for the phase-space 
angle $\theta$ (such that $\phi = \theta/2$ is the orientation of the 
polarization). It takes as input the best estimate of the true 
polarization, $\hat{a}$, and the point estimate of $\theta$ itself (in 
degrees). The output is in the form of the 67\% and 95\% confidence 
interval limits on $\theta$, also in degrees.}

{\tt

\begin{tabbing}

\label{thcl}

\hspace{7ex} \= \hspace{7ex} \= \hspace{7ex} \= \hspace{7ex} \=
\\

\>program thcl\\
\\
\>implicit none\\
\\
\>double precision a0, t0\\
\\
\>integer unit\_no\\
\\
\>character$*$80 my\_filename\\
\\
\>unit\_no = 1\\
\\
\>print $*$, "Welcome to THCL: Theta Confidence Limits"\\
\\
\>print $*$, "(c) Gareth Leyshon, 1996"\\
\\
\>print $*$, " "\\
\\
\>my\_filename = "test.the"\\
\\
$*$ hard-wired filename for output - this could be changed\\
\\
\>open (unit=unit\_no, file=(my\_filename), form="formatted")\\
\\
\>print $*$, " "\\
\\
\>print $*$, "N.B. input 0 for a blank output line, -1 to quit."\\
\\
\>print $*$, " "\\
\\
 145\>continue\\
\\
\>print $*$, "Enter the best estimate of a:"\\
\\
\>read $*$, a0\\
\\
\>if (a0.eq.-1.0d0) then\\
\>\>go to 149\\
\>else\\
\>\>if (a0.eq.0.0d0) then\\
\>\>\>write (unit\_no,$*$), " "\\
\>\>\>go to 145\\
\>\>end if\\
\>endif\\
\\
\>print $*$, "Enter the best estimate of theta:"\\
\\
\>read $*$, t0\\
\\
$*$ t\_0 is the best estimate\\
\\
\>call calculate (unit\_no, a0, t0)\\
\\
\>goto 145\\
\\
 149 \>continue\\
\\
\>close(UNIT=unit\_no, STATUS="KEEP")\\
\\
\>print $*$, " "\\
\\
\>print $*$,"Routine concludes."\\
\\
\>print $*$, " "\\
\\
\>end\\
\\
\\
\>subroutine calculate (unit\_no, a0, t0)\\
\\
\>integer unit\_no\\
\\
\>double precision t0, Cp, a1, a2, a0, tx, getradians, degre\\
\\
\>real ao, tho, b1, b2, c1, c2\\
\\
\\
\>tho = t0\\
\\
\>tx = getradians(t0)\\
\\
\>t0 = tx\\
\\
$*$ convert to radians\\
\\
\>print $*$, " "\\
\>print $*$, "CALCULATE..."\\
\>print $*$, "Best estimate of theta (deg):", tho\\
\>print $*$, "Best estimate of theta (rad):", t0\\
\>print $*$, "Best estimate of a:", a0\\
\\
$*$ calculate for 67
\\
\>Cp = 0.67\\
\\
\>print $*$, " "\\
\>print $*$, "Calling findconf for ", Cp\\
\\
\>call findconf(a0,t0,Cp,a1,a2)\\
\\
\>b1 = degre(a1)\\
\\
\>b2 = degre(a2)\\
\\
$*$ store the results as b1, b2\\
\\
\\
$*$ calculate for 95
\\
\>Cp = 0.95\\
\\
\>print $*$, " "\\
\>print $*$, "Calling findconf for ", Cp\\
\\
\>call findconf(a0,t0,Cp,a1,a2)\\
\\
\>c1=degre(a1)\\
\\
\>c2=degre(a2)\\
\\
$*$ store the results as c1, c2\\
\\
\\
\>ao = a0\\
\\
\>write (unit\_no,$*$), ao,tho,b1,b2,c1,c2\\
\\
\>print $*$, " "\\
\\
\>print $*$, "Results for ", ao,tho\\
\\
\>print $*$, b1,b2,c1,c2\\
\\
\>print $*$, " "\\
\\
\>end\\
\\
\>double precision function degre(alpha)\\
\\
\>double precision confac, alpha\\
\\
\>confac = 180/acos(-1.0d0)\\
\\
\>degre = alpha$*$confac\\
\\
\>end\\
\\
\\
\\
\>double precision function getradians(alpha)\\
\\
\>double precision confac, pi, alpha\\
\\
\>pi = acos(-1.0d0)\\
\\
\>confac = pi/(180.0)\\
\\
\>print $*$, "Degrees: ", alpha\\
\\
\>print $*$, "confac: ", confac\\
\\
\>getradians = alpha$*$confac\\
\\
\>print $*$, "Radians: ", getradians\\
\\
\>end\\
\\
\\
\\
$*$ this is the main subroutine that finds the interval \\
\\
\>subroutine findconf(a0,t0,Cp,a1,a2)\\
\\
\>double precision Cp,a0,alist,blist,rlist,elist,theterr,\\
\> $*$  epsabs,epsrel,result,t0,l,lstep,a1,a2,dstep,abserr\\
\\
\>logical flag\\
\\
\>integer ier,key,limit,neval,iord,last\\
\\
$*$ first set up constants for our integrating\\
\\
\\
\>dstep=2.0d0\\
\>key = 40\\
\>limit = 20000000\\
\>epsabs = 1.0d-20 \\
\>epsrel = 1.0d-20\\
\\
$*$ Cp is the confidence interval we want\\
\\
\>print $*$, " "\\
\\
\>print $*$, "Considering best est pol:", a0\\
\\
\>print $*$, "Considering measured angle (rad):", t0\\
\\
\\
$*$ Now we are going to iterate for theterr.\\
\\
\>\>theterr = t0$*$(0.1d0)\\
\\
\>\>l = 1.0d-3\\
\\
\>\>lstep = 0.4d0\\
\\
 556      continue\\
\\
\>a1 = t0-theterr\\
\\
\>\>a2 = t0+theterr\\
\\
\\
\>print $*$, " "\\
\\
\>print $*$, "t0,err:", t0, theterr\\
\\
\>\>call dqage(a0,t0,a1,a2,epsabs,epsrel,key,limit,result,\\
\>$*$   abserr,neval,ier,alist,blist,rlist,elist,iord,last)\\
\\
$*$ dqage comes from a standard library for integrating under a curve\\
\\
\>print $*$, "Succeeded, area under curve is ", result\\
\\
\>\>if ((abs(result-Cp)).gt.l) then\\
\\
$*$ here we iterate, decreasing our step size\\
\\
\>\>if (result.gt.Cp) then\\
\>\>\>if (flag) then\\
\>\>\>\>continue\\
\>\>\>else\\
\>\>\>\>lstep = lstep/dstep\\
\>\>\>endif\\
\>\>\>theterr = theterr - lstep\\
\>\>\>flag=.true.\\
\>\>else\\
\>\>\>if (flag) then\\
\>\>\>\>lstep = lstep/dstep\\
\>\>\>end if\\
\>\>\>theterr = theterr + lstep\\
\>\>\>flag=.false.\\
\\
\>\>end if\\
\\
\>go to 556 \\
\\
\>\>end if\\
\\
\\
\>print $*$, " "\\
\\
\>print $*$, "Integration error?", ier\\
\\
\>print $*$, "result = ", result\\
\\
\>print $*$, " "\\
\\
\>print $*$, "We have found theterr = ", theterr\\
\\
\>print $*$, " "\\
\\
$*$ depending on result/Cp, modify a and run it again.\\
\\
\>end \\
\\
\\
$*$ HERE IS THE FUNCTION TO BE INTEGRATED, called by dqage\\
\\
\>double precision function f(a0,t0,t)\\
\\
\>double precision a0, t, t0, d, s, c, p, q, r, pi, half\\
\\
\>double precision sq, e, ix, ip\\
\\
\>half = (1.0d0)/(2.0d0)\\
\\
\>pi = acos(-1.0d0)\\
\\
\>ip = (1.0d0)/(pi$*$2.0d0)\\
\\
\>sq = sqrt(ip)\\
\\
\>d = t-t0\\
\\
\>s = a0$*$sin(d)\\
\\
\>c = a0$*$cos(d)\\
\\
\>e = ix(c)\\
\\
\>p = exp(-half$*$(s$*$$*$2))\\
\\
\>q = exp(-half$*$(c$*$$*$2))$*$ip\\
\\
\>r = (half+e)$*$c$*$sq\\
\\
\>f = p$*$(q+r)\\
\\
\>end\\
\\
\\
\\
\>double precision function ix(x)\\
\\
\>double precision x, y, derf, half\\
\\
$*$ derf is the double precision error function, erf(x)\\
\\
\>half = (1.0d0)/(2.0d0)\\
\\
\>y = derf(x$*$sqrt(half))\\
\\
\>ix = half $*$ y\\
\\
\>end\\

\end{tabbing}

}


\nocite{*}


\begin{thebibliography}{}


\bibitem[\protect\citename{Q}0]{Q}
\begin{quote}
A man should keep his little brain-attic stocked with all the furniture
that he is likely to use, and the rest he can put away in the 
lumber-room of his library, where he can get it if he wants it.
\attrib{Sherlock Holmes, {\em The Five Orange Pips}}
\end{quote}


\bibitem[\protect\citename{Aller\ }1987]{Aller-87a}
Aller L. H., 1987, PASP, 99, 1145

\bibitem[\protect\citename{Angel \& Stockman\ }1980]{Angel+80a}Angel J. R.
P., Stockman H. S., 1980, ARAA, 18, 321

\bibitem[\protect\citename{Antonucci\ }1982]{Antonucci-82a}
Antonucci R., 1982, {\em Nature}, 299, 605

\bibitem[\protect\citename{Antonucci\ }1983]{Antonucci-83a}
Antonucci R., 1983, {\em Nature}, 303, 158

\bibitem[\protect\citename{Antonucci\ }1984]{Antonucci-84a}
Antonucci R., 1984, ApJ, 278, 499

\bibitem[\protect\citename{Antonucci\ }1993]{Antonucci-93a}
Antonucci R., 1993, ARAA, 31, 473

\bibitem[\protect\citename{Antonucci \& Barvainis\ 
}1990]{Antonucci+90a}Antonucci R., Barvainis R., 1990, ApJ, 363, L17

\bibitem[\protect\citename{Antonucci, Hurt \& 
Kinney\ }1994]{Antonucci+94a}
Antonucci R., Hurt T., Kinney A., 1994, {\em Nature}, 371, 313

\bibitem[\protect\citename{Antonucci \& Miller\ 
}1985]{Antonucci+85a}Antonucci R. R. J., Miller J. S., 1985, ApJ, 297, 621

\bibitem[\protect\citename{Aspin\ }1994]{Aspin-94a}Aspin C., 1994, IRCAM3 
Operations Manual v1.07, Joint Astronomy Centre, Hilo, Hawaii; available 
on the World Wide Web at {\sc url}\\ 
\verb$http://www.jach.hawaii.edu/~caa/i3/i3.html$

\bibitem[\protect\citename{Aspin\ }1995]{Aspin-95a}Aspin C., 1995, 
IRCAM3\,+\,IRPOL2 polarimetry data acquisition and reduction, Joint 
Astronomy Centre, Hilo, Hawaii; available on the World Wide Web at {\sc 
url}\\ \verb$http://www.jach.hawaii.edu/~caa/pol.doc$

\bibitem[\protect\citename{Axon \& Ellis\ }1976]{Axon+76a}Axon D. J., 
Ellis, 1976, MNRAS, 177, 499

\bibitem[\protect\citename{Bailey et al.\ }1986]{Bailey+86a}
Bailey J., Sparks W., Hough J., Axon D., 1986, {\em Nature}, 322, 150

\bibitem[\protect\citename{Baron \& White\ }1987]{Baron+87a}
Baron E., White S. D. M., 1987, ApJ, 322, 585

\bibitem[\protect\citename{Barthel\ }1989]{Barthel-89a}
Barthel P. D., 1989, ApJ, 336, 606

\bibitem[\protect\citename{Best, Longair \& 
R\"{o}ttgering\ }1996]{Best+96a}Best P. N., Longair M. S.,
R\"{o}ttgering H. J. A., 1996, MNRAS, 280, 9P (BLR-I)

\bibitem[\protect\citename{Best, Longair \& 
R\"{o}ttgering\ }1997]{Best+97a}Best P. N., Longair M. S.,
R\"{o}ttgering H. J. A., 1997, MNRAS, 292, 758 \{astro-ph/9707337\} (BLR-II)

\bibitem[\protect\citename{Best, Longair \& 
R\"{o}ttgering\ }1998]{Best+98a}Best P. N., Longair M. S.,
R\"{o}ttgering H. J. A., 1998, MNRAS, 295, 549 \{astro-ph/9709195\} (BLR-III)

\bibitem[\protect\citename{Binette, Robinson \& Courvoisier\
}1988]{Binette+88a}Binette L., Robinson A., Courvoisier T. J.-L.,
1988, A\&A, 194, 65

\bibitem[\protect\citename{Bithell \& Rees\ }1990]{Bithell+90a}Bithell
M., Rees M. J., 1990, MNRAS, 242, 570

\bibitem[\protect\citename{Blandford \& K\"{o}nigl\
}1979]{Blandford+79a}
Blandford R. D., K\"{o}nigl A., 1979, ApJ, 232, 34

\bibitem[\protect\citename{Boas\ }1983]{Boas-83a}Boas, M. L. {\em 
Mathematical Methods in the Physical Sciences}. John Wiley \& Sons, New 
York, second edition, 1983.

\bibitem[\protect\citename{Brindle et al.\ }1986]{Brindle+86a}Brindle
C., Hough J. H., Bailey J. A., Axon D. J., Hyland A. R., 1986, MNRAS, 221, 739

\bibitem[\protect\citename{Brown \& McLean\ }1977]{Brown+77a}
Brown R. L., McLean I. S., 1977, A\&A, 37, 141

\bibitem[\protect\citename{Bruzual\ }1983]{Bruzual-83a}
Bruzual G., 1983, ApJ, 273, 105

\bibitem[\protect\citename{Bruzual \& Charlot\ }1993]{Bruzual+93a}
Bruzual G., Charlot S., 1993, ApJ, 405, 538

\bibitem[\protect\citename{Burstein \& Heiles\ }1982]{Burstein+82a}Burstein 
D., Heiles C., 1982, AJ, 87, 1165

\bibitem[\protect\citename{Carilli et al.\ }1997]{Carilli+97a}Carilli C. 
L., R\"{o}ttgering H. J. A., van Ojik R., Miley G. K., van Breugel W.
J. M., 1997, ApJS, 109, 1

\bibitem[\protect\citename{Chambers, Miley \& 
van Breugel\ }1987]{Chambers+87a}Chambers K. C., Miley G. K., van 
Breugel W. J. M., 1987, {\em Nature}, 329, 604

\bibitem[\protect\citename{Chambers, Miley \& 
Joyce\ }1988]{Chambers+88a}Chambers K. C., Miley G. K., Joyce R. R., 1988,
ApJ, 329, L75

\bibitem[\protect\citename{Chambers \& Charlot\ }1990]{Chambers+90a}Chambers 
K. C., Charlot S., 1990, ApJ, 348, L1

\bibitem[\protect\citename{Chrysostomou\ }1996]{Chrysostomou-96a}Chrysostomou 
A., 1996, Imaging Polarimetry with IRCAM3, Joint Astronomy Centre, Hilo, 
Hawaii; available on the World Wide Web at {\sc url}\\ 
\verb$http://www.jach.hawaii.edu/UKIRT.new/instruments/irpol/IRCAM/ircampol.html$

\bibitem[\protect\citename{Cimatti \& di Serego Alighieri\
}1995]{Cimatti+95a}Cimatti A., di Serego Alighieri S., 1995, MNRAS, 273, L7

\bibitem[\protect\citename{Cimatti et al.\ }1993]{Cimatti+93a}Cimatti A.,
di Serego Alighieri S., Fosbury R. A. E., Salvati M., Taylor D. N., 1993,
MNRAS, 264, 421

\bibitem[\protect\citename{Cimatti et al.\ }1994]{Cimatti+94a}Cimatti A.,
di Serego Alighieri S., Field G. B., Fosbury R. A. E., 1994, ApJ, 422, 562

\bibitem[\protect\citename{Cimatti et al.\ }1996]{Cimatti+96a}Cimatti
A., Dey A., van Breugel W. J. M., Antonucci R., Spinrad H., 1996, ApJ,
465, 145

\bibitem[\protect\citename{Cimatti et al.\ }1997]{Cimatti+97a}Cimatti A.,
Dey A., van Breugel W. J. M., Hurt T., Antonucci R., 1997, ApJ, 476, 677

\bibitem[\protect\citename{Cimatti et al.\ }1998]{Cimatti+98a}Cimatti A.,
di Serego Alighieri S., Vernet J., Cohen M., Fosbury R. A. E., 1998, ApJ, 499, L21

\bibitem[\protect\citename{Clarke \&
Naghizadeh-Khouei\ }{1994}]{Clarke+94a}Clarke D., Naghizadeh-Khouei J.
1994, AJ, 108, 687

\bibitem[\protect\citename{Clarke \& Stewart\ }1986]{Clarke+86a}Clarke
D., Stewart, B. G., 1986, 
{\em Vistas in Astronomy}, 29, 27

\bibitem[\protect\citename{Clarke et al.\ }1983]{Clarke+83b}Clarke D.,
Stewart B. G., Schwarz H. E., Brooks A., 1983, A\&A, 126, 260
\footnote{In an attempt to use more
consistent notation, my paper uses $\bar{s }$ for the arithmetic mean of a
set of parameters, $\tilde{s}$ for a ratio of means, and $\hat{s}$ for the
best (conservative) errors on certain quantities. Clarke et al.,
however, use $\bar{s}$ for the ratio of non-normalized mean Stokes Paramet
ers, and $\tilde{s}(1)$ for the arithmetic mean.}

\bibitem[\protect\citename{Clarke et al.\ }1993]{Clarke+93a}Clarke D.,
Naghizadeh-Khouei J., Simmons J. F. L., Stewart B. G., 1993, A\&A, 269, 617 

\bibitem[\protect\citename{Clarke \& Cooke\ }1983]{Clarke+83a}Clarke,
G.~M. and Cooke, D. {\em A Basic Course in Statistics}. Edward Arnold,
second edition, 1983. 

\bibitem[\protect\citename{Cohen et al.\ }1977]{Cohen+77a}Cohen 
M. H. et al. (9 co-authors), 1977, {\em Nature}, 268, 405

\bibitem[\protect\citename{Cohen et al.\ }1997]{Cohen+97a}Cohen 
M. H., Vermeulen R. C., Ogle P. M., Tran H. D., Goodrich R. W., 
1997, ApJ, 484, 193

\bibitem[\protect\citename{Coyne, Gehrels \& Serkowski\ }1974]{Coyne+74a}Coyne G.
V., Gehrels T., Serkowski K., 1974, AJ, 79, 581

\bibitem[\protect\citename{Crane \& Vernet\ }1997]{Crane+97a}
Crane P., Vernet J., 1997, A\&AS, 190, 391

\bibitem[\protect\citename{Cristiani \& Vio\ }1990]{Cristiani+90a}
Cristiani S., Vio R., 1990, A\&A, 227, 385

\bibitem[\protect\citename{Daly\ }1990]{Daly-90a}Daly R. A., 1990, ApJ,
355, 416

\bibitem[\protect\citename{Daly\ }1992]{Daly-92a}Daly R. A., 1992, ApJ,
399, 426

\bibitem[\protect\citename{Dey et al.\ }1996]{Dey+96a}Dey A., Cimatti A.,
van Breugel W. J. M., Antonucci R., Spinrad H., 1996, ApJ, 465, 157

\bibitem[\protect\citename{Dey et al.\ }1997]{Dey+97a}Dey A.,
van Breugel W. J. M., Vacca W. D., Antonucci R., 1997, ApJ, 490, 698

\bibitem[\protect\citename{Dickinson, }private
communication]{dpc}Dickinson M., 1996, private communication

\bibitem[\protect\citename{Dickson et al.\ }1995]{Dickson+95a}Dickson 
R., Tadhunter C., Shaw M., Clark N., Morganti R., 1995, MNRAS, 273, 29P

\bibitem[\protect\citename{di Serego Alighieri et al.\ }1988]{Alighieri+88a}di
Serego Alighieri S., Binette L., Courvoisier T., Fosbury R. A. E.,
Tadhunter C. N., 1988, {\em Nature}, 334, 591

\bibitem[\protect\citename{di Serego Alighieri et al.\ }1989]{Alighieri+89a}di
Serego Alighieri S., Fosbury R. A. E., Quinn P. J., Tadhunter C. N., 
1989, {\em Nature}, 341, 307

\bibitem[\protect\citename{di Serego Alighieri, 
Cimatti \& Fosbury\ }{1993}]{Alighieri+93a}di Serego Alighieri S., Cimatti 
A., Fosbury, R. A. E., 1993, ApJ, 404, 584
\footnote{This thesis uses $\eta$ for the instrumental angle which di 
Serego Alighieri et al. call $\phi$.}

\bibitem[\protect\citename{di Serego Alighieri, 
Cimatti \& Fosbury\ }{1994}]{Alighieri+94a}di Serego Alighieri S., Cimatti 
A., Fosbury, R. A. E., 1994, ApJ, 431, 123

\bibitem[\protect\citename{Dolan \& Tapia\ }1986]{Dolan+86a}Dolan J. F., 
Tapia S., 1986, PASP, 98, 792

\bibitem[\protect\citename{Dolan et al. }1994]{Dolan+94a}Dolan J. F. et 
al. (11 co-authors), 1994, ApJ, 432, 560


\bibitem[\protect\citename{Dunlop \& Peacock }1993]{Dunlop+93a}Dunlop J.
S., Peacock J. A., 1993, MNRAS, 263, 936

\bibitem[\protect\citename{Dunlop et al.\ }1996]{Dunlop+96a} Dunlop J.,
Peacock J., Spinrad H., Dey A., Jimenez R., Stern D., Windhorst R., 1996,
{\em Nature}, 381, 581

\bibitem[\protect\citename{Eales\ }1992]{Eales-92a}Eales S.
A., 1992, ApJ, 397, 49

\bibitem[\protect\citename{Eales \& Rawlings\ }1990]{Eales+90a}Eales S.
A., Rawlings S., 1990, MNRAS, 243, 1P

\bibitem[\protect\citename{Eales \& Rawlings\ }1996]{Eales+96a}Eales S.
A., Rawlings S., 1996, ApJ, 460, 68

\bibitem[\protect\citename{Eales et al.\ }1997]{Eales+97a}Eales S. A.,
Rawlings S., Law-Green D., Cotter G., Lacy M., 1997, MNRAS, 291, 593

\bibitem[\protect\citename{Economou et al.\ }1995]{Economou+95a}Economou
F., Lawrence A., Ward M. J., Blanco P. R., 1995, MNRAS, 272, L5

\bibitem[\protect\citename{Eggen, Lynden-Bell \& Sandage\ 
}1962]{Eggen+62a}Eggen O. J., Lynden-Bell D., Sandage A. R., 1962, ApJ,
136, 748

\bibitem[\protect\citename{Eisenhardt \& Chokshi\
}1990]{Eisenhardt+90a}Eisenhardt P., Chokshi A., 1990, ApJ, 351, L9

\bibitem[\protect\citename{Elston \& Jannuzi\ }1999]{Elston+97a}Elston R.,
Jannuzi B., 1999, ApJL, in preparation

\bibitem[\protect\citename{Fabian\ }1989]{Fabian-89a}Fabian A. C., 1989,
MNRAS, 238, 41P

\bibitem[\protect\citename{Fanaroff \& Riley\ }1974]{Fanaroff+74a}Fanaroff 
B. L., Riley J. M., 1974, MNRAS, 167, 31P

\bibitem[\protect\citename{Fernini et al.\ }1993]{Fernini+93a}Fernini 
I., Burns J. O., Bridle A. H., Perley R. A., 1993, AJ, 105, 1690

\bibitem[\protect\citename{Fr\"{o}berg\ }1985]{Froberg-85a}Fr\"{o}berg 
C.-E., {\em Numerical Mathematics}. Addison-Wesley, 1985. 

\bibitem[\protect\citename{Fugmann \& Meisenheimer\ }1988]{Fugmann+88a}Fugmann
W., Meisenheimer K., 1988, A\&AS, 76, 145

\bibitem[\protect\citename{Gehrels\ }1960]{Gehrels-60a}Gehrels T., 1960, AJ,
65, 470

\bibitem[\protect\citename{Clarke, in Gehrels
(ed.)\ }1974]{Clarke-74a}Gehrels T. (ed.), {\em Planets, 
Stars and Nebulae studied with Photopolarimetry}, chapter on Polarimetric
Definitions (by D. Clarke), pages 45--53. The University of Arizona 
Press, 1974. 

\bibitem[\protect\citename{Ghisellini et al.\ }1993]{Ghisellini+93a}Ghisellini 
G., Padovani P., Celotti A., Maraschi L., 1993, ApJ, 407, 65

\bibitem[\protect\citename{Goodrich \& Cohen\ }1992]{Goodrich+92a}Goodrich
R. W., Cohen M. H., 1992, ApJ, 391, 623

\bibitem[\protect\citename{Goodrich et al.\ }1996]{Goodrich+96a}Goodrich
R. W., Miller J. S., Martel A., Cohen M. H., Tran H. D., Ogle P. M., 
Vermeulen R. A., 1996, ApJ, 456, L9

\bibitem[\protect\citename{Grandi et al.\ }1997]{Grandi+97a}Grandi P., 
Sambruna R. M., Maraschi L., Matt G., Urry C. M., Mushotzky R. F., 1997, 
ApJ, 487, 636

\bibitem[\protect\citename{Gunn et al.\ }1981]{Gunn+81a}Gunn J. E.,
Hoessel J. G., Westphal J. A., Perryman M. A. C., Longair M. S., 1981,
MNRAS, 194, 111

\bibitem[\protect\citename{Hammer, LeF\`{e}vre \& 
Angonin\ }1993]{Hammer+93a}Hammer F., LeF\`{e}vre O. N., Angonin
M. C., 1993, {\em Nature}, 362, 324

\bibitem[\protect\citename{Hecht\ }1987]{Hecht-87a}Hecht E., {\em
Optics}. Addison-Wesley, second edition, 1987. 

\bibitem[\protect\citename{Hines \& Wills\ }1993]{Hines+93a}Hines
D. C., Wills B. J., 1993, ApJ, 415, 82

\bibitem[\protect\citename{Hoskin\ }1976]{Hoskin-76a}Hoskin M. A., 
1976, JHistA, 7, 169

\bibitem[\protect\citename{Hough, }private communication]{hpc}Hough J.,
1996, private communication

\bibitem[\protect\citename{Hough et al.\ }1987]{Hough+87a}Hough J., 
Brindle C., Axon D., Bailey J., Sparks W., 1987, MNRAS, 224, 1013

\bibitem[\protect\citename{Hsu \& Breger}{1982}]{Hsu+82a}Hsu J.,
Breger M., 1982, ApJ, 262, 732

\bibitem[\protect\citename{Hurt et al.\ }1999]{Hurt+99a}
Hurt T., Antonucci R., Cohen R., Kinney A., Krolik J., 1999,
ApJ, 514, 579

\bibitem[\protect\citename{Hutchings\ }1987]{Hutchings-87a}Hutchings J. 
B., 1987, ApJ, 311, 526

\bibitem[\protect\citename{Illingworth\ }1994]{Illingworth-94a}Illingworth 
V. (ed.), {\em Collins Dictionary of Astronomy}, HarperCollins, Glasgow, 1994

\bibitem[\protect\citename{Impey, Malkan \& Tapia\ 
}1989]{Impey+89a}Impey C. D., Malkan M. A., Tapia S., 
1989, ApJ, 347, 96

\bibitem[\protect\citename{Impey, Lawrence \& Tapia\ 
}1991]{Impey+91a}Impey C. D., Lawrence C. R., Tapia S., 
1991, ApJ, 375, 46

\bibitem[\protect\citename{Jannuzi \& Elston\ }1991]{Jannuzi+91a}Jannuzi
B., Elston R., 1991, ApJ, 366, L69

\bibitem[\protect\citename{Jannuzi, Green \& French\ }1993]{Jannuzi+93a}
Jannuzi B., Green R., French H., 1993, ApJ, 404, 100

\bibitem[\protect\citename{Jannuzi, }private communication]{jpc}Jannuzi
B., 1996, private communication

\bibitem[\protect\citename{Jenkins, Pooley \& 
Riley\ }1977]{Jenkins+77a}Jenkins C. J., Pooley G. G., Riley J. M., 1977,
MemRAS, 84, 61

\bibitem[\protect\citename{Killeen, Bicknell \& Ekers\ 
}1986]{Killeen+86a}Killeen N. E. B., Bicknell G. V., Ekers R. D., 1986, ApJ, 302, 306

\bibitem[\protect\citename{Kitchin\ }1984]{Kitchin-84a}Kitchin C. R., 
{\em Astrophysical Techniques}. Hilger, Bristol, 1984.

\bibitem[\protect\citename{Kollgaard et al.\ }1992]{Kollgaard+92a}Kollgaard 
R. I., Wardle J. F. C., Roberts D. H., Gabuzda D. C, 1992, AJ, 104, 1687

\bibitem[\protect\citename{Koorneef\ }1983]{Koorneef-83a}Koorneef J.,
1983, A\&A, 128, 84

\bibitem[\protect\citename{Kormendy \& 
Richstone\ }1995]{Kormendy+95a}Kormendy J., Richstone D., 1995, ARAA,
33, 581

\bibitem[\protect\citename{Lacy \& Rawlings\ }1994]{Lacy+94a}Lacy M.,
Rawlings S., 1994, MNRAS, 270, 431

\bibitem[\protect\citename{Lacy et al.\ }1995]{Lacy+95a}Lacy M.,
Rawlings S., Eales S. A., Dunlop J. S., 1995, MNRAS, 273, 821

\bibitem[\protect\citename{Lacy et al.\ }1998]{Lacy+98a}Lacy M., Rawlings
S., Blundell K. M., Ridgway S. K., 1998, MNRAS, 298, 966 
\{astro-ph/9803017\}

\bibitem[\protect\citename{Lacy et al.\ }1999a]{Lacy+99a}Lacy M.,
Kaiser M. E., Hill G. J., Rawlings
S., Leyshon G., 1999a, MNRAS, \{astro-ph/9905357\}

\bibitem[\protect\citename{Lacy et al.\ }1999b]{Lacy+98b}Lacy M.,
Ridgway S. E., Wold M., Lilje P. B., Rawlings
S., 1999b, MNRAS, \{astro-ph/9903314\}

\bibitem[\protect\citename{Laing, Riley \& Longair\ }1983]{Laing+83a}Laing
R. A., Riley J. M., Longair M. S., 1983, MNRAS, 204, 151 


\bibitem[\protect\citename{Lasker et al.\ }1998]{Lasker+90a}Lasker B. M.,
Sturch C. R., McLean B. J., Russell J. L., Jenkner H., Shara M. M., 1990,
AJ, 99, 2019

\bibitem[\protect\citename{Lawrence et al.\ }1993]{Lawrence+93a}Lawrence 
A. et al. (16 co-authors), 1993, MNRAS, 260, 28

\bibitem[\protect\citename{Leahy, Muxlow \& Stephens\ }1989]{Leahy+89a}Leahy
J. P., Muxlow T. W. B., Stephens P. W., 1989, MNRAS, 239, 401 

\bibitem[\protect\citename{LeF\`{e}vre 
et al.\ }1988a]{LeFevre+88a}LeF\`{e}vre O., Hammer F., Nottale L., Mazure A., Christian C., 1988a, ApJ, 324, L1

\bibitem[\protect\citename{LeF\`{e}vre, Hammer \& 
Jones\ }1988b]{LeFevre+88b}LeF\`{e}vre O., Hammer F., Jones J., 1988b, ApJ, 331, L73

\bibitem[\protect\citename{Leyshon\ }1998]{Leyshon-98a}Leyshon G., 1998,
{\em Experimental Astronomy}, 1998, 2, 153 \{astro-ph/9709164\}

\bibitem[\protect\citename{Leyshon \& Eales\ }1998]{Leyshon+98a}Leyshon
G., Eales S. A., 1998, MNRAS, 295, 10 \{astro-ph/9708085\} (L\&E)

\bibitem[\protect\citename{Leyshon, Dunlop \& Eales\ 
}1999]{Leyshon+99a}Leyshon
G., Dunlop J. S., Eales S. A., 1999, MNRAS \{astro-ph/9905282\}

\bibitem[\protect\citename{Lilly \& Longair\ }1984]{Lilly+84a}Lilly S. J.,
Longair M. S., 1984, MNRAS, 211, 833

\bibitem[\protect\citename{Longair\ }1975]{Longair-75a}Longair M. S.,
1975, MNRAS, 173, 309

\bibitem[\protect\citename{Longair, Best \& 
R\"{o}ttgering\ }1995]{Longair+95a}Longair M. S., Best P. N.,
R\"{o}ttgering H. J. A., 1995, MNRAS, 275, 47P

\bibitem[\protect\citename{Manzini \& di Serego 
Alighieri\ }1996]{Manzini+96a}Manzini A., di Serego Alighieri S., 1996, 
A\&A, 311, 79 (MdSA)

\bibitem[\protect\citename{Maronna, Feinstein \& Clocchiatti\ 
}1992]{Maronna+92a}Maronna R.,
Feinstein C., Clocchiatti, A., 1992, A\&A, 260, 525

\bibitem[\protect\citename{Martin \& Whittet\ }1990]{Martin+90a}Martin P.
G., Whittet D. C. B., 1990, ApJ, 357, 113

\bibitem[\protect\citename{Mathewson \& Ford\
}1970]{Mathewson+70a}Mathewson D., Ford V., 1970, MemRAS, 74, 139

\bibitem[\protect\citename{McCarthy\ }1988]{McCarthy-88a}McCarthy P. J.,
1988, PhD thesis, University of California at Berkeley

\bibitem[\protect\citename{McCarthy\ }1993]{McCarthy-93a}McCarthy P. J.,
1993, ARAA, 31, 639

\bibitem[\protect\citename{McCarthy et al.\ }1987a]{McCarthy+87a}McCarthy
P. J., van Breugel W. J. M., Spinrad H., Djorgovski S., 1987a, ApJ, 321, L29

\bibitem[\protect\citename{McCarthy et al.\ }1987b]{McCarthy+87b}McCarthy
P. J., Spinrad H., Djorgovski S., Strauss M. A., van Breugel W. J. M.,
Liebert J., 1987b, ApJ, 319, L39

\bibitem[\protect\citename{McCarthy et al.\ }1988]{McCarthy+88a}McCarthy
P. J., Dickinson M., Filippenko A. V., Spinrad H., van Breugel W. J. M.,
1988, ApJ, 328, L29

\bibitem[\protect\citename{McCarthy et al.\ }1990]{McCarthy+90a}McCarthy
P. J., Kapahi V. K., van Breugel W. J. M., Subrahmanya C. R., 1990, AJ,
100, 1014

\bibitem[\protect\citename{McCarthy, van Breugel \& Kapahi\
}1991]{McCarthy+91a}McCarthy P. J., van Breugel W. J. M., Kapahi V. K., 1991, ApJ, 371, 478

\bibitem[\protect\citename{McCarthy, Persson \& West\
}1992]{McCarthy+92a}McCarthy P. J., Persson S. E., West S. C., 1992, ApJ,
386, 52

\bibitem[\protect\citename{McLean\ }1997]{McLean-97a}McLean I. S.,
{\em Electronic Imaging in Astronomy}, Wiley, Chichester, 1997

\bibitem[\protect\citename{Miley\ }1980]{Miley-80a}Miley G. K., 1980, 
ARAA, 18, 165

\bibitem[\protect\citename{Miller \& Goodrich\ }1990]{Miller+90a}Miller J.
S., Goodrich R. W., 1990, ApJ, 355, 456


\bibitem[\protect\citename{Mood et al.\ }1974]{Mood+74a}Mood A. M.,
Graybill F. A., Boes D. C., {\em Introduction to the Theory of
Statistics}. McGraw-Hill, third edition, 1974. 

\bibitem[\protect\citename{Moore \& Stockman\ }1984]{Moore+84a}Moore R. L., 
Stockman H. S., 1984, ApJ, 279, 465

\bibitem[\protect\citename{Naghizadeh--Khouei \&
Clarke\ }{1994}]{NaghizadehKhouei+93a}Naghizadeh--Khouei J., Clarke D., 1993, A\&A, 274, 968

\bibitem[\protect\citename{Naim, Ratnatunga \& Griffiths\ 
}1997]{Naim+97a}Naim A., Ratnatunga K. U., Griffiths R. E., 1997, ApJ, 
476, 510

\bibitem[\protect\citename{Neeser\ }1996]{Neeser-96a}Neeser M. J., 1996,
PhD thesis, Ruprecht-Karls-Universit\"{a}t, Heidelberg

\bibitem[\protect\citename{NOAO\ }{IRAF}]{IRAFphot}NOAO. {\em On-Line
Documentation for IRAF:} {\sc apphot.phot.} On-line manual page 
for the {\sc phot}
command in the {\sc digiphot.apphot} package of the IRAF data reduction 
system. Software support website: {\tt iraf.noao.edu}.

\bibitem[\protect\citename{Packham et al.\ }1996]{Packham+96a}Packham C., 
Hough J. H., Young S., Chrysostomou A., Bailey J. A., Axon D. J., 
Ward M. J., 1996, MNRAS, 278, 406

\bibitem[\protect\citename{Packham et al.\ }1997]{Packham+97a}Packham C., 
Young S., Hough J. H., Axon D. J., Bailey J. A., 1997, MNRAS, 288, 375

\bibitem[\protect\citename{Padovani \& Urry\ }1992]{Padovani+92a}Padovani P., Urry M. C., 1992, ApJ, 387, 449

\bibitem[\protect\citename{Parsons\ }1974]{Parsons-74a}Parsons R., {\em 
Statistical Analysis: A Decision-Making Approach}. Harper \& Row, New York, 
1974.

\bibitem[\protect\citename{Pentericci et 
al.\ }1998]{Pentericci+98a}Pentericci L., 
R\"{o}ttgering H. J. A., Miley G. K., McCarthy P., Spinrad H.,
van Breugel W. J. M., Macchetto P., 1998, A\&A
\{astro-ph/9809056v2\}


\bibitem[\protect\citename{Peterson\ }1997]{Peterson-97a}Peterson B. M., 
{\em An Introduction to Active Galactic Nuclei}, Cambridge University Press, 
1997

\bibitem[\protect\citename{Prestage \& Peacock\ 
}1988]{Prestage+88a}Prestage R. M., Peacock J. A., 1988, MNRAS, 230, 131

\bibitem[\protect\citename{Rawlings, Lacy \& 
Eales\ }1991]{Rawlings+91a}Rawlings S., Lacy, M., Eales S. A., 1991,
MNRAS, 251, 17P

\bibitem[\protect\citename{Rawlings et al.\ }1995]{Rawlings+95a}Rawlings
S., Lacy, M., Sivia D. S., Eales S. A., 1995, MNRAS, 274, 428

\bibitem[\protect\citename{Rees\ }1966]{Rees-66a}Rees M. J., 1966, {\em
Nature}, 211, 468

\bibitem[\protect\citename{Ridgway \& Stockton\ }1997]{Ridgway+97a}Ridgway
S. K., Stockton A. N., 1997, AJ, 114, 511

\bibitem[\protect\citename{Rieke \& Lebofsky\ }1985]{Rieke+85a}Rieke G.
H., Lebofsky M. J., 1985, ApJ, 288, 618

\bibitem[\protect\citename{Rigler \& Lilly\ }1994]{Rigler+94a}Rigler M.
A., Lilly S. J., 1994, ApJ, 427, L79

\bibitem[\protect\citename{Rigler et al.\ }1992]{Rigler+92a} Rigler M. A.,
Lilly S. J., Stockton A., Hammer F., LeF\`{e}vre O. N.,1992, ApJ, 385, 61

\bibitem[\protect\citename{Riley, Longair \& Gunn\ }1980]{Riley+80a}Riley
J. M., Longair M. S., Gunn J. E., 1980, MNRAS 192, 233

\bibitem[\protect\citename{Robson }1996]{Robson-96a}Robson I., 
{\em Active Galactic Nuclei}, Wiley, Chichester, 1996

\bibitem[\protect\citename{Roche \& Eales\ }1999]{Roche+99a}Roche
N., Eales S. A., 1999, MNRAS {\em in preparation}

\bibitem[\protect\citename{Rudy et al.\ }1983]{Rudy+83a}Rudy R. J., 
Schmidt G. D., Stockman H. S., Moore R. L., 1983, ApJ, 271, 59

\bibitem[\protect\citename{S\'{a}nchez Almeida\
}1995]{Almeida-95a}S\'{a}nchez Almeida J., 1995, A\&ASS, 109, 417

\bibitem[\protect\citename{Sanders et al.\ }1988]{Sanders+88a}Sanders
D. B., Soifer B. T., Elias J. H., Madore B. F., Matthews K., Neugebauer 
G., Scoville N. Z., 1988, ApJ, 325, 74 

\bibitem[\protect\citename{Sansom et al.\ }1987]{Sansom+87a}Sansom
A. E. et al. (8 co-authors), 1987, MNRAS, 229, 15

\bibitem[\protect\citename{Savage \& Mathis\ }1979]{Savage+79a}Savage B.
D., Mathis J. S., 1979, ARAA, 17, 73

\bibitem[\protect\citename{Scarpa \& Falomo\ }1997]{Scarpa+97a}Scarpa 
R., Falomo R., 1997, A\&A, 325, 109

\bibitem[\protect\citename{Scarrott et al.\ }1983]{Scarrott+83a}Scarrott 
S. M., Warren-Smith R. F., Pallister W. S., Axon D. J.,  Bingham R. G., 
MNRAS, 1983, 204, 1163

\bibitem[\protect\citename{Scarrott, Rolph \& Tadhunter\ 
}1990]{Scarrott+90a}Scarrott
S. M., Rolph C.D., Tadhunter C.D., 1990, MNRAS, 243, 5P

\bibitem[\protect\citename{Scarrott et al.\ }1990]{Scarrott+90b}Scarrott
S. M., Rolph C.D., Wolstencroft R. D., Walker H. J., Sekiguchi K., 1990,
MNRAS, 245, 484

\bibitem[\protect\citename{Schilizzi, Kapahi \& Neff\
}1982]{Schilizzi+82a}Schilizzi R., Kapahi V. K., Neff S., 1982, JAp\&A, 3,
173

\bibitem[\protect\citename{Schmidt--Kaler\
}1958]{SchmidtKaler-58a}Schmidt--Kaler T., 1958, Zeitschrift f\"{u}r 
Astrofisik, 46, 145


\bibitem[\protect\citename{Serjeant et al.\
}1998]{Serjeant+98a}Serjeant S., Rawlings S., Maddox S. J., Baker J.
C., Clements D., Lacy M., Lilje P. B., 1998, MNRAS, 292, 494

\bibitem[\protect\citename{Serkowski}{1958}]{Serkowski-58a}Serkowski K., 
1958, {\em Acta Astronomica}, 8, 135 

\bibitem[\protect\citename{Serkowski, Mathewson \& Ford\
}1975]{Serkowski+75a}Serkowski K., Mathewson D., Ford V., 1975, ApJ, 196,
261

\bibitem[\protect\citename{Shaw et al.\ }1995]{Shaw+95a}Shaw M., Tadhunter
C. N., Dickson R., Morganti R., 1995, MNRAS, 275, 703


\bibitem[\protect\citename{Simmons \& Stewart\ }1985]{Simmons+85a}Simmons
J. F. L., Stewart B. G., 1985, A\&A, 142, 100

\bibitem[\protect\citename{Smith et al.\ }1986]{Smith+86a}Smith P. S., 
Balonek T. J., Heckart P. A., Elston R., 1986, ApJ, 305, 484

\bibitem[\protect\citename{Spinrad\ }1982]{Spinrad-82a}
Spinrad H., 1982, PASP, 94, 397

\bibitem[\protect\citename{Spinrad \& Djorgovski\ }1984a]{Spinrad+84a}
Spinrad H., Djorgovski S. G., 1984a, ApJ, 280, L9 

\bibitem[\protect\citename{Spinrad \& Djorgovski\ }1984b]{Spinrad+84b}
Spinrad H., Djorgovski S. G., 1984b, ApJ, 285, L49 

\bibitem[\protect\citename{Spinrad et al.\ }1997]{Spinrad+97a}Spinrad H.,
Dey A., Stern D., Dunlop J., Peacock J., Jimenez R., Windhorst R.,1997,
ApJ, 484, 581

\bibitem[\protect\citename{Sterken \&
Manfroid\ }1992]{Sterken+92a}Sterken Chr., Manfroid J., {\em
Astronomical Photometry: A Guide}. Kluwer Academic, 1992. 

\bibitem[\protect\citename{Stockman, Angel \& 
Miley\ }1979]{Stockman+79a} Stockman H. S., Angel J. R. P., Miley G. K., 
1979, ApJ, 227, L55

\bibitem[\protect\citename{Stockman, Moore \& 
Angel\ }1984]{Stockman+84a} Stockman H. S., Moore R. L., Angel J. R. P., 
1984, ApJ, 279, 485

\bibitem[\protect\citename{Stockton, Kellogg \& 
Ridgway\ }1995]{Stockton+95a} Stockton A., Kellogg M., Ridgway S. E., 1995, 
ApJ, 443, L69

\bibitem[\protect\citename{Strom et al.\ }1990]{Strom+90a}Strom R. G.,
Riley J. M., Spinrad H., van Breugel, W. J. M., Djorgovski S., Liebert J.,
McCarthy P. J., 1990, A\&A, 227, 19

\bibitem[\protect\citename{Tadhunter et al.\
}1992]{Tadhunter+92a}Tadhunter C. N., Scarrott S. M., Draper P., Rolph,
C., 1992, MNRAS, 256, 53P

\bibitem[\protect\citename{Takalo et al.\ }1992]{Takalo+92a}Takalo L. O., 
Kidger M. R., de Diego J. A., Sillanp\"{a}\"{a} A., 1992, A\&A, 261, 415

\bibitem[\protect\citename{Tinbergen\ }1996]{Tinbergen-96a}Tinbergen J.,
{\em Astronomical Polarimetry}. Cambridge University Press, 1996. 


\bibitem[\protect\citename{Tran, Cohen \& Goodrich\ 
}1995]{Tran+95a}Tran H. D., Cohen M. H., Goodrich R. W., 1995, AJ, 110, 2597

\bibitem[\protect\citename{Tran et al.\ }1998]{Tran+98a}Tran H. D., Cohen 
M. H., Ogle P. M., Goodrich R. W., di Serego Alighieri S., 1998, ApJ, 500, 660

\bibitem[\protect\citename{Urry \& Padovani\ }1995]{Urry+95a}Urry 
M. C., Padovani P., 1995, PASP, 107, 803

\bibitem[\protect\citename{Vinokur }1965]{Vinokur-65a}Vinokur M., 1965,
{\em Annales d'Astrophysique}, 28, 412

\bibitem[\protect\citename{Walker\ }1987]{Walker-87a}Walker G., {\em 
Astronomical Observations}.
Cambridge University Press, 1987.

\bibitem[\protect\citename{Walsh et al.\ }1985]{Walsh+85a}Walsh D., 
Lebofsky M. J., Rieke G. H., Shone D., Elston R., 1985, MNRAS, 212, 631

\bibitem[\protect\citename{Wardle \& Kronberg\ }1974]{Wardle+74a}Wardle J.
F. C., Kronberg P. P., 1974, ApJ, 194, 249
\footnote{The Wardle \& Kronberg
paper reproduces Vinokur's equation (my Equation \ref{thetadis}) but
omits the factor `${\mathrm sign}(x)$' from Equation \ref{thetasup} on the
grounds (Wardle, private communication) that the probability of $x$
falling in the domain $x<0$ is negligibly small.}

\bibitem[\protect\citename{Webb et al.\ }1993]{Webb+93a}Webb W., Malkan 
M., Schmidt G., Impey C., 1993, ApJ, 419, 494

\bibitem[\protect\citename{West }1994]{West-94a}West M. J., 1992,
MNRAS, 268, 79

\bibitem[\protect\citename{Whittet et al.\ }1992]{Whittet+92a}Whittet
D. C. B., Martin P. G., Hough J. H., Rouse M. F., Bailey J. A., Axon D.
J., 1992, ApJ, 386, 562

\bibitem[\protect\citename{Wieringa \& 
Katgert\ }1992]{Wieringa+92a}Wieringa M. H., Katgert P, 1992, A\&AS, 93, 399

\bibitem[\protect\citename{Wilking et al.\ }1980]{Wilking+80a}Wilking B.
A., Lebofsky M. J., Martin P. G., Rieke G. H., Kemp J. C., 1980, ApJ, 235,
905

\bibitem[\protect\citename{Willott et al.\
}1998]{Willott+98a}Willott C. J., Rawlings S., Blundell K. M., Lacy
M., 1998, MNRAS, 300, 625

\bibitem[\protect\citename{Willott et al.\
}1999]{Willott+99a}Willott C. J., Rawlings S., Blundell K. M., Lacy
M., 1999, MNRAS, \{astro-ph/9905388\}

\bibitem[\protect\citename{Young et al.\ }1996]{Young+96a}Young S., Packham 
C., Hough J. H., Efstathiou A., 1996, MNRAS, 283, L1

\bibitem[\protect\citename{Zirbel\ }1997]{Zirbel-97a}Zirbel E. L., 1997, 
ApJ, 476, 489

\end{thebibliography}
\end{document}